\documentclass[pdftex,10pt,prd,aps,showpacs,a4paper]{revtex4-1}

\usepackage[latin1]{inputenc}
\usepackage[T1]{fontenc}
\usepackage{graphics}
\usepackage{longtable}
\usepackage{amsfonts,amsmath,amssymb} 
\usepackage{pdfpages}

\usepackage[english]{babel}

\usepackage[pdftitle={Structure Formation independent of Cold Dark Matter},
pdfauthor={P. G. Miedema}, pdfsubject={Gauge
Invariant Cosmological Density Perturbations, Structure Formation,
Population III Stars},
pdfstartview={FitH},
pdfkeywords={Structure Formation, Cosmic Background Fluctuations,
  Population III Stars},
bookmarks, bookmarksnumbered, colorlinks, linkcolor={blue},
citecolor={blue}]{hyperref}

\renewcommand{\vec}[1]{\mbox{\boldmath$#1$}}
\newcommand{\vecs}[1]{\mbox{\scriptsize\boldmath$#1$}}

\newcommand{\nul}{{\mbox{$\scriptscriptstyle(0)$}}}
\newcommand{\een}{{\mbox{$\scriptscriptstyle(1)$}}}
\newcommand{\twee}{{\mbox{$\scriptscriptstyle(2)$}}}
\newcommand{\gi}{{\scriptstyle\mathrm{gi}}}
\newcommand{\dif}{\mathrm{d}}
\newcommand{\me}{\mathrm{e}}
\newcommand{\mi}{\mathrm{i}}

\begin{document}

\title{Structure Formation independent of Cold Dark Matter}


\author{P.\ G.\ Miedema}
\email{pieter.miedema@gmail.com}
\affiliation{Netherlands Defence Academy\\
   Hogeschoollaan 2, \\ NL-4818 CR  Breda, The Netherlands}

\date{January 29, 2011}

\begin{abstract}
  It is shown that a first-order cosmological perturbation theory for
  Friedmann-Lema\^\i tre-Robertson-Walker universes admits one and
  only one gauge-invariant variable which describes the perturbation
  to the energy density and which becomes equal to the usual energy
  density of the Newtonian theory of gravity in the limit that all
  particle velocities are negligible with respect to the speed of
  light.  The same holds true for the perturbation to the particle
  number density.

  A cosmological perturbation theory based on these particular
  gauge-invariant quantities is more precise than any earlier
  first-order perturbation theory. In particular, it explains star
  formation in a satisfactory way, even in the absence of cold dark
  matter. In a baryon-only universe, the earliest stars, the so-called
  Population~III stars, are found to have masses between 400 and
  100,000 solar masses with a peak around 3400 solar masses. If cold
  dark matter, with particle mass 10 times heavier than the proton
  mass, is present then the star masses are between 16 and 4000 solar
  masses with a peak around 140 solar masses. They come into existence
  between 100~Myr and 1000~Myr.  At much later times, star formation
  is possible only in high density regions, for example within
  galaxies. Late time stars may have much smaller masses than early
  stars. The smallest stars that can be formed have masses of 0.2--0.8
  solar mass, depending on the initial internal relative pressure
  perturbation.

  It is demonstrated that the Newtonian theory of gravity cannot be
  used to study the evolution of cosmological density perturbations.
\end{abstract}


\maketitle




\tableofcontents

\baselineskip=1.0\baselineskip

\section{Introduction}

It is well-known, or formulated more precisely, it is generally
accepted, that in a universe filled with only `ordinary matter,' i.e.,
elementary particles and photons but not \emph{cold dark matter}
(\textsc{cdm}), the linear perturbation theory predicts a too small
growth rate to account for star formation in the universe. In this
article we establish that this is not true. The reason brought forward
in all former treatises on first-order cosmological density
perturbations is that the growth of a relative density perturbation in
the era after decoupling of radiation and matter, given by
\begin{equation}\label{eq:standard-growth}
\delta(t)=\delta(t_\mathrm{dec})
\left(\dfrac{t}{t_\mathrm{dec}}\right)^{2/3}, \quad t\ge t_\mathrm{dec},
\end{equation}
is insufficient for relative density perturbations as small as the
observed value $\delta(t_\mathrm{dec})\approx10^{-5}$ to reach the
non-linear phase for times $t\le t_\mathrm{p}$, where
$t_\mathrm{p}=13.7\,\mathrm{Gyr}$, the present age of the universe,
and $t_\mathrm{dec}=380\,\mathrm{kyr}$, the time of decoupling of
matter and radiation. This generally accepted conclusion is suggested
by the solutions (\ref{eq:gen-sol-stand-w}), with $w=0$, of the
standard relativistic evolution equation (\ref{eq:stand-w}) for linear
density perturbations in the era after decoupling of matter and
radiation: this equation yields a growth rate which is much too low to
allow for star formation within $13.7\,\mathrm{Gyr}$.  Therefore,
researchers in the field of structure formation have to assume that a
significant amount of \textsc{cdm} had contracted already before
decoupling in order to explain in their simulations the formation of
large-scale structure after
decoupling~\cite{frenk-2002-300,springel-2006-440}.

The purpose of this article is to show that the formation of structure
can be explained whether or not \textsc{cdm} is present. Our treatise
is independent of a particular system of reference and yields results
which describe the evolution of small density perturbations in the
radiation-dominated era and in the era after decoupling of matter and
radiation. These remarkable and most satisfactory results are a direct
consequence of two facts only. Firstly, we use gauge-invariant
expressions for the first-order perturbations to the energy density,
$\varepsilon^\gi_\een$, and particle number density,
$n^\gi_\een$. Secondly, in the \emph{dark ages} of the universe (i.e.,
the epoch between decoupling and the ignition of the first stars) a
density perturbation from which stars will eventually be formed, has
to cool down~\cite{loeb-2008,greif-2008} in its linear phase in order to
grow. Consequently, the growth of density perturbations can only be
described by a realistic equation of state for the pressure in
combination with the combined First and Second Laws of thermodynamics
(\ref{eq:sec-law-thermo}). Therefore, we use an equation of state for
the pressure of the form $p=p(n,\varepsilon)$, where $n$ is the
particle number density and $\varepsilon$ is the energy density.

Our more sophisticated treatise, which in first-order perturbation
theory is explicitly gauge-invariant, and which uses an equation of
state of the form $p=p(n,\varepsilon)$ rather than of the form
$p=p(\varepsilon)$, makes it possible to explain structure formation
even in the absence of \textsc{cdm}.

\section{Main Results}

In this article no assumptions or approximations (other than linearization of
the Einstein equations and conservation laws) have been made in order to reach
our conclusions. The only assumption we have made is
that Einstein's General Theory of Relativity is the correct theory that
describes gravitation in our universe on all scales and from the onset of the
radiation-dominated era up to the present time.

In order to study structure formation in the universe, one needs the
linearized Einstein equations. The derivation of the evolution
equations (\ref{subeq:final}) for relative density fluctuations in a
Friedmann-Lema\^\i tre-Robertson-Walker (\textsc{flrw}) universe
filled with a perfect fluid with an equation of state
$p=p(n,\varepsilon)$, is one of the main subjects of this article.
Since we use a more general equation of state we are forced to derive
all basic equations from scratch, instead of taking them from
well-known textbooks or renowned articles. This has the advantage that
our article is self-contained and that our results can easily be
checked.

In Section~\ref{sec:eqs-sgi} we show, using the background (i.e.,
zeroth-order) Einstein equations and conservation laws connected with
the combined First and Second Laws of thermodynamics, that the
universe as a whole expands adiabatically. This is a well-known
result. One of the new results of our treatise is that \emph{local}
density perturbations evolve \emph{diabatically}. This has been made
clear in Section~\ref{sec:ad-pert-uni}\@. Only in the non-relativistic
limit, where $\varepsilon=\varepsilon(n)$ and $p=0$, local density
perturbations evolve adiabatically.

In the literature about the subject, all efforts to construct a
gauge-invariant cosmological perturbation theory yield a second-order
differential equation for the density contrast function
$\delta\equiv\varepsilon_\een/\varepsilon_\nul$, with or without an
entropy related source term. For that matter, our treatise is no
exception, as (\ref{sec-ord}) demonstrates. In contrast to the
perturbation theories developed in the literature, we find also a
first-order differential equation, namely (\ref{fir-ord}). This
equation results from the incorporation of the equation of state
$p=p(n,\varepsilon)$ for the pressure and the particle number
conservation law $(nu^\mu)_{;\mu}=0$, (\ref{pncl}). The consequences
of equation (\ref{fir-ord}) are radical: this equation implies that
density perturbations in the total energy density are gravitationally
coupled to density perturbations in the particle number density. This
is the case for ordinary matter as well as \textsc{cdm} throughout
the history of the universe from the onset of the radiation-dominated
era until the present.  As a consequence, perturbations in
\textsc{cdm} evolve gravitationally in exactly the same way as
perturbations in ordinary matter do. The assumption that \textsc{cdm}
would have clustered already before decoupling and thus would have
formed seeds for baryon contraction after decoupling is, therefore,
questionable. This conclusion has, on different grounds, also been
reached by Nieuwenhuizen
\textit{et~al.}\@~\cite{2009arXiv0906.5087N}. This may rule out
\textsc{cdm} as a means to facilitate structure formation in the
universe.

\subsection{Manifestly Gauge-invariant Perturbation Theory and its Non-Relativistic Limit}

In order to solve the structure formation problem of cosmology, we
first develop in Sections~\ref{frommat-mies}--\ref{thirdstep} a
\emph{manifestly} gauge-invariant perturbation theory, i.e., both the
evolution equations and their solutions are independent of the choice
of a system of reference.  In Section~\ref{frommat-mies} we show
that there exist two and only two \emph{unique} and gauge-invariant
first-order quantities $\varepsilon^\gi_\een$ and $n^\gi_\een$ for the
perturbations to the energy density and the particle number
density. The evolution equations for the contrast functions
$\delta_\varepsilon\equiv\varepsilon^\gi_\een/\varepsilon_\nul$ and
$\delta_n\equiv n^\gi_\een/n_\nul$ are given by
(\ref{subeq:final}). From their derivation it follows that these
equations include the ---background as well as first-order---
$G_{00}$- and $G_{0i}$-constraint equations; the $G_{ij}$-evolution
equations; the conservation laws for the energy density; for the
particle number density and for the momentum; and, finally, the
combined First and Second Laws of thermodynamics. Taking into account
the applicability of equations (\ref{subeq:final}) to the open closed
and flat \textsc{flrw} universes filled with a perfect fluid described
by an equation of state $p=p(n,\varepsilon)$, they are rather simple.

In Section~\ref{nrl} we show that in the non-relativistic limit
$v/c\rightarrow0$ the quantities $\varepsilon^\gi_\een$ and
$n^\gi_\een$ survive, and coincide with the usual energy density and
particle number density [see (\ref{eq:poisson}) and
(\ref{eq:rho-Newton})], whereas their gauge dependent counterparts
$\varepsilon_\een$ and $n_\een$ (which are also gauge dependent in the
non-relativistic limit) disappear completely from the scene. Finally,
we show that, in first-order, the \emph{global} expansion of the
universe is not affected by \emph{local} perturbations in the energy
and particle number densities.

\subsection{Large-Scale Perturbations: Confirmation of the Standard Knowledge}

In order to compare our treatise on cosmological density perturbations
with the standard knowledge, we consider a flat \textsc{flrw} universe
in the radiation-dominated era and in the era after decoupling of
matter and radiation. For \emph{large-scale} perturbations these two
cases have been thoroughly studied by a large number of researchers
from 1946 up till now, using the full set of linearized Einstein
equations and conservation laws. Consequently, our refinement of their
work cannot be expected to give results that differ much from those of
the standard theory.  Indeed, we have found that for large-scale
perturbations our manifestly gauge-invariant treatise corroborates the
outcomes of the standard theory in the large-scale limit of both eras,
with the exception that we do \emph{not} find, of course, the
non-physical gauge mode $\delta_\mathrm{gauge}\propto t^{-1}$ which
plagues the standard theory.

For example, our perturbation theory yields in the radiation-dominated
era the well-known solutions (\ref{delta-H-rad})
$\delta_\varepsilon\propto t$ and $\delta_\varepsilon\propto t^{1/2}$
\cite{c15,adams-canuto1975,olson1976,c11,kolb,C12} and in the era
after decoupling of matter and radiation we get the well-known
solutions (\ref{eq:new-dust-53-adiabatic}) $\delta_\varepsilon\propto
t^{2/3}$ \cite{c15,adams-canuto1975,olson1976,c11,kolb,C12} and
$\delta_\varepsilon\propto t^{-5/3}$ \cite{c13,mfb1992}.  A new result
is, however, that the solutions (\ref{delta-H-rad}) as well as
(\ref{eq:new-dust-53-adiabatic}) follow from \emph{one} second-order
differential equation, namely (\ref{eq:delta-rad}) and
(\ref{eq:delta-dust}) respectively.

\subsection{Small-Scale Perturbations and Star Formation: New Results}

The major difference between our treatise and the standard treatise on
the subject lays in the evolution of small-scale density
perturbations. For a radiation-dominated universe, the standard theory
yields oscillating density perturbations (\ref{eq:peacock-sol}) with a
\emph{decaying} amplitude. In contrast, our theory yields
oscillating density perturbations with an \emph{increasing}
(\ref{dc-small}) amplitude. This difference is entirely due to the
presence of the spurious gauge modes in the solutions of the standard
equations, as we will explain in detail in
Section~\ref{sec:rad-stand}.

After decoupling of matter and radiation at $z=1091$, the results of
our treatise and the results found in literature differ
also considerably. Just as is done in previous researches, we
take as equations of state for the energy density and the pressure
$\varepsilon=nm_{\text{H}}c^2+\tfrac{3}{2}nk_{\text{B}}T$ and
$p=nk_{\text{B}}T$, respectively, in the background as well as in the
perturbed universe. Since $m_{\text{H}}c^2\gg k_{\text{B}}T$
throughout the matter-dominated era after decoupling, it follows that
one may neglect the pressure $nk_{\text{B}}T$ and kinetic energy
density $\tfrac{3}{2}nk_{\text{B}}T$ with respect to the rest-mass
energy density $nm_{\text{H}}c^2$ in the unperturbed universe and that
in the perturbed universe one has
$\delta_\varepsilon\approx\delta_n$. Therefore, one takes in the
literature $\delta_\varepsilon=\delta_n$ and solves the (homogeneous)
second-order evolution equation for $\delta_\varepsilon$. As we have
shown in Section~\ref{sec:eq-exact-sol}\@, this yields slowly growing
density perturbations with $\delta_p=\delta_\varepsilon$, i.e., the
relative pressure perturbation is equal to the relative energy density
perturbation, and a vanishing relative \emph{matter} temperature
perturbation $\delta_T$.

In contrast to the standard method, our perturbation theory yields
next to the usual second-order evolution equation
(\ref{eq:delta-dust}) for $\delta_\varepsilon$ also a first-order
evolution equation (\ref{eq:entropy-dust}) for the difference
$\delta_n-\delta_\varepsilon$. Therefore, we need not take $\delta_n$
exactly equal to $\delta_\varepsilon$, so that in our treatise we may
have $\delta_p\neq\delta_\varepsilon$ and $\delta_T\neq0$. As a
consequence, our resulting second-order evolution equation
(\ref{eq:dust-dimless}) becomes inhomogeneous and the initial relative
matter temperature perturbation $\delta_T(t_{\text{dec}},\vec{x})$
enters the source term. This proves to be crucial for star
formation. Although in a linear perturbation theory
$|\delta_T(t,\vec{x})|\le1$, this quantity is not constrained to be as
small as $\delta_\varepsilon(t_{\text{dec}},\vec{x})\approx
\delta_n(t_{\text{dec}},\vec{x})\approx10^{-5}$, as is demanded by
\textsc{wmap}-observations~\cite{2009ApJS..180..306D,Spergel:2003cb,komatsu-2008,hinshaw-2008,2010arXiv1001.4538K}. Since
the gas pressure $p=nk_{\text{B}}T$ is very low, its relative
perturbation $\delta_p\equiv p^\gi_\een/p_\nul$ and, accordingly, the
matter temperature perturbation $\delta_T(t_{\text{dec}},\vec{x})$
could be large. We have shown that just after decoupling at $z=1091$
negative relative matter temperature perturbations as small as
$-0.5\%$ yields massive stars within $13.7\,\text{Gyr}$. The very
first stars, the so-called Population \textsc{iii}
stars~\cite{johnson-2009,glover-2008,norman-2008}, come into existence
between $10^2\,\text{Myr}$ and $10^3\,\text{Myr}$ and have masses
between $4\times10^2\,\text{M}_\odot$ and $10^5\,\text{M}_\odot$, with
a peak around $3.4\times10^3\,\text{M}_\odot$. Stars lighter than
$3.4\times10^3\,\text{M}_\odot$ come into existence at later times,
because their internal gravity is weaker. On the other hand, stars
heavier than $3.4\times10^3\,\text{M}_\odot$ also develop later, since
they do not cool down so fast due to their large scale. These
conclusions, which are valid only in a universe filled with a baryonic
fluid, are outlined in Figure~\ref{fig:collapse}. However, if
\textsc{cdm} with particle mass 10 times the proton mass is present
then the peaks in Figure~\ref{fig:collapse} are at
$140\,\text{M}_\odot$, whereas for \emph{hot dark matter}
(\textsc{hdm}) the peaks are found to be at
$4.8\times10^4\,\text{M}_\odot$. The relation between the particle
mass and the mass of a star will be explained in
Section~\ref{sec:ster-CDM}\@.  The peaks in Figure~\ref{fig:collapse}
can be considered as the relativistic counterparts of the classical
\emph{Jeans mass}.

At much later times, star formation is still possible, however the
mass of the stars may be much smaller. For example, if star formation
starts at $z=1$ or later then the smallest stars that can be formed
have masses of $0.2\,\text{M}_\odot$--$0.8\,\text{M}_\odot$, depending
on the initial internal relative matter temperature perturbation. Also
the initial density perturbations must be considerable at late times
in order to make star formation feasible: for star formation starting
at $z=1$ one must have $0.7\lesssim
\delta_n\approx\delta_\varepsilon<1$.  This shows that late time star
formation is possible only in high density regions within galaxies,
but not in intergalactic space. These findings are summarized in
Figures~\ref{fig:collapse-10} and~\ref{fig:collapse-2}. In contrast to
the relativistic theory developed in this article the standard
Newtonian theory of linear perturbations, which does \emph{not} follow
from Einstein's gravitational theory, predicts, as can be seen from
Figure~\ref{fig:collapse-2-oud}, a lower limit for star formation of
$1.7\,\text{M}_\odot$, implying that our Sun could not exist at all.
This failure of the standard Newtonian perturbation theory can be
attributed to the gauge mode which is present in the solution
(\ref{eq:matter-non-physical}) of the standard equation
(\ref{eq:delta-dust-standard}).

We conclude that density perturbations in ordinary matter can account
for star formation. There is no need to make use of alternative
gravitational theories nor of the inclusion of \textsc{cdm}: the
Theory of General Relativity can be used to explain structure in our
universe. The important conclusion must be that Einstein's
gravitational theory not only describes the \emph{global}
characteristics of the universe, but is also \emph{locally}
successful.

\section{Linear Perturbations in the General Theory of Relativity}
\label{sec:lin-pert-hist}

Perturbation theories for \textsc{flrw} universes are, in general, constructed
along the following lines. First, all quantities relevant to the problem are
divided into two parts: `a background part' and `a perturbation part.' The
background parts are chosen to satisfy the Einstein equations for an isotropic
universe, i.e., one chooses for the background quantities the
\textsc{flrw}-solution. Because of the homogeneity, the background quantities
depend on the time coordinate~$t$ only. The perturbation parts are supposed to
be small compared to their background counterparts, and to depend on the
space-time coordinate $x=(ct,\vec{x})$. The background and perturbations are
often referred to as `zeroth-order' and `first-order' quantities respectively
and we will use this terminology also in this article. After substituting the
sum of the zeroth-order and first-order parts of all relevant quantities into the
Einstein equations, all terms that are products of two or more quantities of
first-order are neglected. This procedure leads, by construction, to a set of
\emph{linear} differential equations for the quantities of first-order. The
solution of this set of linear differential equations is then reduced to a
standard problem of the theory of ordinary, linear differential equations.

\subsection{History}

The first systematic study of cosmological density perturbations is
due to Lifshitz~\cite{lifshitz1946,I.12} (1946) and Lifshitz and
Khalatnikov~\cite{c15} (1963). They considered small variations in the
\emph{metric tensor} to study density perturbations in the
radiation-dominated ($p=\tfrac{1}{3}\varepsilon$) and matter-dominated
($p=0$) universe. The use of metric tensor fluctuations makes their
method vulnerable to spurious solutions, the so called gauge
modes. Adams and Canuto (1975) \cite{adams-canuto1975} extended the
work of Lifshitz to a more general equation of state $p=w\varepsilon$,
where $w$ is a constant. In 1966 Hawking~\cite{hawking1966} presented
a perturbation theory which is explicitly
coordinate-independent. Instead of using the perturbed metric tensor,
he considered small variations in the \emph{curvature} due to density
perturbations. In 1976, Olson \cite{olson1976} corrected and further
developed the work of Hawking. He defined the density perturbation
relative to co-moving proper time. The advantage of Olson's method is
that the gauge mode, still present in his solutions, can be readily
identified since gauge modes yield, in his theory, a vanishing
curvature perturbation. Bardeen \cite{c13} (1980), was one of the
first who realized that one should work with variables which are
themselves gauge-invariant. He used in his work two different
definitions of gauge-invariant density perturbations, which, in the
small-scale limit, coincide with the usual density perturbation which
is gauge dependent outside the horizon. Bardeen assumed that a gauge
dependent perturbation becomes gauge-invariant as soon as the
perturbation becomes smaller than the horizon. In Section~\ref{nrl} we
show that this assumption is invalid. Kodama and Sasaki
\cite{kodama1984} (1984) elaborated and clarified the pioneering work
of Bardeen.  Ellis, Bruni and Hwang \emph{et~al}.~\cite{Ellis1,Ellis2}
(1989) and Ellis and van Elst \cite{ellis-1998} (1998) criticized the
work of Hawking and Bardeen and gave an alternative and elegant
representation of density fluctuations. Their method is both fully
covariant and gauge-invariant. Although the standard equations
(\ref{eq:delta-rad-peacock-ls}) and (\ref{eq:delta-standard}) are,
according to Einstein's General Theory of Relativity, closely related
via (\ref{eq:stand-w}), equation (\ref{eq:delta-standard}) follows
from the theory of Ellis~\textit{et~al.}, but equation
(\ref{eq:delta-rad-peacock-ls}) cannot be derived from their method.
Moreover, they did not take into account the perturbed constraint
equations.  Mukhanov, Feldman and Brandenberger, in their 1992 review
article~\cite{mfb1992} entitled `Theory of Cosmological Perturbations'
mentioned or discussed more than 60 articles on the subject, and,
thereupon, suggested their own approach to the problem. Their method
is also discussed in the textbook of Mukhanov \cite{Mukhanov-2005}.  A
disadvantage of their method is that their perturbation theory does
not yield the usual Poisson equation in the non-relativistic limit
$v/c\rightarrow0$, implying that their gauge-invariant quantities
cannot be linked to their Newtonian counterparts. For an overview of
the literature we refer to Mukhanov~\emph{et~al.}\ \cite{mfb1992} and
the summer course given by Bertschinger \cite{Bertschinger1996}. A
recent and integral overview of the construction of cosmological
perturbation theories for flat \textsc{flrw} universes is given by
Malik and Wands~\cite{malik-2009}.

The fact that so many studies are devoted to a problem that is nothing
but obtaining the solution of a set of ordinary, linear differential
equations is due to the fact that there are several complicating
factors, not regarding the mathematics involved, but with respect to
the physical interpretation of the solutions. As yet there is no
consensus about which solution is the best. In this article we will
demonstrate that there is one and only one solution to the problem how
to construct gauge-invariant quantities. This then enables us to solve
the problem of structure formation.

\subsection{Origin of the Interpretation Problem}\label{intro-noot}

At the very moment that one has divided a physical quantity into a
zeroth-order and a first-order part, one introduces an ambiguity.  Let
us consider the energy density of the universe, $\varepsilon(x)$, and
the particle number density of the universe, $n(x)$. The linearized
Einstein equations contain as \emph{known} functions the zeroth-order
functions $\varepsilon_\nul(t)$ and $n_\nul(t)$, which describe the
evolution of the background, i.e., they describe the evolution of the
unperturbed universe and they obey the unperturbed Einstein equations,
and as \emph{unknown} functions the perturbations
$\varepsilon_\een(x)$ and $n_\een(x)$. The latter are the solutions to
be obtained from the linearized Einstein equations. The
sub-indexes~$0$ and~$1$, which indicate the order, have been put
between round brackets, in order to distinguish them from tensor
indices. In all calculations, products of a zeroth-order and a
first-order quantity are considered to be of first-order, and are
retained, whereas products of first-order quantities are neglected.

The ambiguity is that the linearized Einstein equations do not fix the
quantities $\varepsilon_\een(x)$ and $n_\een(x)$ uniquely. In fact, it turns out
that next to any solution for $\varepsilon_\een$ and $n_\een$ of the linearized
Einstein equations, there exist solutions of the form
\begin{subequations}
\label{subeq:split-e-n}
\begin{align}
  \hat{\varepsilon}_\een(x) & = 
\varepsilon_\een(x)+\psi(x)\partial_0\varepsilon_\nul(t),
     \label{e-ijk} \\
   \hat{n}_{\een} (x) & =  n_{\een}(x)+\psi(x) \partial_0 n_{\nul}(t),
         \label{n-ijk}
\end{align}
\end{subequations}
which also satisfy the linearized Einstein equations. Here the
symbol $\partial_0$ stands for the derivative with respect to
$x^0=ct$. The function $\psi(x)$ is an arbitrary but `small' function of the
space-time coordinate $x=(ct,\vec{x})$, i.e., we consider~$\psi(x)$ to be of
first-order. We will derive (\ref{subeq:split-e-n}) at a later point
in this article; here it is sufficient to note that the perturbations
$\varepsilon_\een$ and $n_\een$ are fixed by the linearized Einstein
equations up to terms that are proportional to an arbitrary, small
function $\psi(x)$, usually called a gauge function in this context.
Since a physical quantity, i.e., a directly measurable property of a
system, may not depend on an arbitrary function, the quantities
$\varepsilon_\een$ and $n_\een$ cannot be interpreted as the real
physical values of the perturbations in the energy density or the
particle number density. But if $\varepsilon_\een$ and $n_\een$ are
not the physical perturbations, what \emph{are} the real
perturbations? This is the notorious `gauge problem' encountered in
any treatise on cosmological perturbations. Many different answers
to this question can be found in the literature, none of which is
completely satisfactory; a fact which explains the ongoing
discussion on this subject. In this article we show that there is a
definitive answer to the gauge problem of cosmology.

\section{Gauge-invariant Quantities}\label{frommat-mies}

In the existing literature on cosmological perturbations, one has attempted to
solve the problem that corresponds to the gauge dependence of the perturbations
$\varepsilon_\een$ and $n_\een$ (\ref{subeq:split-e-n}) in two, essentially
different, ways. The first way is to impose an extra condition on the gauge
field $\psi(x)$~\cite{Bertschinger1996,hwang-noh-1997,Hwang-Noh-2005,
noh-hwang-2005a}. Another way to get rid of the gauge field $\psi(x)$ is to
choose linear combinations of the matter variables $\varepsilon_\een$, $n_\een$
and other gauge dependent variables to construct gauge-invariant quantities. The
latter method is generally considered better than the one where one fixes a
gauge, because it not only leads to quantities that are independent of an
undetermined function, as should be the case for a physical quantity, but it
also does not rely on any particular choice for the gauge function.

The newly constructed gauge-invariant quantities are then shown to obey a set of
linear equations, not containing the gauge function $\psi(x)$ anymore. These
equations follow, by elimination of the gauge dependent quantities in favor of
the gauge-invariant ones, in a straightforward way from the usual linearized
Einstein equations, which did contain $\psi(x)$. In this way, the theory is no
longer plagued by the gauge freedom that is inherent to the original equations
and their solutions: $\psi(x)$ has disappeared completely, as it should, not
with brute force, but as a natural consequence of the definitions of the
perturbations to the energy and the particle number densities. This method is
elaborated by Bardeen~\cite{c13} and Mukhanov \emph{et~al}.~\cite{mfb1992}. From
these two treatises on linear perturbation theory, which differ significantly
from each other, one is tempted to conclude that gauge-invariant quantities can
be constructed in many different ways, and that there is no way to tell which of
these theories describes the evolution of density perturbations correctly. This,
however, is not the case, as we will show in this article.

We follow the method advocated by Mukhanov \emph{et~al.}\ and Bardeen, but
with gauge-invariant quantities which differ substantially from those used by
these researchers. In fact, we will show that there exist \emph{unique}
gauge-invariant quantities
\begin{subequations}
\label{subeq:gi-en}
\begin{align}
 \varepsilon^\gi_\een & \equiv \varepsilon_\een -
   \frac{\partial_0\varepsilon_\nul}{\partial_0
   \theta_\nul}\theta_\een,       \label{gien} \\
 n^\gi_\een & \equiv n_\een - \frac{\partial_0
     n_\nul}{\partial_0 \theta_\nul}
      \theta_\een,       \label{gidi}
\end{align}
\end{subequations}
that describe the perturbations to the energy density and particle number
density. In these expressions $\theta_\nul$ and $\theta_\een$ are the background
and perturbation part of the covariant four-divergence
$\theta=c^{-1}U^\mu{}_{;\mu}$ of the cosmological fluid velocity field
$U^{\mu}(x)$.

In Section~\ref{frommat-zus}, we will show that the quantities
$\varepsilon^\gi_\een$ and $n^\gi_\een$ do \emph{not} change if we
switch from the old coordinates $x^\mu$ to new coordinates
$\hat{x}^\mu$ according to
\begin{equation}
   \hat{x}^\mu = x^\mu - \xi^\mu(x), \label{func}
\end{equation}
where the $\xi^\mu (x)$ ($\mu=0,1,2,3$) are four arbitrary
functions, considered to be of first-order, of the old coordinates
$x^\mu$, i.e., (\ref{func}) is an infinitesimal coordinate transformation, or
gauge transformation. In other words, we will show that
\begin{subequations}
\begin{align}
   \hat{\varepsilon}^\gi_\een (x) & = \varepsilon^\gi_\een (x), \\
   \hat{n}^\gi_\een (x) & = n^\gi_\een (x),
\end{align}
\end{subequations}
i.e., the perturbations (\ref{subeq:gi-en}) are independent of
$\xi^\mu(x)$, i.e., gauge-invariant.

Since the background quantities depend on time, but not on the spatial
coordinates, it will turn out that only the zero component of the gauge
functions $\xi^\mu(x)$ occurs in the transformation of the first-order gauge
dependent variables. We will call it $\psi(x)$:
\begin{equation}
\psi(x)\equiv \xi^0(x). \label{defpsi}
\end{equation}
In the perturbation theory, the gauge function $\psi(x)$ is to be
treated as a first-order quantity, i.e., as a small (or
`infinitesimal') change of the coordinates.

As yet, the quantities (\ref{subeq:gi-en}) are new: they have never
been used before. \emph{The fact that these quantities are unique
  follows immediately from the linearized Einstein equations for
  scalar perturbations, and, therefore, cannot be chosen
  arbitrarily}. Using the quantities (\ref{subeq:gi-en}), our
theory reduces to the usual Newtonian theory (\ref{eq:poisson}) and
(\ref{eq:rho-Newton}) in the limit that the spatial part of the
cosmological fluid velocity four-vector $U^\mu$ is small compared to
the velocity of light.

\subsection{Construction of Gauge-invariant First-order Perturbations}
\label{frommat-zus}

We now proceed with the proof that $\varepsilon^\gi_\een$ and
$n^\gi_\een$ are gauge-invariant, i.e., invariant under the general
infinitesimal coordinate transformation (\ref{func}). To that end, we start by
recalling
the defining expression for the Lie derivative of an arbitrary
tensor field $A^{\alpha\cdots\beta}{}_{\mu\cdots\nu}$ with respect
to a vector field $\xi^{\tau}(x)$. It reads
\begin{align}
\left(\mathcal{L}_\xi A\right)&^{\alpha\cdots\beta}{}_{\mu\cdots\nu}=
  A^{\alpha\cdots\beta}{}_{\mu\cdots\nu;\tau} \xi^{\tau}  \nonumber \\
 & -\,A^{\tau\cdots\beta}{}_{\mu\cdots\nu}\xi^{\alpha}{}_{;\tau}-\cdots-
    A^{\alpha\cdots\tau}{}_{\mu\cdots\nu}\xi^\beta{}_{;\tau}   \nonumber \\
 & +\,A^{\alpha\cdots\beta}{}_{\tau\cdots\nu}\xi^{\tau}{}_{;\mu}+\cdots+
    A^{\alpha\cdots\beta}{}_{\mu\cdots\tau}\xi^{\tau}{}_{;\nu},  
\label{lie1}
\end{align}
where the semi-colon denotes the covariant derivative. At the
right-hand side, there is a term with a plus sign for each lower
index and a term with a minus sign for each upper index. Recall
also, that the covariant derivative in the expression for the Lie
derivative may be replaced by an ordinary derivative, since the
Lie derivative is, by definition, independent of the connection.
This fact simplifies some of the calculations below.

Now, let $\{ x^{\mu} \}$ and $\{ \hat{x}{}^{\mu} = x^{\mu} -
\xi^{\mu} (x) \}$ be two sets of coordinate systems, where
$\xi^{\mu} (x)$ is an arbitrary ---but infinitesimal, i.e., in
this article, of first-order--- vector field. Then the components
$\hat{A}^{\alpha\cdots\beta}{}_{\mu\cdots\nu}(x)$ of the tensor
$A$ with respect to the new coordinates $\hat{x}{}^{\mu}$ can be
related to the components of the tensor
$A^{\alpha\cdots\beta}{}_{\mu\cdots\nu}(x)$, defined with respect
to the old coordinates $\{ x^{\mu} \}$ with the help of the Lie
derivative. Up to and including terms containing first-order
derivatives one has
\begin{equation}
  \hat{A}^{\alpha\cdots\beta}{}_{\mu\cdots\nu}(x) =
  A^{\alpha\cdots\beta}{}_{\mu\cdots\nu}(x)+
\left({\mathcal{L}_{\xi}}A\right)^{\alpha\cdots\beta}{}_{\mu\cdots\nu}
(x)+\cdots.
     \label{lie2}
\end{equation}
For a derivation of this expression, see Weinberg~\cite{c8},
Chapter~10, Section~9.

Note that $x$ in the left-hand side corresponds to a point, $P$ say,
of space-time with coordinates $x^\mu$ in the coordinate frame
\{$x$\}, while in the right-hand side $x$ corresponds to another
point, $Q$ say, with exactly the same coordinates $x^\mu$, but now
with respect to the coordinate frame \{$\hat{x}$\}. Thus,
(\ref{lie2}) is an expression that relates one tensor field $A$ at
two different points of space-time, points that are related via the
relation (\ref{func}).

The following observation is crucial. Because of the general
covariance of the Einstein equations, they are invariant under
general coordinate transformations $x\rightarrow\hat{x}$ and, in
particular, under coordinate transformations given
by~(\ref{func}). Hence, if some tensorial quantity $A(x)$ of
rank~$n$ ($n=0,1,\ldots$) satisfies the Einstein equations with as
source term the energy-momentum tensor $T$, the quantity
$\hat{A}(x)=A(x)+\mathcal{L}_\xi A(x)$ satisfies the Einstein
equations with source term $\hat{T}(x)=T(x)+\mathcal{L}_\xi T(x)$,
for a universe \emph{with precisely the same physical content}.
Because of the linearity of the linearized Einstein equations, a
linear combination of any two solutions is also a solution. In
particular, $\mathcal{L}_\xi A$, being the difference of ${A}$ and
$\hat{A}$, is a solution of the linearized Einstein equations with
source term $\mathcal{L}_\xi T$. In first-order, $\mathcal{L}_\xi
A(x)$ may be replaced by $\mathcal{L}_\xi A_\nul(t)$, where
$A_\nul(t)$ is the solution for $A(t)$ of the zeroth-order Einstein
equations. The freedom to add a term of the form $\mathcal{L}_\xi
A_\nul(t)$, with $\xi^\mu$ ($\mu=0,1,2,3$) four arbitrary
functions of first-order, to any solution of the Einstein
equations of the first-order, is the reason that none of the first-order
solutions is uniquely defined, and, hence, does not
correspond in a unique way to a measurable property of the
universe. This is the notorious gauge problem. The additional terms
$\mathcal{L}_\xi A_\nul(t)$ are called `gauge modes.'

Combining (\ref{lie1}) and (\ref{lie2}) we have
\begin{align}
  \hat{A}^{\alpha\cdots\beta}&{}_{\mu\cdots\nu}(x) =
 A^{\alpha\cdots\beta}{}_{\mu\cdots\nu}(x)+
  A^{\alpha\cdots\beta}{}_{\mu\cdots\nu;\tau}\xi^{\tau}  \nonumber \\
  & -\, A^{\tau\cdots\beta}{}_{\mu\cdots\nu}\xi^{\alpha}{}_{;\tau}-\cdots-
     A^{\alpha\cdots\tau}{}_{\mu\cdots\nu}\xi^{\beta}{}_{;\tau}   \nonumber \\
  &  +\,A^{\alpha\cdots\beta}{}_{\tau\cdots\nu} \xi^{\tau}{}_{;\mu}+\cdots+
     A^{\alpha\cdots\beta}{}_{\mu\cdots\tau} \xi^{\tau}{}_{;\nu}.  
\label{dakjeserop}
\end{align}
We now apply expression (\ref{dakjeserop}) to the case that $A$ is a
scalar $\sigma$, a four-vector $V^\mu$ and a tensor $A_\mu{}_\nu$
respectively,
\begin{subequations}
\label{sca-vec-ten}
\begin{align}
  \hat{\sigma}(x)&=\sigma(x)+\xi^\tau(x) \partial_\tau\sigma(x), \label{sigma}\\
  \hat{V}^\mu&=V^\mu+V^\mu{}_{;\tau}\xi^\tau-V^\tau\xi^\mu{}_{;\tau},
     \label{lievec}\\
\hat{A}_{\mu\nu}&= A_{\mu\nu}+
   A_{\mu\nu;\tau}\xi^\tau+A_{\tau\nu}\xi^\tau{}_{;\mu}+
   A_{\mu\tau}\xi^\tau{}_{;\nu}.
\label{lieder}
\end{align}
\end{subequations}
For the metric tensor, $g_{\mu\nu}$ we find in particular, from
expression (\ref{lieder}),
\begin{equation}
   \hat{g}_{\mu\nu} = g_{\mu\nu} + \xi_{\mu;\nu} + \xi_{\nu;\mu},
           \label{killing}
\end{equation}
where we have used that the covariant derivative of the metric
vanishes.

Our construction of gauge-invariant perturbations totally rest upon
these expressions for hatted quantities. In case $\sigma(x)$ is some
scalar quantity obeying the Einstein equations, $\sigma(x)$ can be
divided in the usual way into a zeroth-order and a first-order part:
\begin{equation}
    \sigma(x) \equiv \sigma_\nul(t) + \sigma_\een(x), \label{sigma1}
\end{equation}
where $\sigma_\nul (t) $ is some background quantity, and hence,
not dependent on the spatial coordinates. Then (\ref{sigma})
becomes
\begin{equation}
   \hat{\sigma}(x)=\sigma_\nul(t) + \sigma_\een(x)+
   \xi^0 (x)\partial_0 \sigma_\nul(t) + \xi^\mu(x) \partial_\mu
  \sigma_\een(x).   \label{sigmahat1}
\end{equation}
The last term, being a product of the first-order quantity
$\xi^{\mu} (x)$ and the first-order quantity $\partial_{\mu}
\sigma_\een$, will be neglected. We thus find
\begin{equation}
  \hat{\sigma} (x) = \sigma_\nul (t) + \hat{\sigma}_\een(x),
     \label{sigmahat2}
\end{equation}
with
\begin{equation}
  \hat{\sigma}_\een(x) \equiv \sigma_\een(x) +
    \psi(x) \partial_0\sigma_\nul(t),   \label{sigmahat3}
\end{equation}
where we used (\ref{defpsi}). Thus, in gauge transformations of scalar
quantities, only the zero component of the gauge functions need to be taken into
account. Similarly, we find from
(\ref{lievec}) and (\ref{killing})
\begin{equation}
  \hat{V}^\mu_\een(x)=V^\mu_\een+
      V_{\nul;\tau}^\mu\xi^\tau-V_\nul^\tau\xi^\mu{}_{;\tau},
 \label{transvec}
\end{equation}
and
\begin{equation}
  \hat{g}_{\een\mu\nu}(x)=g_{\een\mu\nu}(x)+\xi_{\mu;\nu}+\xi_{\nu;\mu}.
    \label{transmetric}
\end{equation}
The latter two expressions will be used later.

We are now in a position that we can conclude the proof of the
statement that $\varepsilon^\gi_\een$ and $n^\gi_\een$ are
gauge-invariant. To that end, we now write down expression
(\ref{sigmahat3}) once again, for another arbitrary scalar quantity
$\omega(x)$ obeying the Einstein equations. We then find the analogue
of expression (\ref{sigmahat3})
\begin{equation}
   \hat{\omega}_\een(x) = \omega_\een(x) + \psi(x)\partial_0
     \omega_\nul(t). \label{omegahat3}
\end{equation}
The left-hand sides of (\ref{sigmahat3}) and (\ref{omegahat3})
give the value of the perturbation at the point with coordinates
$x$ with respect to the old coordinate system $\{x\}$; the
right-hand sides of (\ref{sigmahat3}) and (\ref{omegahat3})
contains quantities with the same values of the coordinates, $x$,
but now with respect to the new coordinate system $\{\hat{x}\}$.
Eliminating the function~$\psi(x)$ from expressions (\ref{sigmahat3})
and~(\ref{omegahat3}) yields
\begin{equation}
\hat{\sigma}_\een(x) - \frac{\partial_0 \sigma_\nul(t)}{\partial_0
\omega_\nul(t)} \hat{\omega}_\een(x) = \sigma_\een(x) -
\frac{\partial_0\sigma_\nul(t)}{\partial_0 \omega_\nul(t)}
\omega_\een(x). \label{invariance1}
\end{equation}
In other words, the particular linear combination occurring in the
right-hand side of (\ref{invariance1}) of any two scalar quantities
$\omega$ and $\sigma$ is gauge-invariant, and, hence, a possible candidate for a
physical quantity.

The expressions (\ref{subeq:gi-en}) are precisely of the form
(\ref{invariance1}). As a consequence, $\varepsilon^\gi_\een$ and $n^\gi_\een$
are indeed invariant under the general infinitesimal coordinate transformation
(\ref{func}), i.e., they are gauge-invariant.

\subsection{Unique Gauge-invariant Density Perturbations}
\label{sec:unique-gi}

Expression (\ref{invariance1}) is the key expression of this article
as far as the scalar quantities $\varepsilon_\een$ and $n_\een$ are
concerned. It tells us how to combine the scalar quantities
occurring in the linearized Einstein equations in such a way that
they become gauge independent. Expression (\ref{invariance1}) can
be used to immediately derive the expressions (\ref{subeq:gi-en})
for the gauge-invariant energy and particle number densities.

In fact, let $U^{\mu}(x)$ be the four-velocity of the cosmological
fluid. In Section~\ref{sec:unique} it is shown that in the linear
theory of cosmological perturbations, defined by the background
equations (\ref{subeq:einstein-flrw}) and the perturbation equations
(\ref{subeq:pertub-flrw}) \emph{only three independent scalars} play a
role, namely
\begin{subequations}
\label{subeq:ent}
\begin{align}
    \varepsilon(x) & = c^{-2} T^{\mu\nu}(x) U_\mu(x) U_\nu(x), \label{eps1} \\
    n(x) & = c^{-2} N^\mu(x) U_\mu(x), \label{en1} \\
    \theta(x) & = c^{-1} U^\mu{}_{;\mu}(x), \label{exp1}
\end{align}
\end{subequations}
where
\begin{equation}\label{eq:current}
    N^\mu\equiv nU^\mu,
\end{equation}
is the cosmological particle current four-vector normalized
according to $U^\mu U_\mu=c^2$. These scalars are divided
according to
\begin{subequations}
\label{subeq:ent-split}
\begin{align}
   \varepsilon (x) & = \varepsilon_\nul (t) + \varepsilon_\een (x), \label{eps2}
\\
   n(x) & = n_\nul(t) + n_\een(x), \label{en2} \\
   \theta(x) & = \theta_\nul(t) + \theta_\een(x), \label{exp2}
\end{align}
\end{subequations}
where the background quantities $\varepsilon_\nul(t)$, $n_\nul(t)$
and $\theta_\nul(t)$ are solutions of the unperturbed Einstein
equations. These quantities depend on the time coordinate $t$ only. The relation
(\ref{invariance1}) inspires us to consider the gauge-invariant
combinations
\begin{subequations}
\label{subeq:ent-gi}
\begin{align}
   \varepsilon^\gi_\een(x) & \equiv \varepsilon_\een(x) - \frac{\partial_0
     \varepsilon_\nul(t)}{\partial_0 \omega_\nul(t)} \omega_\een(x), 
\label{eps3} \\
   n^\gi_\een(x) & \equiv n_\een(x) - \frac{\partial_0 n_\nul(t)}
     {\partial_0 \omega_\nul(t)} \omega_\een(x), \label{en3} \\
   \theta^\gi_\een(x) & \equiv \theta_\een(x) - \frac{\partial_0
     \theta_\nul(t)}{\partial_0 \omega_\nul(t)} \omega_\een(x).  \label{exp3}
\end{align}
\end{subequations}
The question remains what to choose for $\omega$ in these three
cases. In principle, we could choose for $\omega$ any of the
following three scalar functions available in the theory, i.e., we
could choose $\varepsilon$, $n$ or $\theta$. As we will show in
Section~\ref{nrl}, the only choice which satisfies the perturbed energy
density constraint equation (\ref{con-sp-1}) in the non-relativistic
limit $v/c\rightarrow0$ is 
\begin{equation}
  \omega = \theta. \label{omega1}
\end{equation}
This implies the expressions~(\ref{gien}) and (\ref{gidi}) for the
energy and particle number density perturbations, as was to be
shown. Using (\ref{omega1}), we find from (\ref{exp3})
\begin{equation}
   \theta^\gi_\een \equiv \theta_\een-
      \dfrac{\partial_0\theta_\nul}{\partial_0\theta_\nul}\theta_\een=0,
      \quad \theta_\een\neq0.
\label{thetagi}
\end{equation}
The physical interpretation of (\ref{thetagi}) is that, in first-order, the
\emph{global} expansion (\ref{exp1}) is not affected by a
\emph{local} perturbation in the energy density and particle number density. It
should be emphasized here that (\ref{thetagi}) is \emph{not} equivalent to the
`uniform Hubble constant gauge' of Bardeen \cite{c13}, i.e., we do
\emph{not} impose the gauge condition $\theta_\een=0$. In contrast,
$\theta^\gi_\een\equiv0$ follows from the linearized Einstein equations, and,
due to its gauge-invariance, \emph{holds true in arbitrary systems of
reference}. In other words, a `uniform Hubble function' is inherent in a
relativistic cosmological perturbation theory.

The fact that the expressions (\ref{subeq:gi-en}) for the
gauge-invariant quantities $\varepsilon^\gi_\een$ and $n^\gi_\een$ are
unique follows immediately from the background Einstein equations
(\ref{subeq:einstein-flrw}) and their perturbed counterparts
(\ref{subeq:pertub-flrw}). Consequently, these quantities cannot be
chosen arbitrarily. In Section~\ref{nrl} on the non-relativistic limit
we show that $\varepsilon^\gi_\een$ is the perturbation to the energy
density and $n^\gi_\een$ is the perturbation to the particle number
density.

\section{Einstein Equations and Conservation Laws in Synchronous Coordinates}
\label{synchronous}

The system of evolution equations (\ref{subeq:eerste}) or,
equivalently (\ref{subeq:final}), the main results of this article,
are manifestly gauge-invariant. Therefore, one may use any convenient
and suitable system of reference to derive these results.

The choice of a suitable coordinate system can be made as follows.  It
is well-known that in the Newtonian theory of gravity all possible
space-time coordinate systems are \emph{synchronous}, since time and
space transformations (\ref{eq:gauge-trans-newt}) are decoupled in the
Newtonian theory.  Consequently, in order to show that our
perturbation theory yields the Newtonian theory of gravity in the
non-relativistic limit, it is obligatory to work in a synchronous
system of reference.  A second motivation to use synchronous
coordinates is the fact that the background equations
(\ref{subeq:einstein-flrw}) are already given with respect to
synchronous coordinates.  Therefore, the evolution equations for
scalar perturbations (\ref{subeq:pertub-flrw}) turn out to be ---in
synchronous coordinates--- simple extensions of the background
equations. This fact has helped us to find the scalars
(\ref{subeq:ent}) which play a key role in the construction of our
manifestly gauge-invariant perturbation theory.

The name synchronous stems from the fact that surfaces with
$t=\mathrm{constant}$ are surfaces of simultaneity for observers at rest with
respect to the synchronous coordinates, i.e., observers for which the three
space coordinates $x^i$ ($i=1,2,3$) remain constant. A synchronous system can be
used for an arbitrary space-time manifold, not necessarily a homogeneous or
homogeneous and isotropic one \cite{I.12}.

In a synchronous system of reference the line
element for the metric has the form:
\begin{equation}
  \dif s^2=c^2\dif t^2-g_{ij}(t,\vec{x})\dif x^i \dif x^j.
       \label{line-element}
\end{equation}
In a synchronous system, the coordinate~$t$
measures proper time along lines of constant $x^i$. From
(\ref{line-element}) we can read off that ($x^0=ct$):
\begin{equation}
    g_{00}(t,\vec{x})=1, \quad g_{0i}(t,\vec{x})=0.    \label{sync-cond}
\end{equation}
From the form of the line element in four-space
(\ref{line-element}) it follows that minus $g_{ij}(t,\vec{x})$,
($i=1,2,3$), is the metric of three-di\-men\-sional subspaces with
constant~$t$. Because of (\ref{sync-cond}), knowing the
three-geometry in all hyper-surfaces, is equivalent to knowing the
geometry of space-time. The following abbreviations will prove
useful when we rewrite the Einstein equations with respect to
synchronous coordinates:
\begin{equation}
\label{def-gammas}
    \varkappa_{ij} \equiv -\tfrac{1}{2}\dot{g}_{ij}, \quad
    \varkappa^i{}_j \equiv g^{ik}\varkappa_{kj}, \quad
    \varkappa^{ij}\equiv +\tfrac{1}{2}\dot{g}^{ij},
\end{equation}
where a dot denotes differentiation with respect to~$x^0=ct$. From
(\ref{sync-cond})--(\ref{def-gammas}) it follows that the connection
coefficients of (four-di\-men\-sional) space-time
\begin{equation}
   \Gamma^\lambda{}_{\mu\nu}=\tfrac{1}{2} g^{\lambda\rho}
   \left( g_{\rho\mu,\nu}+g_{\rho\nu,\mu}-g_{\mu\nu,\rho} \right),
        \label{concoef1}
\end{equation}
are, in synchronous coordinates, given by
\begin{subequations}
\label{subeq:con}
\begin{align}
    \Gamma^0{}_{00}&=\Gamma^i{}_{00}=\Gamma^0{}_{i0}=\Gamma^0{}_{0i}=0,
           \label{con1}  \\
    \Gamma^0{}_{ij}&=\varkappa_{ij},   \quad
       \Gamma^i{}_{0j}=\Gamma^i{}_{j0}=-\varkappa^i{}_j,     \label{con2}  \\
    \Gamma^k{}_{ij}&=\tfrac{1}{2} g^{kl}
   \left( g_{li,j}+g_{lj,i}-g_{ij,l} \right).   \label{con3}
\end{align}
\end{subequations}
From (\ref{con3}) it follows that the~$\Gamma^k{}_{ij}$ are also the
connection coefficients of (three-dimensional) subspaces of constant
time.

The Ricci tensor $R_{\mu\nu} \equiv {R^{\lambda}}_{\mu\lambda\nu}$ is,
in terms of the connection coefficients, given by
\begin{equation}
  R_{\mu\nu}=
   \Gamma^\lambda{}_{\mu\nu,\lambda}-
   \Gamma^\lambda{}_{\mu\lambda,\nu}+
   \Gamma^\sigma{}_{\mu\nu}\Gamma^\lambda{}_{\lambda\sigma}-
   \Gamma^\sigma{}_{\mu\lambda} \Gamma^\lambda{}_{\nu\sigma}.  \label{Ricci1}
\end{equation}
Upon substituting (\ref{subeq:con}) into (\ref{Ricci1}) one finds
for the components of the Ricci tensor
\begin{subequations}
\label{subeq:ricci}
\begin{align}
   R_{00} & = \dot{\varkappa}^k{}_k - \varkappa^l{}_k \varkappa^k{}_l,
\label{R00} \\
   R_{0i} & = \varkappa^k{}_{k|i}-\varkappa^k{}_{i|k}, \label{R0i} \\
   R_{ij} & = \dot{\varkappa}_{ij} - \varkappa_{ij} \varkappa^k{}_k +
        2\varkappa_{ik}\varkappa^k{}_j + \mbox{$^3\!R_{ij}$},  \label{Rij}
\end{align}
\end{subequations}
where the vertical bar in (\ref{R0i}) denotes covariant
differentiation with respect to the metric~$g_{ij}$ of a
three-dimensional subspace:
\begin{equation}
   \varkappa^i{}_{j|k} \equiv \varkappa^i{}_{j,k} + \Gamma^i{}_{lk} \varkappa^l{}_j
        -\Gamma^l{}_{jk} \varkappa^i{}_l.   \label{threecov}
\end{equation}
The quantities $\mbox{$^3\!R_{ij}$}$ in (\ref{Rij}) are found to be
given by
\begin{equation}
  \mbox{$^3\!R_{ij}$} = \Gamma^k{}_{ij,k}-\Gamma^k{}_{ik,j}+
   \Gamma^l{}_{ij}\Gamma^k{}_{kl}-\Gamma^l{}_{ik} \Gamma^k{}_{jl}.
         \label{Ricci-drie}
\end{equation}
Hence, $\mbox{$^3\!R_{ij}$}$ is the Ricci tensor of the
three-dimensional subspaces of constant time. For the components
$R^\mu{}_\nu=g^{\mu\tau}R_{\tau\nu}$ of the Ricci
tensor~(\ref{subeq:ricci}), we get
\begin{subequations}
\label{subeq:Rmixed}
\begin{align}
   R^0{}_0 & = \dot{\varkappa}^k{}_k - \varkappa^l{}_k \varkappa^k{}_l,
\label{Rm00} \\
   R^0{}_i & = \varkappa^k{}_{k|i}-\varkappa^k{}_{i|k}, \label{Rm0i}\\
   R^i{}_j & = \dot{\varkappa}^i{}_j - \varkappa^i{}_j \varkappa^k{}_k +
\mbox{$^3\!R^i{}_j$},   \label{Rmij}
\end{align}
\end{subequations}
where we have used expressions (\ref{sync-cond})--(\ref{def-gammas}).

The Einstein equations read
\begin{equation}
   G^{\mu\nu} - \Lambda g^{\mu\nu}=\kappa T^{\mu\nu}, \label{ein-verg}
\end{equation}
where $G^{\mu\nu}$, the Einstein tensor, is given by
\begin{equation}
  G^{\mu\nu} \equiv R^{\mu\nu} - \tfrac{1}{2}R^\alpha{}_\alpha g^{\mu\nu}.
     \label{ein-ten}
\end{equation}
In (\ref{ein-verg}), $\Lambda$ is a positive constant, the
well-known cosmological constant. The constant~$\kappa$ is
given~by
\begin{equation}
   \kappa \equiv \frac{8\pi G}{c^4}=2.0766\times10^{-43}
\,\mathrm{m}^{-1}\,\mathrm{kg}^{-1}\,\mathrm{s}^2,   \label{kappa}
\end{equation}
with $G=6.6742\times10^{-11}\,\mathrm{m}^3\,\mathrm{kg}^{-1}\,\mathrm{s}^{-2}$
Newton's gravitational constant and
$c=2.99792458\times10^8\,\mathrm{m}\,\mathrm{s}^{-1}$ the speed of
light. In view of the Bianchi identities one has $G^{\mu\nu}{}_{;\nu}=0$,
hence, since $g^{\mu\nu}{}_{;\nu}=0$, the source term $T^{\mu\nu}$
of the Einstein equations must fulfill the properties
\begin{equation}
    T^{\mu\nu}{}_{;\nu} = 0.    \label{behoud}
\end{equation}
These equations are the energy-momentum conservation laws.

In order to derive simultaneously the background and first-order equations, we
rewrite the Einstein equations~(\ref{ein-verg}) in an alternative form, using
mixed upper and lower indices:
\begin{equation}
   R^\mu{}_\nu=\kappa(T^\mu{}_\nu-
          \tfrac{1}{2}\delta^\mu{}_\nu T^\alpha{}_\alpha) -
          \Lambda\delta^\mu{}_\nu.     \label{einstein}
\end{equation}
Upon substituting the components (\ref{subeq:Rmixed}) into the
Einstein equations~(\ref{einstein}), and eliminating the time
derivative of $\varkappa^k{}_k$ from the $R^0{}_0$-equation with
the help of the $R^i{}_j$-equations, the Einstein equations can be
cast in the form
\begin{subequations}
\label{subeq:Ein-syn}
\begin{align}
    (\varkappa^k{}_k)^2 - \mbox{$^3\!R$} -
       \varkappa^k{}_l\varkappa^l{}_k & =
        2(\kappa T^0{}_0 + \Lambda),  \label{Ein-syn1}   \\
    \varkappa^k{}_{k|i}-\varkappa^k{}_{i|k} & =
           \kappa T^0{}_i, \label{Ein-syn2} \\
    \dot{\varkappa}^i{}_j - \varkappa^i{}_j \varkappa^k{}_k +
        \mbox{$^3\!R^i{}_{j}$} & =
    \kappa(T^i{}_j - \tfrac{1}{2}\delta^i{}_j T^\mu{}_\mu)-\Lambda\delta^i{}_j,
            \label{Ein-syn3}
\end{align}
\end{subequations}
where
\begin{equation}
  \mbox{$^3\!R$}\equiv g^{ij}\,\mbox{$^3\!R_{ij}$}=\mbox{$^3\!R^k{}_k$},
     \label{drieR}
\end{equation}
is the curvature scalar of the three-dimensional subspaces of
constant time. The (differential) equations (\ref{Ein-syn3}) are
the so-called dynamical Einstein equations: they define the
evolution (of the time derivative) of the (spatial part of the)
metric. The (algebraic) equations (\ref{Ein-syn1})
and~(\ref{Ein-syn2}) are constraint equations: they relate the
initial conditions, and, once these are satisfied at one time,
they are satisfied automatically at all times.

The right-hand side of equations (\ref{subeq:Ein-syn}) contain the
components of the energy momentum tensor $T_{\mu\nu}$, which, for
a perfect fluid, are given by
\begin{equation}\label{Tmunu}
   T^\mu{}_\nu = (\varepsilon+p)u^\mu u_\nu - p \delta^\mu{}_\nu,
\end{equation}
where~$u^\mu(t,\vec{x})=c^{-1}U^\mu(t,\vec{x})$ is the hydrodynamic fluid
four-velocity normalized to unity ($u^\mu u_\mu=1$),
$\varepsilon(t,\vec{x})$ the energy density and~$p(t,\vec{x})$ the pressure at a
point~$(t,\vec{x})$ in space-time. In this expression we neglect terms
containing the shear and volume viscosity, and other terms related
to irreversible processes. The equation of state for the pressure
\begin{equation}
     p=p(n,\varepsilon),     \label{toestand}
\end{equation}
where $n(t,\vec{x})$ is the particle number density at a point~$(t,\vec{x})$ in
space-time, is supposed to be a given function of $n$ and
$\varepsilon$ (see also Appendix~\ref{sec:eq-state} for equations
of state in alternative forms).

As stated above already, the Einstein equations (\ref{Ein-syn1})
and (\ref{Ein-syn2}) are constraint equations to the Einstein
equations (\ref{Ein-syn3}) only: they tell us what relations
should exist between the initial values of the various unknown
functions, in order that the Einstein equations be solvable. In
the following, we shall suppose that these conditions are
satisfied. Thus we are left with the nine equations
(\ref{Ein-syn3}), of which, because of the symmetry of $g_{ij}$,
only six are independent. These six equations, together with the
four equations (\ref{behoud}) constitute a set of ten equations
for the eleven $(6+3+1+1)$ independent quantities $g_{ij}$, $u^i$,
$\varepsilon$ and $n$. The eleventh equation needed to close the
system of equations is the particle number conservation law
$N^\mu{}_{;\mu}=0$. Using (\ref{eq:current}), we get
\begin{equation}
     (nu^\mu)_{;\mu}=0,     \label{pncl}
\end{equation}
where a semicolon denotes covariant differentiation with respect to the metric
tensor $g_{\mu\nu}$. This equation can be rewritten in terms of the fluid
expansion scalar defined by expression (\ref{exp1}). Using (\ref{subeq:con}), we
can rewrite the four-divergence (\ref{exp1}) in the form
\begin{equation}
  \theta = \dot{u}^0 - \varkappa^k{}_k u^0 + \vartheta,  \label{vierdiv}
\end{equation}
where the three-divergence $\vartheta$ is given by
\begin{equation}
    \vartheta \equiv u^k{}_{|k}.    \label{driediv}
\end{equation}
Using now expressions (\ref{exp1}), (\ref{subeq:con}), (\ref{threecov})
and~(\ref{vierdiv}), the four energy-momentum conservation
laws~(\ref{behoud}) and the particle number conservation
law~(\ref{pncl}) can be rewritten as
\begin{subequations}
\label{subeq:cons-laws}
\begin{align}
  \dot{T}^{00}+T^{0k}{}_{|k}+\varkappa^k{}_l T^l{}_k -
       \varkappa^k{}_k T^{00} & = 0,     \label{Tnulnu}  \\
  \dot{T}^{i0}+T^{ik}{}_{|k}-2\varkappa^i{}_k T^{k0} -
               \varkappa^k{}_k T^{i0} & = 0,     \label{Tinu}
\end{align}
\end{subequations}
and
\begin{equation}\label{deeltjes}
    \dot{n}u^0 + n_{,k}u^k + n\theta = 0,
\end{equation}
respectively. Since $T^{0i}$ is a vector and $T^{ij}$ is a tensor
with respect to coordinate transformations in a subspace of
constant time, and, hence, are tensorial quantities in this
three-dimensional subspace, we could use in
(\ref{subeq:cons-laws}) a bar to denote covariant differentiation
with respect to the metric $g_{ij}(t,\vec{x})$ of such a subspace
of constant time~$t$.

The Einstein equations (\ref{subeq:Ein-syn}) and conservation laws
(\ref{subeq:cons-laws}) and~(\ref{deeltjes}) describe a universe
filled with a perfect fluid and with a positive cosmological
constant. The  fluid pressure $p$ is described by an equation of
state of the form~(\ref{toestand}): in this stage we only need
that it is some function of the particle number density $n$ and
the energy density~$\varepsilon$.

We have now rewritten the Einstein equations and conservation laws in such a way
that one can easily derive the background and perturbation equations.

\section{Zeroth- and First-order Equations for the  FLRW Universe}  
\label{backpert}

We will now limit the discussion to a particular class of universes,
namely the collection of universes that, apart from a small, local
perturbation in space-time, are homogeneous and isotropic, the
Friedmann-Lema\^\i tre-Robertson-Walker (\textsc{flrw}) universes.

We expand all quantities $Q$ in the form of series, and derive, recursively,
equations for the successive terms of these series.
We will distinguish the successive terms of a series by a sub-index between
brackets:
\begin{equation}
\label{subeq:exp-scalar}
  Q  = Q_\nul + \eta Q_\een + \eta^2 Q_\twee+\cdots, \\
\end{equation}
where the sub-index zero refers to quantities of the unperturbed,
homogeneous and isotropic \textsc{flrw} universe.

In expression (\ref{subeq:exp-scalar}) $\eta$ ($\eta\equiv1$) is a bookkeeping
parameter, the function of which is to enable us in actual calculations to
easily distinguish between the terms of different orders.

\subsection{Zeroth-order Quantities} \label{backpert-zero}

This section is concerned with the background or zeroth-order
quantities occurring in the Einstein equations. All results of this
section are standard~\cite{c8}, and given here only to fix the
notation unambiguously.

For a \textsc{flrw} universe, the background metric $g_{\nul\mu\nu}$ is given by
\begin{subequations}
\label{subeq:metric}
\begin{align}
   g_{\nul 00}(t,\vec{x}) & =1, \quad g_{\nul0i}(t,\vec{x})=0, \label{m1}   \\
   g_{\nul ij}(t,\vec{x}) & =-a^2(t)\tilde{g}_{ij}(\vec{x}), \quad
      g_\nul^{ij}(t,\vec{x}) = -\frac{1}{a^2(t)}
      \tilde{g}^{ij}(\vec{x}).   \label{m2}
\end{align}
\end{subequations}
where $\tilde{g}_{ij}(\vec{x})$ is the metric of a
three-dimensional maximally symmetric subspace:
\begin{equation}
   \tilde{g}_{ij}=\mathrm{diag}
      \left(\frac{1}{1-kr^2}, r^2, r^2\sin^2\varpi\right), \quad
      k=0,\pm1.
\label{tildegij}
\end{equation}
The minus sign in (\ref{m2}) has been introduced in order to switch from the
conventional four-dimensional space-time with signature $(+,-,-,-)$ to the
conventional three-dimensional spatial metric with signature $(+,+,+)$.

All background scalars depend on time only. Furthermore, four-vectors have
vanishing spatial components. Since $u^\mu$ is a unit vector we have
\begin{equation}
    u^\mu_\nul=\delta^\mu{}_0.  \label{u0}
\end{equation}
The time derivative of the three-part of the
metric $g_{\nul ij}$, $\varkappa_{\nul ij}$, may be
expressed in the Hubble function $\mathcal{H}(t)\equiv (\dif a/\dif t)/a(t)$. We
prefer to use a function
\begin{equation}\label{eq:Hubble-function}
    H(t)=\dfrac{\mathcal{H}(t)}{c},
\end{equation}
which we will call Hubble function also. Recalling that a dot
denotes differentiation with respect to $ct$, we have
\begin{equation}
        H\equiv\frac{\dot{a}}{a}. \label{Hubble}
\end{equation}
Substituting the expansion (\ref{subeq:exp-scalar}) into the definitions
(\ref{def-gammas}), we obtain
\begin{equation}\label{metricFRW}
  \varkappa_{\nul ij} = -Hg_{\nul ij}, \!\!\quad \!
  \varkappa^i_{\nul j} = -H\delta^i{}_j, \!\!\quad \!
  \varkappa_\nul^{ij} = -Hg_\nul^{ij},
\end{equation}
where we considered only terms up to the zeroth-order in the
bookkeeping parameter $\eta$.

Similarly, with (\ref{vierdiv}), (\ref{driediv}), (\ref{subeq:exp-scalar}),
(\ref{u0}) and~(\ref{metricFRW}) we find for the background fluid expansion
scalar, $\theta_\nul$, and the
three-divergence, $\vartheta_\nul$,
\begin{equation}
    \theta_\nul = 3 H, \quad
       \vartheta_\nul = 0.   \label{fes2}
\end{equation}
These quantities and their first-order counterparts will play and important role
in our perturbation theory.

Using (\ref{Tmunu}), (\ref{subeq:exp-scalar}), (\ref{subeq:metric}) and
(\ref{u0}) we find for the components of the energy
momentum tensor
\begin{equation}\label{emt}
  T^0_{\nul 0} = \varepsilon_\nul, \quad
     T^i_{\nul 0} =  0, \quad
      T^i_{\nul j} = -p_\nul\delta^i{}_j,
\end{equation}
where the background pressure $p_\nul$ is given by the equation of
state~(\ref{toestand}), which, for the background pressure, is
defined by
\begin{equation}
   p_\nul = p_\nul(n_\nul,\varepsilon_\nul).
       \label{toestandback}
\end{equation}

The background three-dimensional Ricci
tensor, (\ref{Ricci-drie}), is given by
\begin{equation}
  \mbox{$^3\!R_{\nul ij}$} =
   \Gamma^k_{\nul ij,k}-\Gamma^k_{\nul ik,j}+
   \Gamma^l_{\nul ij} \Gamma^k_{\nul kl}-
   \Gamma^l_{\nul ik}
   \Gamma^k_{\nul jl},      \label{Ricci-drie-back}
\end{equation}
where the connection coefficients $\Gamma^k_{\nul ij}$ are given
by
\begin{equation}
   \Gamma^k_{\nul ij}=\tfrac{1}{2} g^{kl}_\nul
   \left( g_{\nul li,j}+g_{\nul lj,i}-g_{\nul ij,l} \right),  \label{cc0}
\end{equation}
where $g^{ij}_\nul$ and $g_{\nul ij}$ depend on time.
Hence, the connection coefficients $\Gamma^k_{\nul ij}$ are equal
to the connection coefficients $\tilde{\Gamma}^k{}_{ij}$ of the
metric~$\tilde{g}_{ij}$:
\begin{equation}
    \Gamma^k_{\nul ij} = \tilde{\Gamma}^k{}_{ij} \equiv
      \tfrac{1}{2} \tilde{g}^{kl}
   \left( \tilde{g}_{li,j}+\tilde{g}_{lj,i}-\tilde{g}_{ij,l}\right).
      \label{con3FRW}
\end{equation}
Therefore, they do not depend on time.

Substituting~(\ref{subeq:metric}) and (\ref{tildegij}) into
(\ref{Ricci-drie-back}), combined
with (\ref{con3FRW}), we find
\begin{equation}
    \mbox{$^3\!R_{\nul ij}$} = 2k \tilde{g}_{ij}. \label{RFLRW}
\end{equation}
From (\ref{RFLRW}) we have
\begin{equation}
   \mbox{$^3\!R^i_{\nul j}$}(t) = -\frac{2k}{a^2(t)}\delta^i{}_j,
       \label{Rijmixed}
\end{equation}
implying that the zeroth-order curvature scalar
$\mbox{$^3\!R_\nul$}=g^{ij}_\nul\,\mbox{$^3\!R_{\nul ij}$}$ is
given by
\begin{equation}
    \mbox{$^3\!R_\nul$}(t) = -\frac{6k}{a^2(t)}.    \label{spRicci}
\end{equation}
This quantity and its perturbed counterpart will play an important role in our
perturbation theory.
Note, that in view of our choice of the metric $(+,-,-,-)$, spaces of positive
curvature~$k$ have a negative curvature scalar~$\mbox{$^3\!R_\nul$}$.

Thus, we have found all background quantities.

\subsection{First-order Quantities}\label{backpert-first}

In this section we express all quantities occurring in the
Einstein equations in terms of zeroth- and first-order quantities. The
equations of state for the energy and pressure, $\varepsilon(n,T)$
and $p(n,T)$, are not specified yet.

Upon substituting the series~(\ref{subeq:exp-scalar}) into the normalization
condition $u^\mu u_\mu=1$, one finds, equating equal powers of the
bookkeeping parameter~$\eta$,
\begin{equation}
   u^0_\een=0,  \label{du0}
\end{equation}
for the first-order perturbation to the four-velocity. Writing the
inverse of $g_{kl}=g_{\nul kl}+\eta g_{\een kl}+\cdots$ as
\begin{equation}
   g^{kl}=g_\nul^{kl} +
      \eta g_\een^{kl} + \cdots, \label{gup}
\end{equation}
where $g_\nul^{kl}$ is the inverse, of $g_{\nul kl}$,
(\ref{m2}), we find
\begin{equation}
   g_\een^{kl} = -g_\nul^{ki}
         g_\nul^{lj}
         g_{\een ij},      \label{g1up}
\end{equation}
and
\begin{equation}
   g^k_{\een i} =
     -g^{kl}_\nul
     g_{\een li}.  \label{mixedg}
\end{equation}
It is convenient to introduce
\begin{equation}
   h_{ij} \equiv - g_{\een ij}, \label{hij}
\end{equation}
so that
\begin{equation}
  h^{ij}=g_\een^{ij}, \quad
      h^i{}_j=g_\nul^{ik} h_{kj}.   \label{hijgij}
\end{equation}
For the time derivative of the first-order perturbations to the
metric, $\varkappa_{\een ij}$ (\ref{def-gammas}), we get
\begin{equation}
  \varkappa_{\een ij}=\tfrac{1}{2}\dot{h}_{ij},  \quad
  \varkappa^i_{\een j}= \tfrac{1}{2}\dot{h}^i{}_j,  \quad
  \varkappa^{ij}_\een=\tfrac{1}{2}\dot{h}^{ij}.
\label{dgam}
\end{equation}

The first-order perturbation~$\theta_\een$ to the fluid expansion
scalar~$\theta$ (\ref{vierdiv}), can be found in the same way.
Using~(\ref{subeq:exp-scalar}) and (\ref{u0}) one arrives~at
\begin{equation}
  \theta_\een=
  \vartheta_\een-\tfrac{1}{2}\dot{h}^k{}_k, \label{fes5}
\end{equation}
where we used (\ref{du0}) and~(\ref{dgam}). This expression will play an important
role in the derivation of the first-order perturbation equations in Section~\ref{evo-scal}\@.
The first-order perturbation $\vartheta_\een$ to the three-divergence $\vartheta$
(\ref{driediv}), is
\begin{equation}
   \vartheta_\een = u^k_{\een|k},      \label{den1a}
\end{equation}
where we have used that
\begin{equation}
   (u^k{}_{|k})_\een= u^k_{\een|k},  \label{div-1}
\end{equation}
which is a consequence of $\Gamma^k_{\een lk}u_\nul^l=0$ as follows from
(\ref{u0}).

Upon substituting the series expansion (\ref{subeq:exp-scalar}), and into
(\ref{Tmunu}) and equating equal powers
of $\eta$, one finds for the first-order perturbation to the
energy-momentum tensor
\begin{subequations}
\label{pertemt}
\begin{align}
    T^0_{\een 0} & = \varepsilon_\een, \\
    T^i_{\een 0} & = (\varepsilon_\nul+p_\nul) u^i_\een,  \\
    T^i_{\een j} & = -p_\een \delta^i{}_j,
\end{align}
\end{subequations}
where we have used (\ref{u0}) and~(\ref{du0}). The first-order
perturbation to the pressure is related to $\varepsilon_\een$ and
$n_\een$ by the first-order perturbation to the equation of
state~(\ref{toestand}). We have
\begin{equation}
    p_\een=p_n n_\een +
    p_\varepsilon \varepsilon_\een,   \label{perttoes}
\end{equation}
where~$p_n$ and~$p_\varepsilon$ are the partial derivatives
of~$p(n,\varepsilon)$ with respect to~$n$ and~$\varepsilon$,
\begin{equation}
    p_n \equiv \left( \frac{\partial p}{\partial n} \right)_{\!\varepsilon}, \quad
  p_\varepsilon \equiv
        \left( \frac{\partial p}{\partial \varepsilon} \right)_{\!n}.
         \label{perttoes1}
\end{equation}
Since we consider only first-order quantities, the partial
derivatives are functions of the background quantities only, i.e.,
\begin{equation}\label{eq:pn-pe-back}
    p_n=p_n(n_\nul,\varepsilon_\nul), \quad
    p_\varepsilon=p_\varepsilon(n_\nul,\varepsilon_\nul).
\end{equation}

Using (\ref{con3}) and the series~(\ref{subeq:exp-scalar}), we
find for the first-order perturbations of the connection
coefficients
\begin{equation}
  \Gamma^k_{\een ij}=-g^{kl}_\nul g_{\een lm}\Gamma^m_{\nul ij}+
    \tfrac{1}{2} g^{kl}_\nul \left( g_{\een li,j}+ g_{\een lj,i}-
       g_{\een ij,l} \right).  \label{cc1}
\end{equation}
The first-order perturbations $\Gamma^k_{\een ij}$ (\ref{cc1}),
occurring in the non-tensor $\Gamma^k{}_{ij}$, happen to
be expressible as a tensor. Indeed, using (\ref{hij}), one can
rewrite (\ref{cc1}) in the form (the so-called Lifshitz formula)
\begin{equation}
    \Gamma^k_{\een ij}=
   -\tfrac{1}{2} g^{kl}_\nul
     (h_{li|j}+h_{lj|i}-h_{ij|l}). \label{con3pert}
\end{equation}
Using the expansion (\ref{subeq:exp-scalar}) for $\mbox{$^3\!R_{ij}$}$ and
$\Gamma^k{}_{ij}$, one finds for the first-order
perturbation to the Ricci tensor~(\ref{Ricci-drie})
\begin{equation}
  \mbox{$^3\!R_{\een ij}$} =
\Gamma^k_{\een ij,k}-\Gamma^k_{\een ik,j}+
   \Gamma^l_{\nul ij} \Gamma^k_{\een kl}+
   \Gamma^l_{\een ij} \Gamma^k_{\nul kl}-
   \Gamma^l_{\nul ik} \Gamma^k_{\een jl}-
   \Gamma^l_{\een ik}\Gamma^k_{\nul jl}, \label{Ricci-drie-pert}
\end{equation}
which can be rewritten in the compact form (the so-called contracted
Palatini identity)
\begin{equation}
   \mbox{$^3\!R_{\een ij}$} =
\Gamma^k_{\een ij|k}-\Gamma^k_{\een ik|j}.
    \label{palatini}
\end{equation}
By substituting (\ref{con3pert}) into (\ref{palatini}), one can
express the first-order perturbation to the Ricci tensor of the
three-dimensional subspace in terms of the perturbation to the
metric and its covariant derivatives:
\begin{equation}
  \mbox{$^3\!R_{\een ij}$} =
   -\tfrac{1}{2} g^{kl}_\nul
     (h_{li|j|k}+h_{lj|i|k}-h_{ij|l|k}-h_{lk|i|j}). \label{deltaR3}
\end{equation}
The perturbation $\mbox{$^3\!R^i_{\een j}$}$ is given by
\begin{equation}
   \mbox{$^3\!R^i_{\een j}$} \equiv (g^{ip}\,\mbox{$^3\!R_{pj}$})_\een=
      g^{ip}_\nul\,\mbox{$^3\!R_{\een pj}$} +
      \tfrac{1}{3}\,\mbox{$^3\!R_\nul$} h^i{}_j,   \label{mixedRij}
\end{equation}
where we have used (\ref{subeq:metric}), (\ref{RFLRW}),
(\ref{spRicci}) and~(\ref{hijgij}). Upon substituting
(\ref{deltaR3}) into (\ref{mixedRij}) we get
\begin{equation}
  \mbox{$^3\!R^i_{\een j}$} =
      -\tfrac{1}{2}g^{ip}_\nul(h^k{}_{p|j|k}+h^k{}_{j|p|k}-h^k{}_{k|p|j})+
      \tfrac{1}{2}g^{kl}_\nul h^i{}_{j|k|l} +
      \tfrac{1}{3}\,\mbox{$^3\!R_\nul$} h^i{}_j.   \label{R1mixed}
\end{equation}
Taking $i=j$ in (\ref{R1mixed}) and summing over the repeated index,
we find for the first-order perturbation to the curvature scalar of
the three-dimensional spaces
\begin{equation}
\mbox{$^3\!R_\een$} =
   g_\nul^{ij} (h^k{}_{k|i|j}-h^k{}_{i|j|k}) +
     \tfrac{1}{3}\,\mbox{$^3\!R_\nul$} h^k{}_k.   \label{driekrom}
\end{equation}
This expression will play an important role in the evolution of
density perturbations, equations (\ref{subeq:pertub-gi}), and in the
non-relativistic limit in Section~\ref{nrl}\@.

We thus have expressed all quantities occurring in the relevant
dynamical equations, i.e., the system of equations formed by the
Einstein equations combined with the conservation laws, in terms of
zeroth- and first-order quantities to be solved from these
equations. In the Sections~\ref{nulde} and \ref{eerste} below we
derive the background and first-order evolution equations
respectively. To that end we substitute the
series~(\ref{subeq:exp-scalar}) into the Einstein equations
(\ref{subeq:Ein-syn}) and conservation laws (\ref{subeq:cons-laws})
and~(\ref{deeltjes}). By equating the powers of $\eta^0$, $\eta^1$,
\ldots, we obtain the zeroth-order, the first-order and higher order
dynamical equations, constraint equations and conservation laws. We
will carry out this scheme for the zeroth- and first-order equations
only.

\subsection{Zeroth-order Equations}  \label{nulde}

With the help of Section~\ref{backpert-zero} and the
series (\ref{subeq:exp-scalar})
we now can find from the Einstein equations~(\ref{subeq:Ein-syn})
and conservation laws~(\ref{subeq:cons-laws}) and~(\ref{deeltjes})
the zeroth-order Einstein equations and the conservation laws.
Furthermore, in view of the symmetry induced by the isotropy, it
is possible to switch from the six quantities $g_{ij}$ and the six
quantities $\varkappa_{ij}$ to the curvature
$\mbox{$^3\!R_\nul$}(t)$ and the Hubble function $H(t)$ only.

\subsubsection{Einstein Equations}\label{nulde-einstein}

Upon substituting (\ref{metricFRW}) and~(\ref{emt}) into the
$(0,0)$-component of the Einstein equations, (\ref{Ein-syn1}), one
finds
\begin{equation}
   3H^2-\tfrac{1}{2}\,\mbox{$^3\!R_\nul$}=
     \kappa\varepsilon_\nul + \Lambda.        \label{endencon0}
\end{equation}
The $(0,i)$-components of the Einstein equations, (\ref{Ein-syn2}),
are identically fulfilled, as follows from (\ref{metricFRW})
and~(\ref{emt}). We thus are left with the six $(i,j)$-components of
the Einstein equations, (\ref{Ein-syn3}). In view of
(\ref{metricFRW}), (\ref{emt}) and~(\ref{Rijmixed}) we find that
$\varkappa^1_{\nul 1}=\varkappa^2_{\nul 2}=\varkappa^3_{\nul 3}$,
$T^1_{\nul 1}=T^2_{\nul 2}=T^3_{\nul 3}$ and $\mbox{$^3\!R^1_{\nul
1}$}=\mbox{$^3\!R^2_{\nul 2}$}= \mbox{$^3\!R^3_{\nul 3}$}$, whereas
for $i\neq j$ these quantities vanish. Hence, the six
$(i,j)$-components reduce to one equation,
\begin{equation}
    \dot{H} = -3H^2+\tfrac{1}{3}\,\mbox{$^3\!R_\nul$}
   +\tfrac{1}{2}\kappa(\varepsilon_\nul - p_\nul) + \Lambda.    \label{dyn0}
\end{equation}
This equation can be simplified by eliminating $3H^2$ with the help of
the constraint equation~(\ref{endencon0}).  We so find for the
dynamical equation for \textsc{flrw} universes
\begin{equation}
  \label{eq:dyn0-flrw}
  \dot{H}=-\tfrac{1}{6}\,\mbox{$^3\!R_\nul$}-
       \tfrac{1}{2}\kappa(\varepsilon_\nul+p_\nul).
\end{equation}
In equations (\ref{endencon0})~and~(\ref{dyn0}) the background
curvature~$\mbox{$^3\!R_\nul$}$ is given by (\ref{spRicci}). It is,
however, of convenience to determine this quantity from a
differential equation. Eliminating $a(t)$ from expressions
(\ref{Hubble}) and~(\ref{spRicci}) we obtain
\begin{equation}
  \mbox{$^3\!\dot{R}_\nul$} +
    2H\,\mbox{$^3\!R_\nul$} = 0,   \label{momback}
\end{equation}
where the initial value~$\mbox{$^3\!R_\nul(t_0)$}$ is given by
\begin{equation}\label{eq:init-R0}
   \mbox{$^3\!R_\nul(t_0)$} = -\dfrac{6k}{a^2(t_0)},
\end{equation}
in accordance with (\ref{spRicci}). It should be emphasized that
equation (\ref{momback}) is not an Einstein equation, since it is
equivalent to expression (\ref{spRicci}). It will be used here as an
ancillary relation.

\subsubsection{Conservation Laws}
\label{nulde-conservation}

Upon substituting (\ref{metricFRW}) and~(\ref{emt}) into the
$0$-component of the conservation law, (\ref{Tnulnu}), one finds
\begin{equation}
   \dot{\varepsilon}_\nul +
    3H(\varepsilon_\nul+p_\nul)=0,
   \label{energyFRW}
\end{equation}
which is the relativistic background continuity equation. The
background momentum conservation laws (i.e., the background
relativistic Euler equations) are identically satisfied, as follows
by substituting (\ref{metricFRW}) and~(\ref{emt}) into the spatial
components of the conservation laws, (\ref{Tinu}).

The background particle number density conservation law can be
found by substituting (\ref{u0}) and~(\ref{fes2}) into
equation (\ref{deeltjes}). One gets
\begin{equation}
   \dot{n}_\nul + 3H n_\nul = 0.   \label{deel1}
\end{equation}

The system of equations (\ref{endencon0}), (\ref{eq:dyn0-flrw}),
(\ref{momback}), (\ref{energyFRW}) and (\ref{deel1}) are five
equations for the four unknown quantities $H$, $\mbox{$^3\!R_\nul$}$,
$\varepsilon_\nul$ and $n_\nul$.  This system is, however not
overdetermined.  This can be shown as follows.  Differentiating the
constraint equation (\ref{endencon0}) with respect to time, yields
\begin{equation}
  \label{eq:time-diff-constraint}
  6H\dot{H}-\tfrac{1}{2}\,\mbox{$^3\!\dot{R}_\nul$}=\kappa\dot{\varepsilon}_\nul.
\end{equation}
Eliminating the time derivatives $\mbox{$^3\!\dot{R}_\nul$}$ and
$\dot{\varepsilon}_\nul$ with the help of equation (\ref{momback}) and
the conservation law (\ref{energyFRW}), respectively, yields the
dynamical equation (\ref{eq:dyn0-flrw}).  Consequently, the general
solution of the system (\ref{endencon0}), (\ref{momback}),
(\ref{energyFRW}) and (\ref{deel1}) is also a solution of the
dynamical equation (\ref{eq:dyn0-flrw}). Therefore, the dynamical
equation (\ref{eq:dyn0-flrw}) need not be considered anymore. This
concludes the derivation of the background equations.

\subsection{First-order Equations}   \label{eerste}

In this section we derive the first-order perturbation equations
from the Einstein equations~(\ref{subeq:Ein-syn}) and conservation
laws~(\ref{subeq:cons-laws}) and~(\ref{deeltjes}). The procedure is,
by now, completely standard. We use the series expansion
(\ref{subeq:exp-scalar}) in~$\eta$
for the various quantities occurring in the Einstein equations and
conservation laws of energy-momentum, and we equate the coefficients
linear in~$\eta$ to obtain the `linearized' or first-order
equations.

\subsubsection{Einstein Equations}  \label{eerste-einstein}

Using the series expansions (\ref{subeq:exp-scalar}) for $\mbox{$^3\!R$}$,
$\varkappa^i{}_j$ and $T^0{}_0$, in the
$(0,0)$-component of the constraint equation (\ref{Ein-syn1}), one
finds
\begin{equation}
  2\varkappa_{\nul k}^k \varkappa_{\een l}^l - \mbox{$^3\!R_\een$} -
  2\varkappa_{\nul l}^k \varkappa_{\een k}^l=
      2\kappa T_{\een 0}^0. \label{endenconpert}
\end{equation}
With the zeroth-order expressions (\ref{metricFRW}), the abbreviations
(\ref{dgam}) and the expression for $T^0_{\een 0}$, (\ref{pertemt}),
we may rewrite this equation in the form
\begin{equation}
   H\dot{h}^k{}_k + \tfrac{1}{2}\,\mbox{$^3\!R_\een$} =
         -\kappa\varepsilon_\een.         \label{endencon1}
\end{equation}

Using the series expansion (\ref{subeq:exp-scalar}) for $\varkappa^i{}_j$ and
$T^0{}_i$, we find for the $(0,i)$-components of the
constraint equations (\ref{Ein-syn2})
\begin{equation}
   \varkappa^k_{\een k|i} - \varkappa^k_{\een i|k}=\kappa T_{\een i}^0,
        \label{momconspert}
\end{equation}
where we noted that
\begin{equation}
   (\varkappa^i{}_{j|k})_\een = \varkappa^i_{\een j|k},    \label{gamijk1}
\end{equation}
which is a consequence of
    $\Gamma^i_{\een lk}\varkappa^l_{\nul j}-
           \Gamma^l_{\een jk}\varkappa^i_{\nul l}=0$,
which, in turn, is a direct consequence of (\ref{metricFRW}). From
(\ref{dgam}) and~(\ref{pertemt}) we find
\begin{equation}
    \dot{h}^k{}_{k|i}-\dot{h}^k{}_{i|k} =
         2\kappa(\varepsilon_\nul + p_\nul) u_{\een i}.    \label{dR0i2}
\end{equation}

Finally, we consider the $(i,j)$-components of the Einstein
equations (\ref{Ein-syn3}). Using the series expansion (\ref{subeq:exp-scalar})
for $\varkappa^i{}_j$, $T^i{}_j$ and
$\mbox{$^3\!R^i{}_j$}$, we find
\begin{equation}
     \dot{\varkappa}^i_{\een j}-\varkappa^i_{\een j}\varkappa^k_{\nul k}-
     \varkappa^i_{\nul j} \varkappa^k_{\een k} +
        \mbox{$^3\!R^i_{\een j}$} =
       \kappa(T^i_{\een j}-\tfrac{1}{2}\delta^i{}_j T^\mu_{\een\mu}).
\end{equation}
With (\ref{metricFRW}), (\ref{dgam}) and~(\ref{pertemt}), we get
\begin{equation}
   \ddot{h}^i{}_j+3H\dot{h}^i{}_j +
    \delta^i{}_j H\dot{h}^k{}_k + 2\,\mbox{$^3\!R^i_{\een j}$}=
     -\kappa\delta^i{}_j(\varepsilon_\een-p_\een),  \label{ddhij}
\end{equation}
where $\mbox{$^3\!R^i_{\een j}$}$ is given by expression (\ref{R1mixed}).

Note that the first-order equations~(\ref{endencon1})
and~(\ref{ddhij}) are independent of the cosmological
constant~$\Lambda$: the effect of the non-zero cosmological
constant is accounted for by the zeroth-order quantities
[cf.~equations (\ref{endencon0}) and~(\ref{dyn0})].

\subsubsection{Conservation Laws}   \label{eerste-conservation}

We now consider the energy conservation law~(\ref{Tnulnu}). With
the help of the series expansion (\ref{subeq:exp-scalar}) for $\varkappa^i{}_j$
and $T^\mu{}_\nu$, one finds for the
first-order equation
\begin{equation}
   \dot{T}^{00}_\een+T^{0k}_\een{}_{|k}+\varkappa^k_{\nul l} T^l_{\een k} +
    \varkappa^k_{\een l} T^l_{\nul k}-\varkappa^k_{\nul k} T^{00}_\een -
     \varkappa^k_{\een k} T^{00}_\nul =0,
           \label{energypert}
\end{equation}
where we have used that for a three-vector $T^{0k}$ we have
\begin{equation}
    (T^{0k}{}_{|k})_\een = T^{0k}_\een{}_{|k},
\end{equation}
see expression (\ref{div-1}). Employing (\ref{metricFRW}),
(\ref{emt}), (\ref{dgam})--(\ref{den1a}) and~(\ref{pertemt}) we
arrive at the first-order energy conservation law
\begin{equation}
   \dot{\varepsilon}_\een+3H(\varepsilon_\een +p_\een)+
         (\varepsilon_\nul +p_\nul)\theta_\een=0.
               \label{enpert}
\end{equation}

Next, we consider the momentum conservation laws~(\ref{Tinu}).
With the series expansion (\ref{subeq:exp-scalar}) for $\varkappa^i{}_j$ and
$T^{\mu\nu}$, we find for the first-order momentum conservation law
\begin{equation}
  \dot{T}^{i0}_\een+(T^{ik}{}_{|k})_\een-2\varkappa^i_{\nul k}T^{k0}_\een-
    2\varkappa^i_{\een k} T^{k0}_\nul-\varkappa^k_{\nul k} T^{i0}_\een-
   \varkappa^k_{\een k} T^{i0}_\nul=0.         \label{Tinu1}
\end{equation}
Using that
\begin{equation}\label{eq:Tik-k}
   (T^{ik}{}_{|k})_\een=-g_\nul^{ik}p_{\een|k},
\end{equation}
and expressions (\ref{metricFRW}), (\ref{emt}), (\ref{dgam})
and~(\ref{pertemt}) we arrive at
\begin{equation}
   \frac{1}{c}\frac{\dif}{\dif t}
   \Bigl[(\varepsilon_\nul+p_\nul) u^i_\een\Bigr]-
    g^{ik}_\nul p_{\een|k}+5H(\varepsilon_\nul+p_\nul) u^i_\een=0,
            \label{mom1}
\end{equation}
where we have also used that the covariant derivative of
$g^{ij}_\nul$ vanishes: $g^{ij}_{\nul |k}=0$.

Finally, we consider the particle number density conservation
law~(\ref{deeltjes}). With the series expansion (\ref{subeq:exp-scalar}) for
$n$, $\theta$, and $u^\mu$, it follows that the
first-order equation reads
\begin{equation}
   \dot{n}_\nul u^0_\een + \dot{n}_\een u^0_\nul + n_{\nul,k}u^k_\een+
   n_{\een,k}u^k_\nul + n_\nul \theta_\een + n_\een \theta_\nul = 0.
       \label{deel2}
\end{equation}
With the help of (\ref{u0}), (\ref{fes2}) and~(\ref{du0}) we find
for the first-order particle number conservation law
\begin{equation}
   \dot{n}_\een +3Hn_\een + n_\nul\theta_\een = 0.
         \label{deel3}
\end{equation}
This concludes the derivation of the basic perturbation equations.

\subsubsection{Summary} \label{sec:summary}

In the preceding two Sections~\ref{eerste-einstein}
and~\ref{eerste-conservation} we have found the equations which,
basically, describe the perturbations in a \textsc{flrw} universe,
in first-order approximation. They are equations (\ref{endencon1}),
(\ref{dR0i2}), (\ref{ddhij}), (\ref{enpert}), (\ref{mom1})
and (\ref{deel3}). For convenience we repeat them here
\begin{subequations}
\label{subeq:basis}
\begin{alignat}{3}
 \text{Constraints:} \quad &&  &H\dot{h}^k{}_k +
\tfrac{1}{2}\,\mbox{$^3\!R_\een$} =
         -\kappa\varepsilon_\een,     \label{basis-1} \\
  &&  &\dot{h}^k{}_{k|i}-\dot{h}^k{}_{i|k} =
         2\kappa(\varepsilon_\nul + p_\nul) u_{\een i}, \label{basis-2} \\
 \text{Evolution:} \quad  && &\ddot{h}^i{}_j+3H\dot{h}^i{}_j+
       \delta^i{}_j H\dot{h}^k{}_k + 2\, \mbox{$^3\!R^i_{\een j}$}=
     -\kappa\delta^i{}_j(\varepsilon_\een-p_\een),
         \label{basis-3} \\
  \text{Conservation:} \quad && &\dot{\varepsilon}_\een+3H(\varepsilon_\een
+p_\een)+
       (\varepsilon_\nul +p_\nul)\theta_\een=0,
               \label{basis-4}  \\
  && &\frac{1}{c}\frac{\dif}{\dif t}
   \Bigl[(\varepsilon_\nul+p_\nul) u^i_\een\Bigr]-
    g^{ik}_\nul p_{\een|k}+5H(\varepsilon_\nul+p_\nul) u^i_\een=0,
            \label{basis-5} \\
  &&  &\dot{n}_\een+3Hn_\een+n_\nul\theta_\een = 0,  \label{basis-6}
\end{alignat}
\end{subequations}
where $\theta_\een$, $\mbox{$^3\!R^i_{\een j}$}$ and $\mbox{$^3\!R_\een$}$ are
given by  (\ref{fes5}), (\ref{R1mixed}) and~(\ref{driekrom}) respectively. Hence,
the equations~(\ref{subeq:basis}) essentially are fifteen equations
for the eleven ($6+3+1+1$) unknown functions $h^i{}_j$, $u^i_\een$,
$\varepsilon_\een$ and $n_\een$. The pressure~$p_\nul$ is given by
an equation of state~(\ref{toestandback}), and the perturbation to
the pressure, $p_\een$, is given by (\ref{perttoes}). The system of
equations is not over-determined, however, since the four equations
(\ref{basis-1}) and~(\ref{basis-2}) are only conditions on the
initial values. These initial value conditions are fulfilled for all
times $t$ automatically if they are satisfied at some (initial)
time~$t=t_0$.

\section{Classification of the Solutions of First-order}
    \label{klasse}

It is well-known that, for \emph{flat} \textsc{flrw} universes, the set of linear
equations~(\ref{subeq:basis}) can be divided into three \emph{independent} sets
of equations, which, together, are equivalent to the original set. In this
section we show that this can also be done for the open and closed \textsc{flrw}
universes. We will refer to these sets by their usual names of scalar, vector
and tensor perturbation equations. We will show that the vector and tensor
perturbations do not, in first-order, contribute to the physical
perturbations~$\varepsilon_\een^\gi$ and~$n_\een^\gi$. As a consequence, we only
need, for our problem, the set of equations which are related to the scalar
perturbations. By considering only the scalar part of the full set of
perturbation equations we are able to cast the perturbation equations into a set
which is directly related to the physical perturbations~$\varepsilon_\een^\gi$
and~$n_\een^\gi$. This is the subject of Section~\ref{evo-scal}.

At the basis of the replacement of one set~(\ref{subeq:basis}) by
three sets of equations stands a theorem proved by York~\cite{York1974} and
Stewart~\cite{Stewart}, which states that a symmetric
second rank tensor can be divided into three irreducible pieces,
and that a vector can be divided into two irreducible pieces. Here,
we will use this general theorem to obtain equations for the scalar
irreducible parts of the tensors $h^i{}_j$ and $\mbox{$^3\!R^i_{\een
j}$}$ and the vector $\vec{u}_\een$, namely $h^i_{\parallel j}$,
$\mbox{$^3\!R^i_{\een\parallel j}$}$ and~$\vec{u}_{\een\parallel}$.

For the perturbation to the metric, a symmetric second rank
tensor we have in particular
\begin{equation}\label{decomp-symh}
  h^i{}_j=h^i_{\parallel j} + h^i_{\perp j} + h^i_{\ast j},
\end{equation}
where, according to the theorem of Stewart~\cite{Stewart}, the
irreducible constituents $h^i_{\parallel j}$, $h^i_{\perp j}$ and
$h^i_{\ast j}$ have the properties
\begin{subequations}
\label{subeq:dec-hij}
\begin{align}
    h^i_{\parallel j} & =
      \frac{2}{c^2}(\phi\delta^i{}_j+\zeta^{|i}{}_{|j}),
        \label{decomp-hij-par} \\
    h^k_{\perp k} & = 0,   \label{decomp-hij-perp} \\
    h^k_{\ast k} & = 0, \quad h^k_{\ast i|k} = 0,
       \label{decomp-hij-ast}
\end{align}
\end{subequations}
with $\phi(t,\vec{x})$ and $\zeta(t,\vec{x})$ arbitrary functions.
The contravariant derivative $A^{|i}$ is defined as
$g_\nul^{ij}A_{|j}$. The functions~$h^i_{\parallel j}$,
$h^i_{\perp j}$ and~$h^i_{\ast j}$ correspond to scalar,
vector and tensor perturbations respectively.

In the same way, the perturbation to the Ricci tensor can be
decomposed into irreducible components, i.e.,
\begin{equation}\label{decomp-Rij}
   \mbox{$^3\!R^i_{\een j}$} = \mbox{$^3\!R^i_{\een\parallel j}$} +
     \mbox{$^3\!R^i_{\een\perp j}$}+\mbox{$^3\!R^i_{\een\ast j}$}.
\end{equation}
The tensors $\mbox{$^3\!R^i_{\een\parallel j}$}$, $\mbox{$^3\!R^i_{\een\perp j}$}$ and $\mbox{$^3\!R^i_{\een\ast j}$}$
have the properties comparable to~(\ref{subeq:dec-hij}), i.e.,
\begin{subequations}
\label{subeq:prop-Rij}
\begin{align}
   \mbox{$^3\!R^i_{\een\parallel j}$} & =
       \frac{2}{c^2}(\gamma\delta^i{}_j+\pi^{|i}{}_{|j}), \label{Rij-paral} \\
   \mbox{$^3\!R^k_{\een\perp k}$} & = 0, \label{Rij-perp} \\
   \mbox{$^3\!R^k_{\een\ast k}$} & = 0, \quad \mbox{$^3\!R^k_{\een\ast i|k}$}=0,
        \label{Rij-ast}
\end{align}
\end{subequations}
where~$\gamma(t,\vec{x})$ and~$\pi(t,\vec{x})$ are two arbitrary
functions. By now using (\ref{R1mixed}) for each of the irreducible
parts we find
\begin{subequations}
\label{subeq:dec-Rij}
\begin{align}
 \mbox{$^3\!R^i_{\een\parallel j}$} = &\dfrac{1}{c^2}
   \Bigl[\phi^{|i}{}_{|j}+\delta^i{}_j\phi^{|k}{}_{|k}-\zeta^{|k|i}{}_{|j|k}-
   \zeta^{|k}{}_{|j}{}^{|i}{}_{|k} \nonumber\\
    & +\zeta^{|k}{}_{|k}{}^{|i}{}_{|j}+
  \zeta^{|i}{}_{|j}{}^{|k}{}_{|k}+
  \tfrac{2}{3}\, \mbox{$^3\!R_\nul$}(\delta^i{}_j\phi+\zeta^{|i}{}_{|j}) 
\Bigr],
    \label{decomp-Rij-par} \\
  \mbox{$^3\!R^i_{\een\perp j}$} = & -\tfrac{1}{2}g^{ip}_\nul(h_{\perp{p|j|k}}^k
+
   h_{\perp{j|p|k}}^k)+
   \tfrac{1}{2}g^{kl}_\nul h_{\perp{j|k|l}}^i+
      \tfrac{1}{3}\,\mbox{$^3\!R_\nul$} h_{\perp j}^i,  \label{decomp-Rij-perp}
\\
  \mbox{$^3\!R^i_{\een\ast j}$} = & -\tfrac{1}{2}g^{ip}_\nul(h_{\ast{p|j|k}}^k +
   h_{\ast{j|p|k}}^k)+
    \tfrac{1}{2}g^{kl}_\nul h_{\ast{j|k|l}}^i +
      \tfrac{1}{3}\,\mbox{$^3\!R_\nul$} h_{\ast j}^i.
       \label{decomp-Rij-ast}
\end{align}
\end{subequations}
Combining expressions~(\ref{Rij-perp})
and~(\ref{decomp-Rij-perp}) it follows that $h^i_\perp{}_j$ has
the property
\begin{equation}\label{eq:hklkl}
  h_{\perp{|k|l}}^{kl} = 0,
\end{equation}
in addition to the property~(\ref{decomp-hij-perp}). In
Section~\ref{vector} we show that this additional condition is needed
to allow for the decomposition (\ref{decomp-u}).

Combining expressions (\ref{Rij-ast})
and~(\ref{decomp-Rij-ast}), we find that $h^i_\ast{}_j$ must obey
\begin{equation}\label{ast-add}
  \tilde{g}^{kl}(h^m_{\ast{k|i|m|l}}+h^m_{\ast{i|k|m|l}}-
      h^m_{\ast{i|k|l|m}})=0,
\end{equation}
in addition to~(\ref{decomp-hij-ast}). The relations~(\ref{ast-add})
are, however, fulfilled identically for \textsc{flrw} universes.
This can easily be shown. First, we recall the well-known relation
that the difference of the covariant derivatives~$A^{i\cdots
j}{}_{k\cdots l|p|q}$ and~$A^{i\cdots j}{}_{k\cdots l|q|p}$ of an
arbitrary tensor can be expressed in terms of the curvature and the
tensor itself (Weinberg~\cite{c8}, Chapter~6, Section~5)
\begin{align}\label{eq:commu-Riemann}
   A^{i\cdots j}&{}_{k\cdots l|p|q}-A^{i\cdots j}{}_{k\cdots l|q|p} =
\nonumber\\
  & +\, A^{i\cdots j}{}_{s\cdots l}\,\mbox{$^3\!R^s_{\nul kpq}$}+\cdots
    +A^{i\cdots j}{}_{k\cdots s}\,\mbox{$^3\!R^s_{\nul lpq}$} \nonumber \\
  & -\, A^{s\cdots j}{}_{k\cdots l}\,\mbox{$^3\!R^i_{\nul spq}$}-\cdots
    -A^{i\cdots s}{}_{k\cdots l}\,\mbox{$^3\!R^j_{\nul spq}$},
\end{align}
where~$\mbox{$^3\!R^i_{\nul jkl}$}$ is the Riemann tensor for the
spaces of constant time. At the right-hand side, there is a term
with a plus sign for each lower index and a term with a minus sign
for each upper index.

We apply this identity taking for~$A$ the second rank
tensor~$h^i_\ast{}_j$ to obtain
\begin{equation}\label{eq:pieter1}
   h^m_{\ast{k|i|m}}-h^m_{\ast{k|m|i}} =
      h^m_{\ast s}\,\mbox{$^3\!R^s_{\nul kim}$}-
       h^s_{\ast k}\,\mbox{$^3\!R^m_{\nul sim}$}.
\end{equation}
Now note that $h^m_{\ast{k|m|i}}$ vanishes in view
of~(\ref{decomp-hij-ast}). Next, we take the covariant derivative
of~(\ref{eq:pieter1}) with respect to $x^l$, and contract
with~$\tilde{g}^{kl}$
\begin{equation}\label{eq:commu-h-ast}
   \tilde{g}^{kl}h^m_{\ast{k|i|m|l}} =
      \tilde{g}^{kl}(h^m_{\ast s}\,\mbox{$^3\!R^s_{\nul kim}$}-
       h^s_{\ast k}\,\mbox{$^3\!R^m_{\nul sim}$})_{|l}.
\end{equation}
Next, using the expression which one has for the Riemann tensor of a
maximally symmetric three-space,
\begin{equation}\label{eq:Riemann}
    \mbox{$^3\!R^a_{\nul bcd}$}=k\left(\delta^a{}_c \tilde{g}_{bd}-
         \delta^a{}_d\tilde{g}_{bc}\right),
\end{equation}
(where~$k=0,\pm1$ is the curvature constant) we find
\begin{equation}\label{eq:first-term}
    \tilde{g}^{kl}h^m_{\ast{k|i|m|l}}=0,
\end{equation}
i.e., the first term of~(\ref{ast-add}) vanishes. The second and
third term can similarly be expressed in the curvature
\begin{equation}\label{eq:commu-h-ast-2}
    h^m_{\ast{i|k|m|l}}-h^m_{\ast{i|k|l|m}}=
    h^m_{\ast{s|k}}\,\mbox{$^3\!R^s_{\nul iml}$}+
    h^m_{\ast{i|s}}\,\mbox{$^3\!R^s_{\nul kml}$}-
    h^s_{\ast{i|k}}\,\mbox{$^3\!R^m_{\nul sml}$},
\end{equation}
where the general property~(\ref{eq:commu-Riemann}) has been used.
Upon substituting the Riemann tensor~(\ref{eq:Riemann}) and
contracting with~$\tilde{g}^{kl}$, we then arrive at
\begin{equation}\label{eq:sec-third-term}
    \tilde{g}^{kl}(h^m_{\ast{i|k|m|l}}-h^m_{\ast{i|k|l|m}})=0,
\end{equation}
i.e., the second and third term of~(\ref{ast-add}) together vanish.
Hence, for \textsc{flrw} universes, equation (\ref{ast-add}) is
identically fulfilled. Consequently, the decomposition
(\ref{decomp-hij-ast}) imposes no additional condition on the
irreducible part $h^i_{\ast j}$ of the perturbation $h^i{}_j$.

The three-vector $\vec{u}_\een$ can be uniquely divided according
to~\cite{Stewart}
\begin{equation}
  \vec{u}_{\een} = \vec{u}_{\een\parallel} +
               \vec{u}_{\een\perp},      \label{decomp-u}
\end{equation}
where $\vec{u}_{\een\parallel}$ is the \emph{longitudinal} part of
$\vec{u}_{\een}$, with the properties
\begin{equation}
   \vec{\tilde{\nabla}}\wedge(\vec{u}_{\een\parallel})=0, \quad
   \vec{\tilde{\nabla}}\cdot\vec{u}_\een=
      \vec{\tilde{\nabla}}\cdot\vec{u}_{\een\parallel},
       \label{longitudinal}
\end{equation}
and $\vec{u}_{\een\perp}$ is the \emph{transverse} part of
$\vec{u}_{\een}$, with the properties
\begin{equation}
   \vec{\tilde{\nabla}}\cdot\vec{u}_{\een\perp}=0, \quad
   \vec{\tilde{\nabla}}\wedge\vec{u}_\een=
       \vec{\tilde{\nabla}}\wedge\vec{u}_{\een\perp},
       \label{transversal}
\end{equation}
where the divergence of the vector $\vec{u}_\een$ is defined by [see
(\ref{driediv})]
\begin{equation}
     \vec{\tilde{\nabla}}\cdot\vec{u}_\een \equiv u^k_{\een|k}=\vartheta_\een,
             \label{divergence}
\end{equation}
and the rotation of the vector $\vec{u}_\een$ is defined by
\begin{equation}
   (\vec{\tilde{\nabla}}\wedge\vec{u}_\een)_i \equiv
          \epsilon_i{}^{jk}u_{\een j|k}=\epsilon_i{}^{jk}u_{\een j,k},
     \label{rotation}
\end{equation}
where $\epsilon_i{}^{jk}$ is the Levi-Civita tensor with
$\epsilon_1{}^{23}=+1$. In expression (\ref{rotation}) we could replace
the covariant derivative by the ordinary partial derivative
because of the symmetry of $\Gamma^i{}_{jk}$.

Having decomposed the tensors $h^i{}_j$, $\mbox{$^3\!R^i_{\een j}$}$
and $u^i_\een$ in a scalar~$\parallel$, a vector~$\perp$ and a
tensor part~$\ast$, we can now decompose the set of
equations~(\ref{subeq:basis}) into three independent sets. The
recipe is simple: all we have to do is to append a
sub-index~$\parallel\,$, $\perp$ or~$\ast$ to the relevant tensorial
quantities in equations~(\ref{subeq:basis}). This will be the
subject of the Sections~\ref{tensor}, \ref{vector} and~\ref{scalar}
below.

\subsection{Tensor Perturbations}   \label{tensor}

We will show that tensor perturbations are not coupled to, i.e.,
do not give rise to, density perturbations.
Upon substituting $h^i{}_j=h^i_{\ast j}$ and $\mbox{$^3\!R^i_{\een j}$}=
\mbox{$^3\!R^i_{\een\ast j}$}$ into the perturbation
equations~(\ref{subeq:basis}) and using the
properties~(\ref{decomp-hij-ast}) and~(\ref{Rij-ast}), we find
from equations~(\ref{basis-1}), (\ref{basis-2})
and~(\ref{basis-4})
\begin{equation}\label{nul-ast}
  \varepsilon_\een=0, \quad p_\een=0, \quad n_\een=0, \quad
  \vec{u}_\een=0,
\end{equation}
where we have also used (\ref{perttoes}). With (\ref{nul-ast}),
equations (\ref{basis-5}) and~(\ref{basis-6}) are identically
satisfied. The only surviving equation is~(\ref{basis-3}), which now
reads
\begin{equation}
   \ddot{h}_{\ast j}^i + 3H\dot{h}_{\ast j}^i +
     2\, \mbox{$^3\!R^i_{\een\ast j}$} = 0,  \label{ddhij-tensor}
\end{equation}
where~$\mbox{$^3\!R^i_{\een\ast j}$}$ is given by
(\ref{decomp-Rij-ast}). Using (\ref{fes5}), (\ref{den1a}),
(\ref{decomp-hij-ast}) and~(\ref{nul-ast}) it follows from
(\ref{subeq:gi-en}) that
\begin{equation}\label{eq:nul-tensor}
  \varepsilon^\gi_\een=0, \quad n^\gi_\een=0,
\end{equation}
so that tensor perturbations do not, in first-order, contribute to
physical energy density and particle number density perturbations.
Hence, the equations (\ref{ddhij-tensor}) do not play a role in this
context, where we are interested in energy density and particle
number density perturbations only.

The equations~(\ref{ddhij-tensor}) have a wave equation like form
with an extra term. Therefore, these
tensor perturbations are sometimes called \emph{gravitational
waves}. The extra term~$3H\dot{h}_{\ast}^i{}_j$ in these
equations is due to the expansion of the universe. The six components
$h^i_{\ast j}$ satisfy the four conditions (\ref{decomp-hij-ast}), leaving
us with two independent functions~$h^i_{\ast j}$. They are related to linearly and
circularly polarized waves.

\subsection{Vector Perturbations} \label{vector}

We will show that, just like tensor perturbations, vector
perturbations are not coupled to density perturbations.
Upon replacing $h^i{}_j$ by $h^i_{\perp j}$ and
$\mbox{$^3\!R^i_{\een j}$}$ by $\mbox{$^3\!R^i_{\een\perp j}$}$
in the perturbation equations~(\ref{subeq:basis}), and using the
expressions~(\ref{decomp-hij-perp}) and~(\ref{Rij-perp}), we find
from equation (\ref{basis-1}) and the trace of equation (\ref{basis-3})
\begin{equation}\label{nul-perp}
  \varepsilon_\een=0, \quad p_\een=0, \quad n_\een=0,
\end{equation}
where we have also used (\ref{perttoes}).

Since $h^i_{\perp j}$ is traceless and raising the index with
$g^{ij}_\nul$ in equation (\ref{basis-2}) we get
\begin{equation}\label{basis-2-raise}
  \dot{h}^{kj}_{\perp{|k}}+2Hh^{kj}_{\perp{|k}}=
  2\kappa(\varepsilon_\nul+p_\nul)u^j_{\een},
\end{equation}
where we have used (\ref{def-gammas}) and~(\ref{metricFRW}). We now
calculate the covariant derivative of equations
(\ref{basis-2-raise}) with respect to~$x^j$, and
use~(\ref{eq:hklkl}) to obtain
\begin{equation}\label{div-0}
      \vec{\tilde{\nabla}}\cdot\vec{u}_{\een}=0,
\end{equation}
where we made use of the fact that the time derivative and the
covariant derivative commute. With
(\ref{decomp-u})--(\ref{transversal}) we see that only the
transverse part of~$\vec{u}_\een$, namely $\vec{u}_{\een\perp}$,
plays a role in vector perturbations. From (\ref{decomp-hij-perp})
and~(\ref{nul-perp}) it follows that the equations~(\ref{basis-4})
and~(\ref{basis-6}) are identically satisfied. The only surviving
equations are~(\ref{basis-2}), (\ref{basis-3}) and~(\ref{basis-5}),
which now read
\begin{subequations}
\label{subeq:vector}
\begin{align}
   & \dot{h}^k_{\perp{i|k}} =
         -2\kappa(\varepsilon_\nul + p_\nul) u_{\een\perp i},
           \label{const-vector} \\
   & \ddot{h}_{\perp j}^i + 3H\dot{h}_{\perp j}^i +
     2\, \mbox{$^3\!R^i_{\een\perp j}$} = 0, \label{ddhij-vector} \\
   &  \frac{1}{c}\frac{\dif}{\dif t}
     \Bigl[(\varepsilon_\nul+p_\nul)u^i_{\een\perp}\Bigr] +
     5H(\varepsilon_\nul+p_\nul) u^i_{\een\perp}=0,
            \label{basis-5-perp}
\end{align}
\end{subequations}
where $\mbox{$^3\!R^i_{\een\perp j}$}$ is given by
(\ref{decomp-Rij-perp}).

Using (\ref{fes5}), (\ref{den1a}), (\ref{nul-perp})
and~(\ref{div-0}) we get from (\ref{subeq:gi-en})
\begin{equation}\label{eq:nul-vector}
  \varepsilon^\gi_\een=0, \quad n^\gi_\een=0,
\end{equation}
implying that also vector perturbations do not, in first-order,
contribute to physical energy density and particle number density
perturbations. Hence, the equations~(\ref{subeq:vector}) do not
play a role when we are interested in energy density and particle
number density perturbations, as we are here. Vector perturbations
are also called \emph{vortices}.

Since vector perturbations obey
$\vec{\tilde{\nabla}}\cdot\vec{u}_{\een\perp}=0$, they have two
degrees of freedom. As a consequence, the tensor $h^i_{\perp j}$
has also two degrees of freedom. These degrees of freedom are
related to clockwise and counter-clockwise rotation of matter.

\subsection{Scalar Perturbations}\label{scalar}

Differentiation of equations (\ref{basis-2}) covariantly with respect
to~$x^j$ we obtain
\begin{equation}
    \dot{h}_{\parallel{k|i|j}}^k-\dot{h}_{\parallel{i|k|j}}^k =
         2\kappa(\varepsilon_\nul + p_\nul)
         u_{\een\parallel i|j}.
    \label{eq:feiko1}
\end{equation}
Interchanging~$i$ and~$j$ in this equation, and subtracting the
resulting equation from~(\ref{eq:feiko1}) we get
\begin{equation}
    \dot{h}_{\parallel{i|k|j}}^k - \dot{h}_{\parallel{j|k|i}}^k =
         -2\kappa(\varepsilon_\nul + p_\nul)
         (u_{\een\parallel i|j}-u_{\een\parallel j|i}),
    \label{dR0i2-rot}
\end{equation}
where we have used that
$\dot{h}_{\parallel{k|i|j}}^k=\dot{h}_{\parallel{k|j|i}}^k$.
Using that $\vec{\tilde{\nabla}}\wedge\vec{u}_{\een\parallel}=0$, we
find from (\ref{decomp-hij-par}) that the function~$\zeta$ must obey
the equations
\begin{equation}
  \dot{\zeta}^{|k}{}_{|i|k|j}-\dot{\zeta}^{|k}{}_{|j|k|i}=0.
      \label{eq:rest-zeta}
\end{equation}
These equations are fulfilled identically in \textsc{flrw}
universes. This can be seen as follows. We first rewrite these
equations by interchanging the covariant derivatives in the form
\begin{equation}
   (\dot{\zeta}^{|k}{}_{|i|k|j}-\dot{\zeta}^{|k}{}_{|i|j|k})-
    (\dot{\zeta}^{|k}{}_{|j|k|i}-\dot{\zeta}^{|k}{}_{|j|i|k})+
    (\dot{\zeta}^{|k}{}_{|i|j}-\dot{\zeta}^{|k}{}_{|j|i})_{|k}=0.
     \label{eq:verwissel}
\end{equation}
Next, we use expression (\ref{eq:commu-Riemann}) and substitute the Riemann
tensor~(\ref{eq:Riemann}) into the resulting expression. Using that
$\dot{\zeta}_{|i|j}=\dot{\zeta}_{|j|i}$, we find that the left-hand
sides of the equations (\ref{eq:verwissel}) vanish. As a consequence, the
equations (\ref{eq:rest-zeta}) are identities. Therefore, the
decomposition (\ref{decomp-hij-par}) imposes no additional condition
on the irreducible part $h^i_{\parallel j}$ of the perturbation $h^i{}_j$.

We thus have shown that the system of first-order Einstein equations
(\ref{subeq:basis}) is, for scalar perturbations, equivalent to the system
\begin{subequations}
\label{subeq:scalar}
\begin{alignat}{3}
  \text{Constraints:} \quad &&  &  H\dot{h}_{\parallel k}^k +
         \tfrac{1}{2}\,\mbox{$^3\!R_{\een\parallel}$}=
         -\kappa\varepsilon_\een,  \label{basis-1-scal} \\
 &&  &\dot{h}_{\parallel{k|i}}^k-\dot{h}_{\parallel{i|k}}^k =
          2\kappa(\varepsilon_\nul + p_\nul) u_{\een\parallel i},
         \label{basis-2-scal} \\
 \text{Evolution:} \quad  && &   \ddot{h}_{\parallel j}^i+3H\dot{h}_{\parallel
j}^i+
      \delta^i{}_j H\dot{h}_{\parallel k}^k +
        2\,\mbox{$^3\!R^i_{\een\parallel j}$}=
     -\kappa\delta^i{}_j(\varepsilon_\een-p_\een),  \label{basis-3-scal} \\
   \text{Conservation:} \quad && &\dot{\varepsilon}_\een+3H(\varepsilon_\een
+p_\een)+
    (\varepsilon_\nul +p_\nul)\theta_\een=0,  \label{basis-4-scal}  \\
 &&  &\frac{1}{c}\frac{\dif}{\dif t}
   \Bigl[(\varepsilon_\nul+p_\nul) u^i_{\een\parallel}\Bigr]-
    g^{ik}_\nul
    p_{\een|k}+5H(\varepsilon_\nul+p_\nul)u^i_{\een\parallel}=0,
            \label{basis-5-scal} \\
 &&   & \dot{n}_\een+3Hn_\een+n_\nul\theta_\een = 0, 
\label{basis-6-scal}
\end{alignat}
\end{subequations}
where the local perturbations to the expansion, metric and the Ricci tensor are given
by (\ref{fes5}), (\ref{decomp-hij-par}) and (\ref{decomp-Rij-par})
respectively. In the tensorial and vectorial case we found
$\varepsilon_\een=0$ and $n_\een=0$, implying that
$\varepsilon^\gi_\een=0$ and $n^\gi_\een=0$, which made the
tensorial and vectorial equations irrelevant for our purpose. Such a
conclusion cannot be drawn from the equations~(\ref{subeq:scalar}).
Perturbations with $\varepsilon_\een\neq0$ and $n_\een\neq0$ are
usually referred to as scalar perturbations.

Since the perturbation equations (\ref{subeq:scalar}) contain only the
components~$h^i_{\parallel j}$, it follows that
\emph{relativistic} energy density and particle number density
perturbations are characterized by \emph{two} potentials, $\phi$
and~$\zeta$. In the next section we will rewrite the system (\ref{subeq:scalar})
into a form which is suitable to study the evolution of the quantities
(\ref{subeq:gi-en}).

\section{First-order Equations for Scalar Perturbations}
\label{evo-scal}

The system of equations (\ref{subeq:scalar}) can be further simplified by taking
into account the decomposition (\ref{decomp-u}).
In the foregoing section we have shown that only the longitudinal part
$\vec{u}_{\een\parallel}$ of the three-vector $\vec{u}_\een$ is coupled to
density perturbations. Using the properties (\ref{decomp-u})--(\ref{divergence})
we can rewrite the scalar perturbation equations~(\ref{subeq:scalar}) in terms
of quantities $\theta_\een$, $\mbox{$^3\!R_{\een\parallel}$}$, $\vartheta_\een$,
$\varepsilon_\een$ and $n_\een$, which are suitable to describe
exclusively the scalar perturbations. The result is that the first-order
quantities occurring in the definitions~(\ref{subeq:gi-en}) do explicitly occur
in the set of equations. A second important result is that the metric components
$h^i_{\parallel j}$ occur, in the resulting equations, only in
$\mbox{$^3\!R_{\een\parallel}$}$. An evolution equation for this quantity
follows from the $(0,i)$ perturbed constraint equations and will be derived
below.

We now successively simplify all equations of the set
(\ref{subeq:scalar}) by replacing $\vec{u}_{\een\parallel}$ by its
divergence $\vartheta_\een$, (\ref{den1a}), and eliminating
$\dot{h}^k_{\parallel k}$ with the help of (\ref{fes5}) and using that
$\dot{h}^k{}_k\equiv\dot{h}^k_{\parallel k}$, as follows from
(\ref{decomp-symh}) and (\ref{subeq:dec-hij}). For equation
(\ref{basis-1-scal}) we find
\begin{equation}
  2H(\theta_\een-\vartheta_\een)-\tfrac{1}{2}\,\mbox{$^3\!R_{\een\parallel}$}=
       \kappa\varepsilon_\een. \label{theta1}
\end{equation}
Thus the $(0,0)$-component of the constraint equations becomes an
algebraic equation which relates the first-order quantities
$\theta_\een$, $\vartheta_\een$, $\mbox{$^3\!R_{\een\parallel}$}$
and $\varepsilon_\een$. In (\ref{theta1}), $\mbox{$^3\!R_{\een\parallel}$}$ is
given by
\begin{equation}\label{eq:driekrom-scalar}
    \mbox{$^3\!R_{\een\parallel}$} =
   g_\nul^{ij} (h^k_{\parallel k|i|j}-h^k_{\parallel i|j|k}) +
     \tfrac{1}{3}\,\mbox{$^3\!R_\nul$} h^k_{\parallel k},
\end{equation}
as follows from (\ref{driekrom}) with
(\ref{subeq:dec-hij})--(\ref{subeq:prop-Rij}) and (\ref{eq:hklkl}).

We will now derive an evolution equation for \mbox{$^3\!R_{\een\parallel}$} from equations 
(\ref{basis-2-scal}). Firstly,
multiplying both sides of equations (\ref{basis-2-scal}) by~$g^{ij}_\nul$ and taking the covariant
divergence with respect to the index~$j$ we find
\begin{equation}
  g_\nul^{ij} (\dot{h}^k_{\parallel k|i|j}-\dot{h}^k_{\parallel i|k|j}) =
    2\kappa(\varepsilon_\nul+p_\nul) \vartheta_\een,  \label{dR0i4}
\end{equation}
where we have also used (\ref{den1a}). The left-hand side will turn
up as a part of the time derivative of the curvature
$\mbox{$^3\!R_{\een\parallel}$}$. In fact, differentiating
(\ref{eq:driekrom-scalar}) with respect to~$ct$ and recalling that
the connection coefficients $\Gamma^k_{\nul ij}$, (\ref{con3FRW}),
are independent of time, one gets
\begin{equation}
  \mbox{$^3\!\dot{R}_{\een\parallel}$} =
-2H\,\mbox{$^3\!R_{\een\parallel}$} +
   g_\nul^{ij} (\dot{h}^k_{\parallel k|i|j}-
   \dot{h}^k_{\parallel i|k|j})+
     \tfrac{1}{3}\,\mbox{$^3\!R_\nul$} \dot{h}^k_{\parallel k},
   \label{dR0i6}
\end{equation}
where we have used (\ref{def-gammas}), (\ref{metricFRW}),
(\ref{momback}) and
\begin{equation}
    g_\nul^{ij}h^k_{\parallel i|j|k} =
    g_\nul^{ij}h^k_{\parallel i|k|j},
\end{equation}
which is a consequence of $g^{ij}_{\nul|k}=0$ and the symmetry of
$h_\parallel^{ij}$. Next, combining equation (\ref{dR0i4}) with
(\ref{dR0i6}), and, finally, eliminating $\dot{h}^k_{\parallel k}$
with the help of (\ref{fes5}), one arrives at
\begin{equation}
 \mbox{$^3\!\dot{R}_{\een\parallel}$}+
        2H\,\mbox{$^3\!R_{\een\parallel}$}-
   2\kappa(\varepsilon_\nul + p_\nul)\vartheta_\een+
     \tfrac{2}{3}\,\mbox{$^3\!R_\nul$}(\theta_\een-\vartheta_\een)=0.
     \label{mompert}
\end{equation}
In this way we managed to recast the three $(0,i)$-components of
the constraint equations in the form of one ordinary differential
equation for the local perturbation,
$\mbox{$^3\!R_{\een\parallel}$}$, to the spatial curvature.

We now consider the dynamical equations~(\ref{basis-3-scal}). Taking
the trace of these equations and using (\ref{fes5}) to eliminate the
quantity~$\dot{h}^k_{\parallel k}$, we arrive at
\begin{equation}
   \dot{\theta}_\een-\dot{\vartheta}_\een +
      6H(\theta_\een-\vartheta_\een)-\mbox{$^3\!R_{\een\parallel}$}=
       \tfrac{3}{2}\kappa(\varepsilon_\een-p_\een).
        \label{eq:168a}
\end{equation}
Eliminating the second term with the help of the constraint equation
(\ref{theta1}) yields
\begin{equation}
  \label{eq:168a-kort}
  \dot{\theta}_\een-\dot{\vartheta}_\een + \tfrac{1}{2}\,\mbox{$^3\!R_{\een\parallel}$}=
       -\tfrac{3}{2}\kappa(\varepsilon_\een+p_\een).
\end{equation}
Thus, for scalar perturbations, the three dynamical Einstein equations
(\ref{basis-3-scal}) with $i=j$ reduce to one ordinary differential
equation for the difference $\theta_\een-\vartheta_\een$. For $i\neq
j$ the dynamical Einstein equations are not coupled to scalar
perturbations. Therefore, these equations need not be considered.

Taking the covariant derivative of equation (\ref{basis-5-scal})
with respect to the metric $g_{\nul ij}$ and using (\ref{den1a}), we
get
\begin{equation}
  \frac{1}{c}\frac{\dif}{\dif t}
     \Bigl[(\varepsilon_\nul+p_\nul)\vartheta_\een\Bigr]-
    g^{ik}_\nul p_{\een|k|i}+5H(\varepsilon_\nul+p_\nul)\vartheta_\een = 0,  
\label{mom2}
\end{equation}
where we have used that the operations of taking the time derivative
and the covariant derivative commute, since the connection
coefficients $\Gamma^k_{\nul ij}$, (\ref{con3FRW}), are independent
of time. With equation (\ref{energyFRW}), we can rewrite equation (\ref{mom2})
in the form
\begin{equation}
  \dot{\vartheta}_\een +
 H\left(2-3\frac{\dot{p}_\nul}{\dot{\varepsilon}_\nul}\right)\vartheta_\een+
     \frac{1}{\varepsilon_\nul+p_\nul}\dfrac{\tilde{\nabla}^2 p_\een}{a^2}=0,
                \label{mom3}
\end{equation}
where $\tilde{\nabla}^2$ is the generalized Laplace operator
which, for an arbitrary function $f(t,\vec{x})$ and with respect
to an arbitrary three-dimensional metric
$\tilde{g}^{ij}(\vec{x})$, is defined by
\begin{equation}\label{Laplace}
 \tilde{\nabla}^2 f \equiv \tilde{g}^{ij} f_{|i|j}=
\dfrac{1}{\displaystyle\sqrt{\det\tilde{g}}}\dfrac{\partial}{\partial x^i}
\left(\tilde{g}^{ij}\sqrt{\det\tilde{g}}\;\dfrac{\partial f}{\partial x^j}  \right).
\end{equation}
Thus, the three first-order momentum conservation laws
(\ref{basis-5-scal}) reduce to one ordinary differential equation
for the divergence~$\vartheta_\een$.

Finally, the conservation laws~(\ref{basis-4-scal})
and~(\ref{basis-6-scal}) are already written in a suitable form. This
concludes the derivation of the first-order equations for scalar perturbations.

We thus have shown that the system of equations (\ref{subeq:scalar}) is
equivalent to the system
\begin{subequations}
\label{subeq:pertub-flrw-sum}
\begin{alignat}{3}
 \text{Constraint:} \quad  &&  &2H(\theta_\een-\vartheta_\een)-
       \tfrac{1}{2}\,\mbox{$^3\!R_{\een\parallel}$} = \kappa\varepsilon_\een,
\label{con-sp-1-sum} \\
 \text{Evolution:} \quad &&  &\mbox{$^3\!\dot{R}_{\een\parallel}$}+
     2H\,\mbox{$^3\!R_{\een\parallel}$}-
    2\kappa \varepsilon_\nul(1 + w)\vartheta_\een
       +\tfrac{2}{3}\,\mbox{$^3\!R_\nul$}(\theta_\een-\vartheta_\een)=0,
               \label{FRW6-sum} \\
&&  &\dot{\theta}_\een-\dot{\vartheta}_\een + \tfrac{1}{2}\,\mbox{$^3\!R_{\een\parallel}$}=
       -\tfrac{3}{2}\kappa(\varepsilon_\een+p_\een),
        \label{eq:168a-sum} \\
 \text{Conservation:} \quad  &&  &\dot{\varepsilon}_\een + 3H(\varepsilon_\een +
p_\een)+
         \varepsilon_\nul(1 + w)\theta_\een=0,  \label{FRW4-sum} \\
&& &\dot{\vartheta}_\een+H(2-3\beta^2)\vartheta_\een+
   \frac{1}{\varepsilon_\nul(1+w)}\dfrac{\tilde{\nabla}^2p_\een}{a^2}=0, 
\label{FRW5-sum}\\
&&  &\dot{n}_\een + 3H n_\een +
         n_\nul\theta_\een=0. \label{FRW4a-sum}
\end{alignat}
\end{subequations}
The quantities $\beta(t)$ and $w(t)$ occurring in equations
(\ref{subeq:pertub-flrw-sum}) are defined by
\begin{equation}
  \beta(t) \equiv
  \sqrt{\dfrac{\dot{p}_\nul(t)}{\dot{\varepsilon}_\nul(t)}},
   \quad w(t)\equiv \dfrac{p_\nul(t)}{\varepsilon_\nul(t)}.
          \label{begam2}
\end{equation}

The algebraic equation (\ref{con-sp-1-sum}) and the five \emph{ordinary}
differential equations (\ref{FRW6-sum})--(\ref{FRW4a-sum}), is a
system of six equations for the five quantities $\theta_\een$,
$\mbox{$^3\!R_\een$}$, $\vartheta_\een$, $\varepsilon_\een$ and
$n_\een$ respectively. This system is, however, not over-determined,
as we will now show. Differentiating the constraint equation
(\ref{con-sp-1-sum}) with respect to time yields
\begin{equation}
\label{eq:con-sp-1-sum-time}
2\dot{H}(\theta_\een-\vartheta_\een)+2H(\dot{\theta}_\een-\dot{\vartheta}_\een)-
 \tfrac{1}{2}\,\mbox{$^3\!\dot{R}_{\een\parallel}$}=\kappa\dot{\varepsilon}_\een.
\end{equation}
Eliminating the time derivatives $\dot{H}$,
$\mbox{$^3\!\dot{R}_{\een\parallel}$}$ and $\dot{\varepsilon}_\een$
with the help of (\ref{eq:dyn0-flrw}), (\ref{FRW6-sum}) and
(\ref{FRW4-sum}), respectively, yields the dynamical equation
(\ref{eq:168a-sum}).  Consequently, the general solution of the system
(\ref{con-sp-1-sum})--(\ref{FRW6-sum}) and
(\ref{FRW4-sum})--(\ref{FRW4a-sum}) is also a solution of the
dynamical equation (\ref{eq:168a-sum}). Therefore, it is not needed to
consider equation (\ref{eq:168a-sum}) anymore.

\section{Summary of the Basic Equations}  \label{resume}

In this section we will show that there exist unique gauge-invariant density
perturbations (\ref{subeq:gi-en}). To that end we summarize the background and
first-order equations for scalar perturbations.

\subsection{Zeroth-order Equations} \label{zoe}

The Einstein equations and conservation laws for the background
\textsc{flrw} universe are given by (\ref{fes2}), (\ref{endencon0}),
(\ref{momback}) and (\ref{energyFRW})--(\ref{deel1}):
\begin{subequations}
\label{subeq:einstein-flrw}
\begin{alignat}{2}
\text{Constraint:} \quad && 3H^2 &=
\tfrac{1}{2}\,\mbox{$^3\!R_\nul$}+\kappa\varepsilon_\nul+\Lambda,\label{FRW3}\\
\text{Evolution:} \quad && \mbox{$^3\!\dot{R}_\nul$} & =
-2H\,\mbox{$^3\!R_\nul$}, \label{FRW3a}\\
\text{Conservation:} \quad &&
        \dot{\varepsilon}_\nul & = -3H\varepsilon_\nul(1+w), \label{FRW2} \\
   && \vartheta_\nul & =0, \\
   &&    \dot{n}_\nul & = -3Hn_\nul,   \label{FRW2a}
\end{alignat}
\end{subequations}
where the initial value for (\ref{FRW3a}) is given by (\ref{eq:init-R0}). As we
have shown in Section~\ref{nulde-conservation}, the dynamical Einstein equation
(\ref{dyn0}) is not needed.
The set (\ref{subeq:einstein-flrw}) consists of one algebraic and three
differential equations with respect to time for the four unknown quantities
$\varepsilon_\nul$, $n_\nul$, $\theta_\nul=3H$ and $\mbox{$^3\!R_\nul$}$. The
pressure $p_\nul(t)$ is related to the energy density $\varepsilon_\nul(t)$ and
the particle number density $n_\nul(t)$ via the equation of
state~(\ref{toestandback}).

We rewrite the Friedmann equation (\ref{FRW3}) by dividing
both sides by $3H^2$ in the form
\begin{equation}\label{eq:Friedmann-Omega}
    1 = \Omega_\mathrm{curv} + \Omega_\mathrm{bar} + \Omega_\mathrm{cdm}+
\Omega_\mathrm{rad} + \Omega_\Lambda,
\end{equation}
where $\Omega_\mathrm{curv}$, $\Omega_\mathrm{bar}$, $\Omega_\mathrm{cdm}$,
$\Omega_\mathrm{rad}$ and $\Omega_\Lambda$ are the contributions due to
curvature; baryonic (ordinary) matter; \textsc{cdm}; radiation and dark
energy respectively. The time dependence of these contributions is given by
\begin{equation}\label{eq:Omegas}
    \Omega_\mathrm{curv}(t) \equiv -\dfrac{k}{a^2H^2}, \quad
    \Omega_\mathrm{bar}(t) + \Omega_\mathrm{cdm}(t) + \Omega_\mathrm{rad}(t)\equiv
\dfrac{\kappa\varepsilon_\nul}{3H^2}, \quad \Omega_\Lambda(t)\equiv
\dfrac{\Lambda}{3H^2},
\end{equation}
where we have used (\ref{spRicci}). This enables us to link the
observations made with the \textsc{wmap} satellite to our treatise on
density perturbations.  The present day values of the quantities in
(\ref{eq:Friedmann-Omega}) are, for a $\Lambda\mathrm{CDM}$ universe
(also referred to as the \emph{concordance
  model})~\cite{2009ApJS..180..306D,Spergel:2003cb,komatsu-2008,hinshaw-2008,2010arXiv1001.4538K},
given by
\begin{equation}\label{eq:Om-La-H}
    \Omega_\mathrm{bar}(t_\mathrm{p})=0.0441, \quad
\Omega_\mathrm{cdm}(t_\mathrm{p})=0.214, \quad
\Omega_\mathrm{rad}(t_\mathrm{p})=0, \quad \Omega_\Lambda(t_\mathrm{p})=0.742.
\end{equation}
With (\ref{eq:Friedmann-Omega}) we get
\begin{equation}\label{eq:Omega-C}
    \Omega_\mathrm{curv}(t_\mathrm{p})=0.000.
\end{equation}
Thus, \textsc{wmap}-observations indicate that the universe is flat. Moreover,
it follows from \textsc{wmap}-observations that the present value of the Hubble
function (\ref{eq:Hubble-function}) is
\begin{equation}
  \label{eq:Hubble-WMAP}
   \mathcal{H}(t_{\mathrm{p}})=71.9\,\mathrm{km/s/Mpc}.
\end{equation}
Using that $1\,\mathrm{Mpc}=3.0857\times10^{22}\,\mathrm{m}$, we get for the curvature
parameter $k$ and the cosmological constant $\Lambda$, using the observed
values (\ref{eq:Om-La-H})--(\ref{eq:Hubble-WMAP}),
\begin{equation}\label{eq:Lambda-value}
     k=0, \quad \Lambda =1.34\times10^{-52}\,\mathrm{m}^{-2},
\end{equation}
respectively. The cosmological constant represents \emph{dark energy}, also known as
\emph{quintessence}, a constant energy
density filling space homogeneously. The existence of dark energy is postulated
in order to explain recent observations that today the universe appears to be
expanding at an \emph{accelerating} rate. Since accelerated expansion takes
place only at late times, we do not take into account $\Lambda$ in our
calculations of star formation in Sections~\ref{sec:an-exam}--\ref{sec:star-for-res}\@.

In astrophysics and cosmology, \textsc{cdm} is hypothetical matter of
unknown composition that does not emit or reflect enough
electromagnetic radiation to be observed directly, but whose presence
can be inferred from gravitational effects on visible matter. One of
the main candidates for \textsc{cdm} are the so-called \textsc{wimp}s
(Weakly Interacting Massive Particles) with a mass of approximately
$10$--$10^3\,\mathrm{GeV}/c^2$. A recent
estimate~\cite{collaboration-2009} yields a \textsc{wimp} mass of
approximately $70\,\mathrm{GeV}/c^2$.  In comparison, the mass of a
proton is $0.938\,\mathrm{GeV}/c^2$. Because of their large mass,
\textsc{wimp}s move relatively slow and are therefore cold. Since
\textsc{wimp}s do interact only through \textit{weak nuclear force}
with a range of approximately $10^{-17}\,\mathrm{m}$, they are dark
and, as a consequence, invisible through electromagnetic
observations. The only perceptible interaction with ordinary matter is
through gravity. In the literature, \textsc{cdm} is therefore treated
as `dust,' i.e., a substance which interacts only through gravity with
itself and ordinary matter.

\subsection{First-order Equations}\label{foe}

Since the evolution equation (\ref{eq:168a-sum}) is not needed, as we have shown
at the end of Section~\ref{evo-scal}, the first-order equations describing
density perturbations are given by the set of one algebraic equation and four
differential equations~(\ref{con-sp-1-sum})--(\ref{FRW6-sum}) and
(\ref{FRW4-sum})--(\ref{FRW4a-sum}):
\begin{subequations}
\label{subeq:pertub-flrw}
\begin{alignat}{3}
 \text{Constraint:} \quad  &&  &2H(\theta_\een-\vartheta_\een)-
       \tfrac{1}{2}\,\mbox{$^3\!R_{\een\parallel}$} = \kappa\varepsilon_\een,
\label{con-sp-1} \\
 \text{Evolution:} \quad &&  &\mbox{$^3\!\dot{R}_{\een\parallel}$}+
     2H\,\mbox{$^3\!R_{\een\parallel}$}-
    2\kappa \varepsilon_\nul(1 + w)\vartheta_\een
       +\tfrac{2}{3}\,\mbox{$^3\!R_\nul$}(\theta_\een-\vartheta_\een)=0,
               \label{FRW6} \\
 \text{Conservation:} \quad  &&  &\dot{\varepsilon}_\een + 3H(\varepsilon_\een +
p_\een)+
         \varepsilon_\nul(1 + w)\theta_\een=0,  \label{FRW4} \\
&& &\dot{\vartheta}_\een+H(2-3\beta^2)\vartheta_\een+
   \frac{1}{\varepsilon_\nul(1+w)}\dfrac{\tilde{\nabla}^2p_\een}{a^2}=0, 
\label{FRW5}\\
&&  &\dot{n}_\een + 3H n_\een +
         n_\nul\theta_\een=0, \label{FRW4a}
\end{alignat}
\end{subequations}
for the five unknown functions $\varepsilon_\een$, $n_\een$,
$\vartheta_\een$, $\mbox{$^3\!R_{\een\parallel}$}$
and $\theta_\een$ respectively. The first-order perturbation to the pressure is
given by the perturbed equation of state (\ref{perttoes}).

From their derivation it follows that the set of equations
(\ref{subeq:pertub-flrw}) is equivalent to the system of equations
(\ref{subeq:scalar}). The metric is in the set
(\ref{subeq:pertub-flrw}) contained in only one quantity, namely the
local first-order perturbation $\mbox{$^3\!R_{\een\parallel}$}$ to the
global spatial curvature. Using the sets of equations
(\ref{subeq:einstein-flrw}) and (\ref{subeq:pertub-flrw}), we show in
Section~\ref{nrl} that our perturbation theory yields the Newtonian
theory of gravity in the non-relativistic limit of an expanding universe
with (\ref{subeq:gi-en}) as the key quantities.

\subsection{Unique Gauge-invariant Density Perturbations} \label{sec:unique}

The background equations (\ref{subeq:einstein-flrw}) and first-order equations
(\ref{subeq:pertub-flrw}) are now rewritten in such a form that we can draw an
important conclusion.
Firstly, we observe that equations (\ref{subeq:pertub-flrw}) are the first-order
counterparts of the background equations (\ref{subeq:einstein-flrw}). Combined,
these two sets of equations describe the background quantities and their
corresponding first-order quantities:
\begin{equation}\label{eq:nulde-eerste}
    (\varepsilon_\nul,\,\varepsilon_\een), \quad (n_\nul,\,n_\een), \quad
(\vartheta_\nul=0,\,\vartheta_\een), \quad
(\mbox{$^3\!R_\nul$},\,\mbox{$^3\!R_{\een\parallel}$}), \quad
(\theta_\nul=3H,\,\theta_\een).
\end{equation}
Secondly, we remark that the quantities $\vartheta$ and
$\mbox{$^3\!R$}$ are not scalars of space-time, since their
first-order perturbations $\vartheta_\een$ and
$\mbox{$^3\!R_{\een\parallel}$}$ transform according to
(\ref{theta-ijk}) and (\ref{drie-ijk}) respectively. In contrast, the
first-order perturbations $\varepsilon_\een$, $n_\een$ and
$\theta_\een$ of the scalars (\ref{subeq:ent}) transform according to
(\ref{sigmahat3}).  Thus, \emph{only three independent scalars}
$\varepsilon$, $n$ and $\theta$, (\ref{subeq:ent}), play a role in a
density perturbation theory. Consequently, the only non-trivial
gauge-invariant combinations which can be constructed from these
scalars and their first-order perturbations are the combinations
(\ref{subeq:gi-en}). We thus have shown that there exist \emph{unique}
gauge-invariant quantities $\varepsilon^\gi_\een$ and $n^\gi_\een$. In
Section~\ref{nrl} we show that a perturbation theory based on these
quantities yields the Newtonian theory of gravity in the
non-relativistic limit of an expanding universe.

By switching from the variables $\varepsilon_\een$,
$n_\een$, $\vartheta_\een$, $\mbox{$^3\!R_{\een\parallel}$}$
and~$\theta_\een$ to the variables~$\varepsilon_\een^\gi$
and~$n_\een^\gi$, we will arrive at a set of equations
for~$\varepsilon_\een^\gi$ and~$n_\een^\gi$ with a unique, i.e.,
gauge-invariant solution. This will be the subject of
Section~\ref{thirdstep}. First, we derive some auxiliary
expressions related to the entropy, pressure and temperature.

\section{Entropy, Pressure, Temperature and Diabatic Perturbations}
\label{sec:pertub-therm}

In order to study the evolution of density
perturbations, we need the laws of thermodynamics. In this section we
rewrite the combined First and Second Laws of thermodynamics in terms
of the gauge-invariant quantities (\ref{subeq:gi-en}) and the
gauge-invariant entropy per particle $s^\gi_\een$.

\subsection{Gauge-invariant Entropy Perturbations}  \label{sec:eqs-sgi}

Consider a gas of $N$ particles with volume $V$. Let $\mu$ be the
thermodynamic ---or chemical--- potential, $p$ the pressure and $S$
its entropy. Then the internal energy $E$ is given by the relation
\begin{equation}
  \label{eq:internal-energy}
  E = TS -pV + \mu N.
\end{equation}
In terms of the energy per particle $e\equiv E/N$, the entropy per
particle $s\equiv S/N$ and the particle number density $n\equiv N/V$
this relation reads
\begin{equation}
  \label{eq:energy-per-particle}
  e = Ts - pn^{-1} + \mu,
\end{equation}
implying that
\begin{equation}
  \label{eq:de-per-part}
  \dif e = T\dif s + s \dif T-n^{-1}\dif p - p\dif n^{-1}+\dif\mu.
\end{equation}
From (\ref{eq:internal-energy}) we also find
\begin{equation}
  \label{eq:dE}
  \dif E= T\dif S+S\dif T-V\dif p- p\dif V+\mu\dif N+N\dif\mu.
\end{equation}
The combined First and Second Laws of thermodynamics reads
\begin{equation}\label{eq:sec-law-thermo}
    \dif E = T\dif S - p\dif V + \mu\dif N.
\end{equation}
From  (\ref{eq:dE}) and (\ref{eq:sec-law-thermo}) we find after
division by $N$
\begin{equation}
  \label{eq:Sdt}
  s\dif T-n^{-1}\dif p + \dif\mu=0.
\end{equation}
This relation enables us to eliminate $\dif\mu$ from
(\ref{eq:de-per-part}). We so find
\begin{equation}
  \label{eq:Tds-klein}
  T\dif s = \dif e + p \dif n^{-1}.
\end{equation}
This thermodynamic relation is independent of $N$ and $\mu$. In terms
of the energy density defined as $\varepsilon\equiv ne$, we so find,
finally,
\begin{equation}\label{eq:TdS}
  T\dif s = \dif \biggl(\dfrac{\varepsilon}{n}\biggr) +
         p\dif \biggl(\dfrac{1}{n}\biggr).
\end{equation}
This is the relation we shall use in the following.

The thermodynamic relation~(\ref{eq:TdS}) is true for a system in
thermodynamic equilibrium. For a non-equilibrium system that is `not
too far' from equilibrium, the equation~(\ref{eq:TdS}) may be replaced
by
\begin{equation}\label{eq:sec-law-noneq}
    T \dfrac{\dif s}{\dif t}=\dfrac{\dif}{\dif
t}\biggl(\dfrac{\varepsilon}{n}\biggr)+
p\dfrac{\dif}{\dif t}\biggl(\dfrac{1}{n}\biggr),
\end{equation}
where $\dif/\dif t$ is the time derivative in a local co-moving
Lorentz system. Now, using
$\varepsilon=\varepsilon_\nul+\varepsilon_\een$,
$s=s_\nul+s_\een$, $p=p_\nul+p_\een$ and $n=n_\nul+n_\een$, we
find from equation (\ref{eq:sec-law-noneq})
\begin{equation}\label{eq:TdS-back}
 T_\nul\dfrac{\dif s_\nul}{\dif t} =
    \dfrac{\dif}{\dif t} \biggl(\dfrac{\varepsilon_\nul}{n_\nul}\biggr)
+p_\nul\dfrac{\dif}{\dif t} \biggl(\dfrac{1}{n_\nul}\biggr),
\end{equation}
where we neglected time derivatives of first-order quantities.
With the help of equations (\ref{FRW2}), (\ref{FRW2a})
and~(\ref{begam2}) we find that the right-hand side of
equation (\ref{eq:TdS-back}) vanishes. Hence, ${\dot{s}_\nul=0}$,
implying that, in zeroth-order, the expansion takes place without
generating entropy: $s_\nul$ is constant in time. Hence, in view of
(\ref{sigmahat3}), which is valid for any scalar, and
$\dot{s}_\nul=0$, the first-order perturbation~$s_\een$ is
automatically a gauge-invariant quantity, i.e.,
$\hat{s}_\een=s_\een$, in contrast to~$\varepsilon_\een$
and~$n_\een$, which had to be redefined according to expressions
(\ref{subeq:gi-en}). Apparently, the entropy per particle~$s_\een$
is such a combination of~$\varepsilon_\een$ and~$n_\een$ that it
need not be redefined. This can be made explicit by noting that in
the linear approximation we are considering, the combined First and Second Laws of
thermodynamics~(\ref{eq:TdS}) should hold for zeroth-order and first-order
quantities separately. In particular, equation (\ref{eq:TdS}) implies
\begin{equation}\label{eq:TdS-1}
    T_\nul s_\een=\dfrac{1}{n_\nul}\left(\varepsilon_\een-
      \dfrac{\varepsilon_\nul+p_\nul}{n_\nul}n_\een\right),
\end{equation}
where we neglected products of differentials and first-order
quantities, and where we replaced~$\dif\varepsilon$ and~$\dif n$
by~$\varepsilon_\een$ and~$n_\een$ respectively. We now note that
the linear combination in the right-hand side of
equation (\ref{eq:TdS-1}) has the property
\begin{equation}\label{eq:lin-gi}
  \varepsilon_\een-\dfrac{\varepsilon_\nul+p_\nul}{n_\nul}n_\een=
  \varepsilon^\gi_\een-\dfrac{\varepsilon_\nul+p_\nul}{n_\nul}n^\gi_\een,
\end{equation}
as may immediately be verified with the help of (\ref{subeq:gi-en})
and the equations (\ref{FRW2}) and~(\ref{FRW2a}). The right-hand
side of expression (\ref{eq:lin-gi}) being gauge-invariant, the
left-hand side must be gauge-invariant. This observation makes
explicit the gauge-invariance of the first-order approximation to
the entropy per particle, $s_\een$. In order to stress the gauge-invariance
of the correction~$s_\een$ to the (constant) entropy per
particle, $s_\nul$, we will write~$s^\gi_\een$, rather
than~$s_\een$.
From (\ref{eq:TdS-1}), (\ref{eq:lin-gi}) and ${s^\gi_\een \equiv s_\een}$ we then get
\begin{equation}\label{eq:TdS-1-gi}
  T_\nul s^\gi_\een = \dfrac{1}{n_\nul}\left(
     \varepsilon^\gi_\een-\frac{\varepsilon_\nul(1+w)}{n_\nul}n^\gi_\een\right),
\end{equation}
where $w$ is the quotient of zeroth-order pressure and zeroth-order
energy density defined by (\ref{begam2}). With (\ref{eq:TdS-1-gi}) we
have rewritten the
thermodynamic law (\ref{eq:sec-law-thermo}) in terms of the quantities
$\varepsilon^\gi_\een$ and $n^\gi_\een$.

We rewrite equation (\ref{eq:TdS-1-gi}) in the form
\begin{equation}\label{eq:hulp-entropy}
   T_\nul s^\gi_\een =
      -\dfrac{\varepsilon_\nul(1+w)}{n_\nul^2}\sigma^\gi_\een,
\end{equation}
where the gauge-invariant, entropy related
quantity~$\sigma^\gi_\een$ is given by
\begin{equation}\label{eq:entropy-gi}
  \sigma_\een^\gi \equiv n_\een^\gi -
  \frac{n_\nul}{\varepsilon_\nul(1+w)}\varepsilon_\een^\gi.
\end{equation}
The quantity~$\sigma^\gi_\een$ occurs as the source term in the
evolution equations~(\ref{subeq:eerste}) below.

\subsection{Gauge-invariant Pressure Perturbations}

We will now derive a gauge-invariant expression for the physical
pressure perturbations. To that end, we first calculate the time
derivative of the background pressure. From the equation of state
(\ref{toestandback}) we have
\begin{equation}\label{eq:pnul-t}
  \dot{p}_\nul=p_n\dot{n}_\nul +p_\varepsilon
    \dot{\varepsilon}_\nul,
\end{equation}
where $p_\varepsilon$ and $p_n$ are the partial derivatives given by
expressions (\ref{perttoes1}) and~(\ref{eq:pn-pe-back}). Multiplying
both sides of this expression by $\theta_\een/\dot{\theta}_\nul$ and
subtracting the result from (\ref{perttoes}) we get
\begin{equation}\label{eq:press-gi}
    p_\een-\dfrac{\dot{p}_\nul}{\dot{\theta}_\nul}\theta_\een=
    p_n n^\gi_\een + p_\varepsilon \varepsilon^\gi_\een,
\end{equation}
where we have used (\ref{subeq:gi-en}) to rewrite the right-hand
side. Since~$p_n$ and~$p_\varepsilon$ depend on the background
quantities~$\varepsilon_\nul$ and~$n_\nul$ only, the right-hand side
is gauge-invariant. Hence, the quantity~$p^\gi_\een$ defined by
\begin{equation}\label{eq:pgi}
  p^\gi_\een \equiv p_\een -
  \dfrac{\dot{p}_\nul}{\dot{\theta}\nul}\theta_\een,
\end{equation}
is gauge-invariant. We thus obtain the gauge-invariant counterpart
of (\ref{perttoes})
\begin{equation}\label{eq:pgi-2}
  p^\gi_\een=p_\varepsilon\varepsilon_\een^\gi+p_n n_\een^\gi.
\end{equation}
We will now rewrite this expression in a slightly different form.
From (\ref{begam2}) and~(\ref{eq:pnul-t}) we obtain
$\beta^2=p_\varepsilon+p_n(\dot{n}_\nul/\dot{\varepsilon}_\nul)$.
Using equations (\ref{FRW2}) and~(\ref{FRW2a}) we find
\begin{equation}\label{eq:begam3}
  \beta^2 = p_\varepsilon+\frac{n_\nul p_n}{\varepsilon_\nul(1+w)}.
\end{equation}
With this expression and (\ref{eq:entropy-gi}) and~(\ref{eq:begam3})
we can rewrite the pressure perturbation (\ref{eq:pgi-2}) as
\begin{equation}\label{eq:pgi-3}
   p^\gi_\een=\beta^2 \varepsilon^\gi_\een + p_n \sigma^\gi_\een.
\end{equation}
We thus have expressed the pressure perturbation $p_\een^\gi$ in
terms of the energy density perturbation~$\varepsilon^\gi_\een$ and
the entropy related quantity~$\sigma^\gi_\een$ rather
than~$\varepsilon^\gi_\een$ and the particle number density
perturbation~$n^\gi_\een$.

Expression (\ref{eq:pgi-3}) can be rewritten into
an equivalent expression containing the entropy perturbation $s^\gi_\een$
explicitly. For $p_n$ we find
\begin{equation}\label{eq:dp-ds}
    p_n \equiv \left(\dfrac{\partial p}{\partial n}\right)_{\!\varepsilon} =
    \left(\dfrac{\partial p}{\partial
s}\right)_{\!\varepsilon}\left(\dfrac{\partial s}{\partial
n}\right)_{\!\varepsilon}=
    -\dfrac{\varepsilon_\nul(1+w)}{n^2_\nul T_\nul}p_s,
      \quad p_s \equiv \left(\dfrac{\partial p}{\partial s}\right)_{\!\varepsilon},
\end{equation}
where we have used (\ref{eq:TdS-1-gi}). Combining (\ref{eq:hulp-entropy}) and
(\ref{eq:pgi-3}) we arrive at
\begin{equation}\label{eq:pgi-s1-gi}
   p_\een^\gi=\beta^2 \varepsilon^\gi_\een + p_s s^\gi_\een.
\end{equation}
Substituting (\ref{eq:dp-ds}) into (\ref{eq:begam3}), we get
\begin{equation}\label{eq:begam3-s}
  \beta^2 = p_\varepsilon-\dfrac{p_s}{n_\nul T_\nul}.
\end{equation}
Expressions (\ref{eq:begam3})--(\ref{eq:pgi-3}) refer to an equation of state
$p=p(n,\varepsilon)$, whereas expressions
(\ref{eq:pgi-s1-gi})--(\ref{eq:begam3-s}) refer to the equivalent equation of
state $p=p(s,\varepsilon)$.

\subsection{Gauge-invariant Temperature Perturbations}

Finally, we will derive an expression for the gauge-invariant
temperature perturbation~$T^\gi_\een$ with the help of
(\ref{subeq:de-dp-a}). For the time derivative of the energy
density~$\varepsilon_\nul(n_\nul,T_\nul)$ we have
\begin{equation}\label{eq:de-dp-time}
  \dot{\varepsilon}_\nul =
   \left(\dfrac{\partial \varepsilon}{\partial n} \right)_{\!T}\dot{n}_\nul +
   \left(\dfrac{\partial \varepsilon}{\partial T} \right)_{\!n}\dot{T}_\nul.
\end{equation}
Replacing the infinitesimal quantities in (\ref{subeq:de-dp-a}) by
perturbations, we find
\begin{equation}\label{eq:de-dp-gd}
  \varepsilon_\een =
   \left(\dfrac{\partial \varepsilon}{\partial n} \right)_{\!T}n_\een +
   \left(\dfrac{\partial \varepsilon}{\partial T} \right)_{\!n}T_\een.
\end{equation}
Multiplying both sides of (\ref{eq:de-dp-time}) by
$\theta_\een/\dot{\theta}_\nul$ and subtracting the result from
(\ref{eq:de-dp-gd}) we get
\begin{equation}\label{eq:de-dp-gi}
  \varepsilon^\gi_\een =
   \left(\dfrac{\partial \varepsilon}{\partial n} \right)_{\!T}n^\gi_\een +
   \left(\dfrac{\partial \varepsilon}{\partial T}
\right)_{\!n}\left(T_\een-\dfrac{\dot{T}_\nul}{\dot{\theta}_\nul}
\theta_\een\right),
\end{equation}
where we have used (\ref{subeq:gi-en}). Hence, the quantity
\begin{equation}\label{eq:Tgi}
  T^\gi_\een \equiv T_\een -
     \dfrac{\dot{T}_\nul}{\dot{\theta}_\nul}\theta_\een,
\end{equation}
is gauge-invariant. Thus, (\ref{eq:de-dp-gi}) can be written as
\begin{equation}\label{eq:de-dp-dT-gi}
  \varepsilon^\gi_\een =
   \left(\dfrac{\partial \varepsilon}{\partial n} \right)_{\!T}n^\gi_\een +
   \left(\dfrac{\partial \varepsilon}{\partial T}
   \right)_{\!n}T^\gi_\een,
\end{equation}
implying that $T^\gi_\een$ can be interpreted as the gauge-invariant
temperature perturbation. We thus have expressed the perturbation in the
absolute temperature as a function of the perturbations in the energy density
and particle number density for a given equation of state of the form
$\varepsilon=\varepsilon(n,T)$ and $p=p(n,T)$. This expression will be used in
Section~\ref{thirdstep} to derive an expression for the fluctuations in the
background temperature, $\delta_T$, a measurable quantity.

Finally, we give the evolution equation for the background
temperature~$T_\nul(t)$. From (\ref{eq:de-dp-time}) it follows that
\begin{equation}\label{eq:evo-T0}
    \dot{T}_\nul=\dfrac{-3H\left[\varepsilon_\nul(1+w)-
    \left(\dfrac{\partial \varepsilon}{\partial n}\right)_{\!T}n_\nul\right]}
    {\left(\dfrac{\partial \varepsilon}{\partial T}\right)_{\!n}},
\end{equation}
where we have used equations (\ref{FRW2}) and~(\ref{FRW2a}). This
equation will be used to follow the time development of the
background temperature once~$\varepsilon_\nul(t)$ and~$n_\nul(t)$
are found from the zeroth-order Einstein equations.

\subsection{Diabatic Perturbations in a FLRW Universe}
\label{sec:ad-pert-uni}

In Section~\ref{sec:eqs-sgi} we have shown that the universe expands
adiabatically. In this section we investigate under which conditions
\emph{local} density perturbations are adiabatic.

By definition, an \emph{adiabatic}, or \emph{isocaloric} process is a thermodynamic process in
which no heat is transferred to or from the working fluid, i.e., it is a process
for which $\delta Q=0$. For a \emph{reversible} process we have $\delta
Q\equiv T_\nul s^\gi_\een$. Hence, reversible
and adiabatic processes are
characterized by $T_\nul s^\gi_\een=0$. From expression
(\ref{eq:TdS-1-gi}) it follows that for adiabatic perturbations we
have
\begin{equation}
  \label{eq:Tsi-0}
  n_\nul \varepsilon^\gi_\een - \varepsilon_\nul(1+w)n^\gi_\een=0.
\end{equation}
Using the background conservation laws (\ref{FRW2}) and (\ref{FRW2a}),
we get the adiabatic condition for density perturbations in a
\textsc{flrw} universe
\begin{equation}
  \label{eq:adia-cond}
  \dot{n}_\nul\varepsilon^\gi_\een-\dot{\varepsilon}_\nul n^\gi_\een=0.
\end{equation}
In a non-static universe we have $\dot{\varepsilon}_\nul\neq0$ and
$\dot{n}_\nul\neq0$.  In this case, equation (\ref{eq:adia-cond}) is
fulfilled if and only if the energy density is a function of the
particle number density \emph{only}, i.e., if
$\varepsilon=\varepsilon(n)$.  This can be demonstrated as follows.
From thermodynamics it is known that $\varepsilon=\varepsilon(n,T)$
and $p=p(n,T)$, where the particle number density $n$ and the
temperature $T$ are independent quantities.  Substituting
$\varepsilon=\varepsilon(n,T)$ into (\ref{eq:adia-cond}), yields
\begin{equation}
  \label{eq:adia-cond-n-T}
  \left(\dfrac{\partial\varepsilon}{\partial T} \right)_n
  \left[\dot{n}_\nul T^\gi_\een-n^\gi_\een \dot{T}_\nul  \right]=0.
\end{equation}
Since $n$ and $T$ are independent quantities, equation
(\ref{eq:adia-cond-n-T}) if, and only if, 
\begin{equation}
  \label{eq:eps-ind-T}
  \left(\dfrac{\partial\varepsilon}{\partial T}\right)_n=0,
\end{equation}
implying that $\varepsilon=\varepsilon(n)$.  In particular, density
perturbations in a perfect pressureless fluid with $\varepsilon=nmc^2$
are adiabatic.  This is the case in the non-relativistic limit
$v/c\rightarrow0$ of an expanding \textsc{flrw} universe, see
Section~\ref{nrl}\@. In all other cases,
$\varepsilon=\varepsilon(n,T)$ and $p=p(n,T)$, local density
perturbations evolve \emph{diabatically}.

\section{Manifestly Gauge-invariant First-order Equations} \label{thirdstep}

The five perturbation equations~(\ref{subeq:pertub-flrw}) form a set
of five equations for the five unknown quantities~$\varepsilon_\een$,
$n_\een$, $\vartheta_\een$, $\mbox{$^3\!R_{\een\parallel}$}$
and~$\theta_\een$. This system of equations can be further reduced in
the following way. As has been explained in
Section~\ref{sec:unique-gi}, our perturbation theory yields automatically
$\theta^\gi_\een=0$, (\ref{thetagi}). As a consequence, we do not need
the gauge dependent quantity~$\theta_\een$. Eliminating the quantity
$\theta_\een$ from equations~(\ref{subeq:pertub-flrw}) with the help
of equation (\ref{con-sp-1}), we arrive at the set of four first-order
differential equations
\begin{subequations}
\label{subeq:pertub-gi}
\begin{align}
 \dot{\varepsilon}_\een + 3H(\varepsilon_\een + p_\een)+
     \varepsilon_\nul(1 + w)\Bigl[\vartheta_\een +\frac{1}{2H}\left(
 \kappa\varepsilon_\een+\tfrac{1}{2}\,\mbox{$^3\!R_{\een\parallel}$}\right)\Bigr]&=0,
         \label{FRW4gi} \\
 \dot{n}_\een + 3H n_\een+
    n_\nul \Bigl[\vartheta_\een +\frac{1}{2H}\left(\kappa\varepsilon_\een +
      \tfrac{1}{2}\,\mbox{$^3\!R_{\een\parallel}$}\right)\Bigr]&=0, 
\label{FRW4agi} \\
  \dot{\vartheta}_\een+H(2-3\beta^2)\vartheta_\een +
        \frac{1}{\varepsilon_\nul(1+w)}\dfrac{\tilde{\nabla}^2p_\een}{a^2}&=0,   
\label{FRW5gi}\\
  \mbox{$^3\!\dot{R}_{\een\parallel}$} +
     2H\,\mbox{$^3\!R_{\een\parallel}$}
      -2\kappa\varepsilon_\nul(1 + w)\vartheta_\een+
      \frac{\mbox{$^3\!R_\nul$}}{3H} \left(\kappa\varepsilon_\een +
 \tfrac{1}{2}\,\mbox{$^3\!R_{\een\parallel}$} \right)&=0,     \label{FRW6gi}
\end{align}
\end{subequations}
for the four quantities~$\varepsilon_\een$, $n_\een$,
$\vartheta_\een$ and~$\mbox{$^3\!R_{\een\parallel}$}$. From their derivation it
follows that the system of equations (\ref{subeq:pertub-gi}) is, for scalar
perturbations, equivalent to the full set (\ref{subeq:basis}) of first-order
Einstein equations.

The system of equations~(\ref{subeq:pertub-gi}) is now cast in a
suitable form to arrive at a system of manifestly gauge-invariant
equations for the physical quantities~$\varepsilon^\gi_\een$
and~$n^\gi_\een$, since we then can immediately calculate these
quantities. Indeed, eliminating the quantity~$\theta_\een$ from
equations (\ref{subeq:gi-en}) with the help of equation (\ref{con-sp-1}), and
using the background equations (\ref{subeq:einstein-flrw}) to eliminate the time
derivatives $\dot{\varepsilon}_\nul$, $\dot{n}_\nul$ and
$\dot{\theta}_\nul=3\dot{H}$, we get
\begin{subequations}
\label{subeq:pertub-gi-e-n}
\begin{align}
    \varepsilon_\een^\gi & =
     \frac{ \varepsilon_\een \,\mbox{$^3\!R_\nul$} -
   3\varepsilon_\nul(1 + w) (2H\vartheta_\een +
  \tfrac{1}{2}\,\mbox{$^3\!R_{\een\parallel}$}) }
   {\mbox{$^3\!R_\nul$}+3\kappa\varepsilon_\nul(1 + w)},
        \label{Egi} \\
   n_\een^\gi & = n_\een-\dfrac{3n_\nul(\kappa\varepsilon_\een+2H\vartheta_\een+
            \tfrac{1}{2}\,\mbox{$^3\!R_{\een\parallel}$})}
         {\mbox{$^3\!R_\nul$}+3\kappa\varepsilon_\nul(1+w)}.  \label{nu2}
\end{align}
\end{subequations}
The quantities ~$\varepsilon^\gi_\een$ and~$n^\gi_\een$ are now
completely determined by the system of background
equations~(\ref{subeq:einstein-flrw}) and the first-order
equations~(\ref{subeq:pertub-gi}).

\subsection{Evolution Equations for Density Perturbations}

Instead of calculating~$\varepsilon^\gi_\een$ and~$n^\gi_\een$ in
the way described above, we proceed by first making explicit the
gauge-invariance of the theory. To that end, we rewrite the system
of four differential equations~(\ref{subeq:pertub-gi}) for the
gauge dependent variables~$\varepsilon_\een$, $n_\een$,
$\vartheta_\een$ and~$\mbox{$^3\!R_{\een\parallel}$}$ into a system
of equations for the gauge-invariant
variables~$\varepsilon_\een^\gi$ and~$n_\een^\gi$. It is, however,
of convenience to use the entropy related
perturbation~$\sigma_\een^\gi$, defined by (\ref{eq:entropy-gi}),
rather than the particle number density perturbation~$n_\een^\gi$.
The result is
\begin{subequations}
\label{subeq:eerste}
\begin{align}
 & \ddot{\varepsilon}^\gi_\een+a_1\dot{\varepsilon}^\gi_\een+
  a_2\varepsilon^\gi_\een = a_3 \sigma^\gi_\een,
      \label{eq:vondst2}  \\
 & \dot{\sigma}^\gi_\een = -3H\left(1-\frac{n_\nul
  p_n}{\varepsilon_\nul(1+w)}\right) \sigma^\gi_\een.   \label{eq:vondst1}
\end{align}
\end{subequations}
The derivation of these equations is given in detail in
Appendices~\ref{app:hulp2} and~\ref{app:hulp3}. The coefficients
$a_1$, $a_2$ and $a_3$ occurring in equations (\ref{subeq:eerste}) are
given~by
\begin{subequations}
\label{subeq:coeff}
\begin{align}
  a_1 & = \dfrac{\kappa\varepsilon_\nul(1+w)}{H}
  -2\dfrac{\dot{\beta}}{\beta}+H(4-3\beta^2)
    +\mbox{$^3\!R_\nul$}\left(\dfrac{1}{3H} + \dfrac{2H(1+3\beta^2)}
  {\mbox{$^3\!R_\nul$}+3\kappa\varepsilon_\nul(1+w)}\right), \label{eq:alpha-1}
\\
  a_2 & = \kappa\varepsilon_\nul(1+w)-
  4H\dfrac{\dot{\beta}}{\beta}+2H^2(2-3\beta^2)
  +\mbox{$^3\!R_\nul$}\left(\dfrac{1}{2}+
\dfrac{5H^2(1+3\beta^2)-2H\dfrac{\dot{\beta}}{\beta}}
  {\mbox{$^3\!R_\nul$}+3\kappa\varepsilon_\nul(1+w)}\right)
-\beta^2\left(\frac{\tilde{\nabla}^2}{a^2}-\tfrac{1}{2}\,\mbox{$^3\!R_\nul$}
\right), \label{eq:alpha-2} \\
  a_3 & =
\Biggl\{\dfrac{-18H^2}{\mbox{$^3\!R_\nul$}+3\kappa\varepsilon_\nul(1+w)}
  \Biggl[\varepsilon_\nul p_{\varepsilon n}(1+w)+
   \dfrac{2p_n}{3H}\dfrac{\dot{\beta}}{\beta}
   -\beta^2p_n+p_\varepsilon p_n+n_\nul
   p_{nn}\Biggr]+p_n\Biggr\}
\left(\frac{\tilde{\nabla}^2}{a^2}-\tfrac{1}{2}\,\mbox{$^3\!R_\nul$}\right),
       \label{eq:alpha-3}
\end{align}
\end{subequations}
where the functions $\beta(t)$ and $w(t)$ are given by
(\ref{begam2}). In the derivation
of the above results, we used equations (\ref{subeq:einstein-flrw}). The
abbreviations~$p_n$ and~$p_\varepsilon$ are
given by (\ref{perttoes1}). Furthermore, we used the abbreviations
\begin{equation}
    p_{nn} \equiv \frac{\partial^2 p}{\partial n^2}, \quad
    p_{\varepsilon n} \equiv
           \frac{\partial^2 p}{\partial \varepsilon\,\partial n}.
        \label{pne}
\end{equation}

The equations~(\ref{subeq:eerste}) contain only gauge-invariant
quantities and the coefficients are scalar functions. Thus, these
equations are \emph{manifestly gauge-invariant}. In contrast, the equations
(\ref{subeq:pertub-gi}), being linear Einstein equations and conservation
laws, are themselves gauge-invariant, but their solutions are gauge
dependent, see Appendix~\ref{giofoe} for a detailed explanation.

The equations~(\ref{subeq:eerste}) are equivalent to one equation
of the third-order, whereas the four first-order
equations~(\ref{subeq:pertub-gi}) are equivalent to one
equation of the fourth-order. This observation reflects the fact
that the solutions of the four first-order equations (\ref{subeq:pertub-gi}) are gauge
dependent, while the solutions $\varepsilon^\gi_\een$
and~$\sigma^\gi_\een$ of equations (\ref{subeq:eerste}) are gauge-invariant.
One `degree of freedom,' say, the gauge
function~$\psi(\vec{x})$ has disappeared from the scene completely.

The equations~(\ref{subeq:eerste}) constitute the main result of
this article. In view of (\ref{eq:entropy-gi}), they essentially are
two differential equations for the
perturbations~$\varepsilon_\een^\gi$ and~$n_\een^\gi$ to the energy
density~$\varepsilon_\nul(t)$ and the particle number
density~$n_\nul(t)$ respectively, for \textsc{flrw} universes with
$k=-1,0,+1$. They describe the evolution of the energy density
perturbation~$\varepsilon_\een^\gi$ and the particle number density
perturbation~$n_\een^\gi$ for \textsc{flrw} universes filled with a
fluid which is described by an equation of state of the form
${p=p(n,\varepsilon)}$, the precise form of which is left
unspecified.

\subsection{Evolution Equations for Contrast Functions}

In the study of the evolution of density perturbations it is of
convenience to use a quantity which measures the perturbation to
the density relative to the background densities. To that end we
define the gauge-invariant contrast functions $\delta_\varepsilon$
and $\delta_n$ by
\begin{equation}\label{eq:contrast}
  \delta_\varepsilon(t,\vec{x}) \equiv
      \dfrac{\varepsilon^\gi_\een(t,\vec{x})}{\varepsilon_\nul(t)}, \quad
  \delta_n(t,\vec{x}) \equiv
      \dfrac{n^\gi_\een(t,\vec{x})}{n_\nul(t)}.
\end{equation}
Using these quantities, equations (\ref{subeq:eerste}) can be
rewritten as (Appendix~\ref{app:contrast})
\begin{subequations}
\label{subeq:final}
\begin{align}
   \ddot{\delta}_\varepsilon + b_1 \dot{\delta}_\varepsilon +
      b_2 \delta_\varepsilon &=
      b_3 \left(\delta_n - \frac{\delta_\varepsilon}{1+w}\right),
              \label{sec-ord}  \\
   \frac{1}{c}\frac{\dif}{\dif t}
      \left(\delta_n - \frac{\delta_\varepsilon}{1 + w}\right)& =
     \frac{3Hn_\nul p_n}{\varepsilon_\nul(1 + w)}
     \left(\delta_n - \frac{\delta_\varepsilon}{1 + w}\right),
                 \label{fir-ord}
\end{align}
\end{subequations}
where the coefficients $b_1$, $b_2$ and~$b_3$ are given by
\begin{subequations}
\label{subeq:coeff-contrast}
\begin{align}
  b_1 & =  \dfrac{\kappa\varepsilon_\nul(1+w)}{H}
  -2\dfrac{\dot{\beta}}{\beta}-H(2+6w+3\beta^2)
   +\mbox{$^3\!R_\nul$}\left(\dfrac{1}{3H}+
  \dfrac{2H(1+3\beta^2)}
  {\mbox{$^3\!R_\nul$}+3\kappa\varepsilon_\nul(1+w)}\right), \\
  b_2 &= -\tfrac{1}{2}\kappa\varepsilon_\nul(1+w)(1+3w)+
  H^2\left(1-3w+6\beta^2(2+3w)\right) \nonumber \\
  & \phantom{=}\;\,+6H\dfrac{\dot{\beta}}{\beta}\left(w+\dfrac{\kappa\varepsilon_\nul(1+w)}
   {\mbox{$^3\!R_\nul$}+3\kappa\varepsilon_\nul(1+w)}\right)
 -\mbox{$^3\!R_\nul$}\left(\tfrac{1}{2}w+
\dfrac{H^2(1+6w)(1+3\beta^2)}{\mbox{$^3\!R_\nul$}+3\kappa\varepsilon_\nul(1+w)}
\right)
-\beta^2\left(\frac{\tilde{\nabla}^2}{a^2}-\tfrac{1}{2}\,\mbox{$^3\!R_\nul$}
\right), \\
  b_3 &= 
\Biggl\{\dfrac{-18H^2}{\mbox{$^3\!R_\nul$}+3\kappa\varepsilon_\nul(1+w)}
  \Biggl[\varepsilon_\nul p_{\varepsilon n}(1+w)
  +\dfrac{2p_n}{3H}\dfrac{\dot{\beta}}{\beta}
  -\beta^2p_n+p_\varepsilon p_n+n_\nul p_{nn}\Biggr]+
   p_n\Biggr\}\dfrac{n_\nul}{\varepsilon_\nul}
\left(\frac{\tilde{\nabla}^2}{a^2}-\tfrac{1}{2}\,\mbox{$^3\!R_\nul$}\right).
\label{eq:b3}
\end{align}
\end{subequations}
In Sections~\ref{sec:an-exam}--\ref{sec:star-for-res} we use the equations (\ref{subeq:final}) to study
the evolution of small energy density perturbations and
particle number perturbations in \textsc{flrw} universes.

The combined First and Second Laws of thermodynamics~(\ref{eq:TdS-1-gi}) reads,
in terms of the contrast functions~(\ref{eq:contrast}),
\begin{equation}\label{eq:S-contrast}
    T_\nul s^\gi_\een = -\dfrac{\varepsilon_\nul(1+w)}{n_\nul}
  \left( \delta_n - \frac{\delta_\varepsilon}{1 + w} \right)=
  -\dfrac{\varepsilon_\nul}{n_\nul}(\delta_n-\delta_\varepsilon)-
  \dfrac{p_\nul}{n_\nul}\delta_n,
\end{equation}
where $s^\gi_\een$ is the entropy per particle.

Finally, we define the relative temperature
perturbation~$\delta_T$ and the relative pressure
perturbation $\delta_p$ by
\begin{equation}\label{eq:delta-T}
  \delta_T(t,\vec{x}) \equiv \dfrac{T^\gi_\een(t,\vec{x})}{T_\nul(t)},
  \quad \delta_p(t,\vec{x}) \equiv \dfrac{p^\gi_\een(t,\vec{x})}{p_\nul(t)}.
\end{equation}
Using the expressions (\ref{eq:de-dp-dT-gi}), (\ref{eq:contrast}) and
(\ref{eq:delta-T}) we arrive at the relative temperature perturbation
\begin{equation}\label{eq:rel-T-pert}
  \delta_T = \dfrac{\varepsilon_\nul\delta_\varepsilon-
       \left(\dfrac{\partial \varepsilon}{\partial n}\right)_{\!T}n_\nul\delta_n}
        {T_\nul\left(\dfrac{\partial \varepsilon}{\partial T}\right)_{\!n}}.
\end{equation}
The relative pressure perturbation follows directly from
(\ref{eq:pgi-2}). We get
\begin{equation}
  \label{eq:rel-press-pert}
  \delta_p = \dfrac{\varepsilon_\nul}{p_\nul}
     \left(\dfrac{\partial p}{\partial \varepsilon}\right)_{\!n}\delta_\varepsilon+
             \dfrac{n_\nul}{p_\nul} \left(\dfrac{\partial p}
                  {\partial n}\right)_{\!\varepsilon}\delta_n.
\end{equation}
We thus have found the relative temperature and the relative pressure perturbations as
functions of the relative perturbations in the energy density and
particle number density for an equation of state of the form
$\varepsilon=\varepsilon(n,T)$ and $p=p(n,T)$ (see
Appendix~\ref{sec:eq-state}). The quantity
$\delta_T(t,\vec{x})$ is a measurable quantity in the
cosmic background radiation.

\section{Non-Relativistic Limit in an Expanding FLRW Universe} \label{nrl}

In Section~\ref{sec:unique} we have shown that there exist only two
gauge-invariant quantities $\varepsilon^\gi_\een$ and $n^\gi_\een$,
which could be the real energy density and particle number density
perturbations. In this section we show that in the non-relativistic
limit $v/c\rightarrow0$ the quantities $\varepsilon^\gi_\een$ and
$n^\gi_\een$ become equal to their Newtonian counterparts. This
implies that $\varepsilon^\gi_\een$ and $n^\gi_\een$ are indeed the
local perturbations to the energy density and particle number density
perturbations.

It is well known that if the gravitational field is weak and
velocities are small with respect to the velocity of light
($v/c\rightarrow0$), the system of Einstein equations and conservation
laws may reduce to the single field equation of the Newtonian theory
of gravity, namely the Poisson equation~(\ref{eq:poisson}). In the
first-order perturbation theory developed in this article, the
gravitational field is already weak, so that, at first sight, all we
have to do to arrive at the Newtonian theory of gravity is to take the
non-relativistic limit $v/c\rightarrow0$ in all equations. Since in
the Newtonian theory the gravitational field is described by only one,
time-independent, potential $\varphi(\vec{x})$, one cannot obtain the
Newtonian theory by simply taking the non-relativistic limit
$v/c\rightarrow0$, since in a relativistic theory the gravitational
field is described, in general, by six potentials, namely the six
components $h_{ij}(t,\vec{x})$ of the metric.  In this article we have
used the decomposition~(\ref{decomp-symh}). Moreover, we have shown in
Section~\ref{klasse} that only $h^i_{\parallel j}$ given by
(\ref{decomp-hij-par}) is coupled to density perturbations. By using
this decomposition we have reduced the number of potentials to two,
namely $\phi(t,\vec{x})$ and $\zeta(t,\vec{x})$.  We have rewritten
the system of equations (\ref{subeq:scalar}) for scalar perturbations
into an equivalent system (\ref{subeq:pertub-flrw}). As a result, the
perturbation to the metric, $h^i_{\parallel j}$, enters the system
(\ref{subeq:pertub-flrw}) via the trace
\begin{equation}\label{RnabEE}
   \mbox{$^3\!R_{\een\parallel}$} =
   \dfrac{2}{c^2}\Bigl[2\phi^{|k}{}_{|k}-
  \zeta^{|k|l}{}_{|l|k}+\zeta^{|k}{}_{|k}{}^{|l}{}_{|l}+
  \tfrac{1}{3}\,\mbox{$^3\!R_\nul$}(3\phi+\zeta^{|k}{}_{|k})\Bigr],
\end{equation}
of the spatial part of the perturbation to the Ricci tensor
(\ref{decomp-Rij-par}), and via the perturbed expansion scalar
(\ref{fes5})
\begin{equation}
  \label{eq:expan-potentials}
  \theta_\een=\vartheta_\een-\dfrac{1}{c^2}\bigl(3\dot{\phi}+\dot{\zeta}^{|k}{}_{|k}\bigr),
\end{equation}
where we have used (\ref{decomp-hij-par}).  This shows explicitly that
density perturbations are, in \textsc{flrw} universes, described by
two potentials $\phi(t,\vec{x})$ and $\zeta(t,\vec{x})$.  In this
section we show that in the non-relativistic limit of a \emph{flat}
\textsc{flrw} universe, the potential $\zeta$ drops from the
perturbation theory. For a flat (i.e., $k=0$) \textsc{flrw} universe we have
$\mbox{$^3\!R_\nul$}=0$, as follows from (\ref{spRicci}). The
perturbation to the spatial part of the Ricci scalar, (\ref{RnabEE}),
reduces in this case to
\begin{equation}\label{RnabEE-0}
  \mbox{$^3\!R_{\een\parallel}$} =\dfrac{4}{c^2}\phi^{|k}{}_{|k}=
       -\dfrac{4}{c^2}\dfrac{\nabla^2\phi}{a^2},
\end{equation}
where~$\nabla^2$ is the usual Laplace operator see
(\ref{subeq:metric}) and (\ref{Laplace}).  For a flat \textsc{flrw}
universe, the zeroth-order equations~(\ref{subeq:einstein-flrw})
reduce~to
\begin{subequations}
\label{subeq:einstein-flrw-flat}
\begin{alignat}{3}
  \text{Constraint:}\quad &&  3H^2 & = \kappa\varepsilon_\nul + \Lambda, \label{FRW3-flat} \\
  \text{Conservation:}\quad &&   \dot{\varepsilon}_\nul & = -3H\varepsilon_\nul(1+w), \label{FRW2-flat} \\
     &&   \dot{n}_\nul & = -3Hn_\nul.   \label{FRW2a-flat}
\end{alignat}
\end{subequations}
Upon substituting (\ref{RnabEE-0}) into (\ref{subeq:pertub-flrw}) and
putting $\mbox{$^3\!R_\nul$}=0$, we arrive at the set of perturbation
equations for a flat \textsc{flrw} universe:
\begin{subequations}
\label{subeq:pertub-gi-flat}
\begin{alignat}{3}
 \text{Constraint:} \quad  &&  & H(\theta_\een-\vartheta_\een)+
       \dfrac{1}{c^2}\dfrac{\nabla^2\phi}{a^2} = \dfrac{4\pi G}{c^4}
        \biggl(\varepsilon^\gi_\een+
         \dfrac{\dot{\varepsilon}_\nul}{\dot{\theta}_\nul}\theta_\een\biggr),
\label{con-sp-1-flat} \\
 \text{Evolution:} \quad &&  & \dfrac{\nabla^2\dot{\phi}}{a^2}+\dfrac{4\pi G}{c^2}
\varepsilon_\nul(1 + w)\vartheta_\een=0, \label{FRW6gi-flat} \\
 \text{Conservation:} \quad  &&  &\dot{\varepsilon}_\een +
3H(\varepsilon_\een + p_\een)+
         \varepsilon_\nul(1 + w)\theta_\een=0,  \label{FRW4gi-flat} \\
&& &\dot{\vartheta}_\een+H(2-3\beta^2)\vartheta_\een+
   \frac{1}{\varepsilon_\nul(1+w)}\dfrac{\nabla^2p_\een}{a^2}=0, 
  \label{FRW5gi-flat} \\
&&  &\dot{n}_\een + 3H n_\een +
         n_\nul\theta_\een=0, \label{FRW4agi-flat}
\end{alignat}
\end{subequations}
where we have used (\ref{kappa}), and our new definition (\ref{gien})
to eliminate $\varepsilon_\een$ from (\ref{con-sp-1}).  The scale
factor of the universe $a(t)$ follows from the Einstein equations
(\ref{subeq:einstein-flrw-flat}) via $H \equiv \dot{a}/a$.  In the
equations (\ref{subeq:pertub-gi-flat}), the potential $\zeta$ occurs
only in the quantity $\theta_\een$, see (\ref{eq:expan-potentials}).

We now consider the sets of equations (\ref{subeq:einstein-flrw-flat})
and (\ref{subeq:pertub-gi-flat}) in the non-relativistic limit
$v/c\rightarrow0$.  Since the spatial part $u^i_{\een\parallel}$ of
the fluid four-velocity is gauge dependent, (\ref{eq:trans-u0-ui-b}),
with a physical component and a non-physical gauge part, we define the
non-relativistic limit $v/c\rightarrow0$ by
\begin{equation}
  \label{eq:nrl-limit}
  u^i_{\een\parallel\text{physical}} \equiv
    \dfrac{U^i_{\een\parallel\text{physical}}}{c} \rightarrow 0,
\end{equation}
i.e., the \emph{physical} part of the spatial fluid velocity is
negligible with respect to the speed of light. In this limit the
kinetic energy per particle
$\tfrac{1}{2}m\langle{v^2}\rangle=\tfrac{3}{2}k_{\text{B}}T\rightarrow0$
is small compared to the rest energy $mc^2$ per particle, implying
that the pressure $p=nk_{\text{B}}T\rightarrow0$ ($n\neq0$) vanishes
also and that the energy density of the universe is given by
$\varepsilon=nmc^2$. In other words, the non-relativistic limit
(\ref{eq:nrl-limit}) implies 
\begin{equation}\label{eq:rest-energy}
   p=0, \quad  \varepsilon = n m c^2, \quad w\equiv \dfrac{p}{\varepsilon}=0.
\end{equation}
In view of (\ref{eq:rest-energy}), the background equations
(\ref{subeq:einstein-flrw-flat}) take on the simple form in the
non-relativistic limit (\ref{eq:nrl-limit})
\begin{subequations}
\label{subeq:einstein-flrw-newt}
\begin{alignat}{3}
   \text{Constraint:}\quad &&  3H^2 & = \kappa\varepsilon_\nul + \Lambda, \label{FRW3-newt} \\
   \text{Conservation:}\quad && \dot{\varepsilon}_\nul &  = -3H\varepsilon_\nul,  \label{FRW2-newt}\\
    &&  \dot{n}_\nul & = -3H n_\nul.  \label{FRW2a-newt}
\end{alignat}
\end{subequations}
Thus, even in the non-relativistic limit, a non-empty universe
cannot be static, therefore we have $H\neq0$.  Note that in the limit
(\ref{eq:nrl-limit}), equations (\ref{FRW2-newt})
and~(\ref{FRW2a-newt}) are identical, since then
$\varepsilon_\nul=n_\nul mc^2$, in view of (\ref{eq:rest-energy}).

The consequences of the limit (\ref{eq:nrl-limit}) are as follows. 
In the limit $p\rightarrow0$, equation (\ref{basis-5-scal}) reads
\begin{equation}
\frac{1}{c}\frac{\dif}{\dif t}
   \Bigl(\varepsilon_\nul u^i_{\een\parallel}\Bigr)+5H\varepsilon_\nul
u^i_{\een\parallel}=0.
            \label{basis-5-scal-p0}
\end{equation}
Using the background equation (\ref{FRW2-newt}), equation
(\ref{basis-5-scal-p0}) takes on the simple form
\begin{equation}\label{eq:ui-par-p0}
    \dot{u}^i_{\een\parallel}=-2H u^i_{\een\parallel}.
\end{equation}
The general solution of this equation is, using $H\equiv\dot{a}/a$,
\begin{equation}
\label{eq:ui-gauge-mode}
    u^{i}_{\een\parallel\mathrm{gauge}}=-\dfrac{1}{a^2(t)}
\tilde{g}^{ik}\partial_k\psi(\vec{x}).
\end{equation}
Since the physical part of $u^{i}_{\een\parallel}$ vanishes in the
limit (\ref{eq:nrl-limit}), the solution (\ref{eq:ui-gauge-mode}) is a
gauge mode, as follows from (\ref{eq:trans-u0-ui-b}). Thus, in the
limit (\ref{eq:nrl-limit}) we are left with the non-physical quantity
(\ref{eq:ui-gauge-mode}).  If a quantity is proportional to the gauge
function $\psi(\vec{x})$ or its partial derivatives, it may be put
equal to zero without loss of physical information.  If we require
that in the limit~(\ref{eq:nrl-limit})
$u^{i}_{\een\parallel\mathrm{gauge}}=0$ holds true before and after a
gauge transformation we find from (\ref{eq:ui-gauge-mode}) that
$\partial_k\psi(\vec{x})=0$ or, equivalently,
\begin{equation}\label{eq:psi-K}
  \psi(\vec{x})=\psi.
\end{equation}
In view of (\ref{eq:psi-K}), the gauge dependent functions~$\varepsilon_\een$
and~$n_\een$ transform under a gauge transformation in the limit
(\ref{eq:nrl-limit}) according to (\ref{subeq:split-e-n}), with constant $\psi$:
\begin{subequations}
\label{subeq:split-e-n-nrl}
\begin{alignat}{3}
 \varepsilon_\een & \rightarrow \hat{\varepsilon}_\een & \,= &\;
      \varepsilon_\een + \psi\dot{\varepsilon}_\nul, \label{e-ijk-nrl} \\
 n_\een & \rightarrow \hat{n}_{\een} & \,= &\; n_{\een} + \psi\dot{n}_{\nul}.
         \label{n-ijk-nrl}
\end{alignat}
\end{subequations}
Since in an expanding universe the time derivatives
$\dot{\varepsilon}_\nul(t)\neq0$, (\ref{FRW2-newt}), and
$\dot{n}_{\nul}(t)\neq0$, (\ref{FRW2a-newt}), the
functions~$\varepsilon_\een(t,\vec{x})$ and~$n_\een(t,\vec{x})$ do
still depend upon the gauge, so that these quantities have, also in
the limit (\ref{eq:nrl-limit}), no physical significance.  Finally, it
follows from (\ref{xi-syn}) and (\ref{eq:psi-K}) that the gauge
transformation (\ref{func}) reduces in the limit (\ref{eq:nrl-limit})
to the gauge transformation
\begin{equation}\label{eq:gauge-trans-newt}
    x^0 \rightarrow x^0 - \psi, \quad x^i \rightarrow
    x^i-\chi^i(\vec{x}).
\end{equation}
In other words, time and space transformations are decoupled: time
coordinates may be shifted, whereas spatial coordinates may be chosen
arbitrarily.  Thus, in the limit (\ref{eq:nrl-limit}) the general
relativistic infinitesimal coordinate transformation (\ref{func}),
with $\xi^\mu$ given by (\ref{xi-syn}), reduces to the most general
infinitesimal coordinate transformation (\ref{eq:gauge-trans-newt})
which is possible in the Newtonian theory of gravity.  By now, it
should be clear that the gauge problem of cosmology cannot be solved
by `fixing the gauge,' since the constant $\psi$ and the three
functions $\chi^i(\vec{x})$ cannot be determined.

We now consider the perturbation equations
(\ref{subeq:pertub-gi-flat}) in the limit (\ref{eq:nrl-limit}).  Upon
substituting $\vartheta_\een \equiv (u^k_{\een\parallel})_{|k}=0$,
$w\equiv p_\nul/\varepsilon_\nul=0$ and $p_\een=0$ into these
equations we find that equation (\ref{FRW5gi-flat}) is identically
satisfied, whereas the remaining
equations~(\ref{subeq:pertub-gi-flat}) reduce to
\begin{subequations}
\label{subeq:pertub-gi-flat-newt}
\begin{alignat}{3}
 \text{Constraint:} \quad  &&  &
     \nabla^2\phi = \dfrac{4\pi G}{c^2}a^2\varepsilon^\gi_\een,
\label{con-sp-1-flat-newt} \\
 \text{Evolution:} \quad &&  & \nabla^2\dot{\phi}=0, \label{FRW6gi-flat-newt} \\
 \text{Conservation:} \quad  &&  &\dot{\varepsilon}_\een + 3H\varepsilon_\een+
         \varepsilon_\nul\theta_\een=0,  \label{FRW4gi-flat-newt} \\
&&  &\dot{n}_\een + 3H n_\een +
         n_\nul\theta_\een=0. \label{FRW4agi-flat-newt}
\end{alignat}
\end{subequations}
The constraint equation (\ref{con-sp-1-flat-newt}) can be found by
subtracting $\tfrac{1}{6}\theta_\een/\dot{H}$ times the
time-derivative of the background constraint equation
(\ref{FRW3-newt}) from the constraint equation (\ref{con-sp-1-flat})
and using that $\theta_\nul=3H$, (\ref{fes2}). This shows explicitly
that (\ref{gien}) is the only possible choice to construct the
gauge-invariant quantity $\varepsilon^\gi_\een$.  Equations
(\ref{FRW4gi-flat-newt}) and~(\ref{FRW4agi-flat-newt}) are identical
in the non-relativistic limit (\ref{eq:nrl-limit}) since
$\varepsilon_\een=n_\een mc^2$, in view of (\ref{eq:rest-energy}).
The quantities $\varepsilon_\een$ and $n_\een$, which are gauge
dependent in the General Theory of Relativity, are also gauge
dependent in the non-relativistic limit (\ref{eq:nrl-limit}), so that
equations (\ref{FRW4gi-flat-newt}) and~(\ref{FRW4agi-flat-newt}) need
not be considered: the equations (\ref{FRW4gi-flat-newt}) and
(\ref{FRW4agi-flat-newt}) have no physical significance and can be
discarded. Consequently, the perturbed expansion scalar $\theta_\een$
does not occur anymore in the perturbation theory, and we are left with
one potential $\phi$ only.

Equations (\ref{con-sp-1-flat-newt}) and (\ref{FRW6gi-flat-newt}) can
be combined to give
\begin{equation}
  \nabla^2\phi(\vec{x})=
    \dfrac{4\pi G}{c^2}a^2(t)\varepsilon_\een^\gi(t,\vec{x}).
    \label{Egi-poisson}
\end{equation}
Or, equivalently,
\begin{equation}
  \nabla^2\phi(\vec{x})=
    \dfrac{4\pi G}{c^2}a^2(t_\mathrm{p})
    \varepsilon_\een^\gi(t_\mathrm{p},\vec{x}),
    \label{Egi-poisson-present}
\end{equation}
where $t_\mathrm{p}$ indicates the present time. This Einstein
equation can be rewritten in a form which is equivalent to the
Poisson equation, by introducing the potential~$\varphi$
\begin{equation}\label{eq:ident}
   \varphi(\vec{x}) \equiv \frac{\phi(\vec{x})}{a^2(t_\mathrm{p})}.
\end{equation}
Inserting (\ref{eq:ident}) into~(\ref{Egi-poisson-present}) we
obtain the result
\begin{equation}\label{poisson}
    \nabla^2\varphi(\vec{x})=
    4\pi G\dfrac{\varepsilon^\gi_\een(t_\mathrm{p},\vec{x})}{c^2}.
\end{equation}

Finally, we have to check the expressions (\ref{subeq:gi-en}) or,
equivalently, (\ref{subeq:pertub-gi-e-n}) in the limit
(\ref{eq:nrl-limit}). It can easily be verified that in the
non-relativistic limit (\ref{eq:nrl-limit}) of a flat \textsc{flrw}
universe, expression (\ref{Egi}) reduces to
\begin{equation}
  \label{eq:R-eps}
   \varepsilon^\gi_\een=-\dfrac{1}{2\kappa}\mbox{$^3\!R_{\een\parallel}$}.
\end{equation}
In view of (\ref{kappa}) and (\ref{RnabEE-0}), this equation is
equivalent to (\ref{con-sp-1-flat-newt}). Using (\ref{eq:R-eps}), we
find that expression (\ref{nu2}) reduces in the non-relativistic limit
(\ref{eq:nrl-limit}) to
\begin{equation}
  \label{nu2-flat-newt}
  n^\gi_\een=n_\een-\dfrac{n_\nul}{\varepsilon_\nul}(\varepsilon_\een-\varepsilon^\gi_\een).
\end{equation}
Combining $\varepsilon_\nul=n_\nul mc^2$ and $\varepsilon_\een=n_\een
mc^2$, (\ref{eq:rest-energy}),
with (\ref{nu2-flat-newt}) yields
\begin{equation}
  \label{eq:n-eps-gi}
  n^\gi_\een = \dfrac{\varepsilon^\gi_\een}{mc^2},
\end{equation}
which is the gauge-invariant counterpart of $\varepsilon_\een=n_\een
mc^2$. Again, we have to 
conclude that (\ref{gidi}) is the only possible choice to construct
the gauge-invariant quantity $n^\gi_\een$.

Combining equations (\ref{poisson}) and (\ref{eq:n-eps-gi}), we arrive
at the Poisson equation of the Newtonian theory of gravity:
\begin{equation}\label{eq:poisson}
  \nabla^2\varphi(\vec{x})=
    4\pi G \rho_\een(\vec{x}),
\end{equation}
where
\begin{equation}
  \label{eq:rho-Newton}
  \rho_\een(\vec{x})\equiv
  \dfrac{\varepsilon^\gi_\een(t_\mathrm{p},\vec{x})}{c^2}, \quad
   \rho_\een(\vec{x}) =  mn^\gi_\een(t_\mathrm{p},\vec{x}),
\end{equation}
is the mass density of a perturbation.

In this section we have shown that in the non-relativistic limit
(\ref{eq:nrl-limit}) of a flat \textsc{flrw} universe, the first-order
perturbation equations (\ref{subeq:pertub-flrw}) together with the new
definitions (\ref{subeq:gi-en}) reduce to the well-known Newtonian
results (\ref{eq:gauge-trans-newt}), (\ref{eq:n-eps-gi}) and
(\ref{eq:poisson}).  Consequently, the gauge-invariant quantities
$\varepsilon^\gi_\een$ and $n^\gi_\een$ given by (\ref{subeq:gi-en})
are indeed the energy density and particle number density
perturbations.

\section{Perturbation equations for a Flat FLRW Universe}
\label{sec:an-exam}

In this section we derive the perturbation equations for a flat \textsc{flrw} universe
with a vanishing cosmological constant in its radiation-dominated,
plasma-dominated and matter-dominated stages.

The six numerical values that we need in the following are:
\textit{i.}\@ the redshift at time $t_{\text{eq}}$ when the matter
density has become equal to the radiation density, \textit{ii.}\@ the
redshift at time $t_{\text{dec}}$ when matter and radiation decouple,
\textit{iii.}\@ the present value of the Hubble function,
\textit{iv.}\@ the present value of the background radiation
temperature, \textit{v.}\@ the age of the universe and, \textit{vi.}\@
the fluctuations in the background radiation at decoupling. They
follow from
\textsc{wmap}~\cite{2009ApJS..180..306D,Spergel:2003cb,komatsu-2008,hinshaw-2008,2010arXiv1001.4538K}
observations:
\begin{subequations}
\label{subeq:wmap}
\begin{align}
   z(t_{\mathrm{eq}}) &= 3176, \label{eq:z-eq} \\
   z(t_{\mathrm{dec}})&= 1091,  \label{eq:init-wmap-z} \\
   \mathcal{H}(t_{\mathrm{p}})&= 71.9\,\mathrm{km/sec/Mpc}=
      2.33\times10^{-18}\,\mathrm{s}^{-1}, \label{eq:init-wmap-T} \\
  T_{\nul\gamma}(t_\mathrm{p})&= 2.725\,\mathrm{K}, \label{eq:back-T-tp}\\
  t_{\mathrm{p}} & =
   13.7\,\mathrm{Gyr}=4.32\times10^{17}\,\text{sec}, \label{eq:present-age}\\
  \delta_{T_\gamma}(t_{\text{dec}}) & \lesssim  10^{-5}, \label{eq:background-rad}
\end{align}
\end{subequations}
where we have used that $1\,\mathrm{Mpc}=3.0857\times10^{22}\,\mathrm{m}$
($1\,\mathrm{pc}=3.2616\;\mathrm{ly}$).

The cosmological redshift $z(t)$ is given by
\begin{equation}\label{eq:redshift}
    z(t)=\dfrac{a(t_\mathrm{p})}{a(t)}-1, \quad a(t_{\mathrm{p}})=1,
\end{equation}
where we have normalized the scale factor $a(t)$ to unity at
$t=t_{\mathrm{p}}$, which is only allowed in a \emph{flat} \textsc{flrw}
universe.

\subsection{Radiation-dominated Phase}
\label{sec:rad}

We consider the universe after the era of inflation. Let $t_\mathrm{rad}$ be the
moment at which exponential expansion of the universe has come to an end, and
the radiation-dominated era sets in. The radiation-dominated universe starts
with a temperature $T_{\nul\gamma}(t_\mathrm{rad})\approx10^{12}\,\mathrm{K}$
and lasts until matter-energy equality, which occurs at the redshift
$z(t_\mathrm{eq})=3176$. If we assume [see
Weinberg~\cite{c8}, (15.5.7) for an explanation] that during the time that
matter and radiation were in thermal contact the temperature of the radiation,
$T_{\nul\gamma}$, dropped according to the formula $T_{\nul\gamma}(t)=A/a(t)$,
where $A$ is a constant and $a(t)$ the scale factor, we have
\begin{equation}\label{eq:redshift-temp}
\dfrac{T_{\nul\gamma}(t)}{T_{\nul\gamma}(t_\mathrm{p})}=\dfrac{a(t_\mathrm{p})}{a(t)}.
\end{equation}
This relation follows also from the Einstein equations
in Section~\ref{sec:zero-order-rad}.
We suppose that matter and radiation are still
in thermal equilibrium at the end of the radiation-dominated era, i.e., at a
redshift of $z(t_\mathrm{eq})=3176$, yielding
$a(t_\mathrm{p})/a(t_{\mathrm{eq}})=3177$.
In this way we find from (\ref{eq:back-T-tp}) and (\ref{eq:redshift-temp})
\begin{equation}\label{eq:T-eq}
    T_{\nul\gamma}(t_\mathrm{eq})=8657\,\mathrm{K}.
\end{equation}
Hence, as a rough estimate we may conclude that the universe is dominated by
radiation in the temperature interval
\begin{equation}\label{eq:temp-interval-rad}
    10^{12}\,\mathrm{K} \ge T_{\nul\gamma} \ge 10^4\,\mathrm{K}.
\end{equation}
If we neglect, in the radiation-dominated era, the contribution of the electrons
and neutrinos, we have for the photon energy density $\varepsilon$
\begin{equation}\label{eq:state-rad}
    \varepsilon = a_\mathrm{B}T_{\nul\gamma}^4,
\end{equation}
where $a_\mathrm{B}$ is the black body constant
\begin{equation}\label{eq:blackbody-constant}
    a_\mathrm{B}=\dfrac{\pi^2 k_\mathrm{B}^4}{15\hbar^3c^3}=
    7.5658\times10^{-16}\,\mathrm{J}\, \mathrm{m}^{-3}\,\mathrm{K}^{-4},
\end{equation}
with $k_\mathrm{B}$ and $\hbar$ Boltzmann's and Planck's constant
respectively.
Furthermore, we could neglect the cosmological constant $\Lambda$,
(\ref{eq:Lambda-value}), with respect to the energy content of the universe at
the matter-radiation equality time $\kappa
a_\mathrm{B}T^4_{\nul\gamma}(t_\mathrm{eq})=1.0\times10^{-42}\,\mathrm{m}^{-2}$,
see equation (\ref{FRW3}). Finally, we consider a flat universe ($k=0$),
implying, with (\ref{spRicci}), that $\mbox{$^3\!R_\nul$}(t)=0$.

\subsubsection{Zeroth-order Equations} \label{sec:zero-order-rad}

In the radiation-dominated era, the contributions to the total energy
density due to baryons, $nm_{\text{H}}c^2$, can be
neglected, so that the equations of state read
\begin{equation}
\label{eq:rad-p-3e}
   \varepsilon(n,T)=a_{\mathrm{B}}T_\gamma^4, \quad
        p(n,T)=\tfrac{1}{3}a_{\mathrm{B}}T_\gamma^4.
\end{equation}
Moreover, we can put $\Lambda=0$ for reasons mentioned at the end of
Section~\ref{zoe}\@. The zeroth-order equations  (\ref{subeq:einstein-flrw}) then
reduce to
\begin{subequations}
\label{subeq:rad-R0}
\begin{align}
     H^2 & = \tfrac{1}{3}\kappa\varepsilon_\nul, \label{con}\\
    \dot{\varepsilon}_\nul & = -4H
         \varepsilon_\nul,  \label{eq:behoud} \\
    \dot{n}_\nul & = -3Hn_\nul.  \label{eq:n-behoud}
\end{align}
\end{subequations}
These equations can easily be solved, and we get the well-known results
\begin{subequations}
\label{subeq:rad-R0-sol}
\begin{align}
   H(t) & = \tfrac{1}{2} \left(ct\right)^{-1}=
      H(t_\mathrm{rad})\left(\frac{t}{t_\mathrm{rad}}\right)^{-1}, 
\label{sol1a} \\
   \varepsilon_{\scriptscriptstyle(0)}(t) & =
       \frac{3}{4\kappa} \left(ct\right)^{-2}=
\varepsilon_\nul(t_\mathrm{rad})\left(\frac{t}{t_\mathrm{rad}}\right)^{-2}, 
\label{sol1c}  \\
   n_\nul(t) & = n_\nul(t_\mathrm{rad})
\left(\frac{a(t)}{a(t_\mathrm{rad})}\right)^{-3}. \label{sol1d}
\end{align}
\end{subequations}
The initial values $H(t_\mathrm{rad})$ and $\varepsilon_\nul(t_\mathrm{rad})$ are related
by the constraint equation (\ref{con}) taken at $t_\mathrm{rad}$, the time at
which the radiation-dominated era sets in.

Using the definition of the Hubble function $H \equiv \dot{a}/a$ we find
from (\ref{sol1a}) that
\begin{equation}\label{sol1b}
   a(t) = a(t_\mathrm{rad})
\left(\frac{t}{t_\mathrm{rad}}\right)^{\tfrac{1}{2}}.
\end{equation}
Combining expressions (\ref{eq:state-rad}), (\ref{sol1c}) and (\ref{sol1b}), we
arrive at (\ref{eq:redshift-temp}), which can be rewritten in the form
\begin{equation}
  \label{eq:T-gamma-z}
  T_{\nul\gamma}(t)=T_{\nul\gamma}(t_{\text{p}}) \bigl[z(t)+1\bigr],
\end{equation}
where we have used (\ref{eq:redshift}).

With (\ref{subeq:rad-R0-sol})--(\ref{sol1b}) the coefficients $b_1$, $b_2$ and
$b_3$, defined by (\ref{subeq:coeff-contrast}), occurring in the equations
(\ref{subeq:final}) can be calculated. This will be the subject of the next
section.

\subsubsection{First-order Equations}

In view of (\ref{perttoes1}), we find from (\ref{eq:rad-p-3e})
\begin{equation}
  \label{eq:pn-pe-rad}
  p_n = 0, \quad p_\varepsilon=\tfrac{1}{3},
\end{equation}
so that, according to (\ref{begam2}),
we have $w=\tfrac{1}{3}$ and $\beta=1/\sqrt{3}$, see (\ref{eq:begam3}).
The zeroth-order solutions (\ref{subeq:rad-R0-sol}) can now be
substituted into the coefficients (\ref{subeq:coeff-contrast}) of
the equations (\ref{subeq:final}). Since $\vec{\tilde{\nabla}}=\vec{\nabla}$ for a flat
universe, the coefficients $b_1$, $b_2$ and $b_3$ reduce to
\begin{equation}
\label{subeq:coeff-contrast-flat}
  b_1 = -H, \quad
  b_2 = -\frac{1}{3}\frac{\nabla^2}{a^2}+
     \tfrac{2}{3}\kappa\varepsilon_\nul, \quad
  b_3  = 0,
\end{equation}
where we have used (\ref{con}). For the first-order
equations~(\ref{subeq:final}) this yields the simple forms
\begin{subequations}
\label{subeq:final-rad}
\begin{align}
   \ddot{\delta}_\varepsilon-H\dot{\delta}_\varepsilon-
  \left(\frac{1}{3}\frac{\nabla^2}{a^2}-
   \tfrac{2}{3}\kappa\varepsilon_\nul\right)
   \delta_\varepsilon & = 0,   \label{eq:delta-rad} \\
    \frac{1}{c}\frac{\dif}{\dif t}
        \left(\delta_n-\tfrac{3}{4}\delta_\varepsilon\right) & = 0,
            \label{eq:entropy-rad}
\end{align}
\end{subequations}
where $H\equiv\dot{a}/a$, (\ref{Hubble}). Since the right-hand side of
(\ref{eq:delta-rad}) vanishes, the evolution of density perturbations
is independent of the actual value of
$\delta_n-\tfrac{3}{4}\delta_\varepsilon$, see (\ref{subeq:final}) and
(\ref{eq:S-contrast}). In other words, the evolution of density
perturbations is not affected by perturbations in the entropy.
Equation (\ref{eq:entropy-rad}) implies that
$\delta_n-\tfrac{3}{4}\delta_\varepsilon$ is constant in
time. Consequently, during the radiation-dominated era, perturbations
in the particle number density are gravitationally coupled to
radiation perturbations:
\begin{equation}\label{eq:coupled-rad}
    \delta_n(t,\vec{x})-\tfrac{3}{4}\delta_\varepsilon(t,\vec{x})=
\delta_n(t_\mathrm{rad},\vec{x})-\tfrac{3}{4}\delta_\varepsilon(t_\mathrm{rad},
\vec{x}),
\end{equation}
where the right-hand side is constant with respect to time. From
(\ref{eq:S-contrast}) we find for the perturbation of the entropy per
particle:
\begin{equation}
  \label{eq:entropy-rad-constant}
  s_\een^\gi(t,\vec{x})=s_\een^\gi(t_{\text{rad}},\vec{x})=
  -\dfrac{a_{\text{B}}T^3_{\nul\gamma}(t_{\text{rad}})}{n_\nul(t_{\text{rad}})}
  \Bigl[ \delta_n(t_\mathrm{rad},\vec{x})-\tfrac{3}{4}\delta_\varepsilon(t_\mathrm{rad})\Bigr],
\end{equation}
where we have used that $T_{\nul\gamma}\propto a^{-1}$, $n_\nul\propto
a^{-3}$ and (\ref{eq:coupled-rad}). Thus, throughout the
radiation-dominated era the entropy per particle is constant with
respect to time. Since $n_\nul\neq0$ and $n_\een^\gi\neq0$, we have
$s^\gi_\een\not\equiv0$, see (\ref{eq:adia-cond}).

Equation (\ref{eq:delta-rad}) may be solved by Fourier expansion of
the function $\delta_\varepsilon$. Writing
\begin{equation}\label{pw12}
    \delta_\varepsilon(t,\vec{x}) =
    \delta_\varepsilon(t,\vec{q})\me^{\mi {\vecs{q}} \cdot {\vecs{x}}},
\end{equation}
with $q=|\vec{q}|=2\pi/\lambda$, where $\lambda$ is the wavelength
of the perturbation and ${\mi^2=-1}$, we find
\begin{equation}\label{Fourier12}
   \nabla^2\delta_\varepsilon(t,\vec{x}) =
       -q^2\delta_\varepsilon(t,\vec{q}),
\end{equation}
so that the evolution equation (\ref{eq:delta-rad}) for the
amplitude $\delta_\varepsilon(t,\vec{q})$ reads
\begin{equation}
  \ddot{\delta}_\varepsilon -
    H(t_\mathrm{rad})\left(\frac{t}{t_\mathrm{rad}}\right)^{-1}
\dot{\delta}_\varepsilon
+\left[\frac{1}{3}\frac{q^2}{a^2(t_\mathrm{rad})}\left(\frac{t}{t_\mathrm{rad}}
\right)^{-1} +
2H^2(t_\mathrm{rad})\left(\frac{t}{t_\mathrm{rad}}\right)^{-2}\right]
\delta_\varepsilon =0,  \label{delta-pnue}
\end{equation}
where we have used (\ref{con})--(\ref{subeq:rad-R0-sol}). This
equation will be rewritten in such a way that the coefficients
become dimensionless. To that end a dimensionless time variable is
introduced, defined by
\begin{equation}
   \tau \equiv \frac{t}{t_\mathrm{rad}}, \quad t \ge t_\mathrm{rad},  \label{tau}
\end{equation}
with $t_\mathrm{rad}$ the time immediately after the inflationary era.
This definition implies
\begin{equation}
   \frac{\dif^n}{c^n\dif
t^n}=\left(\frac{1}{ct_\mathrm{rad}}\right)^n\frac{\dif^n}{\dif\tau^n}=
   \left[2H(t_\mathrm{rad})\right]^n
   \frac{\dif^n}{\dif\tau^n}, \quad n=1,2,\ldots\,,  \label{dtau-n}
\end{equation}
where we have used (\ref{sol1a}). Using (\ref{sol1a}), (\ref{tau})
and (\ref{dtau-n}), equation (\ref{delta-pnue}) for the density
contrast $\delta_\varepsilon(\tau,\vec{q})$ can be written as
\begin{equation}
    \delta_\varepsilon^{\prime\prime} - \frac{1}{2\tau}\delta_\varepsilon^\prime+
  \left(\frac{\mu_\mathrm{r}^2}{4\tau} +
\frac{1}{2\tau^2}\right)\delta_\varepsilon=0,
         \label{delta-pnue-tau}
\end{equation}
where a prime denotes differentiation with respect to $\tau$. The
constant $\mu_\mathrm{r}$ is given by
\begin{equation}
     \mu_\mathrm{r} \equiv
\frac{q}{a(t_\mathrm{rad})}\frac{1}{H(t_\mathrm{rad})}\frac{1}{\sqrt{3}}\,.  
\label{xi}
\end{equation}
The general solution of equation (\ref{delta-pnue-tau}) is a linear
combination of the functions
$J_{\pm\frac{1}{2}}(\mu_\mathrm{r}\sqrt{\tau})\tau^{3/4}$, where
$J_{+\frac{1}{2}}(x)=\sqrt{2/(\pi x)}\sin x$ and
$J_{-\frac{1}{2}}(x)=\sqrt{2/(\pi x)}\cos x$ are Bessel functions of
the first kind:
\begin{equation}
 \delta_\varepsilon(\tau,\vec{q}) =
      \Bigl[A_1(\vec{q})\sin\left(\mu_\mathrm{r}\sqrt{\tau}\right) +
A_2(\vec{q})\cos\left(\mu_\mathrm{r}\sqrt{\tau}\right)\Bigr]\sqrt{\tau},
\label{nu13}
\end{equation}
where the functions $A_1(\vec{q})$ and $A_2(\vec{q})$ are given by
\begin{subequations}
\label{subeq:C1-C2}
\begin{align}
   A_1(\vec{q}) & = \delta_\varepsilon(t_\mathrm{rad},\vec{q})\sin\mu_\mathrm{r}
- \frac{\cos\mu_\mathrm{r}}{\mu_\mathrm{r}}
\left[\delta_\varepsilon(t_\mathrm{rad},\vec{q})-\frac{\dot{\delta}
_\varepsilon(t_\mathrm{rad},\vec{q})}{H(t_\mathrm{rad})}\right],
        \label{C1} \\
   A_2(\vec{q}) & = \delta_\varepsilon(t_\mathrm{rad},\vec{q})\cos\mu_\mathrm{r}
+ \frac{\sin\mu_\mathrm{r}}{\mu_\mathrm{r}}
\left[\delta_\varepsilon(t_\mathrm{rad},\vec{q})-
\frac{\dot{\delta}_\varepsilon(t_\mathrm{rad},\vec{q})}{H(t_\mathrm{rad})}\right],
      \label{C2}
\end{align}
\end{subequations}
where we have used that
\begin{equation}
     \delta_\varepsilon(t_\mathrm{rad},\vec{q}) =
     \delta_\varepsilon(\tau=1,\vec{q}), \quad
     \dot{\delta}_\varepsilon(t_\mathrm{rad},\vec{q}) =
         2H(t_\mathrm{rad}) \delta_\varepsilon^\prime(\tau=1,\vec{q}),
\end{equation}
as follows from (\ref{dtau-n}). The relative perturbations in the particle
number density, $\delta_n$, evolve by virtue of
(\ref{eq:coupled-rad}) also according to (\ref{nu13}).

We consider the contribution of the terms of the Fourier expansion  of the
energy density perturbation [see (\ref{pw12})] in two limiting cases, namely the
case of small $\lambda$ (large $q$) and the case of large $\lambda$ (small $q$).

For large-scale perturbations, ${\lambda\rightarrow\infty}$, the
magnitude of the wave vector $|\vec{q}|=2\pi/\lambda$ vanishes.
Writing ${\delta_\varepsilon(t)\equiv\delta_\varepsilon(t,q=0)}$ and
${\dot{\delta}_\varepsilon(t)\equiv\dot{\delta}_\varepsilon(t,q=0)}$,
we find from (\ref{xi})--(\ref{subeq:C1-C2}) that, for ${t\ge t_\mathrm{rad}}$,
\begin{equation}
    \delta_\varepsilon(t) = -\left[\delta_\varepsilon(t_\mathrm{rad})-
       \frac{\dot{\delta}_\varepsilon(t_\mathrm{rad})}{H(t_\mathrm{rad})}\right]
       \frac{t}{t_\mathrm{rad}}
    +\left[2\delta_\varepsilon(t_\mathrm{rad})
    - \frac{\dot{\delta}_\varepsilon(t_\mathrm{rad})}{H(t_\mathrm{rad})}\right]
     \left(\frac{t}{t_\mathrm{rad}}\right)^{\tfrac{1}{2}}. \label{delta-H-rad}
\end{equation}
It is seen that the energy density contrast has two contributions to
the growth rate, one proportional to $t$ and one proportional to
$t^{1/2}$. These have been found by a large number of authors. See
Lifshitz and Khalatnikov \cite{c15}, (8.11), Adams and Canuto
\cite{adams-canuto1975}, (4.5b), Olson \cite{olson1976}, page 329,
Peebles \cite{c11}, (86.20), Kolb and Turner \cite{kolb}, (9.121) and
Press and Vishniac \cite{C12}, (33). The precise factors of
proportionality, however, have not been published earlier. From the
first of them we may conclude, in particular, that large-scale
perturbations only grow if the initial growth rate is large enough,
i.e.,
\begin{equation}\label{eq:ls-grow}
    \dot{\delta}_\varepsilon(t_\mathrm{rad}) \ge
    \delta_\varepsilon(t_\mathrm{rad})H(t_\mathrm{rad}) \quad
    \Rightarrow \quad
    \delta^\prime_\varepsilon(\tau=1,\vec{q})\ge\tfrac{1}{2}
       \delta_\varepsilon(\tau=1,\vec{q}),
\end{equation}
otherwise the perturbations are decaying. For \textsc{cdm} and ordinary matter
the same growth rate $\delta_n\propto t$ is found in the literature for
super-horizon perturbations, see, for example, the textbook of
Padmanabhan~\cite{paddy1993}, Section~4.4. Thus, for large-scale perturbations,
our treatise corroborates the outcomes found in the literature on the subject.

We now come to the second case. In the small-scale limit ${\lambda\rightarrow0}$
(or, equivalently,
${|\vec{q}|\rightarrow\infty}$) we find, using
(\ref{xi})--(\ref{subeq:C1-C2}), that
\begin{equation}\label{dc-small}
   \delta_\varepsilon(t,\vec{q}) \approx
\delta_\varepsilon(t_\mathrm{rad},\vec{q})
\left(\frac{t}{t_\mathrm{rad}}\right)^{\tfrac{1}{2}}\cos\left[\mu_\mathrm{r}-
\mu_\mathrm{r}\left(\frac{t}{t_\mathrm{rad}}\right)^{\tfrac{1}{2}}\right].
\end{equation}
We see that in the limit of small $\lambda$, the contribution to the growth rate
is smaller than the leading term in the expression (\ref{delta-H-rad}).
Physically, this can be understood: on small scales, the pressure gradients
$|\vec{\nabla}p|\approx p/\lambda$ are much higher than on large scales.

In Section~\ref{sec:stan-th} we review the standard results. Comparing
our result (\ref{dc-small}) with the standard result (\ref{eq:peacock-sol})
found in the literature, we observe that we obtain oscillating solutions with an
increasing amplitude, whereas the standard equation (\ref{eq:delta-rad-peacock})
yields oscillating solutions (\ref{eq:peacock-sol}) with a decreasing amplitude.

The above calculated behavior of density perturbations in the
radiation-dominated universe is important for star formation in the
era after decoupling of matter and radiation: the oscillating growth
shows up in the cosmic background radiation as random fluctuations on
different scales and amplitudes (i.e., intensities in the background
radiation).  As we will show in Section~\ref{sec:star-for-res},
density perturbations yield massive stars.

\subsection{Plasma Era} 
\label{sec:plasma-era}

The so-called plasma era sets in at time $t_{\mathrm{eq}}$, when the energy
density of ordinary matter equals the energy density of radiation,
i.e., when $n_\nul(t_{\text{eq}})mc^2=a_{\text{B}}T_{\nul\gamma}^4(t_{\text{eq}})$, and ends at time $t_{\mathrm{dec}}$, the time
of decoupling of matter and radiation. In the plasma era the
matter-radiation mixture can be characterized by the equations of
state, see Kodama and Sasaki~\cite{kodama1984} Chapter~V,
\begin{equation}
  \label{eq:plasma-state} \varepsilon(n,T)=nmc^2 +
a_{\mathrm{B}} T_\gamma^4, \quad p(n,T)=\tfrac{1}{3} a_{\mathrm{B}}
T_\gamma^4,
\end{equation}
where we have not taken into account the contributions to the pressure
from ordinary matter, $m=m_{\text{H}}$, or \textsc{cdm}, $m=m_{\text{cdm}}$, since these contributions are
vanishingly small in comparison to the radiation energy density. Using
(\ref{subeq:final-pn-pe}), we find
\begin{equation}
  \label{eq:pn-pe-plasma} p_n = -\tfrac{1}{3}mc^2,
\quad p_\varepsilon=\tfrac{1}{3}.
\end{equation}
Substituting the expressions (\ref{eq:plasma-state})
and (\ref{eq:pn-pe-plasma}) into the entropy equation (\ref{fir-ord})
and integrating the resulting equation yields
\begin{equation}
  \label{eq:int-entropy-rad}
\delta_n(t,\vec{x})-\dfrac{\delta_\varepsilon(t,\vec{x})}{1+w(t)}=
     \left[\delta_n(t_{\mathrm{eq}},\vec{x})-\dfrac{\delta_\varepsilon(t_{\mathrm{eq}},\vec{x})}
      {1+w(t_{\mathrm{eq}})}\right] \exp\left[-\int_{ct_{\mathrm{eq}}}^{ct}
     \dfrac{H(\tau)n_\nul(\tau)mc^2}{n_\nul(\tau)mc^2+
    \tfrac{4}{3}a_{\mathrm{B}}T^4_{\nul\gamma}(\tau)}\dif\tau\right].
\end{equation}
It is well-known that ordinary matter perturbations are coupled to
perturbations in the radiation density.  This coupling is attributed
to the high radiation pressure in the radiation-dominated and plasma
eras.  In addition to this coupling, expression
(\ref{eq:int-entropy-rad}) shows that perturbations in the particle
number density, $\delta_n$, are also \emph{gravitationally} coupled to
perturbations in the total energy density, $\delta_\varepsilon$.  This
has considerable consequences for the growth of perturbations in
\textsc{cdm}. Since the energy density of \textsc{cdm} can be written
as $n_\nul m_{\text{cdm}}c^2$, (\ref{eq:int-entropy-rad}) implies that
also density perturbations in \textsc{cdm} are gravitationally coupled
to perturbations in the radiation energy density. Consequently,
\textsc{cdm} perturbations can, just as perturbations in ordinary
matter, start to grow only after decoupling. In other words,
\textsc{cdm} and ordinary matter behave gravitationally in exactly the
same way.  This may rule out \textsc{cdm} as a means to facilitate the
formation of structure in the universe. The same conclusion, on
different grounds, has also been reached by Nieuwenhuizen
\textit{et~al.}\@~\cite{2009arXiv0906.5087N}.

\subsection{Era after  Decoupling of Matter and Radiation}
\label{matter}

In the era after decoupling of matter and radiation, we have to distinguish
between the matter temperature and the radiation temperature. The radiation
temperature $T_{\nul\gamma}$ evolves as (\ref{eq:redshift-temp}), whereas the
matter temperature $T_\nul$ evolves according to (\ref{e-T3}).
Once protons and electrons recombine to yield hydrogen, the radiation pressure
becomes negligible, and the equations of state reduce to those of a
non-relativistic monatomic perfect gas [Weinberg \cite{c8},
equations (15.8.20) and (15.8.21)]
\begin{equation}
  \varepsilon(n,T) = nm_\mathrm{H}c^2+\tfrac{3}{2}nk_\mathrm{B}T, \quad
  p(n,T) = nk_{\mathrm{B}}T,    \label{state-mat}
\end{equation}
where $k_\mathrm{B}=1.3806504\times10^{-23}\,\mathrm{J}\,\mathrm{K}^{-1}$ is
Boltzmann's constant, $m_\mathrm{H}$ the mass of a proton, and $T$ the
temperature of the \emph{matter}. Since the energy density in
(\ref{state-mat}) is not of the form $\varepsilon=\varepsilon(n)$, see
Section~\ref{sec:ad-pert-uni}, density perturbations cannot be adiabatic.

\subsubsection{Zeroth-order Equations}

The maximum gas temperature occurs around time $t_\mathrm{dec}$ of the
decoupling of matter and radiation and is equal to the radiation
temperature $T_{\nul\gamma}(t_\mathrm{dec})$.  Using
(\ref{subeq:wmap})--(\ref{eq:redshift-temp}), we get for the
temperature at the time of decoupling
\begin{equation}\label{eq:decoup-temp}
    T_\nul(t_\mathrm{dec})=T_{\nul\gamma}(t_\mathrm{dec})=2976\,\mathrm{K}.
\end{equation}
Since the universe cools down during its expansion, it follows from
(\ref{state-mat}) that

\begin{equation}\label{eq:p-neg-e}
    \frac{p}{\varepsilon}\approx\frac{k_\mathrm{B}T_\nul(t)}{m_\mathrm{H}c^2} \le
       \frac{k_\mathrm{B}T_\nul(t_{\text{dec}})}{m_\mathrm{H}c^2}=
 2.73\times 10^{-10}, \quad t \ge t_{\text{dec}}.
\end{equation}
Hence, the pressure is negligible with respect to the energy
density. This implies that, to a good approximation,
$\varepsilon_{\scriptscriptstyle(0)}\pm p_{\scriptscriptstyle(0)}
\approx \varepsilon_{\scriptscriptstyle(0)}$ and
$\varepsilon_\nul\approx n_\nul m_\mathrm{H} c^2$. Hence, in an
unperturbed flat \textsc{flrw} universe the pressure can, in the
background equations, be neglected with respect to the energy
density. The above facts yield that the Einstein equations and
conservation laws (\ref{subeq:einstein-flrw}) for a flat \textsc{flrw}
universe reduce~to
\begin{subequations}
\label{subeq:matp0}
\begin{align}
  H^2 & = \tfrac{1}{3}\kappa\varepsilon_\nul, \label{mat3p0}\\
   \dot{\varepsilon}_\nul & = -3H\varepsilon_\nul, \label{mat2p0} \\
   \dot{n}_\nul & = -3Hn_\nul,   \label{mat4p0}
\end{align}
\end{subequations}
where we have put the cosmological constant $\Lambda$ equal to zero.
The solutions of (\ref{subeq:matp0}) are
\begin{subequations}
\label{subeq:matp0-sol}
\begin{align}
 H(t) & = \tfrac{2}{3} \left(ct\right)^{-1}=
   H(t_\mathrm{mat})\left(\frac{t}{t_\mathrm{mat}}\right)^{-1}, \label{matsol1a}
\\
   \varepsilon_\nul(t) & =
     \frac{4}{3\kappa} \left(ct\right)^{-2}=
     \varepsilon_\nul(t_\mathrm{mat})\left(\frac{t}{t_\mathrm{mat}}\right)^{-2},
     \label{matsol1c} \\
     n_\nul(t) & =
     n_\nul(t_\mathrm{mat})\left(\frac{a(t)}{a(t_\mathrm{mat})}\right)^{-3}.
\label{matsol1d}
\end{align}
\end{subequations}
The scale factor $a(t)$ after decoupling, however, has a time
dependence which differs from that of the radiation-dominated era (\ref{sol1b}).
From (\ref{matsol1a}) we find
\begin{equation}\label{matsol1b}
    a(t) = a(t_\mathrm{mat})
\left(\frac{t}{t_\mathrm{mat}}\right)^{\tfrac{2}{3}},
\end{equation}
where $t_{\text{mat}}$ is some initial time after decoupling of matter and
radiation:
\begin{equation}\label{eq:tmat}
   t_{\text{dec}} \le t_{\text{mat}} \le t_{\text{p}}.
\end{equation}
The initial values $H(t_\mathrm{mat})$ and $\varepsilon_\nul(t_\mathrm{mat})$
are related
by the constraint equation (\ref{mat3p0}) taken at $t=t_\mathrm{mat}$.
With (\ref{subeq:matp0-sol})--(\ref{matsol1b}) the coefficients
(\ref{subeq:coeff-contrast}) of the perturbation equations (\ref{subeq:final})
are known functions of time.

\subsubsection{First-order Equations}

We first remark that, in the study of the evolution of density
perturbations, we may not neglect the pressure with respect to the
energy density. The case of a pressureless perfect fluid is already
thoroughly discussed in Section~\ref{nrl} on the non-relativistic
limit. We neglect $k_{\text{B}}T_\nul/(m_{\text{H}}c^2)$ with respect
to terms of the order unity in the final expressions.

Using equations (\ref{subeq:final-pn-pe}) we find from the equations of
state~(\ref{state-mat})
\begin{equation}
   p_\varepsilon = \tfrac{2}{3}, \quad p_n=-\tfrac{2}{3} m_\mathrm{H}c^2,
     \label{pepn-mat}
\end{equation}
From (\ref{eq:begam3}) it follows that
\begin{equation}
  \label{eq:non-ad-speed}
  \beta(t)=\sqrt{\dfrac{2}{3}
   \left[1-\dfrac{m_{\mathrm{H}}c^2}{m_{\mathrm{H}}c^2+\tfrac{5}{2}k_{\mathrm{B}}T_\nul(t)}\right]}\approx
\sqrt{\dfrac{2}{3}\left[1-\left(1-\dfrac{5}{2}\dfrac{k_{\text{B}}T_\nul(t)}{m_{\text{H}}c^2}\right)\right]},
\end{equation}
where it is used that $k_{\mathrm{B}}T_\nul(t) \ll m_{\mathrm{H}}c^2$,
(\ref{eq:p-neg-e}).  Therefore, we have to a good approximation
\begin{equation}
   \beta(t) \approx \frac{v_\mathrm{s}(t)}{c}=\sqrt{\frac{5}{3}
        \frac{k_\mathrm{B}T_\nul(t)}{m_\mathrm{H}c^2}},
\label{coef-nu1}
\end{equation}
where $v_\mathrm{s}$ is the speed of sound. Differentiating
(\ref{coef-nu1}) with respect to time yields
\begin{equation}
   \frac{\dot{\beta}}{\beta} = \frac{\dot{T}_\nul}{2T_\nul}.
     \label{coef-a}
\end{equation}
For the time development of the matter temperature it is found
from (\ref{eq:evo-T0}) and~(\ref{state-mat}) that
\begin{equation}\label{eq:bd-b}
    \dot{T}_\nul=-2HT_\nul.
\end{equation}
Combining (\ref{coef-a}) and (\ref{eq:bd-b}) results in
 \begin{equation}
   \frac{\dot{\beta}}{\beta} = -H.
     \label{eq:db-b-H}
\end{equation}
For the evolution of the matter temperature it is found from
(\ref{eq:bd-b}) that
\begin{equation}
    T_\nul(t)= T_\nul(t_\mathrm{mat})
\left(\frac{a(t)}{a(t_\mathrm{mat})}\right)^{-2}=
   T_\nul(t_{\text{mat}}) \left(\dfrac{z(t)+1}{z(t_{\text{mat}})+1}\right)^2,
       \label{e-T3}
\end{equation}
where we have used that $H\equiv\dot{a}/a$ and (\ref{eq:redshift}).

We now consider equation (\ref{fir-ord}).
Using (\ref{pepn-mat}) it is found for this equation
\begin{equation}\label{eq:entropy-dust}
      \frac{1}{c}\frac{\dif}{\dif t}
        \left(\delta_n-\delta_\varepsilon\right)=
        -2H\left(\delta_n-\delta_\varepsilon\right).
\end{equation}
The general solution of equation (\ref{eq:entropy-dust}) is, using also
(\ref{matsol1b}),
\begin{equation}
  \delta_n(t,\vec{x})-\delta_\varepsilon(t,\vec{x})=
    \bigl[\delta_n(t_\mathrm{mat},\vec{x})-\delta_\varepsilon(t_\mathrm{mat},\vec{x})\bigr]
  \left(\dfrac{a(t)}{a(t_\mathrm{mat})}\right)^{-2}=
\bigl[\delta_n(t_\mathrm{mat},\vec{x})-\delta_\varepsilon(t_\mathrm{mat},\vec{x})\bigr]
    \left(\dfrac{t}{t_\mathrm{mat}}\right)^{-\tfrac{4}{3}}.
      \label{eq:entropy-dust-sol}
\end{equation}
We now calculate the perturbed counterpart of the equations of state
(\ref{state-mat}). For the first-order perturbation to the pressure we find
\begin{equation}
  \label{eq:pert-press}
  p^\gi_\een=n^\gi_\een k_{\text{B}}T_\nul + n_\nul k_{\text{B}}T^\gi_\een.
\end{equation}
Dividing this expression by $p_\nul$ yields the relative pressure perturbation
\begin{equation}
  \label{eq:rel-press-pert-matter}
  \delta_p\equiv\dfrac{p^\gi_\een}{p_\nul}=
   \dfrac{n^\gi_\een k_{\text{B}}T_\nul + n_\nul
     k_{\text{B}}T^\gi_\een}
       {n_\nul k_{\text{B}}T_\nul}=\delta_n+\delta_T.
\end{equation}
The first-order perturbation to the energy density reads
\begin{equation}
  \label{eq:pert-energy}
  \varepsilon^\gi_\een=n^\gi_\een m_{\text{H}}c^2+\tfrac{3}{2}n^\gi_\een k_{\text{B}}T_\nul+
       \tfrac{3}{2}n_\nul k_{\text{B}}T^\gi_\een.
\end{equation}
Dividing this expression by $\varepsilon_\nul$ one finds the energy
density contrast
\begin{equation}
  \label{eq:rel-energy-pert}
  \delta_\varepsilon\equiv\dfrac{\varepsilon^\gi_\een}{\varepsilon_\nul}=
      \dfrac{n^\gi_\een m_{\text{H}}c^2+\tfrac{3}{2}n^\gi_\een k_{\text{B}}T_\nul+
       \tfrac{3}{2}n_\nul k_{\text{B}}T^\gi_\een}
       {n_\nul m_{\text{H}}c^2+\tfrac{3}{2}n_\nul k_{\text{B}}T_\nul}.
\end{equation}
This expression can be rewritten in the form
\begin{equation}
  \label{eq:rel-pert-energy-exact}
  \delta_\varepsilon=\dfrac{\delta_n\left[1+\dfrac{3}{2}\dfrac{k_{\text{B}}T_\nul}{m_{\text{H}}c^2}\right]+
        \dfrac{3}{2}\dfrac{k_{\text{B}}T_\nul}{m_{\text{H}}c^2}\delta_T}
     {1+\dfrac{3}{2}\dfrac{k_{\text{B}}T_\nul}{m_{\text{H}}c^2}}=
  \delta_n+\dfrac{\dfrac{3}{2}\dfrac{k_{\text{B}}T_\nul}{m_{\text{H}}c^2}\delta_T}
     {1+\dfrac{3}{2}\dfrac{k_{\text{B}}T_\nul}{m_{\text{H}}c^2}}\approx
   \delta_n+\dfrac{3}{2}\dfrac{k_{\text{B}}T_\nul}{m_{\text{H}}c^2}
    \left(1-\dfrac{3}{2}\dfrac{k_{\text{B}}T_\nul}{m_{\text{H}}c^2}\right)\delta_T.
\end{equation}
Since $k_{\text{B}}T_\nul\ll m_{\text{H}}c^2$ expression
(\ref{eq:rel-pert-energy-exact}) can to a very good approximation be
written as
\begin{equation}
  \label{eq:rel-press-final}
  \delta_\varepsilon\approx\delta_n+\dfrac{3}{2}\dfrac{k_{\text{B}}T_\nul}
     {m_{\text{H}}c^2}\delta_T=
   \delta_n+\dfrac{9}{10}\dfrac{v_{\text{s}}^2}{c^2}\delta_T,
\end{equation}
where the second equality follows from (\ref{coef-nu1}).
Combining (\ref{eq:entropy-dust-sol}) and (\ref{eq:rel-press-final}),
we find that in the linear regime $\delta_T$ is constant to a very
good approximation, i.e.,
\begin{equation}
  \label{eq:delta-T-constant}
  \delta_T(t,\vec{x})\approx\delta_T(t_{\text{mat}},\vec{x}).
\end{equation}
The perturbed equations of state can now be written as
\begin{equation}
  \label{eq:pert-eq-state}
  \delta_n(t,\vec{x})-\delta_\varepsilon(t,\vec{x})\approx-\dfrac{3}{2}\dfrac{k_{\text{B}}T_\nul(t)}
     {m_{\text{H}}c^2}\delta_T(t_{\text{mat}},\vec{x}), \quad
      \delta_p(t,\vec{x})\approx\delta_n(t,\vec{x})+\delta_T(t_{\text{mat}},\vec{x}).
\end{equation}
These expressions are the perturbed counterparts of the equations of
state (\ref{state-mat}).  Finally, we calculate the heat exchange of a
perturbation with its environment.  Using (\ref{eq:S-contrast}), we
get
\begin{equation}
  \label{eq:heat-exch}
  T_\nul(t)s^\gi_\een(t,\vec{x})\approx\tfrac{1}{2}k_{\text{B}}T_\nul(t)
     \bigl[3\delta_T(t_{\text{mat}},\vec{x})-2\delta(t,\vec{x})\bigr].
\end{equation}
This concludes the discussion of equation (\ref{fir-ord}).

We now consider (\ref{sec-ord}). After substituting
(\ref{pepn-mat}), (\ref{coef-nu1}) and (\ref{eq:db-b-H}) into the
coefficients (\ref{subeq:coeff-contrast}) it is found that
\begin{equation}\label{eq:coeff-dust}
    b_1=3H, \quad b_2=-\tfrac{5}{6}\kappa\varepsilon_\nul-
      \frac{v_\mathrm{s}^2}{c^2}\frac{\nabla^2}{a^2}, \quad
      b_3=-\frac{2}{3}\frac{\nabla^2}{a^2},
\end{equation}
considering that for a flat universe $\tilde{\nabla}^2=\nabla^2$. In the
derivation of $b_3$ we have used that ${\beta^2\ll 1}$, as follows from
(\ref{eq:p-neg-e}) and (\ref{coef-nu1}).
For the evolution equation for density perturbations,
(\ref{sec-ord}), this results in the simple form
\begin{equation}\label{eq:delta-dust}
  \ddot{\delta}_\varepsilon + 3H\dot{\delta}_\varepsilon-
  \left(\frac{v_\mathrm{s}^2}{c^2}\frac{\nabla^2}{a^2}+
   \tfrac{5}{6}\kappa\varepsilon_\nul\right)
   \delta_\varepsilon=\dfrac{3}{5}
   \dfrac{v^2_{\text{s}}}{c^2}\dfrac{\nabla^2}{a^2}\delta_T(t_{\text{mat}},\vec{x}),
\end{equation}
where we have used the results (\ref{coef-nu1}) and (\ref{eq:pert-eq-state}).
Using (\ref{pw12}) and (\ref{Fourier12}) the evolution equation for the
amplitude $\delta_\varepsilon(t,\vec{q})$ can be rewritten as
\begin{equation}
  \ddot{\delta}_\varepsilon+
3H(t_\mathrm{mat})\left(\frac{t}{t_\mathrm{mat}}\right)^{-1}\dot{\delta}_\varepsilon
  + H^2(t_\mathrm{mat})\left[\mu_\mathrm{m}^2\left(\frac{t}{t_\mathrm{mat}}
\right)^{-\tfrac{8}{3}}-
  \frac{5}{2}\left(\frac{t}{t_\mathrm{mat}}\right)^{-2}
 \right]\delta_\varepsilon=
-\dfrac{3}{5}H^2(t_{\text{mat}})\mu^2_{\text{m}}\left(\dfrac{t}{t_\mathrm{mat}}
\right)^{-\tfrac{8}{3}}\delta_T(t_\mathrm{mat},\vec{q}),
 \label{eq:delta-dust-amp}
\end{equation}
where we have incorporated (\ref{subeq:matp0})--(\ref{matsol1b}),
(\ref{coef-nu1}) and (\ref{e-T3}). The constant $\mu_\mathrm{m}$ is
given by
\begin{equation}\label{eq:const-mu}
\mu_\mathrm{m}\equiv\frac{q}{a(t_\mathrm{mat})}\frac{1}{H(t_\mathrm{mat})}\frac{
v_\mathrm{s}(t_\mathrm{mat})}{c}, \quad
    v_\mathrm{s}(t_\mathrm{mat})=
\sqrt{\frac{5}{3}\frac{k_\mathrm{B}T_\nul(t_\mathrm{mat})}{m_\mathrm{H}}}.
\end{equation}
Using the dimensionless time variable
\begin{equation}\label{eq:tau-mat}
    \tau \equiv \dfrac{t}{t_\mathrm{mat}}, \quad
     t_{\text{mat}} \le t \le t_{\text{p}},
\end{equation}
it is found from (\ref{matsol1a}) that
\begin{equation}
   \frac{\dif^n}{c^n\dif
t^n}=\left(\frac{1}{ct_\mathrm{mat}}\right)^n\frac{\dif^n}{\dif\tau^n}=
   \left[\frac{3}{2}H(t_\mathrm{mat})\right]^n
   \frac{\dif^n}{\dif\tau^n}, \quad n=1,2,\ldots\,.    \label{dtau-n-dust}
\end{equation}
Using this expression, equation (\ref{eq:delta-dust-amp}) can be
rewritten in the form
\begin{equation}\label{eq:dust-dimless}
    \delta_\varepsilon^{\prime\prime}+\frac{2}{\tau}\delta_\varepsilon^\prime+
\left(\dfrac{4}{9}\dfrac{\mu_\mathrm{m}^2}{\tau^{8/3}}-\frac{10}{9\tau^2}
\right)\delta_\varepsilon=
-\dfrac{4}{15}\dfrac{\mu^2_\mathrm{m}}{\tau^{8/3}}
\delta_T(t_\mathrm{mat},\vec{q}),
\end{equation}
where we have also used (\ref{coef-nu1}) and
(\ref{eq:pert-eq-state}). In equation (\ref{eq:dust-dimless}), a prime
denotes differentiation with respect to~$\tau$.

It is of convenience for the numerical integration of equation
(\ref{eq:dust-dimless}) to express the dimensionless time variable
$\tau$,
(\ref{eq:tau-mat}), in the cosmological redshift $z(t)$. Using
(\ref{eq:redshift}) and (\ref{matsol1b}), we get
\begin{equation}
  \label{eq:tau-max}
  \tau=\left(\dfrac{a(t)}{a(t_{\text{mat}})}\right)^{\tfrac{3}{2}}=
    \left(\dfrac{z(t_{\text{mat}})+1}{z(t)+1}\right)^{\tfrac{3}{2}}.
\end{equation}
The integration will then be halted if either the time variable $\tau$
has reached the value $\tau_{\text{end}}$ for which $z=0$, i.e.,
\begin{equation}
  \label{eq:present-tau}
  \tau_{\text{end}}=\bigl[z(t_{\text{mat}})+1\bigr]^{3/2},
\end{equation}
or when $|\delta_\varepsilon(\tau,\vec{q})|=1$ has been reached for
$\tau<\tau_{\text{end}}$.  In Section~\ref{sec:star-for-res}, we solve
numerically the inhomogeneous equation (\ref{eq:dust-dimless}). First,
however, we consider the homogeneous part of the evolution equation
(\ref{eq:dust-dimless}).

\subsubsection{General Solution of the Evolution Equation}
\label{sec:eq-exact-sol}

In this section we calculate the exact solution of
equation~(\ref{eq:dust-dimless}). To that end, we replace the
independent variable $\tau$ in this equation by the new independent
variable 
\begin{equation}
  \label{eq:t-y}
  x(\tau) \equiv 2\mu_{\text{m}}\tau^{-1/3},
\end{equation}
so that
\begin{equation}
  \label{eq:diff-x}
  \dfrac{\dif}{\dif\tau}=-\tfrac{2}{3}\mu_{\text{m}}\tau^{-4/3}\dfrac{\dif}{\dif x},
  \qquad  \dfrac{\dif^2}{\dif\tau^2}=\tfrac{8}{9}\mu_{\text{m}}\tau^{-7/3}\dfrac{\dif}{\dif x}+
     \tfrac{4}{9}\mu_{\text{m}}^2\tau^{-8/3}\dfrac{\dif^2}{\dif x^2}.
\end{equation}
Using the variable (\ref{eq:t-y}), equation~(\ref{eq:dust-dimless})
can be rewritten in a much more tractable form
\begin{equation}
  \label{eq:tractable}
  \dfrac{\dif^2\delta_\varepsilon}{\dif x^2}-
      \dfrac{2}{x}\dfrac{\dif\delta_\varepsilon}{\dif x}+
      \left(1-\dfrac{10}{x^2}\right)\delta_\varepsilon=
      -\tfrac{3}{5}\delta_T(t_{\text{mat}},\vec{q}).
\end{equation}
The general solution of this equation is
\begin{equation}
  \label{eq:gen-sol-dust-Bessel}
  \delta_\varepsilon(x)=\left[A_1J_{+\tfrac{7}{2}}(x)+A_2J_{-\tfrac{7}{2}}(x)\right]x^{3/2}
      -\dfrac{3}{5}\left(1+\dfrac{10}{x^2}\right)\delta_T(t_{\text{mat}},\vec{q}),
\end{equation}
where $A_1$ and $A_2$ are the constants of integration and
$J_{\pm\nu}(x)$ are the Bessel functions of the first kind:
\begin{subequations}
\label{subeq:Bessel-72}
\begin{align}
  J_{+\tfrac{7}{2}}(x) & =\sqrt{\dfrac{2}{\pi}}\Bigl[(x^3-15x)\cos x -
  (6x^2-15)\sin x \Bigr]x^{-7/2}, \\
  J_{-\tfrac{7}{2}}(x) & =\sqrt{\dfrac{2}{\pi}}\Bigl[(x^3-15x)\sin x +
  (6x^2-15)\cos x \Bigr]x^{-7/2},
\end{align}
\end{subequations}
with the asymptotic expressions for $x\rightarrow0$
\begin{equation}
  \label{eq:prop-Bessel72}
  J_{+\tfrac{7}{2}}(x)\approx \sqrt{\dfrac{2}{\pi}}\dfrac{x^{7/2}}{105}, \qquad
     J_{-\tfrac{7}{2}}(x)\approx-\sqrt{\dfrac{2}{\pi}}\dfrac{15}{x^{7/2}}.
\end{equation}
Transforming back from $x$ to $\tau$, we arrive at the general
solution of equation~(\ref{eq:dust-dimless})
\begin{equation}\label{eq:matter-physical}
 \delta_\varepsilon(\tau,\vec{q}) =
    \left[B_1(\vec{q}) J_{+\tfrac{7}{2}}\bigl(2\mu_\mathrm{m}\tau^{-1/3}\bigr)+
       B_2(\vec{q})J_{-\tfrac{7}{2}}\bigl(2\mu_\mathrm{m}\tau^{-1/3}\bigr)\right]\tau^{-1/2}
      -\dfrac{3}{5}\left(1+
   \dfrac{5\tau^{2/3}}{2\mu_{\text{m}}^2}\right)\delta_T(t_{\text{mat}},\vec{q}),   
\end{equation}
where $B_1(\vec{q})$ and $B_2(\vec{q})$ are arbitrary functions. The
first two terms are the solution of the homogeneous equation and the
last term is the particular solution. The solution of the homogeneous
equation follows from $\delta_T(t_{\text{mat}},\vec{q})=0$ and has the property
\begin{equation}
  \label{eq:homo-de-dn}
   \delta_T(t,\vec{x})=0 \quad \Leftrightarrow
 \quad   \delta_n(t,\vec{x})=\delta_\varepsilon(t,\vec{x})=\delta_p(t,\vec{x}),
\end{equation}
where we have used (\ref{eq:delta-T-constant}) and
(\ref{eq:pert-eq-state}). Thus, if $\delta_\varepsilon=\delta_n$ then
pressure perturbations counteract the growth or decay of a density
perturbation in such a way that $\delta_n=\delta_p$ and temperature
perturbations in the matter do not occur. This is always the case in
the standard Newtonian perturbation theory given by equation
(\ref{eq:delta-dust-standard}), since in this theory one has
$\delta_n=\delta_\varepsilon$. In the literature about linear
cosmological perturbations, equation (\ref{eq:entropy-dust}) does not
exist, and the standard second-order evolution equation
(\ref{eq:delta-dust-standard}) is homogeneous. Since
$m_{\text{H}}c^2\gg k_{\text{B}}T_\nul$, one is forced to take
$\delta_n=\delta_\varepsilon$. It follows from (\ref{eq:heat-exch})
and (\ref{eq:homo-de-dn}) that, in this case, the heat transfer to a
perturbation is given by
\begin{equation}
  \label{eq:diabatic} T_\nul s^\gi_\een=-k_{\mathrm{B}}T_\nul\delta_n.
\end{equation}
This implies that a density perturbation with $\delta_n>0$ loses some
of its internal heat energy to its surroundings, so that it may
grow. The heat transfer $T_\nul s^\gi_\een$ from the perturbation to
its surroundings is very small. In contrast to the standard theory,
our treatise allows $\delta_n\neq\delta_\varepsilon$, which may result
in a larger heat loss and, hence, a faster growth rate. This will be
studied in Section~\ref{sec:star-for-res}\@.

The evolution of density perturbations can be studied by imposing
initial conditions $\delta_\varepsilon(t_{\text{mat}},\vec{q})$ and
$\dot{\delta}_\varepsilon(t_{\text{mat}},\vec{q})$ on the general solution (\ref{eq:matter-physical}). Since the
resulting expression is far too complicated, we investigate the
evolution of density perturbations by solving equation
(\ref{eq:dust-dimless}) numerically in
Section~\ref{sec:star-for-res}\@. In this section we only consider the
two limiting cases of large-scale and small-scale perturbations.

In the large-scale limit $|\vec{q}|\rightarrow0$, $\lambda\rightarrow\infty$ (i.e., larger than the horizon,
see Appendix~\ref{app:horizon}), it is found that, transforming back from $\tau$
to $t$,
\begin{equation}
    \delta_\varepsilon(t) =
    \frac{1}{7}\left[5\delta_\varepsilon(t_\mathrm{mat})+
       \dfrac{2\dot{\delta}_\varepsilon(t_{\text{mat}})}{H(t_{\text{mat}})}\right]
 \left(\frac{t}{t_\mathrm{mat}}\right)^{\tfrac{2}{3}}
  +\frac{2}{7}\left[\delta_\varepsilon(t_\mathrm{mat})-
    \dfrac{\dot{\delta}_\varepsilon(t_{\text{mat}})}{H(t_{\text{mat}})} \right]
      \left(\frac{t}{t_\mathrm{mat}}\right)^{-\tfrac{5}{3}}.
  \label{eq:new-dust-53-adiabatic}
\end{equation}
Thus, for large-scale perturbations, the initial value
$\delta_T(t_{\text{mat}},\vec{q})$ does not play a role during the
evolution: large-scale perturbations evolve only under the influence
of gravity.  The solution proportional to $t^{2/3}$ is a standard
result. Since $\delta_\varepsilon$ is gauge-invariant, the standard
non-physical gauge mode proportional to $t^{-1}$ is absent from our
theory. Instead, a physical mode proportional to $t^{-5/3}$ is found.
This mode has also been found by Bardeen~\cite{c13}, Table~I, and by
Mukhanov \emph{et~al.}~\cite{mfb1992}, expression (5.33). In order to
arrive at the $t^{-5/3}$ mode, Bardeen has to use the `uniform Hubble
constant gauge.' In our treatise, however, the Hubble function is
automatically uniform, without any additional gauge condition, see
(\ref{thetagi}).
  
In the small-scale limit $\lambda\rightarrow0$ or, equivalently,
$|\vec{q}|\rightarrow\infty$, we find, transforming back from $\tau$
to $t$,
\begin{equation}
  \label{eq:evo-small-case}
   \delta_\varepsilon(t,\vec{q})\approx-\tfrac{3}{5}\delta_T(t_{\text{mat}},\vec{q})+
   \left(\dfrac{t}{t_{\text{mat}}}\right)^{-\tfrac{1}{3}}
    \Bigl[\tfrac{3}{5}\delta_T(t_{\text{mat}},\vec{q})+\delta_\varepsilon(t_{\text{mat}},\vec{q})\Bigr]
   \cos\left[2\mu_{\text{m}}-2\mu_{\text{m}}\left(\dfrac{t}{t_{\text{mat}}}\right)^{-\tfrac{1}{3}}  \right].
\end{equation}
Thus, small-scale density perturbations oscillate with a decaying
amplitude which is smaller than unity so that the non-linear regime
will never be reached. 

For scales between the two extremes discussed above the growth of the perturbations can
be considerable as we will show in Section~\ref{sec:star-for-res}\@.

\section{Star Formation: Basic Equations}
\label{sec:galaxy}

This section is a preparation for the numerical solution of the
evolution equation (\ref{eq:dust-dimless}). We express the constant
$\mu_{\text{m}}$ and the star mass $M(t_{\text{mat}})$ in the
observable quantities $z(t_{\text{dec}})$, $\mathcal{H}(t_\mathrm{p})$
and $T_\nul(t_{\text{dec}})$, the initial redshift
$z(t_{\text{mat}})$, where $t_{\text{mat}}$ is some initial time
(\ref{eq:tmat}) at which a density perturbation starts to contract,
and the physical dimensions $\lambda_{\text{mat}}$ of a density
perturbation. Finally, we study the influence of the particle mass $m$
and the initial time $t_{\text{mat}}$ on the mass of a star.

\subsection{Initial Values and the Mass of a Star}

Writing $q=2\pi/\lambda$, where $\lambda a(t_\mathrm{mat})$ is the
physical scale of a perturbation at time $t_\mathrm{mat}$ and $\lambda
a(t_\mathrm{p})\equiv\lambda$ is the physical scale as measured at the
present time $t_\mathrm{p}$, we get for (\ref{eq:const-mu})
\begin{equation}\label{eq:nu-m-lambda-phys}
    \mu_\mathrm{m}=\dfrac{2\pi}{\lambda_\mathrm{mat}}
\dfrac{1}{H(t_\mathrm{mat})}
\sqrt{\dfrac{5}{3}\dfrac{k_{\mathrm{B}}T_\nul(t_{\mathrm{mat}})}{m_{\mathrm{H}}c^2}},
\quad \lambda_{\mathrm{mat}} \equiv \lambda a(t_{\mathrm{mat}}).
\end{equation}
The Hubble function $H(t_\mathrm{mat})$ follows from
(\ref{eq:redshift}), (\ref{matsol1a}) and (\ref{matsol1b}). We find
\begin{equation}
\label{eq:H-mat}
H(t_\mathrm{mat})=H(t_{\mathrm{p}})\bigl[z(t_{\mathrm{mat}})+1\bigr]^{3/2}.
\end{equation}
From (\ref{e-T3}), we get
\begin{equation}
  \label{eq:Tmat-z}
  T_\nul(t_{\text{mat}})= T_\nul(t_{\text{dec}})
  \left(\dfrac{z(t_{\text{mat}})+1}{z(t_{\text{dec}})+1}\right)^2.
\end{equation}
Using (\ref{eq:H-mat}) and (\ref{eq:Tmat-z}), expression
(\ref{eq:nu-m-lambda-phys}) can be rewritten as
\begin{equation}
  \label{eq:H-dec-wmap}
  \mu_{\mathrm{m}}=\dfrac{2\pi}{\lambda_{\mathrm{mat}}}
     \dfrac{1}{\mathcal{H}(t_\mathrm{p})\bigl[z(t_{\mathrm{dec}})+1\bigr]\sqrt{z(t_{\text{mat}})+1}}
\sqrt{\dfrac{5}{3}
    \dfrac{k_{\mathrm{B}}T_{\nul}(t_{\mathrm{dec}})}{m_{\mathrm{H}}}},
\end{equation}
where we have used (\ref{eq:Hubble-function}). With
(\ref{eq:H-dec-wmap}) we have expressed $\mu_{\mathrm{m}}$ in the
observable quantities $\mathcal{H}(t_{\mathrm{p}})$,
$z(t_{\mathrm{dec}})$ and $T_\nul(t_{\text{dec}})$. The matter
temperature just after decoupling, $T_\nul(t_{\mathrm{dec}})$, is
given by (\ref{eq:decoup-temp}). Using also (\ref{subeq:wmap}) we get
\begin{equation}\label{eq:nu-m-lambda}
    \mu_\mathrm{m}=\dfrac{512.0}{\lambda_{\mathrm{mat}}\sqrt{z(t_{\text{mat}})+1}}, \quad
\lambda_{\mathrm{mat}} \mbox{ in pc},
\end{equation}
where we have used that $1\,\mathrm{pc}=3.0857\times10^{16}\,\mathrm{m}$
($1\,\mathrm{pc}=3.2616\;\mathrm{ly}$).

The mass $M(t_{\mathrm{mat}})$ of a spherical density perturbation with
radius  $\tfrac{1}{2}\lambda_{\mathrm{mat}}$ is given by
\begin{equation}
  \label{eq:M-dec}
  M(t_{\mathrm{mat}})=
     \dfrac{4\pi}{3}\left(\tfrac{1}{2}\lambda_{\mathrm{mat}}\right)^3
     n_\nul(t_{\mathrm{mat}})m_{\mathrm{H}},
\end{equation}
where $m_{\mathrm{H}}$ is the proton mass.
The particle number density can be calculated from the value at the
end of the radiation-dominated era. 
By definition, at the end of the radiation-domination era the
matter energy density $n_\nul m_\mathrm{H}c^2$ equals the energy density of the radiation. Hence
\begin{equation}\label{eq:mat-en-eq}
n_\nul(t_\mathrm{eq})m_\mathrm{H}c^2=a_\mathrm{B}T_{\nul\gamma}^4(t_\mathrm{eq}).
\end{equation}
Since $n_\nul\propto a^{-3}$ and $T_{\nul\gamma}\propto a^{-1}$, we
find, using (\ref{eq:redshift}), the particle number density at
time $t_{\text{mat}}$
\begin{equation}
  \label{eq:n-nul-t-dec}
  n_\nul(t_{\mathrm{mat}})=
   \dfrac{a_{\mathrm{B}}T_{\nul\gamma}^4(t_{\mathrm{p}})}{m_{\mathrm{H}}c^2}
   \bigl[z(t_{\mathrm{eq}})+1\bigr]\bigl[z(t_{\mathrm{mat}})+1\bigr]^3.
\end{equation}
Combining (\ref{eq:M-dec}) and (\ref{eq:n-nul-t-dec}), we get
\begin{equation}
  \label{eq:M-dec-n-dec}
  M(t_{\text{mat}})=
  \dfrac{4\pi}{3}\left(\tfrac{1}{2}\lambda_{\mathrm{mat}}\right)^3
 \dfrac{a_{\mathrm{B}}T_{\nul\gamma}^4(t_{\mathrm{p}})}{c^2}
   \bigl[z(t_{\mathrm{eq}})+1\bigr]\bigl[z(t_{\mathrm{mat}})+1\bigr]^3.
\end{equation}
Using that one solar mass is $1.98892\times10^{30}\,\text{kg}$, we
find from (\ref{subeq:wmap})
\begin{equation}
  \label{eq:M-dec-solar}
  M(t_{\text{mat}})=1.141\times10^{-8}\lambda^3_{\text{mat}}
      \bigl[z(t_{\mathrm{mat}})+1\bigr]^3\,\text{M}_\odot, \quad
  \lambda_{\text{mat}} \text{ in pc}.
\end{equation}
This expression will be used to convert the scale
$\lambda_{\text{mat}}$ of a perturbation, which starts to contract at
a redshift of $z(t_{\text{mat}})$, into its mass. The latter is
expressed in units of the solar mass. It should be stressed here that
the numeric factors in expressions (\ref{eq:nu-m-lambda}) and
(\ref{eq:M-dec-solar}) hold true only for a fluid consisting of
protons. In Section~\ref{sec:ster-CDM} we investigate the influence of
the particle mass on the mass of a star.

In order to solve equation (\ref{eq:dust-dimless}), we need the
initial values $\delta_\varepsilon(t_{\text{mat}},\vec{q})$,
$\delta^\prime_\varepsilon(t_{\text{mat}},\vec{q})$ and
$\delta_T(t_{\text{mat}},\vec{q})$. From (\ref{eq:pert-eq-state}) it
follows that, since $|\delta_T(t,\vec{q})|\le1$, we must have
\begin{equation}
  \label{eq:heel-klein}
  |\delta_n(t,\vec{q})-\delta_\varepsilon(t,\vec{q})|\approx
   \tfrac{3}{2}|\delta_T(t_{\text{mat}},\vec{q})|
   \dfrac{k_{\mathrm{B}}T_\nul(t)}{m_{\mathrm{H}}c^2}\le
 \tfrac{3}{2}|\delta_T(t_{\text{mat}},\vec{q})|
   \dfrac{k_{\mathrm{B}}T_\nul(t_{\text{dec}})}{m_{\mathrm{H}}c^2}=
    4.10\times 10^{-10}|\delta_T(t_{\text{mat}},\vec{q})|,
\end{equation}
implying that
\begin{equation}
  \label{eq:bijna-gelijk}
  \delta_n(t,\vec{q})\approx\delta_\varepsilon(t,\vec{q}), \quad
  t \ge t_{\text{dec}}.
\end{equation}
We take, however,
$\delta_n(t_{\text{mat}},\vec{q})\neq\delta_\varepsilon(t_{\text{mat}},\vec{q})$. The
case
$\delta_n(t_{\text{mat}},\vec{q})=\delta_\varepsilon(t_{\text{mat}},\vec{q})$
has been discussed in Section~\ref{sec:eq-exact-sol} on the general
solution of the evolution equation (\ref{eq:dust-dimless}).

An initial value for the relative matter temperature perturbation
$\delta_T(t_{\text{mat}},\vec{q})$ can be found as follows.  A weak
condition for growth is that a perturbation must lose some of its
internal heat energy, i.e., $T_\nul s^\gi_\een<0$ or, equivalently. A
larger growth can be achieved if we take the initial values such that
the heat loss of a density perturbation is larger than given by
(\ref{eq:diabatic}), i.e.,
\begin{equation}
  \label{eq:heat-loss-more}
  -k_{\text{B}}T_\nul(t)\delta_n(t,\vec{q}) >
    \tfrac{1}{2}k_{\text{B}}T_\nul(t)
    \bigl[3\delta_T(t_{\text{mat}},\vec{q})-2\delta_n(t,\vec{q})\bigr]
    \quad \Rightarrow
     \quad \delta_T(t_{\text{mat}},\vec{q})<0,
\end{equation}
where we have used the combined First and Second Laws of
thermodynamics in the form (\ref{eq:heat-exch}). The criterion
(\ref{eq:heat-loss-more}), yields a positive source term of the
evolution equation (\ref{eq:dust-dimless}) and, therefore, a larger
growth than the homogeneous equation.

Finally, there are no observations of the growth rates of density
perturbations. Therefore, we take the initial growth rates equal to
zero, i.e.,
\begin{equation}
  \label{eq:zero-init-rate}
    \delta^\prime_\varepsilon(t_{\text{mat}},\vec{q})=
\delta^\prime_n(t_{\text{mat}},\vec{q})=0, \quad
t_{\text{dec}} \le t_{\text{mat}} \le t_{\text{p}}.
\end{equation}

In the next two subsections we investigate the influence of both the
particle mass and the initial time of star formation on the mass of a
particular star. In Section~\ref{sec:star-for-res} we present the
results of our new perturbation theory for structure formation.

\subsection{The Influence of  the Particle  Mass on the Mass of  a Star}

\label{sec:ster-CDM}

Up till now we have assumed that the matter content of the universe
after decoupling consists of protons. However, \textsc{wmap}
observations suggest that there may exist a considerable amount of, as
yet unknown, dark matter (\textsc{dm}) particles.  In this section we
investigate the influence of the mean particle mass on the evolution
of density perturbations.  We assume that after decoupling, a gas
consisting of particles with mean mass $\tilde{m}$, given by
($\alpha(t)>0$)
\begin{equation}
  \label{eq:cdm-mass-H}
  \tilde{m}=\alpha(t_{\text{mat}}) m_{\text{H}},
\end{equation}
can be described by an equation of state of the form (\ref{state-mat})
with the mass $m_{\text{H}}$ replaced by (\ref{eq:cdm-mass-H}).  As a
crude estimate for $\tilde{m}$, we take the mean mass of the baryons
and the \textsc{dm} particles, i.e.,
\begin{equation}
  \label{eq:mean-mass}
  \alpha(t)=\dfrac{\Omega_{\text{bar}}(t)m_{\text{H}}+\Omega_{\text{dm}}(t)m_{\text{dm}}}
    {\bigl[\Omega_{\text{bar}}(t)+\Omega_{\text{dm}}(t)\bigr]m_{\text{H}}},
\end{equation}
where $\Omega_{\text{bar}}$ and $\Omega_{\text{dm}}$ are the baryon and
dark matter particle densities respectively, given in units of the
critical density, see the
Friedmann equation (\ref{eq:Friedmann-Omega}).

By $\tilde{\mu}_{\text{m}}$, we denote the parameter $\mu_{\text{m}}$
(\ref{eq:H-dec-wmap}) in which $m_{\text{H}}$ is replaced by
$\alpha(t_{\text{mat}}) m_{\text{H}}$.  Density perturbations in a gas
with mean particle mass $\tilde{m}$ evolve in exactly the same way as
perturbations in a gas of which the particles have mass $m_{\text{H}}$
if $\mu_{\text{m}}=\tilde{\mu}_{\text{m}}$: in this case we have
$\delta_\varepsilon(t,\vec{q})=\tilde{\delta}_\varepsilon(t,\vec{q})$. From
(\ref{eq:H-dec-wmap}) it follows that
\begin{equation}
  \label{eq:lambda-cdm-m}
  \tilde{\lambda}_{\text{mat}}=\dfrac{\lambda_{\text{mat}}}{\sqrt{\alpha(t_{\text{mat}})}},
\end{equation}
where $\tilde{\lambda}_{\text{mat}}$ is the scale of a density
perturbation in a gas with mean particle mass $\tilde{m}$.  Using
(\ref{eq:M-dec-n-dec}) we find
\begin{equation}
  \label{eq:mass-cdm}
  \tilde{M}(t_{\text{mat}})=M(t_{\text{mat}}) \alpha^{-3/2}(t_{\text{mat}}),
\end{equation}
where $\tilde{M}(t_{\text{mat}})$ refers to a perturbation mass in a
gas with mean particle mass $\tilde{m}$. In the derivation of
(\ref{eq:mass-cdm}) we have assumed that the right-hand side of
(\ref{eq:mat-en-eq}) is independent of the particle mass, i.e., the
total mass of the universe is independent of the particle mass.  We
thus have found that perturbations in a gas consisting of particles
which are heavier (i.e., $\alpha(t_{\text{mat}})>1$) than protons the
collapse takes place at a smaller total mass than perturbations in
ordinary matter. In other words, heavier particles yield lighter
stars. We will illustrate (\ref{eq:mass-cdm}) with two extreme cases,
namely a mixture of heavy \textsc{wimp}s (\textsc{cdm}) with a mass of
$m_{\text{cdm}}\approx10\,m_{\text{H}}$~\cite{collaboration-2009} and
baryons, and a mixture of ordinary matter and \emph{hot dark matter}
(\textsc{hdm}) with a mass of $m_{\text{hdm}}=1.5\,\text{eV}/c^2$, as
suggested by Nieuwenhuizen
\textit{et~al.}\@~\cite{2009arXiv0906.5087N,2009EL.....8659001N}. Using
that $m_{\text{H}}=0.938\,\text{GeV}/c^2$, we can calculate with the
help of (\ref{eq:Om-La-H}) and (\ref{eq:mean-mass}) the mean particle
mass and, hence, $\alpha$. Assuming that
$\Omega_{\text{bar}}(t)/\Omega_{\text{dm}}(t)\approx
\Omega_{\text{bar}}(t_{\text{p}})/\Omega_{\text{dm}}(t_{\text{p}})$
for both types of particles and for $t>t_{\text{dec}}$, we find
$\alpha(t)\approx\alpha(t_{\text{p}})$, so that
$\alpha^{-3/2}_{\text{cdm}}\approx4.1\times10^{-2}$ and
$\alpha^{-3/2}_{\text{hdm}}\approx14$.  This implies that if the dark
matter particles are heavy \textsc{wimp}s then the stars formed are
much lighter than the stars formed in a universe filled with baryons
only. On the other hand, if light neutrinos are the dark matter
(\textsc{hdm}), then the stars formed will be heavier.

\subsection{The Influence of Initial Time on the Mass of a Star}
\label{sec:star-late}
In this section we show that density fluctuations which start to
contract at late times yield stars with smaller masses than early
density perturbations. To show this, we consider the evolution
equation (\ref{eq:dust-dimless}).

It follows from equation (\ref{eq:dust-dimless}) that perturbations
starting to grow at $t_{\text{mat}}>t_{\text{dec}}$ and obeying the same
initial conditions as fluctuations starting to contract at $t_{\text{dec}}$, i.e.,
\begin{equation}
  \label{eq:gelijke-beginw}
  \delta_\varepsilon(t_{\text{mat}},\vec{x}) =\delta_\varepsilon(t_{\text{dec}},\vec{x}),\quad
  \delta^\prime_\varepsilon(t_{\text{mat}},\vec{x}) =\delta^\prime_\varepsilon(t_{\text{dec}},\vec{x}),\quad
  \delta_T(t_{\text{mat}},\vec{x}) =\delta_T(t_{\text{dec}},\vec{x}),
\end{equation}
evolve in exactly the same way, provided that
\begin{equation}
  \label{eq:mumat-mudec}
 \mu_{\text{m}}(t_{\text{mat}})=\mu_{\text{m}}(t_{\text{dec}}), 
\end{equation}
where we have written $\mu_{\text{m}}(t_0)$ to denote the value of the
parameter $\mu_{\text{m}}$ taken at time $t=t_0$.  Using the relation
(\ref{eq:mumat-mudec}) we can relate the masses of density
perturbations starting at different times.  Replacing the initial time
$t_{\text{mat}}$ in equation (\ref{eq:dust-dimless}) by the initial
time $t_{\text{dec}}$, we get for the parameter $\mu_{\mathrm{m}}$,
(\ref{eq:H-dec-wmap}), at $t=t_{\text{dec}}$
\begin{equation}
  \label{eq:star-mu-m-obs}
    \mu_{\mathrm{m}}(t_{\text{dec}})=\dfrac{2\pi}{\lambda_{\text{dec}}}
     \dfrac{1}{\mathcal{H}(t_\mathrm{p})\bigl[z(t_{\mathrm{dec}})+1\bigr]^{3/2}}
\sqrt{\dfrac{5}{3}
    \dfrac{k_{\mathrm{B}}T_{\nul}(t_{\mathrm{dec}})}{m_{\mathrm{H}}}}.
\end{equation}
Equating (\ref{eq:H-dec-wmap}) and
(\ref{eq:star-mu-m-obs}), we arrive at
\begin{equation}
  \label{eq:lambda-0-lambda-dec}
  \lambda_{\text{mat}}=\lambda_{\text{dec}}\sqrt{\dfrac{z(t_{\text{dec}})+1}{z(t_{\text{mat}})+1}},
\end{equation}
i.e., the evolution of a density perturbation starting at
$t=t_{\text{dec}}$ with scale $\lambda_{\text{dec}}$ is exactly equal
to the evolution of a density perturbation starting at $t=t_{\text{mat}}$ with
scale $\lambda_{\text{mat}}$, if and only if the relation between the scales
obeys (\ref{eq:lambda-0-lambda-dec}).
Since $n_\nul\propto a^{-3}\propto(z+1)^3$, we have
\begin{equation}
  \label{eq:n-0-zt}
  n_\nul(t_{\text{mat}})=n_\nul(t_{\text{dec}})\left(\dfrac{z(t_{\text{mat}})+1}
     {z(t_{\text{dec}})+1} \right)^3.
\end{equation}
Using these expressions and (\ref{eq:M-dec}), we finally arrive at
\begin{equation}
  \label{eq:Mt0-Mtdec}
    M(t_{\text{mat}})=M(t_{\text{dec}})\left(\dfrac{z(t_{\text{mat}})+1}
         {z(t_{\text{dec}})+1} \right)^{\tfrac{3}{2}}.
\end{equation}
This expression relates the masses of density perturbations, which
start to evolve at different times $t_\text{dec}$ and $t_{\text{mat}}$
with the same initial values (\ref{eq:gelijke-beginw}) and
(\ref{eq:mumat-mudec}). Consequently, stars formed from late time
fluctuations have smaller masses.

\section{Star Formation: Results}
\label{sec:star-for-res}

The standard cosmological theory of small perturbations is
characterized by $\delta_n=\delta_\varepsilon$,
(\ref{eq:homo-de-dn}). This is, however, too restrictive. Although
$|\delta_n-\delta_\varepsilon|$ is very small, it need not be zero, as
follows from (\ref{eq:heel-klein}). Since $m_{\mathrm{H}}c^2\gg
k_{\mathrm{B}}T_\nul$, it follows from (\ref{eq:pert-eq-state}) that
the quantities $\delta_T$, $\delta_p$ and $\delta_n$ are not,
beforehand, confined to small values: small changes in
$\Delta\equiv\delta_n-\delta_\varepsilon$ may lead to large changes in
$\delta_T$, $\delta_p$ and $\delta_n$. These quantities must only
fulfill the linearity conditions $|\delta_T|\le1$, $|\delta_p|\le1$
and $|\delta_n|\le1$. The fact that $\delta_n$ need not be exactly
equal to $\delta_\varepsilon$ is paramount for the formation of
structure in the universe.

In this section we will solve the evolution equation
(\ref{eq:dust-dimless}) numerically. To that end we use the
differential equation solver \texttt{lsodar} with root finding
capabilities, included in the package \texttt{deSolve}, which, in
turn, is included in~\textsf{R}, a system for statistical computation
and graphics~\cite{R}. We investigate star formation which starts at
cosmological redshifts $z=1091$ and $z=1$.

\subsection{Star Formation starting at $z=1091$: Population III Stars}
\label{sec:star-decoup}

At the moment of decoupling of matter and radiation photons could not ionize
matter any more and the two constituents fell out of thermal
equilibrium. As a consequence, the pressure drops from a very high
radiation pressure $p=\tfrac{1}{3}a_{\mathrm{B}}T^4_\gamma$
just before decoupling to a very low gas pressure
$p=nk_{\mathrm{B}}T$ after decoupling. This fast transition from a
high pressure epoch to a very low pressure era may result in large
relative pressure perturbations. In this subsection we study the
influence of these relative pressure perturbations on the formation of stars. 

\subsubsection{Initial Values}
\label{sec:init-val-1091}

In order to integrate equation (\ref{eq:dust-dimless}), we need
initial values.  The initial value
$\delta_\varepsilon(t_{\text{dec}},\vec{q})$ is related to the
relative perturbation $\delta_{T_\gamma}(t_{\text{dec}},\vec{q})$ in
the \emph{background radiation}. Using (\ref{eq:rel-T-pert}) and
(\ref{eq:plasma-state}), we get at the end of the plasma era
\begin{equation}\label{eq:temp-eps-fluct}
   \delta_{T_\gamma}(t_{\text{dec}},\vec{q})= \dfrac{1}{4}
  \left[\dfrac{n_\nul(t_{\text{dec}})m_{\text{H}}c^2}
      {a_{\text{B}}T^4_{\nul\gamma}(t_{\text{dec}})}
      \bigl[\delta_\varepsilon(t_{\text{dec}},\vec{q})-\delta_n(t_{\text{dec}},\vec{q})\bigr]+
  \delta_\varepsilon(t_{\text{dec}},\vec{q})\right].
\end{equation}
Using that $n_\nul\propto a^{-3}$ and $T_{\nul\gamma}\propto a^{-1}$,
we get from (\ref{eq:redshift}) and (\ref{eq:mat-en-eq})
\begin{equation}
  \label{eq:plasma-end}
  n_\nul(t_{\text{dec}})m_{\text{H}}c^2=a_{\text{B}}T^4_{\nul\gamma}(t_{\text{dec}})
  \dfrac{z(t_{\text{eq}})+1}{z(t_{\text{dec}})+1},
\end{equation}
so that (\ref{eq:temp-eps-fluct}) can be rewritten as
\begin{equation}
  \label{eq:temp-fluct-zz}
    \delta_{T_\gamma}(t_{\text{dec}},\vec{q})= \dfrac{1}{4}
  \left[\dfrac{z(t_{\text{eq}})+1}{z(t_{\text{dec}})+1}
      \bigl[\delta_\varepsilon(t_{\text{dec}},\vec{q})-\delta_n(t_{\text{dec}},\vec{q})\bigr]+
  \delta_\varepsilon(t_{\text{dec}},\vec{q})\right]. 
\end{equation}
In order to eliminate $\delta_n(t_{\text{dec}},\vec{q})$, we use that
at the end of the plasma era we have, according to
(\ref{eq:int-entropy-rad}),
\begin{equation}
  \label{eq:dn-de-nul}
  \delta_n(t_{\text{dec}},\vec{q})-
     \dfrac{\delta_\varepsilon(t_{\text{dec}},\vec{q})}
     {1+w(t_{\text{dec}})}\approx0,
\end{equation}
where, using (\ref{eq:plasma-state}) and (\ref{eq:plasma-end}),
\begin{equation}
  \label{eq:w-end-plasma}
  w(t_{\text{dec}})\equiv\dfrac{p_\nul(t_{\text{dec}})}
          {\varepsilon_\nul(t_{\text{dec}})}=
\dfrac{z(t_{\text{dec}})+1}{3z(t_{\text{eq}})+3z(t_{\text{dec}})+6}\approx0.085.
\end{equation}
Combining (\ref{eq:dn-de-nul}) and (\ref{eq:w-end-plasma}), we find
\begin{equation}
  \label{eq:dn-de-gelijk}
  \delta_n(t_{\text{dec}},\vec{q})=\delta_\varepsilon(t_{\text{dec}},\vec{q})
\dfrac{3z(t_{\text{eq}})+3z(t_{\text{dec}})+6}{3z(t_{\text{eq}})+4z(t_{\text{dec}})+7}
\approx0.92\;\delta_\varepsilon(t_{\text{dec}},\vec{q}).
\end{equation}
We can now rewrite (\ref{eq:temp-fluct-zz}) as
\begin{equation}
  \label{eq:delta-eps-num}
   \delta_{T_\gamma}(t_{\text{dec}},\vec{q})=
   \delta_\varepsilon(t_{\text{dec}},\vec{q})
   \dfrac{z(t_{\text{eq}})+z(t_{\text{dec}})+2}{3z(t_{\text{eq}})+4z(t_{\text{dec}})+7}\approx
       0.31\;\delta_\varepsilon(t_{\text{dec}},\vec{q}).
\end{equation}
From the \textsc{wmap} observation (\ref{eq:background-rad}) we find,
using (\ref{eq:delta-eps-num}),
\begin{equation}
  \label{eq:initval-delta-dot-delta}
  \delta_\varepsilon(t_{\text{dec}}, \vec{q})\approx 3.3\times10^{-5}, \quad 
 \delta^\prime_\varepsilon(t_{\text{dec}}, \vec{q})\approx0,
\end{equation}
where we have assumed, for lack of observations, that during the
decoupling of matter and radiation the growth rate $\delta^\prime_\varepsilon$ is very small, see
(\ref{eq:zero-init-rate}). With (\ref{eq:initval-delta-dot-delta}) the
condition (\ref{eq:bijna-gelijk}) implies that
\begin{equation}
  \label{eq:init-delta-n}
  \delta_n(t_{\text{dec}},\vec{q})\approx3.3\times10^{-5}, \quad
   \delta^{\prime}_n(t_{\text{dec}},\vec{q})\approx0,
\end{equation}
i.e., \textsc{wmap} observations demand that, just after decoupling,
also the relative particle number density perturbations are very
small.

\begin{center}
\begin{figure}
\includegraphics[scale=0.75]{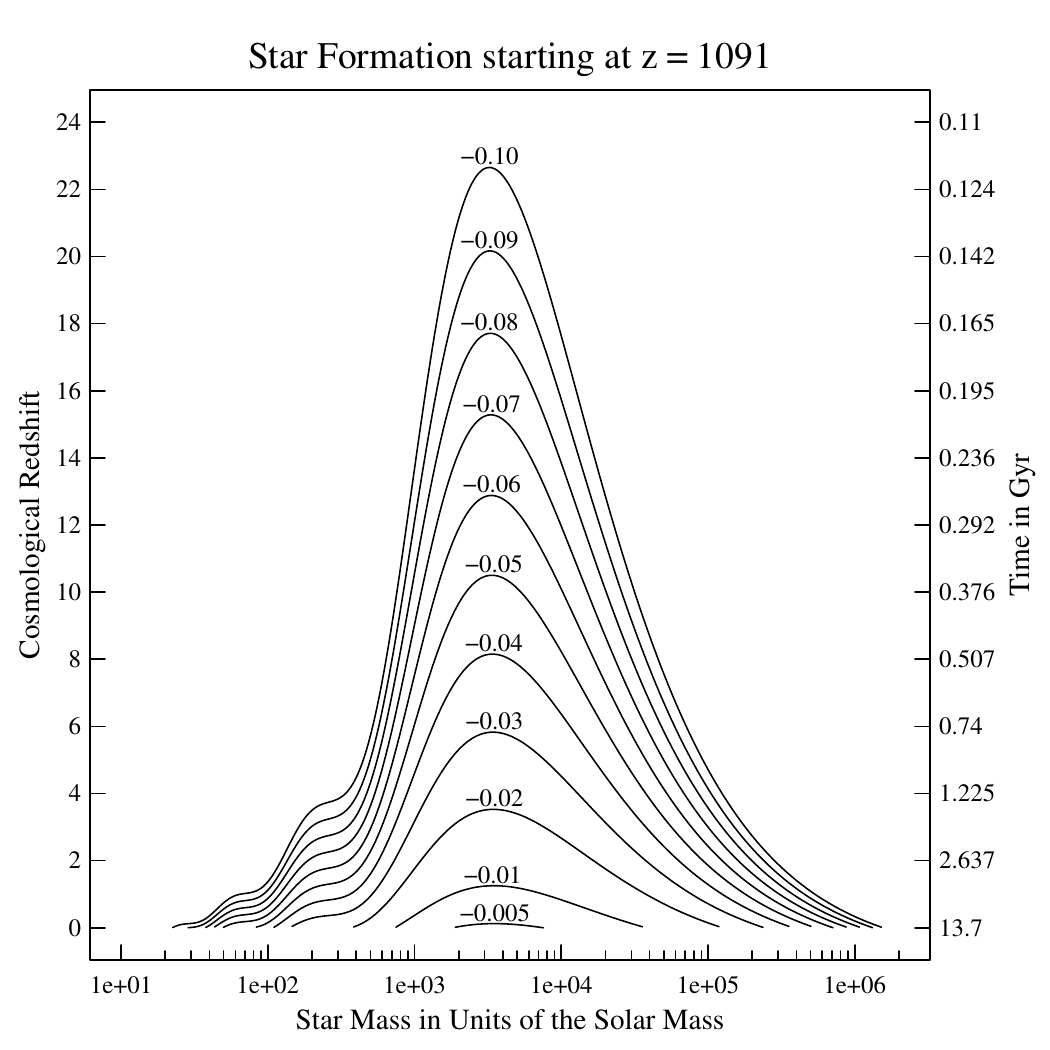}
\caption{The curves give the redshift at which a linear perturbation
  in the particle number density with initial values
  $\delta_n(t_\text{dec},\vec{q})\approx10^{-5}$ and
  $\delta^{\prime}_n(t_{\text{dec}},\vec{q})=0$ starting to grow at an
  initial redshift of $z(t_{\mathrm{dec}})=1091$ becomes non-linear,
  i.e., $\delta_n\approx1$. During the evolution we have
  $\delta_p(t,\vec{q})=\delta_T(t_{\text{dec}},\vec{q})+\delta_n(t,\vec{q})$.
  The numbers at each of the curves are the initial relative
  perturbations in the matter temperature
  $\delta_T(t_{\mathrm{dec}},\vec{q})$. For each curve, the maximum is
  at $3.4\times10^3\,\mathrm{M}_\odot$. If \textsc{cdm} is present
  then the peak mass is at $4.5\times10^2\,\mathrm{M}_\odot$.}
\label{fig:collapse}
\end{figure}
\end{center}

As we have shown, the relative matter temperature perturbation
$\delta_T$ is very nearly constant for a contracting or expanding
density perturbation. From the fact that the pressure is given by
$p=nk_{\text{B}}T$ and (\ref{eq:init-delta-n}) it follows that large
pressure perturbations can, just after decoupling, only be realized by
large matter temperature perturbations.  In our calculations we take,
therefore, $|\delta_T(t_{\mathrm{dec}},\vec{x})|$ in the range
$0.5\%$--$10\%$. With (\ref{eq:pert-eq-state}) and (\ref{eq:init-delta-n})
this implies that, initially,
\begin{equation}
  \label{eq:delta-p-delta-T}
\delta_p(t_{\text{dec}},\vec{q})\approx\delta_T(t_{\text{dec}},\vec{q}).
\end{equation} 
In other words, the perturbation in the pressure is,
initially, mainly determined by a perturbation in the matter
temperature. In view of (\ref{eq:heat-loss-more}) and
(\ref{eq:init-delta-n}) we have to take for the initial value of the
relative matter temperature perturbation
\begin{equation}
  \label{eq:dp-negative}
  \delta_T(t_{\text{dec}},\vec{q})<0,
\end{equation}
in order to find growing density perturbations. We now have gathered
the necessary ingredients to integrate the evolution equation
(\ref{eq:dust-dimless}) numerically.

\subsubsection{Results}
\label{sec:result-1091}

Figure~\ref{fig:collapse} has been constructed \cite{R} as follows.  For
each choice of $\delta_T(t_{\text{dec}},\vec{q})$ we integrate
equation (\ref{eq:dust-dimless}) for a large number of values for
$\lambda_{\text{dec}}$, using the initial values
(\ref{eq:initval-delta-dot-delta}) and (\ref{eq:init-delta-n}). The
integration starts at $\tau\equiv t/t_{\text{dec}}=1$, i.e., at
$z=z(t_{\text{dec}})$ and will be halted if either $z=0$ [i.e.,
$\tau=\tau_{\text{end}}$, (\ref{eq:present-tau})] or
$\delta_\varepsilon(t,\vec{q})=1$ has been reached. One integration
run yields one point on the curve for a particular choice of
$\lambda_{\text{dec}}$ if $\delta_\varepsilon(t,\vec{q})=1$ has been
reached for $z>0$. If the integration halts at $z=0$ and
$\delta_\varepsilon(t,\vec{q})<1$, then the
perturbation belonging to that particular $\lambda_{\text{dec}}$ has not yet
reached its non-linear phase today, i.e., at $t=13.7\,\text{Gyr}$. On
the other hand, if the integration is stopped at
$\delta_\varepsilon(t,\vec{q})=1$, then the perturbation has become
non-linear within $13.7\,\text{Gyr}$. In Figure~\ref{fig:collapse} we
have used the star mass $M(t_{\text{dec}})$, expressed in solar
masses, instead of the scale $\lambda_{\text{dec}}$ of a density
perturbation. To that end we have used expression
(\ref{eq:M-dec-solar}).

\begin{figure}
\begin{center}
\includegraphics[scale=0.75]{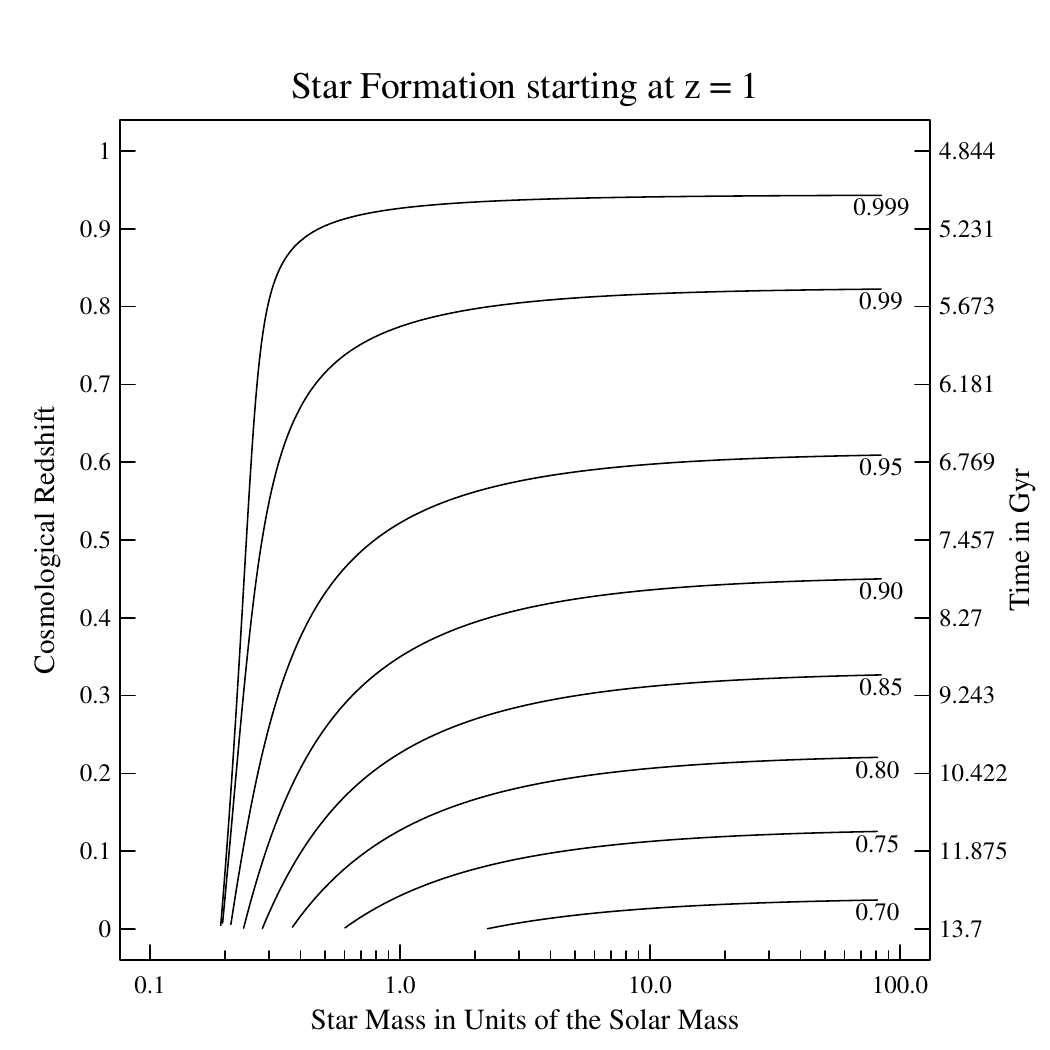}
\caption{Star formation for $\delta_T(t,\vec{q})=-1$.  The curves give
  the redshift at which a linear density perturbation starting to grow
  at an initial redshift of $z(t_{\mathrm{mat}})=1$ becomes
  non-linear, i.e., $\delta_n\approx\delta_\varepsilon\approx1$. The
  numbers at each of the curves are the initial relative density
  perturbations $\delta_n(t_{\text{mat}},\vec{q})\approx
  \delta_\varepsilon(t_{\text{mat}},\vec{q})$. During the evolution we
  have $\delta_p(t,\vec{q})=-1+\delta_n(t,\vec{q})$.  For masses in
  excess of $100\,\text{M}_\odot$, the growth becomes independent of
  the scale of a perturbation, i.e., the growth is proportional to
  $t^{2/3}$.}
\label{fig:collapse-10}
\end{center}
\end{figure}

The above described procedure is repeated for
$\delta_T(t_{\text{dec}},\vec{q})$ in the range $-0.005, -0.01, -0.02,
\ldots, -0.1$. During the evolution, the relative pressure
perturbation evolves as
\begin{equation}
  \label{eq:evo-const-T}
  \delta_p(t,\vec{q})=\delta_T(t_{\text{dec}},\vec{q})+\delta_n(t,\vec{q}).
\end{equation}
The fastest growth is seen for perturbations with a mass
of approximately $3.4\times10^3\,\text{M}_\odot$. This value is nearly
independent of the initial value of the matter temperature perturbation
$\delta_T(t_{\text{dec}},\vec{q})$. Even density perturbations with an
initial matter temperature perturbation as small as
$\delta_T(t_{\text{dec}},\vec{q})=-0.5\%$ reach their non-linear phase
at $z=0.13$ ($T_{\nul\gamma}=3.1\,\text{K}$, $t=11.5\,\text{Gyr}$)
provided that its mass is around $3.4\times10^3\,\text{M}_\odot$.
Perturbations with masses smaller than $3.4\times10^3\,\text{M}_\odot$ reach
their non-linear phase at a later time, because their internal gravity
is weaker. On the other hand, perturbations with masses larger than
$3.4\times10^3\,\text{M}_\odot$ have larger scales so that they cool down
slower, resulting also in a smaller growth rate.  Since the growth
rate decreases rapidly for perturbations with masses below
$3.4\times10^3\,\text{M}_\odot$, the latter may be considered as the
relativistic counterpart of the \emph{Jeans mass}.

Figure~\ref{fig:collapse} has been calculated for a baryonic cosmic
fluid, without \textsc{cdm} or \textsc{hdm}. If \textsc{cdm} is
present, then it follows from (\ref{eq:mass-cdm}) that the graphs
shift towards smaller masses such that the Jeans masses are at
$140\,\text{M}_\odot$. For \textsc{hdm} the peaks are at
$4.8\times10^4\,\text{M}_\odot$.

Finally, using (\ref{eq:M-dec-solar}), we find that all star masses in
Figure~\ref{fig:collapse} start to contract at decoupling from density
perturbations with diameters less than $52\,\text{pc}$, which is much
smaller than the particle horizon size (Appendix~\ref{app:horizon})
$d_{\text{H}}(t_{\text{dec}})=349\,\text{kpc}$.

\begin{figure}
\begin{center}
\includegraphics[scale=0.75]{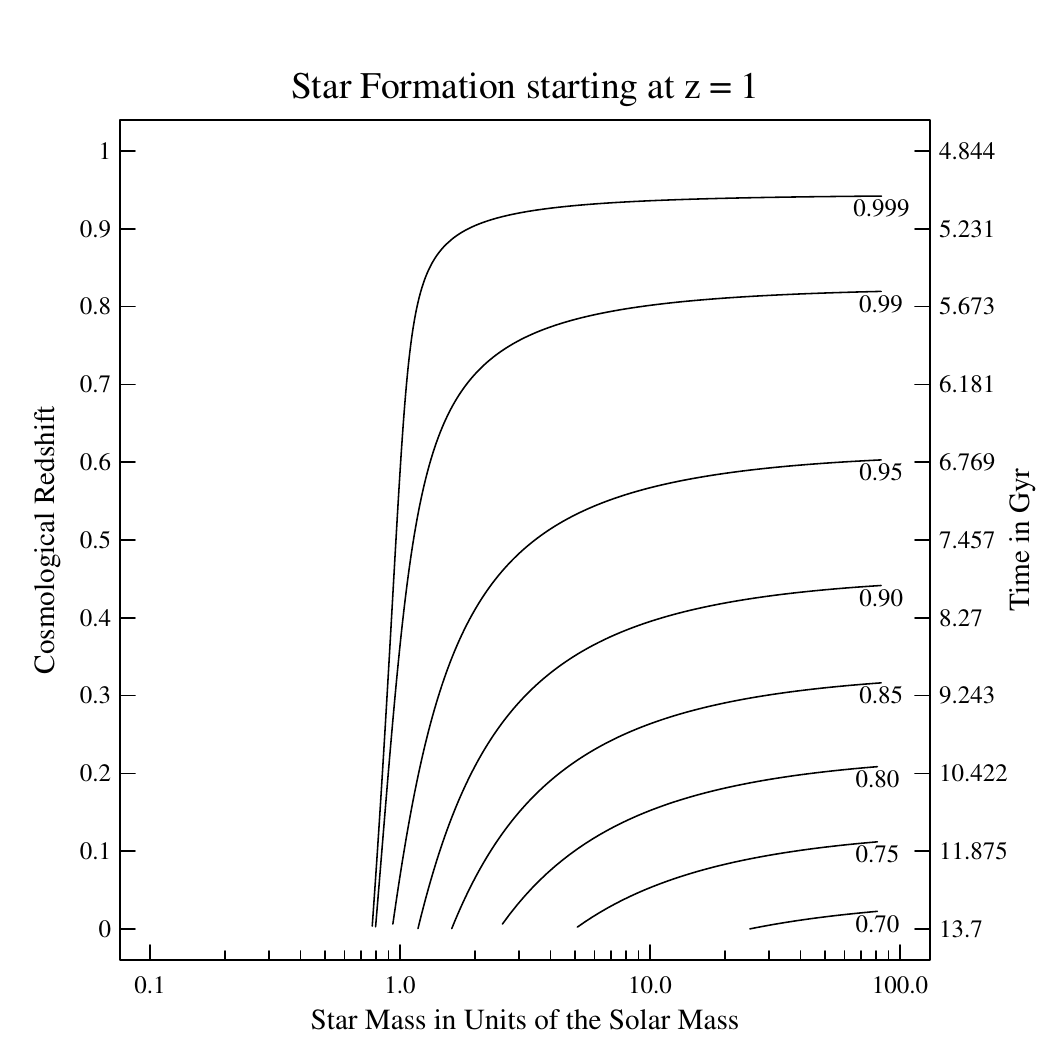}
\caption{Star formation for $\delta_T(t,\vec{q})=0$. The curves give
  the redshift at which a linear density perturbation starting to grow
  at an initial redshift of $z(t_{\mathrm{mat}})=1$ becomes
  non-linear, i.e., $\delta_n=\delta_\varepsilon\approx1$.  The
  numbers at each of the curves are the initial relative perturbations
  $\delta_n(t_{\mathrm{mat}},\vec{q})=
  \delta_\varepsilon(t_{\text{mat}},\vec{q})$.  During the evolution
  we have
  $\delta_p(t,\vec{q})=\delta_n(t,\vec{q})=\delta_\varepsilon(t,\vec{q})$. For
  masses in excess of $100\,\text{M}_\odot$, the growth becomes
  independent of the scale of a perturbation, i.e., the growth is
  proportional to $t^{2/3}$.}
\label{fig:collapse-2}
\end{center}
\end{figure}

\subsection{Star Formation starting at $z=1$}
\label{sec:star-z10}

At $z=1$ the interstellar gas has been diluted so much that star
formation is not possible anymore in the interstellar gas with only
the negative matter temperature perturbation
$\delta_T(t_{\text{mat}},\vec{q})$ as the driving force. Consequently,
late time star formation can only take place in regions which have a
higher density with respect to the intergalactic space. Therefore,
late time star formation takes place mainly within galaxies.  As can
be seen from Figures~\ref{fig:collapse-10} and~\ref{fig:collapse-2}, the
initial density perturbation should be at least of the order of
$\delta_n\approx\delta_\varepsilon\approx0.70$, in order to yield
eventually a gravitational collapse for $z>0$.

We have considered two extreme cases of star formation starting at
$z(t_{\text{mat}})=1$ or, equivalently,
$t_{\text{mat}}\approx4.8\,\text{Gyr}$. In both cases, we assume that
the initial growth rate vanishes, (\ref{eq:zero-init-rate}).  In the
first case, depicted in Figure~\ref{fig:collapse-10}, we have chosen
$\delta_T(t_{\text{mat}},\vec{q})=-1$, the smallest value that a
relative perturbation can have without violating the linearity
conditions. During the evolution of a density perturbation, we have,
according to (\ref{eq:pert-eq-state}),
\begin{equation}
  \label{eq:deltaT-1}
     \delta_p(t,\vec{q})=-1+\delta_n(t,\vec{q}),
\end{equation}
so that during the evolution the relative pressure perturbation is
negative. This is the most favorable situation for a perturbation to
grow: pressure cooperates with gravitation. The most conspicuous
feature in Figure~\ref{fig:collapse-10} is the sharp lower limit of
star formation. Density perturbations with masses below
$0.2\,\text{M}_\odot$, do not become non-linear before
$z=0$ is reached. Apparently, the gravitational field is, for masses below
$0.2\,\text{M}_\odot$, too weak to collapse in due time and,
eventually, become a star. Another notable characteristic of star
formation is that it is slow: even if the initial density perturbation
is as high as $\delta_n(t_{\text{mat}},\vec{q})=0.999$ it takes, for a
$100\,\text{M}_\odot$ perturbation, approximately $214\,\text{Myr}$ to
reach the value $\delta_n(t,\vec{q})=1$.

The second case, characterized by $\delta_T(t_{\text{mat}},\vec{q})=0$
and summarized in Figure~\ref{fig:collapse-2}, is the most detrimental
for star formation, because during the evolution of a density perturbation we
have
\begin{equation}
  \label{eq:deltaT-0}
   \delta_p(t,\vec{q})=\delta_n(t,\vec{q}),
\end{equation}
so that the pressure opposes the contraction of a density
perturbation. However, if the internal gravity of a perturbation is
strong enough, then gravity will overcome the pressure and the
perturbation will eventually collapse to form a star. Due to the
counteracting pressure, the lower limit of star formation is much
larger than in the case of cooperating pressure, namely
$0.8\,\text{M}_\odot$.

Both extreme cases have one common characteristic. For perturbations
with masses larger than $100\,\text{M}_\odot$, the collapse time is nearly the
same, as can be seen from Figures~\ref{fig:collapse-10}
and~\ref{fig:collapse-2}. Apparently, the internal gravitational field
is so strong that opposing or cooperating pressure perturbations do not
play a role anymore.

\section{Standard Newtonian Theory of Cosmological Perturbations} 
\label{sec:stan-th}

The new evolution equations (\ref{eq:delta-rad}) and (\ref{eq:delta-dust})
are different from their standard counterparts
(\ref{eq:delta-rad-peacock}) and (\ref{eq:delta-dust-standard})
respectively. In this section we explain the differences and we show
why equations (\ref{eq:delta-rad-peacock}) and
(\ref{eq:delta-dust-standard}) should not be used anymore in the study
of cosmological density perturbations.

Padmanabhan had already in $1993$ the supposition that the Newtonian theory of
cosmological density perturbations is questionable. On page~$136$ of his
textbook~\cite{paddy1993} he states:
\begin{quote}
  To avoid any misunderstanding, we emphasize the following fact: it
  is not possible to study cosmology using Newtonian gravity without
  invoking procedures which are mathematically dubious.
\end{quote}
Padmanabhan does not conclude, however, that the Newtonian theory of
cosmological perturbations is incorrect, since he subsequently states:
\begin{quote}
  However, if we are only interested in regions much smaller than the
  characteristic length scale set by the curvature of space-time, then
  one can introduce a valid approximation to general relativity.
\end{quote}
Up till now, this point of view is considered as standard
knowledge. This is due to the fact that the standard equation
(\ref{eq:delta-dust-standard}) of the Newtonian theory for small-scale
perturbations is, in the low velocity limit
$v_\mathrm{s}/c\rightarrow0$, similar to the relativistic equation
(\ref{eq:delta-standard}) derived from the General Theory of
Relativity. This similarity has led to much confusion in the
literature (see, for example,
\cite{Hwang-Noh-2005,hwang-noh-1997,noh-hwang-2005a,lima-1997-291,dent-2008}),
as we will now explain. First, we remark that gauge problems, although
time and space coordinates may be chosen freely according to
(\ref{eq:gauge-trans-newt}), do \emph{not} occur in the Newtonian
theory of gravity.  This is because in the Newtonian theory itself the
universe is \emph{static}, i.e., $\dot{\varepsilon}_\nul=0$ and
$\dot{n}_\nul=0$, implying with (\ref{subeq:split-e-n-nrl}) that both
$\varepsilon_\een$ and $n_\een$ are independent of the choice of a
system of reference. Now, the reasoning in the literature is as
follows. The quantity $\delta\equiv\varepsilon_\een/\varepsilon_\nul$
occurring in the Newtonian equation (\ref{eq:delta-dust-standard}) is,
according to the standard knowledge, independent of the gauge choice,
since this equation is derived from the Newtonian theory, in which
gauge problems concerning perturbed scalar quantities do not
occur. Furthermore, the Newtonian equation
(\ref{eq:delta-dust-standard}) is valid only for density perturbations
with scales smaller than the horizon size. By virtue of the
resemblance of the Newtonian equation (\ref{eq:delta-dust-standard})
and the relativistic equation (\ref{eq:delta-standard}) it is,
therefore, put forward that
\begin{quote}
  a gauge dependent quantity, such as
  $\delta\equiv\varepsilon_\een/\varepsilon_\nul$, which is initially
  larger than the horizon, becomes automatically gauge-invariant as
  soon as the perturbation becomes smaller than the horizon.
\end{quote}
This viewpoint is, however, incorrect as we will now show in detail.
Firstly, we remark that the universe \emph{cannot be static} in the
non-relativistic \emph{limit} (see Section~\ref{nrl}) of the General
Theory of Relativity, since $H\rightarrow0$ violates the Einstein
equations (\ref{subeq:einstein-flrw-newt}).  Secondly, it has been
shown in Section~\ref{nrl} that there remains some gauge freedom in
the non-relativistic limit, namely the freedom to shift time
coordinates, $x^0\rightarrow x^0-\psi$, and the freedom to choose
spatial coordinates, $x^i\rightarrow x^i-\chi^i(\vec{x})$. This
coordinate freedom is a well-known and natural property of Newtonian
physics and, therefore, should follow from a relativistic perturbation
theory in the low velocity limit $v_{\text{s}}/c\rightarrow0$.
Thirdly, we will show that, just because of the resemblance of the
relativistic and Newtonian equation, the gauge function $\psi$ also
occurs in the solution (\ref{eq:matter-non-physical}) of the Newtonian
equation (\ref{eq:delta-dust-standard}). Consequently, the standard
equation of the Newtonian theory has no physical significance: due to
the appearance of the gauge function $\psi$ in one of the two
independent solutions, one cannot impose initial conditions to arrive
at a physical solution. From the fact that, in the non-relativistic
limit of the General Theory of Relativity, the universe is not static,
combined with the freedom in time coordinates implies that if a
quantity, such as $\delta\equiv\varepsilon_\een/\varepsilon_\nul$, is
gauge dependent in the General Theory of Relativity it is also gauge
dependent in the non-relativistic limit, as follows from
(\ref{subeq:split-e-n-nrl}). The important conclusion is, therefore,
that \emph{a quantity which has no physical significance outside the
  horizon, does not become a physical quantity inside the
  horizon}. This conclusion is consistent with the facts that, in the
non-relativistic limit of our perturbation theory, the gauge-invariant
quantities $\varepsilon^\gi_\een$, (\ref{poisson}), and $n^\gi_\een$,
(\ref{eq:n-eps-gi}), survive, while the gauge dependent quantities
$\varepsilon_\een$ and $n_\een$ disappear from the scene, as we have
shown in detail in Section~\ref{nrl}\@. Hence, with
$\varepsilon^\gi_\een$ and $n^\gi_\een$ surviving in the
non-relativistic limit, there is indeed no gauge problem in the
Newtonian theory of gravity.

The fact that linear perturbation theory is plagued by the gauge solutions
(\ref{eq:gauge-rad}) and (\ref{eq:gauge-matter}) has already been pointed
out by Lifschitz~\cite{c15,lifshitz1946,I.12} as early as $1946$. In
1980, thirty-four years later, Press and Vishniac \cite{C12} called attention to
the same issue. In spite of these warnings, the standard equations
(\ref{eq:delta-rad-peacock}) and (\ref{eq:delta-dust-standard}) are still
ubiquitous in the cosmological literature. Apparently, the
cosmological gauge problem is quite persevering.

In the next two subsections, we will elucidate the gauge problem. We
go one step further than Lifschitz and Press and Vishniac. These
researchers have shown that only for large scales the solutions of the
standard equations are infected by spurious gauge modes. We show, in
addition, that the solutions (\ref{eq:peacock-sol}) and
(\ref{eq:matter-non-physical}) of the standard equations
(\ref{eq:delta-rad-peacock}) and (\ref{eq:delta-dust-standard})
contain the gauge function $\psi(\vec{x})$, \emph{independent} of the
scale of a perturbation. Consequently,
the important conclusion must be that
\begin{quote}
  the Newtonian theory of gravity is \emph{not} suitable to study the
  evolution of cosmological density perturbations.
\end{quote}
In order to show that the Newtonian theory of cosmological
perturbations is invalid, we consider a flat ($k=0$) \textsc{flrw}
universe with a vanishing cosmological constant ($\Lambda=0$) in the
radiation-dominated era and the era after decoupling of matter and
radiation.

\subsection{Radiation-dominated Universe} 
\label{sec:rad-stand}

The standard equation for the density contrast function $\delta$
which can be found, for example, in the textbook of Peacock
\cite{peacock1999}, equation (15.25), is given by
\begin{equation}\label{eq:delta-rad-peacock}
  \ddot{\delta}+2H\dot{\delta}-
  \left(\frac{1}{3}\frac{\nabla^2}{a^2}+
   \tfrac{4}{3}\kappa\varepsilon_\nul\right) \delta=0.
\end{equation}
Compare this equation with equation (\ref{eq:delta-rad}) for the
gauge-invariant contrast function $\delta_\varepsilon$. Equation
(\ref{eq:delta-rad-peacock}) is derived by using special relativistic
fluid mechanics and the Newtonian theory of gravity with a
relativistic source term. In agreement with the text under equation
(15.25) of this textbook, the term $-\frac{1}{3}\nabla^2\delta/a^2$
has been added. The same result, equation
(\ref{eq:delta-rad-peacock}), can be found in Weinberg's classic
\cite{c8}, equation (15.10.57) with $p=\frac{1}{3}\rho$ and
$v_\mathrm{s}=1/\sqrt{3}$. Note, that equation
(\ref{eq:delta-rad-peacock}) cannot be derived from the General Theory
of Relativity.

Using (\ref{subeq:rad-R0-sol}), (\ref{sol1b}), (\ref{pw12}),
(\ref{tau}) and (\ref{dtau-n}), we can rewrite equation
(\ref{eq:delta-rad-peacock}) in the form
\begin{equation}
  \label{eq:fout-rad}
  \delta^{\prime\prime}+\dfrac{1}{\tau}\delta^\prime+
   \left(\dfrac{\mu^2_{\text{r}}}{4\tau}-\dfrac{1}{\tau^2}\right)\delta=0,
\end{equation}
where the constant $\mu_\mathrm{r}$ is given by (\ref{xi}). A prime
denotes differentiation with respect to $\tau$, (\ref{tau}).  The
general solution of this equation is found to be
\begin{equation}\label{eq:peacock-sol}
    \delta(\tau,\vec{q})=\frac{8C_1(\vec{q})}{\mu_\mathrm{r}^2}
            J_2\bigl(\mu_\mathrm{r}\sqrt{\tau}\bigr)+
           \psi(\vec{q})\pi \mu_\mathrm{r}^2 H(t_{\text{rad}})
               Y_2\bigl(\mu_\mathrm{r}\sqrt{\tau}\bigr),
\end{equation}
where $C_1(\vec{q})$ and $\psi(\vec{q})$ are arbitrary functions
(the integration `constants') and $J_\nu(x)$ and
$Y_\nu(x)$ are Bessel functions of the first and second kind
respectively. The factors $8/\mu_\mathrm{r}^2$ and
$\pi\mu_\mathrm{r}^2H(t_{\text{rad}})$ have been inserted for convenience. Thus, the standard
equation (\ref{eq:delta-rad-peacock}) yields oscillating density
perturbations with a \emph{decaying} amplitude.

For large-scale
perturbations ($|\vec{q}|\rightarrow0$ or, equivalently,
$\mu_\mathrm{r}\rightarrow0$), the asymptotic expressions for the
Bessel functions $J_2$ and $Y_2$ are given by
\begin{equation}\label{eq:J2Y2-0}
    J_2\bigl(\mu_\mathrm{r}\sqrt{\tau}\bigr)\approx
             \frac{\mu_\mathrm{r}^2}{8}\tau, \quad
    Y_2\bigl(\mu_\mathrm{r}\sqrt{\tau}\bigr)
             \approx-\frac{4}{\pi\mu_\mathrm{r}^2}\tau^{-1}.
\end{equation}
Substituting these expressions into (\ref{eq:peacock-sol}), it is
found for large-scale perturbations that
\begin{equation}\label{eq:delta-rad-peacock-sol}
    \delta(\tau)=C_1 \tau - 4H(t_{\text{rad}})\psi\tau^{-1},
\end{equation}
where we have used that $C_1(|\vec{q}|\rightarrow0)=C_1$ and
$\psi(|\vec{q}|\rightarrow0)=\psi$ become constants in the large-scale limit.
Large-scale perturbations can also be obtained from the standard equation
(\ref{eq:delta-rad-peacock}) by substituting
${\nabla^2\delta=0}$, i.e.,
\begin{equation}\label{eq:delta-rad-peacock-ls}
  \ddot{\delta}+2H\dot{\delta}-
   \tfrac{4}{3}\kappa\varepsilon_\nul \delta=0.
\end{equation}
The general solution of this equation is, using (\ref{eq:fout-rad})
with $\mu_{\text{r}}=0$, given by (\ref{eq:delta-rad-peacock-sol}).
Thus far, the functions $C_1(\vec{q})$ and $\psi(\vec{q})$ are the
integration `constants' which can be determined by the initial values
$\delta(t_{\text{rad}},\vec{q})$ and
$\dot{\delta}(t_{\text{rad}},\vec{q})$.

However, equation (\ref{eq:delta-rad-peacock-ls}) can, in contrast to
equation (\ref{eq:delta-rad-peacock}), also be derived from the
General Theory of Relativity: see the derivation in
Appendix~\ref{app:standard-equation}\@. As a consequence, equation
(\ref{eq:delta-rad-peacock-ls}) is found to be also a
\emph{relativistic} equation, implying that the quantity
${\delta=\varepsilon_\een/\varepsilon_\nul}$ is gauge dependent.
Therefore, the second term in the solution
(\ref{eq:delta-rad-peacock-sol}) is not a physical mode, but equal to
the gauge mode
\begin{equation}\label{eq:gauge-rad}
    \delta_\mathrm{gauge}(\tau)=
    \psi\frac{\dot{\varepsilon}_\nul}{\varepsilon_\nul}=
    -4H(t_{\text{rad}})\psi\tau^{-1},
\end{equation}
as follows from (\ref{e-ijk}), (\ref{eq:behoud}) and (\ref{sol1a}).
Consequently, the constant $\psi$ in (\ref{eq:delta-rad-peacock-sol})
and, hence, the function $\psi(\vec{q})$ in (\ref{eq:peacock-sol})
should \emph{not} be interpreted as an integration constant, but as a
gauge function, which cannot be determined by imposing initial value
conditions, see Appendix~\ref{giofoe} for a detailed
explanation. Thus, the general solution (\ref{eq:peacock-sol}) of the
standard equation (\ref{eq:delta-rad-peacock}) depends on the gauge
function $\psi(\vec{q})$ and has, as a consequence, no physical
significance.  This, in turn, implies that the standard equation
(\ref{eq:delta-rad-peacock}) does \emph{not} describe the evolution of
density perturbations.

Here the negative effect of the gauge function is clearly seen: as yet it was
commonly accepted that small-scale perturbations in the radiation-dominated era
of a flat \textsc{flrw} universe oscillate with a \emph{decaying} amplitude,
according to (\ref{eq:peacock-sol}). The treatise presented in this article
reveals, however, that small-scale density perturbations oscillate with an
\emph{increasing} amplitude, according to (\ref{dc-small}). This is the real
behavior of a small-scale density perturbation.

\subsection{Era after Decoupling of Matter and Radiation}
\label{sec:matter}

\begin{figure}
\begin{center}
\includegraphics[scale=0.75]{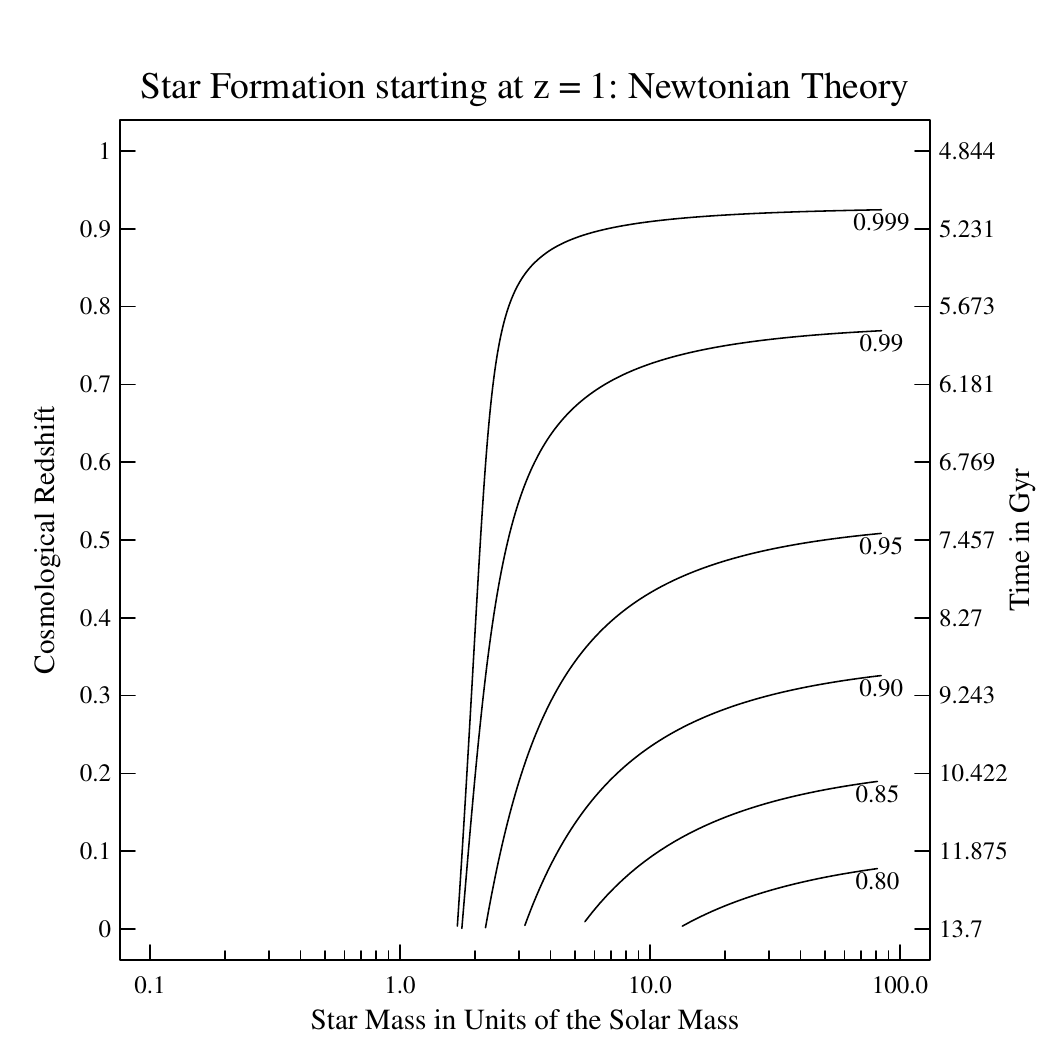}
\caption{Star formation according to the standard equation
  (\ref{eq:delta-dust-standard}) with $\delta_T(t,\vec{q})=0$.  The
  curves give the redshift at which a linear density perturbation
  starting to grow at an initial redshift of $z(t_{\mathrm{mat}})=1$
  becomes non-linear, i.e.,
  $\delta_n=\delta_\varepsilon=\delta_p\approx1$. The numbers at each
  of the curves are the initial relative density perturbations
  $\delta_n(t_{\mathrm{mat}},\vec{q})=
  \delta_\varepsilon(t_{\text{mat}},\vec{q})=\delta_p(t_{\text{mat}},\vec{q})$.
  During the evolution we have
  $\delta_p(t,\vec{q})=\delta_n(t,\vec{q})=\delta_\varepsilon(t,\vec{q})$. For
  masses in excess of $100\,\text{M}_\odot$, the growth becomes
  independent of the scale of a perturbation, i.e., the growth is
  proportional to $t^{2/3}$.}
\label{fig:collapse-2-oud}
\end{center}
\end{figure}

The standard perturbation equation of the Newtonian theory of
gravity is derived from \emph{approximate}, \emph{non-relativistic}
equations. It reads
\begin{equation}\label{eq:delta-dust-standard}
  \ddot{\delta} + 2H\dot{\delta}-
  \left(\frac{v_\mathrm{s}^2}{c^2}\frac{\nabla^2}{a^2}+
   \tfrac{1}{2}\kappa\varepsilon_\nul\right)
   \delta =0,
\end{equation}
where $v_\mathrm{s}$ is the speed of sound. (See, for example, Weinberg
\cite{c8}, Section~15.9, or Peacock \cite{peacock1999}, Section~15.2.)
Compare this equation with equation (\ref{eq:delta-dust}) for the gauge-invariant
contrast function~$\delta_\varepsilon$. Just as equation
(\ref{eq:delta-rad-peacock}), the standard equation
(\ref{eq:delta-dust-standard}) cannot be derived from Einstein's
gravitation theory.

Using (\ref{pw12}), (\ref{subeq:matp0-sol}), (\ref{matsol1b}),
(\ref{eq:tau-mat}) and (\ref{dtau-n-dust}), equation
(\ref{eq:delta-dust-standard}) can rewritten in the form
\begin{equation}
  \label{eq:fout-mat}
  \delta^{\prime\prime}+\dfrac{4}{3\tau}\delta^\prime+
   \left(\dfrac{4}{9}\dfrac{\mu^2_{\text{m}}}{\tau^{8/3}}-\dfrac{2}{3\tau^2}\right)\delta=0,  
\end{equation}
where the constant $\mu_\mathrm{m}$ is given by (\ref{eq:const-mu}). A
prime denotes differentiation with respect to $\tau$,
(\ref{eq:tau-mat}).  The general solution of equation
(\ref{eq:fout-mat}) is found to~be
\begin{equation}\label{eq:matter-non-physical}
  \delta(\tau,\vec{q}) =
    \Biggl[\dfrac{4}{3}D_1(\vec{q})\textstyle{\sqrt{\pi\mu_\mathrm{m}^5}}
        J_{-\tfrac{5}{2}}\bigl(2\mu_\mathrm{m}\tau^{-1/3}\bigr)
\displaystyle{-\frac{45}{8}\psi(\vec{q})H(t_{\text{mat}})\sqrt{\frac{\pi}{\mu_\mathrm{m}^5}}}
    J_{+\tfrac{5}{2}}\bigl(2\mu_\mathrm{m}\tau^{-1/3}\bigr)\Biggr]\tau^{-1/6},
\end{equation}
where $D_1(\vec{q})$ and $\psi(\vec{q})$ are arbitrary functions (the
`constants' of integration) and $J_{\pm\nu}(x)$ is the Bessel function
of the first kind. The factors $\frac{4}{3}\sqrt{\pi\mu_\mathrm{m}^5}$
and $\frac{45}{8}H(t_{\text{mat}})\sqrt{\pi/\mu_\mathrm{m}^5}$ have
been inserted for convenience.

We now consider large-scale perturbations characterized by
$\nabla^2\delta=0$ (i.e., $|\vec{q}|\rightarrow0$) or perturbations of
all scales in the limit $v_\mathrm{s}/c\rightarrow0$. Both limits
imply $\mu_\mathrm{m}\rightarrow0$, as follows from
(\ref{eq:const-mu}).  The asymptotic expressions for the Bessel
functions in the limit $\mu_\mathrm{m}\rightarrow0$ are given by
\begin{equation}\label{eq:limit-J5/2}
    J_{-\tfrac{5}{2}}\bigl(2\mu_\mathrm{m}\tau^{-1/3}\bigr)\approx
       \frac{3}{4\sqrt{\pi\mu_\mathrm{m}^5}}\tau^{5/6}, \quad
  J_{+\tfrac{5}{2}}\bigl(2\mu_\mathrm{m}\tau^{-1/3}\bigr)\approx
       \frac{8}{15}\sqrt{\frac{\mu_\mathrm{m}^5}{\pi}}\tau^{-5/6}.
\end{equation}
Substituting these expressions into the general solution
(\ref{eq:matter-non-physical}), results in
\begin{equation}\label{eq:delta-standard-sol}
    \delta(\tau)=D_1 \tau^{2/3} -3 H(t_{\text{mat}})\psi\tau^{-1}.
\end{equation}
where we have used that $D_1(|\vec{q}|\rightarrow0)=D_1$ and
$\psi(|\vec{q}|\rightarrow0)=\psi$ become constants in either the large-scale
limit $|\vec{q}|\rightarrow0$ or in the limit $v_\mathrm{s}/c\rightarrow0$.
In the limit $\mu_\mathrm{m}\rightarrow0$, equation
(\ref{eq:delta-dust-standard}) reduces to
\begin{equation}\label{eq:delta-standard}
    \ddot{\delta}+2H\dot{\delta}-\tfrac{1}{2}\kappa\varepsilon_\nul\delta=0.
\end{equation}
Using (\ref{eq:fout-mat}) with $\mu_{\text{m}}=0$, we find that the
general solution of equation (\ref{eq:delta-standard}) is given by
(\ref{eq:delta-standard-sol}). Thus far, the functions $D_1(\vec{q})$
and $\psi(\vec{q})$ are the integration `constants' which can be
determined by the initial values $\delta(t_{\text{mat}},\vec{q})$ and
$\dot{\delta}(t_{\text{mat}},\vec{q})$.

However, equation (\ref{eq:delta-standard}) can, unlike equation
(\ref{eq:delta-dust-standard}), also be derived from the General
Theory of Relativity, and is, as a consequence, a \emph{relativistic}
equation: see Appendix~\ref{app:standard-equation} for
a derivation. In this case, however, it is based on the gauge
dependent quantity $\delta=\varepsilon_\een/\varepsilon_\nul$.  As a
consequence, the second term of (\ref{eq:delta-standard-sol}) is equal
to the gauge mode
\begin{equation}\label{eq:gauge-matter}
    \delta_\mathrm{gauge}(\tau)=
    \psi\frac{\dot{\varepsilon}_\nul}{\varepsilon_\nul}=
    -3H(t_{\text{mat}})\psi\tau^{-1},
\end{equation}
as follows from (\ref{e-ijk}), (\ref{mat2p0}) and (\ref{matsol1a}).
Therefore, the constant $\psi$ in (\ref{eq:gauge-matter}) and, hence,
the function $\psi(\vec{q})$ in (\ref{eq:matter-non-physical}), should
\emph{not} be interpreted as an integration constant, but as a gauge
function, which cannot be determined by imposing initial value
conditions (see Appendix~\ref{giofoe}).  Since the solution
(\ref{eq:matter-non-physical}) of equation
(\ref{eq:delta-dust-standard}) depends on the gauge function
$\psi(\vec{q})$ it has no physical significance. Consequently, the
standard equation (\ref{eq:delta-dust-standard}) does \emph{not}
describe the evolution of density perturbations.

Again, we encounter the negative effect of the gauge function: up till
now it was commonly accepted that for small-scale density
perturbations (i.e., density perturbations with wave lengths much
smaller than the particle horizon, Appendix~\ref{app:horizon}) the
Newtonian theory suffices and gauge ambiguities do not occur and that
the evolution of density perturbations in the Newtonian regime is
described by the standard equation (\ref{eq:delta-dust-standard}). The
treatise presented in this article reveals, however, that in a fluid
with an equation of state (\ref{state-mat}), the evolution of density
perturbations is described by the \emph{relativistic} equation
(\ref{eq:delta-dust}), for small-scale as well as large-scale
perturbations. In fact, equation (\ref{eq:delta-dust}) explains the
formation of stars in the universe.

For pedagogical reasons only, we have calculated the `formation of
stars,' starting at $z=1$, with the help of the standard equation
(\ref{eq:delta-dust-standard}) for which the conditions
(\ref{eq:homo-de-dn}) hold true. The result is depicted in
Figure~\ref{fig:collapse-2-oud}. This figure should be compared with
Figure~\ref{fig:collapse-2}, which is calculated with the help of the
new equation (\ref{eq:delta-dust}) under the conditions
(\ref{eq:homo-de-dn}).  The standard Newtonian theory underestimates
the internal gravitational field of a perturbation by a factor of
$\tfrac{5}{3}$ with respect to our perturbation theory. Compare the
gravitational fields $\tfrac{1}{2}\kappa\varepsilon_\nul$ in equation
(\ref{eq:delta-dust-standard}) and its relativistic counterpart
(\ref{eq:delta-standard}) with $\tfrac{5}{6}\kappa\varepsilon_\nul$ in
the new equation (\ref{eq:delta-dust}). As a result, the standard
theory predicts a lower limit of $1.7\,\text{M}_\odot$ for star
formation, whereas our theory yields a lower limit of
$0.8\,\text{M}_\odot$ under the conditions (\ref{eq:homo-de-dn}). As a
consequence, our Sun could not exist at all according to the standard
theory. This fact can be considered as an `experimental proof' that
the standard Newtonian theory has indeed no physical significance.


\appendix

\section{Equations of State for the Energy Density and Pressure}
\label{sec:eq-state}

We have used an equation of state for the pressure of the form
$p=p(n,\varepsilon)$. In general, however, this equation of state is
given in the form of two equations for the energy
density~$\varepsilon$ and the pressure~$p$ which contain also the
absolute temperature~$T$:
\begin{equation}\label{eq:state-e-p}
  \varepsilon=\varepsilon(n,T), \quad  p=p(n,T).
\end{equation}
In principle it is possible to eliminate $T$ from the two
equations~(\ref{eq:state-e-p}) to get $p=p(n,\varepsilon)$, so that
our choice of the form $p=p(n,\varepsilon)$ is justified. In
practice, however, it may in general be difficult to eliminate the
temperature~$T$ from the equations~(\ref{eq:state-e-p}). However,
this is not necessary, since the partial derivatives~$p_\varepsilon$
and~$p_n$ (\ref{perttoes1}), the only quantities that are actually
needed, can be found in an alternative way. From equations
(\ref{eq:state-e-p}) it follows
\begin{subequations}
\label{subeq:de-dp}
\begin{align}
\dif \varepsilon & =
   \left(\dfrac{\partial \varepsilon}{\partial n} \right)_{\!T}\dif n+
   \left(\dfrac{\partial \varepsilon}{\partial T} \right)_{\!n}\dif T,
      \label{subeq:de-dp-a}  \\
 \dif p &=
   \left(\dfrac{\partial p}{\partial n} \right)_{\!T}\dif n +
   \left(\dfrac{\partial p}{\partial T} \right)_{\!n}\dif T.
      \label{subeq:de-dp-b}
\end{align}
\end{subequations}
From (\ref{subeq:de-dp-b}) it follows that the partial
derivatives~(\ref{perttoes1}) are
\begin{subequations}
\label{subeq:pn-pe}
\begin{align}
 p_n & =
   \left(\dfrac{\partial p}{\partial n} \right)_{\!T} +
   \left(\dfrac{\partial p}{\partial T} \right)_{\!n}
   \left(\dfrac{\partial T}{\partial n} \right)_{\!\varepsilon},
      \label{subeq:pn-pe-a}  \\
 p_\varepsilon & =
   \left(\dfrac{\partial p}{\partial T} \right)_{\!n}
   \left(\dfrac{\partial T}{\partial \varepsilon} \right)_{\!n}.
      \label{subeq:pn-pe-b}
\end{align}
\end{subequations}
From (\ref{subeq:de-dp-a}) it follows
\begin{subequations}
\label{eq:dT-dn-de}
\begin{align}
  \left(\dfrac{\partial T}{\partial n} \right)_{\!\varepsilon} & =
  -\left(\dfrac{\partial \varepsilon}{\partial n} \right)_{\!T}
  \left(\dfrac{\partial \varepsilon}{\partial T} \right)_{\!n}^{-1}, \\
  \left(\dfrac{\partial T}{\partial \varepsilon}\right)_{\!n} & =
  \left(\dfrac{\partial \varepsilon}{\partial T} \right)_{\!n}^{-1}.
\end{align}
\end{subequations}
Upon substituting the expressions~(\ref{eq:dT-dn-de}) into
(\ref{subeq:pn-pe}), we find for the partial derivatives defined by
(\ref{perttoes1})
\begin{subequations}
\label{subeq:final-pn-pe}
\begin{align}
 p_n & \equiv \left(\dfrac{\partial p}{\partial n}  \right)_{\!\varepsilon} =
   \left(\dfrac{\partial p}{\partial n} \right)_{\!T} -
   \left(\dfrac{\partial p}{\partial T} \right)_{\!n}
   \left(\dfrac{\partial \varepsilon}{\partial n} \right)_{\!T}
   \left(\dfrac{\partial \varepsilon}{\partial T} \right)_{\!n}^{-1},
      \label{subeq:final-pn-pe-a}  \\
 p_\varepsilon & \equiv \left(\dfrac{\partial p}{\partial\varepsilon}  \right)_{\!n} =
   \left(\dfrac{\partial p}{\partial T} \right)_{\!n}
   \left(\dfrac{\partial \varepsilon}{\partial T}
   \right)_{\!n}^{-1},
      \label{subeq:final-pn-pe-b}
\end{align}
\end{subequations}
where $\varepsilon$ and~$p$ are given by~(\ref{eq:state-e-p}). In
order to calculate the second-order derivative $p_{nn}$ replace $p$
in (\ref{subeq:final-pn-pe-a}) by $p_n$. For
$p_{\varepsilon\varepsilon}$ replace $p$ in
(\ref{subeq:final-pn-pe-b}) by $p_\varepsilon$. Finally, for
$p_{\varepsilon n}\equiv p_{n\varepsilon}$, replace $p$ in
(\ref{subeq:final-pn-pe-a}) by $p_\varepsilon$ or, equivalently,
replace $p$ in (\ref{subeq:final-pn-pe-b}) by~$p_n$.

\section{Derivation of the Manifestly Gauge-invariant  Perturbation Equations}
\label{sec:deriv-egi}

In this appendix we derive the perturbation
equations~(\ref{subeq:eerste}) and the evolution
equations~(\ref{subeq:final}).

\subsection{Derivation of the Evolution Equation for the Entropy}
\label{app:hulp2}

With the help of equations (\ref{FRW4gi}) and~(\ref{FRW4agi}) and
equations (\ref{begam2})--(\ref{subeq:einstein-flrw}) one may verify
that
\begin{equation}\label{hulp2}
 \frac{1}{c}\frac{\dif}{\dif t}
      \left(n_\een-\frac{n_\nul}{\varepsilon_\nul(1+w)}\varepsilon_\een\right)=
     -3H\left(1-\frac{n_\nul p_n}{\varepsilon_\nul(1+w)}\right)
\left(n_\een-\dfrac{n_\nul}{\varepsilon_\nul(1+w)}\varepsilon_\een\right).
\end{equation}
In view of (\ref{eq:lin-gi}) one may replace~$\varepsilon_\een$
and~$n_\een$ by~$\varepsilon^\gi_\een$ and~$n^\gi_\een$. Using
(\ref{eq:entropy-gi}) yields equation (\ref{eq:vondst1}) of the main text.

\subsection{Derivation of the Evolution Equation for the Energy  Density Perturbation}
\label{app:hulp3}

\begin{table}
\renewcommand{\arraystretch}{2.25}
\caption{\label{eq:aij}The coefficients $\alpha_{ij}$ figuring in
the equations~(\ref{subeq:nieuw}).}  \normalsize   
\[  \begin{array}{|c|c|c|c|} \hline
  3H(1+p_\varepsilon)+\dfrac{\kappa\varepsilon_\nul(1+w)}{2H} &
      3Hp_n & \varepsilon_\nul(1+w) &
      \dfrac{\varepsilon_\nul(1+w)}{4H} \\ \hline
  \dfrac{\kappa n_\nul}{2H} & 3H & n_\nul &
  \dfrac{n_\nul}{4H} \\ \hline
  \dfrac{p_\varepsilon}{\varepsilon_\nul(1+w)}\dfrac{\tilde{\nabla}^2}{a^2} &
       \dfrac{p_n}{\varepsilon_\nul(1+w)}\dfrac{\tilde{\nabla}^2}{a^2} &
       H(2-3\beta^2) & 0 \\ \hline
  \dfrac{\kappa\,\mbox{$^3\!R_\nul$}}{3H} & 0 &
  -2\kappa\varepsilon_\nul(1+w) &
        2H+\dfrac{\mbox{$^3\!R_\nul$}}{6H} \\ \hline
  \dfrac{-\,\mbox{$^3\!R_\nul$}}{\mbox{$^3\!R_\nul$}+
         3\kappa\varepsilon_\nul(1+w)} & 0 &
     \dfrac{6\varepsilon_\nul
H(1+w)}{\mbox{$^3\!R_\nul$}+3\kappa\varepsilon_\nul(1+w)} &
     \dfrac{\tfrac{3}{2}\varepsilon_\nul(1+w)}
     {\mbox{$^3\!R_\nul$}+3\kappa\varepsilon_\nul(1+w)}  \\  \hline
\end{array}  \]
\end{table}

We will now derive equation (\ref{eq:vondst2}). To that end, we rewrite
the system (\ref{subeq:pertub-gi}) and expression (\ref{Egi}) in the form
\begin{subequations}
\label{subeq:nieuw}
\begin{align}
  \dot{\varepsilon}_\een+\alpha_{11}\varepsilon_\een+
    \alpha_{12}n_\een+
   \alpha_{13}\vartheta_\een+\alpha_{14}\,\mbox{$^3\!R_{\een\parallel}$} & = 0,
\label{nieuw1} \\
   \dot{n}_\een+\alpha_{21}\varepsilon_\een +
    \alpha_{22}n_\een+
   \alpha_{23}\vartheta_\een+\alpha_{24}\,\mbox{$^3\!R_{\een\parallel}$} & = 0,
\label{nieuw2} \\
  \dot{\vartheta}_\een +\alpha_{31}\varepsilon_\een +
     \alpha_{32}n_\een+
   \alpha_{33}\vartheta_\een+\alpha_{34}\,\mbox{$^3\!R_{\een\parallel}$} & = 0,
\label{nieuw3} \\
  \mbox{$^3\!\dot{R}_{\een\parallel}$}+\alpha_{41}\varepsilon_\een+
     \alpha_{42}n_\een+
   \alpha_{43}\vartheta_\een+\alpha_{44}\,\mbox{$^3\!R_{\een\parallel}$} & = 0,
\label{nieuw4} \\
  \varepsilon^\gi_\een+\alpha_{51}\varepsilon_\een+\alpha_{52}n_\een+
    \alpha_{53}\vartheta_\een+\alpha_{54}\,\mbox{$^3\!R_{\een\parallel}$} & = 0,
\label{nieuw5}
\end{align}
\end{subequations}
where the coefficients~$\alpha_{ij}(t)$ are given in
Table~\ref{eq:aij}.

In calculating the coefficients~$a_1$, $a_2$ and~$a_3$,
(\ref{subeq:coeff}) in the main text, we use that the time
derivative of the quotient $w$, defined by (\ref{begam2}) is given
by
\begin{equation}\label{eq:td-w}
  \dot{w} = 3H(1+w)(w-\beta^2),
\end{equation}
as follows from equation (\ref{FRW2}) and the expression
(\ref{begam2}). Moreover, it is of convenience \emph{not} to expand
the function~$\beta(t)$ defined by (\ref{begam2}) since this will
considerably complicate the expressions for the coefficients~$a_1$,
$a_2$ and~$a_3$.

\paragraph{Step 1.} We first eliminate the
quantity~$\mbox{$^3\!R_{\een\parallel}$}$ from
equations (\ref{subeq:nieuw}). Differentiating equation (\ref{nieuw5}) with
respect to time and eliminating the time derivatives
~$\dot{\varepsilon}_\een$, $\dot{n}_\een$, $\dot{\vartheta}_\een$
and~$\mbox{$^3\!\dot{R}_{\een\parallel}$}$ with the help of
equations (\ref{nieuw1})--(\ref{nieuw4}), we arrive at the equation
\begin{equation}\label{eq:equiv}
   \dot{\varepsilon}^\gi_\een + p_1\varepsilon_\een+p_2 n_\een+
   p_3\vartheta_\een+p_4\,\mbox{$^3\!R_{\een\parallel}$}=0,
\end{equation}
where the coefficients $p_1(t),\ldots,p_4(t)$ are given by
\begin{equation}\label{eq:coef-pi}
  p_i=\dot{\alpha}_{5i}-\alpha_{51}\alpha_{1i}-
         \alpha_{52}\alpha_{2i}-\alpha_{53}\alpha_{3i}-\alpha_{54}\alpha_{4i}.
\end{equation}
From equation (\ref{eq:equiv}) it follows that
\begin{equation}\label{eq:sol-3R1}
  \mbox{$^3\!R_{\een\parallel}$}=-\dfrac{1}{p_4}\dot{\varepsilon}^\gi_\een-
     \dfrac{p_1}{p_4}\varepsilon_\een-\dfrac{p_2}{p_4}n_\een-
     \dfrac{p_3}{p_4}\vartheta_\een.
\end{equation}
In this way we have expressed the
quantity~$\mbox{$^3\!R_{\een\parallel}$}$ as a linear combination of
the quantities~$\dot{\varepsilon}^\gi_\een$, $\varepsilon_\een$,
$n_\een$ and~$\vartheta_\een$. Upon replacing
$\mbox{$^3\!R_{\een\parallel}$}$ given by~(\ref{eq:sol-3R1}), in
equations (\ref{subeq:nieuw}), we arrive at the system of equations
\begin{subequations}
\label{subeq:tweede}
\begin{align}
 \dot{\varepsilon}_\een+q_1\dot{\varepsilon}_\een^\gi+
   \beta_{11}\varepsilon_\een+\beta_{12}n_\een+
   \beta_{13}\vartheta_\een & = 0, \label{tweede1} \\
 \dot{n}_\een+q_2\dot{\varepsilon}_\een^\gi+
    \beta_{21}\varepsilon_\een+\beta_{22}n_\een+
    \beta_{23}\vartheta_\een & = 0, \label{tweede2} \\
 \dot{\vartheta}_\een+q_3\dot{\varepsilon}_\een^\gi+
   \beta_{31}\varepsilon_\een+\beta_{32}n_\een+
   \beta_{33}\vartheta_\een & = 0, \label{tweede3} \\
 \mbox{$^3\!\dot{R}_{\een\parallel}$}+
   q_4\dot{\varepsilon}^\gi_\een+\beta_{41}\varepsilon_\een+\beta_{42}n_\een+
   \beta_{43}\vartheta_\een & = 0, \label{tweede4} \\
 \varepsilon^\gi_\een+
   q_5\dot{\varepsilon}^\gi_\een+\beta_{51}\varepsilon_\een+\beta_{52}n_\een+
   \beta_{53}\vartheta_\een & = 0, \label{tweede5}
\end{align}
\end{subequations}
where the coefficients~$q_i(t)$ and~$\beta_{ij}(t)$ are given by
\begin{equation}\label{eq:betaij}
  q_i=-\dfrac{\alpha_{i4}}{p_4}, \quad
   \beta_{ij}=\alpha_{ij}+q_i p_j.
\end{equation}
We now have achieved that the
quantity~$\mbox{$^3\!R_{\een\parallel}$}$ occurs only in
equation (\ref{tweede4}). Since we are not interested in the
non-physical quantity~$\mbox{$^3\!R_{\een\parallel}$}$, we do not
need this equation any more.

\paragraph{Step 2.} We now proceed in the same way as in step~1:
eliminating this time the quantity~$\vartheta_\een$ from the system
of equations~(\ref{subeq:tweede}). Differentiating
equation (\ref{tweede5}) with respect to time and eliminating the time
derivatives~$\dot{\varepsilon}_\een$, $\dot{n}_\een$
and~$\dot{\vartheta}_\een$ with the help of
equations (\ref{tweede1})--(\ref{tweede3}), we arrive at
\begin{equation}
\label{eq:ddot-egi}
  q_5\ddot{\varepsilon}^\gi_\een+r\dot{\varepsilon}^\gi_\een+
     s_1\varepsilon_\een+s_2n_\een+s_3\vartheta_\een=0,
\end{equation}
where the coefficients~$r(t)$ and~$s_i(t)$ are given by
\begin{subequations}
\label{eq:coef-qi}
\begin{align}
  s_i & = \dot{\beta}_{5i}-\beta_{51}\beta_{1i}-\beta_{52}\beta_{2i}-
       \beta_{53}\beta_{3i}, \\
  r & = 1+\dot{q}_5-\beta_{51}q_1-\beta_{52}q_2-\beta_{53}q_3.
\end{align}
\end{subequations}
From equation (\ref{eq:ddot-egi}) it follows that
\begin{equation}\label{eq:sol-theta1}
  \vartheta\een=-\dfrac{q_5}{s_3}\ddot{\varepsilon}^\gi_\een-
     \dfrac{r}{s_3}\dot{\varepsilon}^\gi_\een-
     \dfrac{s_1}{s_3}\varepsilon_\een-\dfrac{s_2}{s_3}n_\een.
\end{equation}
In this way we have expressed the quantity~$\vartheta_\een$ as a
linear combination of the quantities~$\ddot{\varepsilon}^\gi_\een$,
$\dot{\varepsilon}^\gi_\een$, $\varepsilon_\een$ and~$n_\een$. Upon
replacing~$\vartheta_\een$ given by~(\ref{eq:sol-theta1}) in
equations (\ref{subeq:tweede}), we arrive at the system of equations
\begin{subequations}
\label{subeq:derde}
\begin{align}
\dot{\varepsilon}_\een-\beta_{13}\dfrac{q_5}{s_3}\ddot{\varepsilon}^\gi_\een+
   \left(q_1-\beta_{13}\dfrac{r}{s_3}\right)\dot{\varepsilon}^\gi_\een
  +\left(\beta_{11}-\beta_{13}\dfrac{s_1}{s_3}\right)\varepsilon_\een
  +\left(\beta_{12}-\beta_{13}\dfrac{s_2}{s_3}\right)n_\een & =0, \label{derde1}
\\
\dot{n}_\een-\beta_{23}\dfrac{q_5}{s_3}\ddot{\varepsilon}^\gi_\een+
   \left(q_2-\beta_{23}\dfrac{r}{s_3}\right)\dot{\varepsilon}^\gi_\een
    +\left(\beta_{21}-\beta_{23}\dfrac{s_1}{s_3}\right)\varepsilon_\een
  +\left(\beta_{22}-\beta_{23}\dfrac{s_2}{s_3}\right)n_\een & =0, \label{derde2}
\\
\dot{\vartheta}_\een-\beta_{33}\dfrac{q_5}{s_3}\ddot{\varepsilon}^\gi_\een+
   \left(q_3-\beta_{33}\dfrac{r}{s_3}\right)\dot{\varepsilon}^\gi_\een
   +\left(\beta_{31}-\beta_{33}\dfrac{s_1}{s_3}\right)\varepsilon_\een
  +\left(\beta_{32}-\beta_{33}\dfrac{s_2}{s_3}\right)n_\een & =0, \label{derde3}
\\
\mbox{$^3\!\dot{R}_{\een\parallel}$}-\beta_{43}\dfrac{q_5}{s_3}\ddot{\varepsilon
}^\gi_\een+
   \left(q_4-\beta_{43}\dfrac{r}{s_3}\right)\dot{\varepsilon}^\gi_\een
  + \left(\beta_{41}-\beta_{43}\dfrac{s_1}{s_3}\right)\varepsilon_\een
  +\left(\beta_{42}-\beta_{43}\dfrac{s_2}{s_3}\right)n_\een & =0, \label{derde4}
\\
\varepsilon^\gi_\een-\beta_{53}\dfrac{q_5}{s_3}\ddot{\varepsilon}^\gi_\een+
   \left(q_5-\beta_{53}\dfrac{r}{s_3}\right)\dot{\varepsilon}^\gi_\een
    +\left(\beta_{51}-\beta_{53}\dfrac{s_1}{s_3}\right)\varepsilon_\een
  +\left(\beta_{52}-\beta_{53}\dfrac{s_2}{s_3}\right)n_\een & =0. \label{derde5}
\end{align}
\end{subequations}
We have achieved now that the quantities $\vartheta_\een$ and
$\mbox{$^3\!R_{\een\parallel}$}$ occur only in equations
(\ref{derde3}) and (\ref{derde4}), so that these equations will not
be needed anymore. We are left, in principle, with equations
(\ref{derde1}), (\ref{derde2}) and (\ref{derde5}) for the three
unknown quantities $\varepsilon_\een$, $n_\een$ and
$\varepsilon^\gi_\een$, but we first proceed with all five
equations.

\paragraph{Step 3.} At first sight, the next steps would be to
eliminate, successively, the quantities~$\varepsilon_\een$
and~$n_\een$ from equation (\ref{derde5}) with the help of
equations (\ref{derde1}) and~(\ref{derde2}). We then would end up with
a fourth-order differential equation for the unknown
quantity~$\varepsilon^\gi_\een$.

However, it is possible to extract a second-order equation for the
gauge-invariant energy density from the
equations~(\ref{subeq:derde}). This will now be shown.
Equation (\ref{derde5}) can be rewritten
\begin{equation}\label{eq:eindelijk}
  \ddot{\varepsilon}^\gi_\een+a_1\dot{\varepsilon}^\gi_\een+
    a_2\varepsilon^\gi_\een=
    a_3\left(n_\een+\dfrac{\beta_{51}s_3-\beta_{53}s_1}
    {\beta_{52}s_3-\beta_{53}s_2} \varepsilon_\een\right),
\end{equation}
where the coefficients $a_1(t)$, $a_2(t)$ and~$a_3(t)$ are given
by
\begin{equation}
\label{subeq:vierde}
  a_1  = -\dfrac{s_3}{\beta_{53}}+\dfrac{r}{q_5}, \quad
  a_2  = -\dfrac{s_3}{\beta_{53}q_5}, \quad
  a_3  = \dfrac{\beta_{52}s_3}{\beta_{53}q_5}-\dfrac{s_2}{q_5}.
\end{equation}
These are precisely the coefficients
(\ref{eq:alpha-1})--(\ref{eq:alpha-3}) of the main text.
Furthermore, we find
\begin{equation}\label{eq:check}
  \dfrac{\beta_{51}s_3-\beta_{53}s_1}{\beta_{52}s_3-\beta_{53}s_2}=
      -\dfrac{n_\nul}{\varepsilon_\nul(1+w)}.
\end{equation}
Hence,
\begin{equation}\label{eq:hulp3}
  n_\een+\dfrac{\beta_{51}s_3-\beta_{53}s_1}{\beta_{52}s_3-\beta_{53}s_2}
    \varepsilon_\een=
    n_\een-\dfrac{n_\nul}{\varepsilon_\nul(1+w)}\varepsilon_\een.
\end{equation}
With the help of this expression and (\ref{eq:lin-gi}) we can
rewrite equation (\ref{eq:eindelijk}) in the
form~(\ref{eq:vondst2}).

The derivation of the expressions~(\ref{subeq:coeff})
from~(\ref{subeq:vierde}) and the proof of the
equality~(\ref{eq:check}) is straightforward, but extremely
complicated. We used \textsc{Maple~14}~\cite{MapleV} to perform
this algebraic task.

\subsection{Evolution Equations for the Contrast Functions}
\label{app:contrast}

In this section we derive equations (\ref{subeq:final}). We start
off with equation (\ref{fir-ord}). From (\ref{eq:entropy-gi}) and
the definitions~(\ref{eq:contrast}) it follows that
\begin{equation}\label{eq:sgi-contrast}
  \sigma^\gi_\een=n_\nul\left(\delta_n-\dfrac{\delta_\varepsilon}{1+w}\right).
\end{equation}
Differentiating this equation with respect to~$ct$ yields
\begin{equation}\label{eq:diff-sgi}
  a_4
\sigma^\gi_\een=\dot{n}_\nul\left(\delta_n-\dfrac{\delta_\varepsilon}{1+w}
\right)+
  n_\nul\dfrac{1}{c}\dfrac{\dif}{\dif t}
  \left(\delta_n-\dfrac{\delta_\varepsilon}{1+w}\right),
\end{equation}
where we have used equation (\ref{eq:vondst1}). Using equation
(\ref{FRW2a}) and the expression
(\ref{eq:sgi-contrast}) to eliminate~$\sigma^\gi_\een$, we arrive at
equation (\ref{fir-ord}) of the main text.

Finally, we derive equation (\ref{sec-ord}). Upon substituting the
expressions
\begin{equation}
\label{subeq:afgeleiden}
\varepsilon^\gi_\een  = \varepsilon_\nul\delta_\varepsilon, \quad
\dot{\varepsilon}^\gi_\een =\dot{\varepsilon}_\nul\delta_\varepsilon+
      \varepsilon_\nul\dot{\delta}_\varepsilon, \quad
\ddot{\varepsilon}^\gi_\een=\ddot{\varepsilon}_\nul\delta_\varepsilon+
      2\dot{\varepsilon}_\nul\dot{\delta}_\varepsilon+
      \varepsilon_\nul\ddot{\delta}_\varepsilon,
\end{equation}
into equation (\ref{eq:vondst2}), and dividing by~$\varepsilon_\nul$,
we find
\begin{equation}
\label{subeq:tilde-alpha}
 b_1=2\dfrac{\dot{\varepsilon}_\nul}{\varepsilon_\nul}+a_1, \quad
 b_2=\dfrac{\ddot{\varepsilon}_\nul}{\varepsilon_\nul}+
      a_1\dfrac{\dot{\varepsilon}_\nul}{\varepsilon_\nul}+a_2, \quad
 b_3=a_3\dfrac{n_\nul}{\varepsilon_\nul}.
\end{equation}
where we have also used (\ref{eq:sgi-contrast}). These are the
coefficients~(\ref{subeq:coeff-contrast}) of the main text.

\section{Gauge-invariance of the First-order Equations}
\label{giofoe}

If we go over from one synchronous system of reference with
coordinates $x$ to another synchronous system of reference with
coordinates $\hat{x}$ given by expression (\ref{func}), we have
\begin{equation}
   \xi_{\mu;0} + \xi_{0;\mu} = 0,    \label{kil-syn}
\end{equation}
as follows from the transformation rule (\ref{killing}) and the
conditions (\ref{sync-cond}). From this equation we find, using
(\ref{def-gammas}), (\ref{con1}), (\ref{con2}) and~(\ref{metricFRW})
that $\xi^\mu(t,\vec{x})$ must be of the form
\begin{equation}
   \xi^0=\psi(\vec{x}), \quad
     \xi^i=\tilde{g}^{ik}\partial_k \psi(\vec{x})
       \int^{ct} \!\! \frac{\dif\tau}{a^2(\tau)} + \chi^i(\vec{x}),
               \label{xi-syn}
\end{equation}
where $\psi(\vec{x})$ and $\chi^i(\vec{x})$ are \emph{arbitrary}
functions ---of the first-order---  of the spatial coordinates
$\vec{x}$. The fact that the gauge function $\psi$ does not depend
on the time coordinate $x^0=ct$ anymore, as it did in general coordinates, see
(\ref{defpsi}), is a consequence of the choice of synchronous
coordinates for the original coordinates as well as for the
transformed system of reference.

The energy density perturbation transforms according to
(\ref{e-ijk}), where $\varepsilon_\nul$ is a solution of equation
(\ref{FRW2}). Similarly, the particle number density transforms
according to (\ref{n-ijk}) where $n_\nul$ is a solution of equation
(\ref{FRW2a}). Finally, as follows from (\ref{sigmahat3}), the fluid
expansion scalar $\theta$, (\ref{exp1}), transforms as
\begin{equation}
  \hat{\theta}_\een(t,\vec{x}) = \theta_\een(t,\vec{x}) +
        \psi(\vec{x})\dot{\theta}_\nul(t) , \label{tr-theta}
\end{equation}
where $\theta_\nul=3H$ is a solution of the set (\ref{subeq:einstein-flrw}).

From~(\ref{sigmahat3}) with $\sigma=p$, $\varepsilon$, or~$n$
and~(\ref{perttoes}) we find for the transformation rule for the
first-order perturbations to the pressure
\begin{equation}\label{perttoes-hat}
  \hat{p}_\een = p_\varepsilon \hat{\varepsilon}_\een +
                        p_n \hat{n}_\een.
\end{equation}

The transformation rule~(\ref{transvec}) with $V^\mu$ the
four-velocity $u^\mu$ implies
\begin{equation}\label{eq:trans-alg-umu}
    \hat{u}^\mu_\een=u^\mu_\een-\xi^\mu{}_{,0},
\end{equation}
where we have used that $u^\mu_\nul=\delta^\mu{}_0$, expression
(\ref{u0}).  From the transformation rule (\ref{eq:trans-alg-umu}) it
follows that $u^\mu_\een$ transforms under transformations (\ref{xi-syn})
between synchronous coordinates as
\begin{subequations}
\label{eq:trans-u0-ui}
\begin{align}
   \hat{u}^0_\een(t,\vec{x}) & =u^0_\een(t,\vec{x}) =0, \\
        \hat{u}^i_{\een\parallel}(t,\vec{x}) & =u^i_{\een\parallel}(t,\vec{x})-
        \frac{1}{a^2(t)}\tilde{g}^{ik}(\vec{x})
        \partial_k\psi(\vec{x}). \label{eq:trans-u0-ui-b}
\end{align}
\end{subequations}

We want to determine the transformation rules for~$\vartheta_\een$
and~$\mbox{$^3\!R_{\een\parallel}$}$. Since the
quantities~$\mbox{$^3\!R$}$, (\ref{drieR}), and $\vartheta$,
(\ref{driediv}), are both non-scalars under general space-time
transformations, the transformation rule~(\ref{sigmahat3}) is not
applicable to determine the transformation of their first-order
perturbations under infinitesimal space-time transformations
$x^\mu\rightarrow\hat{x}^\mu$, (\ref{func}).
Since~$u^i_{\een\parallel}$ satisfies equation (\ref{basis-5-scal}),
and since~$u^i_{\een\parallel}$ transforms according
to~(\ref{eq:trans-u0-ui}), and since we know
that~$\hat{u}^i_{\een\parallel}$ satisfies
equation (\ref{basis-5-scal}) with hats, one may verify,
using~(\ref{perttoes-hat}), that
\begin{equation}
   \hat{\vartheta}_\een(t,\vec{x}) \equiv
   \vartheta_\een(t,\vec{x})-\frac{\tilde{\nabla}^2\psi(\vec{x})}{a^2(t)},
          \label{theta-ijk}
\end{equation}
satisfies equation (\ref{FRW5}) with hatted quantities. The quantity
$\hat{\vartheta}_\een$ is defined in analogy to $\vartheta_\een$
in~(\ref{den1a})
\begin{equation}
    \hat{\vartheta}_\een=(\hat{u}_{\een\parallel}^k){}_{|k}.
\end{equation}
Apparently, $\vartheta_\een$ transforms according to~(\ref{theta-ijk})
under arbitrary infinitesimal space-time transformations between
synchronous coordinates. Similarly, one may verify that
\begin{equation}
       \mbox{$^3\!\hat{R}_{\een\parallel}(t,\vec{x})$}\equiv
       \mbox{$^3\!R_{\een\parallel}(t,\vec{x})$} +
      4H(t)\left(\frac{\tilde{\nabla}^2\psi(\vec{x})}{a^2(t)} -
       \tfrac{1}{2}\,\mbox{$^3\!R_\nul(t)$}\psi(\vec{x})\right),
            \label{drie-ijk}
\end{equation}
satisfies equation (\ref{con-sp-1}). Apparently, expression
(\ref{drie-ijk}) is the transformation rule for
$\mbox{$^3\!R_{\een\parallel}$}$ under arbitrary infinitesimal
space-time transformations between synchronous coordinates. An
alternative way to find the results~(\ref{theta-ijk})
and~(\ref{drie-ijk}) is to write
$\hat{\vartheta}_\een=\vartheta_\een-f$ and
$\mbox{$^3\!\hat{R}_{\een\parallel}$}=\mbox{$^3\!R_{\een\parallel}$}-g$,
where $f$ and $g$ are unknown functions, to substitute, thereupon,
$\hat{\vartheta}_\een$ and $\mbox{$^3\!\hat{R}_{\een\parallel}$}$ into
equations (\ref{FRW5}) and~(\ref{con-sp-1}), and to determine $f$ and
$g$ such that the old equations (\ref{FRW5}) and~(\ref{con-sp-1})
reappear. In fact, our method to define $\hat{\vartheta}_\een$,
(\ref{theta-ijk}), and $\mbox{$^3\!\hat{R}_{\een\parallel}$}$,
(\ref{drie-ijk}), is nothing but a shortcut to this procedure.

It may now easily be verified by substitution that if
$\varepsilon_\een$, $n_\een$, $\theta_\een$, $\vartheta_\een$, and
\mbox{$^3\!R_{\een\parallel}$} are solutions of the systems
(\ref{subeq:pertub-flrw-sum}) and (\ref{subeq:pertub-gi}), then the
quantities $\hat{\varepsilon}_\een$, (\ref{e-ijk}), $\hat{n}_\een$,
(\ref{n-ijk}), $\hat{\theta}_\een$, (\ref{tr-theta}),
$\hat{\vartheta}_\een$, (\ref{theta-ijk}),
and~\mbox{$^3\!\hat{R}_{\een\parallel}$}, (\ref{drie-ijk}), are, for
an arbitrary function~$\psi(\vec{x})$, also solutions of these
systems. In other words, the systems (\ref{subeq:pertub-flrw-sum}) and
(\ref{subeq:pertub-gi}) are gauge-invariant under gauge
transformations between synchronous coordinates. The solutions
$\varepsilon_\een$, $n_\een$, $\theta_\een$, $\vartheta_\een$, and
\mbox{$^3\!R_{\een\parallel}$}, however, contain an arbitrary function
$\psi(\vec{x})$ and are, therefore, gauge dependent.

\section{Horizon Size after Decoupling of Matter and Radiation}
\label{app:horizon}

As is well-known, a density perturbation can only grow if all its particles are
in causal contact with each other, so that gravity can act in such a way that a
density perturbation may eventually collapse. In this appendix we calculate the
horizon size at time $t_{\text{mat}}$ between the decoupling time
$t_{\text{dec}}$ and the present time $t_{\text{p}}$.
The horizon size at time $t$ is given by
\begin{equation}\label{eq:horizon}
    d_\mathrm{H}(t) = c a(t) \int_0^t \dfrac{\dif t^\prime}{a(t^\prime)}.
\end{equation}
Using  (\ref{matsol1b}), we get
\begin{equation}\label{eq:horizon-flat}
    d_\mathrm{H}(t)=3ct.
\end{equation}
For the horizon size at time $t_{\text{mat}}$, we find
\begin{equation}\label{eq:horizon-dec}
    d_\mathrm{H}(t_\mathrm{mat})=3ct_\mathrm{mat}=
\dfrac{3ct_\mathrm{p}}{\bigl[z(t_{\text{mat}})+1\bigr]^{3/2}}=
    \dfrac{1.260\times10^7}{\bigl[z(t_{\text{mat}})+1\bigr]^{3/2}} \;\mathrm{kpc},
\end{equation}
where we have used (\ref{subeq:wmap}), (\ref{eq:redshift}) and
(\ref{matsol1b}). At decoupling, $z(t_{\text{dec}})=1091$, the horizon
size is $d_{\text{H}}(t_{\text{dec}})=349\,\text{kpc}$.

\section{Derivation of the Relativistic Standard Equation for Density Perturbations}
\label{app:standard-equation}

In this appendix equations (\ref{eq:delta-rad-peacock-ls}) and
(\ref{eq:delta-standard}) of the main text are derived, for a flat
\textsc{flrw} universe, $\mbox{$^3\!R_\nul$}=0$, with vanishing
cosmological constant, $\Lambda=0$, using the background equations
(\ref{subeq:einstein-flrw}) and the linearized
Einstein equations and conservation laws for scalar perturbations
(\ref{subeq:pertub-gi}).

From (\ref{eq:td-w}) it follows that $w$ is constant if and only if
$w=\beta^2$ for all times. Using (\ref{eq:begam3}) it is found for
constant $w$ that $p_n=0$ and $p_\varepsilon=w$, i.e., the pressure
does not depend on the particle number density. Consequently, in the
derivation of equations (\ref{eq:delta-rad-peacock-ls}) and
(\ref{eq:delta-standard}) the equations (\ref{FRW2a}) for
$n_\nul(t)$ and (\ref{FRW4agi}) for $n_\een(t,\vec{x})$ are not
needed. In this case, the equation of state is given by
\begin{equation}\label{eq:state-simple}
    p = w \varepsilon.
\end{equation}
To derive the standard equations (\ref{eq:delta-rad-peacock-ls}) and
(\ref{eq:delta-standard}), it is required that
$u^i_{\een\parallel}=0$, implying that
\begin{equation}\label{eq:theta1-is-0}
    \vartheta_\een(t,\vec{x})=0, \quad \psi(\vec{x})=\psi,
\end{equation}
where we have used (\ref{eq:psi-K}). The first of
(\ref{eq:theta1-is-0}) implies, using, equation
(\ref{FRW5gi}) that
\begin{equation}
  \label{eq:press-grad-0}
  \nabla^2p_\een=0,
\end{equation}
i.e., pressure gradients should vanish in order to derive the
\emph{relativistic} standard equation (\ref{eq:stand-w}).
Substituting $\varepsilon_\een=\varepsilon_\nul\delta$ into equation
(\ref{FRW4gi}) and eliminating $\dot{\varepsilon}_\nul$ with the help
of equation (\ref{FRW2}), it is found that
\begin{equation}\label{eq:delta-eerste}
    \dot{\delta}+\frac{1+w}{2H}\left(\kappa\varepsilon_\nul\delta+
        \tfrac{1}{2}\,\mbox{$^3\!R_{\een\parallel}$}\right)=0.
\end{equation}
Differentiating equation (\ref{eq:delta-eerste}) with respect to
$x^0=ct$ and using equations (\ref{subeq:einstein-flrw}) and (\ref{FRW6gi}),
yields
\begin{equation}\label{eq:delta-tweede}
    \ddot{\delta}+\tfrac{3}{2}(1+w)H\dot{\delta}-
    \tfrac{3}{4}(1+w)^2\kappa\varepsilon_\nul\delta-
    \tfrac{1}{8}(1+w)(1-3w)\,\mbox{$^3\!R_{\een\parallel}$}=0.
\end{equation}
Eliminating $\mbox{$^3\!R_{\een\parallel}$}$ from equation
(\ref{eq:delta-tweede}) with the help of equation
(\ref{eq:delta-eerste}), yields the standard equation for
large-scale perturbations in a flat \textsc{flrw} universe:
\begin{equation}\label{eq:stand-w}
    \ddot{\delta}+2H\dot{\delta}-
         \tfrac{1}{2}\kappa\varepsilon_\nul\delta(1+w)(1+3w)=0.
\end{equation}
This equation has been derived by Weinberg \cite{c8}, equation (15.10.57)
and Peebles~\cite{c11}, equation (86.11). For $w=\tfrac{1}{3}$ (the
radiation-dominated era) equation (\ref{eq:delta-rad-peacock-ls}) is
found, whereas for $w=0$ (the era after decoupling of matter and radiation)
equation
(\ref{eq:delta-standard}) applies.

Using that the general solution of the background equations
(\ref{subeq:einstein-flrw}) for
$\mbox{$^3\!R_\nul$}=0$, $\Lambda=0$ and constant~$w$ is given by
\begin{subequations}
\begin{align}
  H(t) &= \dfrac{2}{3(1+w)}(ct)^{-1}=H(t_0)\left(\dfrac{t}{t_0}\right)^{-1},\label{eq:Ht-dust-k0} \\
  \varepsilon_\nul(t) &= \dfrac{4}{3\kappa(1+w)^2}(ct)^{-2}
  =\varepsilon_\nul(t_0)\left(\dfrac{t}{t_0}\right)^{-2},
\end{align}
\end{subequations}
we find for the general solution of equation (\ref{eq:stand-w}), with
$\tau\equiv t/t_0$,
\begin{equation}\label{eq:gen-sol-stand-w}
    \delta(\tau) = E_1\tau^{(2+6w)/(3+3w)} -
       3(1+w)H(t_0)\psi \tau^{-1},
\end{equation}
where $E_1$ is an arbitrary integration constant.

The only surviving gauge mode is, using
(\ref{e-ijk}), (\ref{FRW2}) and $\delta\equiv\varepsilon_\een/\varepsilon_\nul$
[cf.\ (\ref{eq:contrast})],
\begin{equation}\label{eq:gauge-mode-e}
    \varepsilon_{\een\mathrm{gauge}}(t)=\psi\dot{\varepsilon}_\nul(t),
    \quad \delta_\mathrm{gauge}(t)=\dfrac{\varepsilon_{\een\mathrm{gauge}}(t)}{\varepsilon_\nul(t)}=
       \psi\dfrac{\dot{\varepsilon}_\nul(t)}{\varepsilon_\nul(t)}=
       -3(1+w)H(t)\psi.
\end{equation}
It follows from (\ref{eq:Ht-dust-k0}) and
(\ref{eq:gauge-mode-e}) that the second term in the right-hand side of
(\ref{eq:gen-sol-stand-w}) is a gauge mode
and $\psi$ is the gauge constant. For $w=\tfrac{1}{3}$ we find
(\ref{eq:delta-rad-peacock-sol}) with gauge mode (\ref{eq:gauge-rad}) and for
$w=0$ we get (\ref{eq:delta-standard-sol}) with gauge mode
(\ref{eq:gauge-matter}). The solution~(\ref{eq:gen-sol-stand-w}) is exactly
equal to the result found by Peebles~\cite{c11}, \S86, expression (86.12).

Finally, we note that in the derivation of
(\ref{eq:delta-rad-peacock-ls}) [i.e., (\ref{eq:stand-w}) with
$w=\tfrac{1}{3}$] Peebles uses $\vartheta_\een=0$ (in his notation:
$\theta=0$), see (\ref{eq:theta1-is-0}). In this case, Peebles'
method yields a physical mode $\delta\propto\tau$ and a gauge mode
$\delta\propto\tau^{-1}$. For $\vartheta_\een\neq0$ Peebles finds a
physical mode $\delta\propto\tau^{1/2}$. However, in the treatise
presented in this article both physical modes $\delta\propto\tau$ and
$\delta\propto\tau^{1/2}$ [see (\ref{delta-H-rad})] follow from
\emph{one} second-order differential equation
(\ref{eq:delta-rad}). Moreover, in our treatise neither the pressure
gradient $\vec{\nabla}p^\gi_\een$, nor the velocity divergence
$\vartheta_\een\equiv(u^k_{\een||})_{|k}$ are neglected in our evolution equations
(\ref{subeq:final}).


\section{Symbols and their Meaning}\label{notations}
\vspace{-0.7cm}

\begin{longtable}{lll}
\caption{Symbols and their meaning of all quantities, except for those occurring
in the appendices.} \\
\hline\hline \textbf{Symbol} & \textbf{Meaning} &
\textbf{Reference Equation} \\ \hline \vspace{1ex}
\endfirsthead
\caption[]{(continued)} \\ \hline \hline
\endhead
\hline
\endfoot
$\vec{\nabla}f$& $(\partial_1f,\partial_2f,\partial_3f)$ & --- \\
$(\vec{\tilde{\nabla}} f)^i$ & $\tilde{g}^{ij}\partial_j f$ &
(\ref{longitudinal})--(\ref{rotation})\\
$\vec{\tilde{\nabla}}\cdot\vec{u}$ & $u^k{}_{|k}$ & (\ref{longitudinal}),
(\ref{transversal}) \\
$(\vec{\tilde{\nabla}}\wedge\vec{u})_i$ &
          $\epsilon_i{}^{jk}u_{j|k}=\epsilon_i{}^{jk}u_{j,k}$ &
(\ref{rotation}) \\
$\tilde{\nabla}^2 f $ &
      $\vec{\tilde{\nabla}}\cdot(\vec{\tilde{\nabla}}f)=\tilde{g}^{ij}f_{|i|j}
      $ & (\ref{Laplace})\\
$\partial_{0}$& derivative with respect to $x^{0}=ct$& ---\\
$\partial_{i}$& derivative with respect to $x^{i}$& ---\\
hat: $\hat{}$&   computed with respect to
$\hat x$& (\ref{subeq:split-e-n})\\
dot: $\dot{}$ & derivative with respect to $x^0=ct$&
(\ref{def-gammas})\\
prime: $^\prime$ & derivative with respect to $\tau$ &  (\ref{delta-pnue-tau}), (\ref{eq:dust-dimless}) \\ 
tilde: $\tilde{}$ & computed with respect to three-metric $\tilde{g}_{ij}$&
(\ref{m2}), (\ref{con3FRW}) \\
super-index: ${}^\gi$ &   gauge-invariant& (\ref{subeq:gi-en})\\
super-index: ${}^{|k}$ & contravariant derivative with respect to
      $x^k$: $\zeta^{|k}=g^{kj}_\nul\zeta_{|j}$ & (\ref{decomp-hij-par}) \\
sub-index: ${}_\nul$ & background quantity& (\ref{subeq:ent-split})\\
sub-index: ${}_\een$& perturbation of first-order & (\ref{subeq:ent-split}) \\
sub-index: ${}_\twee$& perturbation of second-order&  (\ref{subeq:exp-scalar}) \\
sub-index: ${}_{;\lambda}$& covariant derivative with respect to $x^\lambda$&
(\ref{lie1})\\
sub-index: ${}_{|k}$ & covariant derivative with respect to $x^k$ &
(\ref{threecov})\\
sub-index: ${}_{,\mu}$ & derivative with respect to $x^\mu$& (\ref{concoef1}) \\
sub-index: ${}_{\parallel}$ & longitudinal part of a vector or tensor&
(\ref{decomp-symh})\\
sub-index: ${}_{\perp}$ & perpendicular part of vector or  tensor &
(\ref{decomp-symh}) \\
sub-index: ${}_{\ast}$ & transverse and traceless part of a tensor &
(\ref{decomp-symh}) \\
$\beta$ & $c^{-1}$ times speed of sound:
$\textstyle{\sqrt{\dot{p}_{(0)}/\dot{\varepsilon}_{(0)}}}$& (\ref{begam2})\\
$\Gamma^\lambda{}_{\mu\nu}$ & connection coefficients & (\ref{concoef1})\\
$\gamma$ & arbitrary function & (\ref{Rij-paral}) \\
$\delta_\varepsilon$ & energy contrast function & (\ref{eq:contrast})\\
$\delta^\mu{}_\nu$ & Kronecker delta & (\ref{u0})\\
$\delta_n$ & particle number contrast function & (\ref{eq:contrast})\\
$\delta_p$ & pressure contrast function &  (\ref{eq:rel-press-pert})\\
$\delta_T$ &  (matter) temperature contrast function & (\ref{eq:rel-T-pert})\\
$\delta_{T_\gamma}$ &  background radiation temperature contrast function &
(\ref{eq:rel-T-pert}), (\ref{eq:temp-eps-fluct})\\
$\varepsilon$& energy density& (\ref{eps1})\\
$e$ & energy per particle & (\ref{eq:energy-per-particle}) \\
$\epsilon_i{}^{jk}$ & Levi-Civita tensor,
    $\epsilon_1{}^{23}=+1$ & (\ref{rotation})  \\
$\zeta,\phi$ & potentials due to
      relativistic density perturbations & (\ref{decomp-hij-par}),
(\ref{RnabEE})  \\
$\eta$ & bookkeeping parameter equal to 1& (\ref{subeq:exp-scalar}) \\
$\theta$ &  expansion scalar in four-space& (\ref{exp1}) \\
$\vartheta$ & expansion scalar in three-space & (\ref{driediv})\\
$\kappa$ &  $8\pi G/c^4$& (\ref{kappa}) \\
$\varkappa_{ij}$ & time derivative metric coefficients & (\ref{def-gammas})\\
$\Lambda$& cosmological constant & (\ref{ein-verg}) \\
$\lambda$ & wavelength, physical scale at time $t_\mathrm{p}$ & (\ref{pw12}) \\
$\mu$ & thermodynamic potential & (\ref{eq:sec-law-thermo})\\
$\mu_\mathrm{r}$ & reduced wave-number (radiation)& (\ref{xi}) \\
$\mu_\mathrm{m}$ & reduced wave-number (matter)& (\ref{eq:const-mu}) \\
$\xi^\mu$& first-order space-time translation& (\ref{func}) \\
$\pi$ & arbitrary function & (\ref{Rij-paral}) \\
$\varpi$ & \textsc{flrw}-coordinate & (\ref{tildegij}) \\
$\rho_\een$  & $\varepsilon^\gi_\een/c^2$ & (\ref{eq:poisson}) \\
$\sigma$ & arbitrary scalar &  (\ref{sigma}) \\
$\sigma^\gi_\een$ &  abbreviation for
      $n_\een^\gi-n_\nul\varepsilon^\gi_\een/[\varepsilon_\nul(1+w)]$ &
      (\ref{eq:hulp-entropy}), (\ref{eq:entropy-gi})\\
$\tau$ & dimensionless time &  (\ref{tau}), (\ref{eq:tau-mat}) \\
$\phi, \zeta$ & potentials due to relativistic density perturbations &
(\ref{decomp-hij-par}),  (\ref{RnabEE}) \\
$\varphi$ & Newtonian potential  & (\ref{poisson}), (\ref{eq:ident})\\
$\chi^i$  & arbitrary three-vector & (\ref{xi-syn})  \\
$\psi$& first-order time translation& (\ref{subeq:split-e-n}),  (\ref{func}),
(\ref{defpsi})\\
$\Omega$ & components in the Friedmann equation &
(\ref{eq:Friedmann-Omega}) \\
$\omega$& arbitrary scalar &  (\ref{omegahat3})\\
$A^{\alpha\ldots\beta}{}_{\mu\ldots\nu}$& arbitrary tensor&  (\ref{lie1})\\
$A_{\mu\nu}$& arbitrary rank two tensor&  (\ref{lieder})\\
$a$ & scale factor or radius of universe& (\ref{m2}) \\
$a_\mathrm{B}$ & black body constant & (\ref{eq:state-rad}) \\
$a_1,a_2,a_3,a_4$& coefficients in perturbation equations& (\ref{subeq:coeff})\\
$b_1,b_2,b_3$& coefficients in perturbation equations&
(\ref{subeq:coeff-contrast})\\
$c$  &   speed of light & ---\\
$\dif s^2$& line element in four-space& (\ref{line-element}) \\
$E$ & energy within volume $V$ & (\ref{eq:sec-law-thermo}) \\
$G$ & Newton's gravitation constant & --- \\
$g_{\mu\nu}$& metric tensor& (\ref{killing})\\
$\hat g_{\mu\nu}$& metric with respect to $\hat x^\mu$& (\ref{killing})\\
$\tilde{g}_{ij}$& time-independent metric of three-space & (\ref{m2})\\
$H$ &  $c^{-1}\mathcal{H}$ & (\ref{Hubble}) \\
$\mathcal{H}$& Hubble function: $\mathcal{H}=(\dif a/\dif t)/a$ & (\ref{eq:Hubble-function}) \\
$h_{ij}$& minus first-order perturbation of metric & (\ref{hij})\\
$k$ & $k=-1$ (open), $0$ (flat), $+1$ (closed) & (\ref{tildegij})\\
$k_\mathrm{B}$ & Boltzmann's constant & (\ref{state-mat}) \\
$\mathcal{L}_\xi$& Lie derivative with respect to $\xi^\mu$& (\ref{lie1}) \\
$m$ & mass of particle of cosmological fluid & (\ref{eq:rest-energy})\\
$\text{M}_\odot$ & solar mass, $1.98892\times10^{30}\,\mathrm{kg}$ & --- \\
$m_\mathrm{H}$ & baryonic (proton) mass & (\ref{state-mat}) \\
$N$ & number of particles within volume $V$ & (\ref{eq:sec-law-thermo})\\
$N^\mu$& particle density four-flow& (\ref{eq:current})\\
$n$ & particle number density &  (\ref{en1})\\
$p$ & pressure & (\ref{toestand}) \\
$p_\varepsilon, p_n$& partial derivatives of pressure & (\ref{perttoes1})\\
$p_{nn}, p_{\varepsilon n}, p_{\varepsilon \varepsilon}$ & partial derivatives of pressure & (\ref{pne}) \\
$q$ & circular wave number: $2\pi/\lambda$ & (\ref{pw12})\\
$r$ & \textsc{flrw}-coordinate & (\ref{tildegij}) \\
$\mbox{$^3\!R$}$ & Ricci scalar in three-space& (\ref{drieR})\\
$R_{\mu\nu}$& Ricci tensor in four-space & (\ref{Ricci1})\\
$\mbox{$^3\!R_{ij}$}$ & Ricci tensor in three-space & (\ref{Ricci-drie})\\
$S$ & entropy within a volume $V$ & (\ref{eq:sec-law-thermo}) \\
$s$ & entropy per particle $s\equiv S/N$ & (\ref{eq:energy-per-particle})\\
$s^\gi_\een$ & gauge-invariant entropy perturbation &
(\ref{eq:TdS-1})--(\ref{eq:lin-gi})\\
$T$ & absolute temperature & (\ref{eq:sec-law-thermo})\\
$T_\gamma$ & photon temperature & (\ref{eq:decoup-temp}) \\
$T^{\mu\nu}$& energy momentum tensor& (\ref{Tmunu})\\
$T^\gi_\een$ & gauge-invariant temperature perturbation  & (\ref{eq:Tgi})\\
$t$ &  cosmological time & ---\\
$t_0$ & initial cosmological time & --- \\
$t_\mathrm{dec}$ & decoupling time & (\ref{subeq:wmap}) \\
$t_\mathrm{eq}$ & matter-radiation equality (onset of matter-dominated era) &
(\ref{subeq:wmap}) \\
$t_{\text{mat}}$ & initial time between decoupling and the present &
(\ref{eq:tmat}) \\
$t_\mathrm{p}$ & present cosmological time ($13.7\,\mathrm{Gyr}$) &
(\ref{subeq:wmap}) \\
$t_\mathrm{rad}$ & onset of radiation-dominated era & (\ref{subeq:rad-R0-sol}) \\
$U^\mu$& cosmological four-velocity $U^\mu U_\mu=c^2$   & (\ref{subeq:ent}) \\
$u^\mu$ & $c^{-1}U^\mu$ & --- \\
$\vec{U}$ & spatial velocity & (\ref{eq:nrl-limit}) \\
$\vec{u}$ & $c^{-1}\vec{U}$ & (\ref{decomp-u}) \\
$V^\mu$& arbitrary four-vector& (\ref{lievec})\\
$v_{\text{s}}$ & speed of sound & (\ref{coef-nu1})  \\
$w$ & pressure divided by energy density & (\ref{begam2}) \\
$x$& space-time point $x^\mu=(ct,\vec{x})$& ---\\
$\hat{x}^\mu$& locally transformed coordinates&  (\ref{func})\\
$\vec{x}$ &  spatial point $\vec{x}=(x^1,x^2,x^3)$ &  --- \\
$z(t)$ & redshift at time $t$ & (\ref{eq:redshift})
\end{longtable}



\section*{Acknowledgments}

The present article is a continuation and extension of an earlier
article~\cite{Miedema-Leeuwen-2003}, written in collaboration with
Willem van Leeuwen: \url{http://cosmology.minerva83.nl}

The author should like to take the opportunity of thanking
Marie-Louise Pennings \textsc{ma} (lecturer of English at the
\textsc{nlda}) for her assistance in discovering the intricacies of
the English language.

\bibliographystyle{unsrtnat}
\bibliography{/home/pieter/Dropbox/Documents/artikel/literatuur-database/diss}





\section*{Supplementary material}

```
> ###############################################################
> #         Structure Formation independent of Cold Dark Matter
> #
> #                          P.G.Miedema
> ###############################################################
> #                          Maple 14
> #
> # Consistency check of the perturbation equations (203) and (207).
> # 1) To check Eqs.(203) with coefficients (204), substitute the expressions
> #    (202) into Eqs.(203) and use the background equations (158) and the
> #    perturbation equations (201) to evaluate all time derivatives.
> # 2) To check Eqs.(207) with coefficients (208), substitute (206) into Eqs.(207)
> #    and use the background equations (158) and the perturbation equations
> #    (201) to evaluate all time derivatives.
> ###############################################################
> restart;
> # Constants:
> #     kappa = 8*pi*G/c^4 (38),
> #     Delta is the nabla operator (154),
> #     Lambda is the cosmological constant, see (158a).
> ###############################################################
> # Coefficients a_1, a_2, a_3, Eqs.(204):
> a1:=kappa*e_0(t)*(1+w(t))/H(t)-2*diff(beta(t),t)/beta(t)+H(t)*(4-3*beta(t)^2)+R_0(t)*
  (1/(3*H(t))+2*H(t)*(1+3*beta(t)^2)/(R_0(t)+3*kappa*e_0(t)*(1+w(t))));
```

$$a1 := \frac{\kappa\, e\_0(t)\,(1+w(t))}{H(t)} - \frac{2\left(\frac{d}{dt}\,\beta(t)\right)}{\beta(t)} + H(t)\,(4-3\,\beta(t)^2) + R\_0(t)\left(\frac{1}{3\,H(t)}\right.$$
$$\left. + \frac{2\,H(t)\,(1+3\,\beta(t)^2)}{R\_0(t)+3\,\kappa\,e\_0(t)\,(1+w(t))}\right) \tag{1}$$

```
> a2:=kappa*e_0(t)*(1+w(t))-4*H(t)*diff(beta(t),t)/beta(t)+2*H(t)^2*(2-3*beta(t)^2)+R_0
  (t)*(1/2+((5*H(t)^2)*(1+3*beta(t)^2)-2*H(t)*diff(beta(t),t)/beta(t))/(R_0(t)+3*kappa*
  e_0(t)*(1+w(t))))-beta(t)^2*(Delta/a(t)^2-(1/2)*R_0(t));
```

$$a2 := \kappa\, e\_0(t)\,(1+w(t)) - \frac{4\,H(t)\left(\frac{d}{dt}\,\beta(t)\right)}{\beta(t)} + 2\,H(t)^2\,(2-3\,\beta(t)^2) + R\_0(t)\left(\frac{1}{2}\right.$$
$$\left. + \frac{5\,H(t)^2\,(1+3\,\beta(t)^2) - \frac{2\,H(t)\left(\frac{d}{dt}\,\beta(t)\right)}{\beta(t)}}{R\_0(t)+3\,\kappa\,e\_0(t)\,(1+w(t))}\right) - \beta(t)^2\left(\frac{\Delta}{a(t)^2} - \frac{1}{2}\,R\_0(t)\right) \tag{2}$$

```
> a3:=(-18*H(t)^2*(e_0(t)*p_en(t)*(1+w(t))+2*p_n(t)*diff(beta(t),t)/beta(t)/(3*H(t))-beta
  (t)^2*p_n(t)+p_e(t)*p_n(t)+n_0(t)*p_nn(t))/(R_0(t)+3*kappa*e_0(t)*(1+w(t)))+p_n(t))*
  (Delta/a(t)^2-(1/2)*R_0(t));
```

$$a3 := \left(\vphantom{\frac{1}{2}}\right.$$
$$-\frac{1}{R\_0(t)+3\,\kappa\,e\_0(t)\,(1+w(t))}\left(18\,H(t)^2\left(e\_0(t)\,p\_en(t)\,(1+w(t)) + \frac{2}{3}\,\frac{p\_n(t)\left(\frac{d}{dt}\,\beta(t)\right)}{\beta(t)\,H(t)}\right.\right.$$
$$\left.\left. - \beta(t)^2\,p\_n(t) + p\_e(t)\,p\_n(t) + n\_0(t)\,p\_nn(t)\right)\right) + p\_n(t)\left)\left(\frac{\Delta}{a(t)^2} - \frac{1}{2}\,R\_0(t)\right)\right. \tag{3}$$

```
> # Coefficients b_1, b_2, b_3, Eqs.(208):
> b1:=kappa*e_0(t)*(1+w(t))/H(t)-2*diff(beta(t),t)/beta(t)-H(t)*(2+6*w(t)+3*beta(t)^2)+
  R_0(t)*(1/(3*H(t))+2*H(t)*(1+3*beta(t)^2)/(R_0(t)+3*kappa*e_0(t)*(1+w(t))));
```

$$b1 := \frac{\kappa\, e\_0(t)\,(1+w(t))}{H(t)} - \frac{2\left(\dfrac{d}{dt}\,\beta(t)\right)}{\beta(t)} - H(t)\left(2+6\,w(t)+3\,\beta(t)^2\right) + R\_0(t)\left(\frac{1}{3\,H(t)}\right.$$
$$\left. + \frac{2\,H(t)\,(1+3\,\beta(t)^2)}{R\_0(t)+3\,\kappa\,e\_0(t)\,(1+w(t))}\right) \tag{4}$$

```
> b2:=-(1/2)*kappa*e_0(t)*(1+w(t))*(1+3*w(t))+H(t)^2*(1-3*w(t)+6*beta(t)^2*(2+3*w(t)))+6*
  H(t)*(diff(beta(t),t)/beta(t))*(w(t)+kappa*e_0(t)*(1+w(t))/(R_0(t)+3*kappa*e_0(t)*(1+w
  (t))))-R_0(t)*((1/2)*w(t)+H(t)^2*(1+6*w(t))*(1+3*beta(t)^2)/(R_0(t)+3*kappa*e_0(t)*(1+w
  (t))))-beta(t)^2*(Delta/a(t)^2-(1/2)*R_0(t));
```

$$b2 := -\frac{1}{2}\,\kappa\,e\_0(t)\,(1+w(t))\,(1+3\,w(t)) + H(t)^2\left(1-3\,w(t)+6\,\beta(t)^2\,(2+3\,w(t))\right)$$
$$+ \frac{6\,H(t)\left(\dfrac{d}{dt}\,\beta(t)\right)\left(w(t)+\dfrac{\kappa\,e\_0(t)\,(1+w(t))}{R\_0(t)+3\,\kappa\,e\_0(t)\,(1+w(t))}\right)}{\beta(t)} - R\_0(t)\left(\frac{1}{2}\,w(t)\right.$$
$$\left. + \frac{H(t)^2\,(1+6\,w(t))\,(1+3\,\beta(t)^2)}{R\_0(t)+3\,\kappa\,e\_0(t)\,(1+w(t))}\right) - \beta(t)^2\left(\frac{\Delta}{a(t)^2}-\frac{1}{2}\,R\_0(t)\right) \tag{5}$$

```
> b3:=a3*(n_0(t)/e_0(t));
```

$$b3 := \frac{1}{e\_0(t)}\left(\left(\vphantom{\frac11}\right.\right.$$
$$- \frac{1}{R\_0(t)+3\,\kappa\,e\_0(t)\,(1+w(t))}\left(18\,H(t)^2\left(e\_0(t)\,p\_en(t)\,(1+w(t))+\frac{2}{3}\,\frac{p\_n(t)\left(\dfrac{d}{dt}\,\beta(t)\right)}{\beta(t)\,H(t)}\right.\right.$$
$$\left.\left. -\beta(t)^2\,p\_n(t)+p\_e(t)\,p\_n(t)+n\_0(t)\,p\_nn(t)\right)\right) + p\_n(t)\left(\frac{\Delta}{a(t)^2}-\frac{1}{2}\,R\_0(t)\right)n\_0(t)\right) \tag{6}$$

```
> ###############################################################################
> #                          Background equations
> ###############################################################################
> # Eq.(156):
> w(t):=p_0(t)/e_0(t);
```

$$w(t) := \frac{p\_0(t)}{e\_0(t)} \tag{7}$$

```
> # Scale factor Eq.(55):
> define(a,diff(a(t),t)=H(t)*a(t));
> diff(a(t),t);
```

$$H(t)\,a(t) \tag{8}$$

```
> # Define H(t) by the dynamical equation (90):
> define(H,diff(H(t),t)=-(1/6)*R_0(t)-(1/2)*kappa*(e_0(t)+p_0(t)));
> diff(H(t),t);
```

$$-\frac{1}{6}\,R\_0(t) - \frac{1}{2}\,\kappa\,(e\_0(t)+p\_0(t)) \tag{9}$$

```
> # Either define the three-space curvature by Eq.(158b):
> define(R_0,diff(R_0(t),t)=-2*H(t)*R_0(t));
> # or by Eq.(65) with k=+1, 0, -1:
> k:=+1; R_0(t):=-6*k/a(t)^2;
> diff(R_0(t),t);
```

$$-2\,H(t)\,R\_0(t) \tag{10}$$

```
> # Energy conservation law (158c):
> define(e_0,diff(e_0(t),t)=-3*H(t)*e_0(t)*(1+w(t)));
> diff(e_0(t),t);
```

$$\tag{11}$$

$$-3\,H(t)\,e\_0(t)\left(1+\frac{p\_0(t)}{e\_0(t)}\right)$$

**(11)**

```
> # Particle number conservation law (158e):
> define(n_0,diff(n_0(t),t)=-3*H(t)*n_0(t));
> diff(n_0(t),t);
```

$$-3\,H(t)\,n\_0(t)$$

**(12)**

```
> # Time derivative of p_0(t), Eq.(59):
> define(p_0,diff(p_0(t),t)=p_e(t)*diff(e_0(t),t)+p_n(t)*diff(n_0(t),t));
> diff(p_0(t),t);
```

$$-3\,p\_e(t)\,H(t)\,e\_0(t)\left(1+\frac{p\_0(t)}{e\_0(t)}\right)-3\,p\_n(t)\,H(t)\,n\_0(t)$$

**(13)**

```
> # Partial derivative p_e(t) of the pressure p, Eq.(79):
> define(p_e,diff(p_e(t),t)=p_ee(t)*diff(e_0(t),t)+p_en(t)*diff(n_0(t),t));
> diff(p_e(t),t);
```

$$-3\,p\_ee(t)\,H(t)\,e\_0(t)\left(1+\frac{p\_0(t)}{e\_0(t)}\right)-3\,p\_en(t)\,H(t)\,n\_0(t)$$

**(14)**

```
> # Partial derivative p_n(t) of the pressure p, Eq.(79):
> define(p_n,diff(p_n(t),t)=p_en(t)*diff(e_0(t),t)+p_nn(t)*diff(n_0(t),t));
> diff(p_n(t),t);
```

$$-3\,p\_en(t)\,H(t)\,e\_0(t)\left(1+\frac{p\_0(t)}{e\_0(t)}\right)-3\,p\_nn(t)\,H(t)\,n\_0(t)$$

**(15)**

```
> # Quantity beta(t), Eq.(156):
> beta(t):=sqrt(diff(p_0(t),t)/diff(e_0(t),t));
```

$$\beta(t):=\frac{1}{3}\sqrt{-\frac{-9\,p\_e(t)\,H(t)\,e\_0(t)\left(1+\frac{p\_0(t)}{e\_0(t)}\right)-9\,p\_n(t)\,H(t)\,n\_0(t)}{H(t)\,e\_0(t)\left(1+\frac{p\_0(t)}{e\_0(t)}\right)}}$$

**(16)**

```
> ##############################################################################
> #                     First order perturbation equations
> ##############################################################################
> # First order perturbation to the pressure, Eq.(77):
> p_1(t):=p_e(t)*e_1(t)+p_n(t)*n_1(t);
```

$$p\_1(t):=p\_e(t)\,e\_1(t)+p\_n(t)\,n\_1(t)$$

**(17)**

```
> # Energy conservation law (201a):
> define(e_1,diff(e_1(t),t)=-3*H(t)*(e_1(t)+p_1(t))-e_0(t)*(1+w(t))*(theta(t)+(kappa*e_1
  (t)+(1/2)*R_1(t))/(2*H(t))));
> diff(e_1(t),t);
```

$$-3\,H(t)\,(e\_1(t)+p\_e(t)\,e\_1(t)+p\_n(t)\,n\_1(t))-e\_0(t)\left(1+\frac{p\_0(t)}{e\_0(t)}\right)\left(\theta(t)+\frac{1}{2}\,\frac{\kappa e\_1(t)+\frac{1}{2}\,R\_1(t)}{H(t)}\right)$$

**(18)**

```
> # Particle number conservation law (201b):
> define(n_1,diff(n_1(t),t)=-3*H(t)*n_1(t)-n_0(t)*(theta(t)+(kappa*e_1(t)+(1/2)*R_1(t))/
  (2*H(t))));
> diff(n_1(t),t);
```

$$-3\,H(t)\,n\_1(t)-n\_0(t)\left(\theta(t)+\frac{1}{2}\,\frac{\kappa e\_1(t)+\frac{1}{2}\,R\_1(t)}{H(t)}\right)$$

**(19)**

```
> # Momentum conservation law (201c):
> define(theta,diff(theta(t),t)=-H(t)*(2-3*beta(t)^2)*theta(t)-Delta/a(t)^2*p_1(t)/(e_0
  (t)*(1+w(t))));
```

```
> diff(theta(t),t);
```

$$-H(t)\left(2 + \frac{1}{3}\, \frac{-9\,p\_e(t)\,H(t)\,e\_0(t)\left(1+\frac{p\_0(t)}{e\_0(t)}\right) - 9\,p\_n(t)\,H(t)\,n\_0(t)}{H(t)\,e\_0(t)\left(1+\frac{p\_0(t)}{e\_0(t)}\right)}\right)\theta(t)$$
$$-\frac{\Delta\left(p\_e(t)\,e\_1(t) + p\_n(t)\,n\_1(t)\right)}{a(t)^2\,e\_0(t)\left(1+\frac{p\_0(t)}{e\_0(t)}\right)} \qquad \textbf{(20)}$$

```
> # Evolution equation for the local perturbation to the
> # global spatial curvature, (201d):
> define(R_1,diff(R_1(t),t)=-2*(H(t)*R_1(t)-kappa*e_0(t)*(1+w(t))*theta(t))-R_0(t)/(3*H
  (t))*(kappa*e_1(t)+(1/2)*R_1(t)));
> diff(R_1(t),t);
```

$$-2\,H(t)\,R\_1(t) + 2\,\kappa\,e\_0(t)\left(1+\frac{p\_0(t)}{e\_0(t)}\right)\theta(t) - \frac{1}{3}\,\frac{R\_0(t)\left(\kappa\,e\_1(t) + \frac{1}{2}\,R\_1(t)\right)}{H(t)} \qquad \textbf{(21)}$$

```
> ###############################################################################
> #                    Gauge-invariant density perturbations
> ###############################################################################
> # Gauge-invariant perturbation to the energy density, Eq.(202a):
> e_gi(t):=(e_1(t)*R_0(t)-3*e_0(t)*(1+w(t))*(2*H(t)*theta(t)+(1/2)*R_1(t)))/(R_0(t)+3*
  kappa*e_0(t)*(1+w(t)));
```

$$e\_gi(t) := \frac{e\_1(t)\,R\_0(t) - 3\,e\_0(t)\left(1+\frac{p\_0(t)}{e\_0(t)}\right)\left(2\,H(t)\,\theta(t) + \frac{1}{2}\,R\_1(t)\right)}{R\_0(t) + 3\,\kappa\,e\_0(t)\left(1+\frac{p\_0(t)}{e\_0(t)}\right)} \qquad \textbf{(22)}$$

```
> # Gauge-invariant perturbation to the particle number density, Eq.(202b):
> n_gi(t):=n_1(t)-3*n_0(t)*(kappa*e_1(t)+2*H(t)*theta(t)+(1/2)*R_1(t))/(R_0(t)+3*kappa*
  e_0(t)*(1+w(t)));
```

$$n\_gi(t) := n\_1(t) - \frac{3\,n\_0(t)\left(\kappa\,e\_1(t) + 2\,H(t)\,\theta(t) + \frac{1}{2}\,R\_1(t)\right)}{R\_0(t) + 3\,\kappa\,e\_0(t)\left(1+\frac{p\_0(t)}{e\_0(t)}\right)} \qquad \textbf{(23)}$$

```
> # Definition of sigma^gi: Eq.(181):
> sigma_gi(t):=simplify(n_gi(t)-n_0(t)*e_gi(t)/(e_0(t)*(1+w(t))));
```

$$sigma\_gi(t) := \frac{n\_1(t)\,e\_0(t) + n\_1(t)\,p\_0(t) - n\_0(t)\,e\_1(t)}{e\_0(t) + p\_0(t)} \qquad \textbf{(24)}$$

```
> ###############################################################################
> #                         Consistency checks
> ###############################################################################
> # Consistency check for Eq.(203b) [left-hand side minus right-hand side must be zero]:
> simplify(diff(sigma_gi(t),t)-(-3*H(t)*(1-n_0(t)*p_n(t)/(e_0(t)*(1+w(t)))))*sigma_gi(t))
  ;
```

$$0 \qquad \textbf{(25)}$$

```
> # Consistency check for Eq.(203a) [left-hand side minus right-hand side must be zero]:
> simplify(diff(e_gi(t),t$2)+a1*diff(e_gi(t),t)+a2*e_gi(t)-a3*sigma_gi(t));
```

$$0 \qquad \textbf{(26)}$$

```
> # Gauge-invariant contrast functions, Eq.(206):
> delta_e(t):=e_gi(t)/e_0(t); delta_n(t):=n_gi(t)/n_0(t);
```

$$delta\_e(t) := \frac{e\_1(t)\, R\_0(t) - 3\, e\_0(t) \left(1 + \dfrac{p\_0(t)}{e\_0(t)}\right) \left(2\, H(t)\, \theta(t) + \dfrac{1}{2}\, R\_1(t)\right)}{\left(R\_0(t) + 3\, \kappa\, e\_0(t) \left(1 + \dfrac{p\_0(t)}{e\_0(t)}\right)\right) e\_0(t)}$$

$$delta\_n(t) := \frac{n\_1(t) - \dfrac{3\, n\_0(t) \left(\kappa\, e\_1(t) + 2\, H(t)\, \theta(t) + \dfrac{1}{2}\, R\_1(t)\right)}{R\_0(t) + 3\, \kappa\, e\_0(t) \left(1 + \dfrac{p\_0(t)}{e\_0(t)}\right)}}{n\_0(t)}$$

**(27)**

```
> # Consistency check for Eq.(207b) [left-hand side minus right-hand side must be zero]:
> simplify(diff(delta_n(t)-delta_e(t)/(1+w(t)),t)-3*H(t)*n_0(t)*p_n(t)/(e_0(t)*(1+w(t)))*
  (delta_n(t)-delta_e(t)/(1+w(t))));
```

$$0$$

**(28)**

```
> # Consistency check for Eq.(207a) [left-hand side minus right-hand side must be zero]:
> simplify(diff(delta_e(t),t$2)+b1*diff(delta_e(t),t)+b2*delta_e(t)-b3*(delta_n(t)-
  delta_e(t)/(1+w(t))));
```

$$0$$

**(29)**

```
>
```



\end{document}